\documentclass[11pt,letterpaper]{article}

\usepackage[utf8]{inputenc}
\DeclareTextCommand{\DJ}{OT1}{\leavevmode\rlap{\raisebox{.15ex}{\kern-.08em-}}D}
\DeclareTextCommand{\dj}{OT1}{\leavevmode d\kern-.1em\raisebox{.55ex}{\rule{.25em}{.3pt}}}
\DeclareTextCommand{\k}{OT1}[1]{#1\kern-.05em\raisebox{-.5ex}{\scriptsize,}}
\usepackage[letterpaper,top=1in,bottom=1in,left=1in,right=1in]{geometry}
\usepackage{amsmath,amssymb}
\usepackage{setspace}
\usepackage[table]{xcolor}
\usepackage{tikz}
\usepackage{placeins}
\usepackage{graphicx,adjustbox,booktabs,array,multirow,textcomp,ragged2e,titlesec,longtable,needspace}
\usepackage[super]{cite}
\usepackage[colorlinks=true,citecolor=blue,linkcolor=blue,urlcolor=blue,pdftitle={Single-electron Fluorine Bonding. I. The SEFB-15 and CSFB-14 Benchmark Datasets, Reference Binding Energies and Intermolecular Bond Distances: Assessment of M06-2X density functional},pdfauthor={Govinda Prasad Khanal, Elangannan Arunan},pdfsubject={Single-electron fluorine bonding; benchmark datasets},pdfkeywords={fluorine bonding, single-electron bond, SEFB-15, CSFB-14, M06-2X}]{hyperref}
\usepackage[protrusion=true,expansion=true]{microtype}
\newcommand{\orcidicon}[1]{\href{https://orcid.org/#1}{\raisebox{-0.2ex}{\begin{tikzpicture}[baseline=-0.5ex]\fill[color={rgb,255:red,166;green,206;blue,57}] (0,0) circle (0.9ex);\node[white,font=\sffamily\bfseries\fontsize{5}{5}\selectfont,inner sep=0pt] at (0,0) {iD};\end{tikzpicture}}}}
\renewcommand{\topfraction}{0.9}
\renewcommand{\bottomfraction}{0.9}
\renewcommand{\textfraction}{0.08}
\renewcommand{\floatpagefraction}{0.7}
\titleformat{\section}[block]{\normalfont\bfseries\fontsize{14}{17}\selectfont}{\thesection.\space}{0pt}{}
\titleformat{\subsection}[block]{\normalfont\bfseries\fontsize{12}{14.5}\selectfont}{\thesubsection.\space}{0pt}{}
\titlespacing*{\section}{0pt}{16pt}{6pt}
\titlespacing*{\subsection}{0pt}{12pt}{5pt}
\newcommand{\GPKfig}[3]{\par\vspace{2pt}\noindent\hfil\adjustbox{max width=\linewidth,max totalheight=0.42\textheight}{\includegraphics[width=#1,height=#2]{#3}}\hfil\null\par\vspace{1pt}}
\usepackage{caption}
\renewcommand{\thetable}{\Roman{table}}\renewcommand{\LTcaptype}{LTdummy}
\renewcommand{\figurename}{FIG.}
\renewcommand{\tablename}{TABLE}
\DeclareCaptionFormat{gpk}{\setstretch{1}\fontsize{9.5}{11.5}\selectfont\setlength{\parskip}{0pt}\justifying\setlength{\parindent}{0pt}\noindent #1#2#3\par}
\newcommand{\Tref}[1]{\mbox{\textbf{TABLE \ref{#1}}}}
\newcommand{\Fref}[1]{\mbox{\textbf{FIG. \ref{#1}}}}
\newcommand{\Sref}[1]{\textbf{Section \ref*{#1}}}
\usepackage{fancyhdr}
\renewcommand{\headrulewidth}{0pt}
\makeatletter

\renewcommand\@biblabel[1]{\textsuperscript{#1}}
\makeatother
\AtBeginDocument{\renewcommand{\textperiodcentered}{\ensuremath{\cdot}}
\renewcommand{\textdegree}{\ensuremath{^{\circ}}}
\renewcommand{\texttimes}{\ensuremath{\times}}
\renewcommand{\textless}{\ensuremath{<}}}

\begin{document}
\nocite{MullerDethlefs2000,Hobza2010,Kolar2016,Rezac2016,Desiraju2001,Johnson2010,Stone2013,Neese2019,AlHamdani2021,Villot2022,Ballesteros2021,AlHamdani2019,Guthrie1863,Remsen1896,Benesi1949,Mulliken1950,Hassel1962,Bent1968,Hassel1970,Blackstock1987,Andrews1988,Desiraju1995,Legon1999,Metrangolo2005,Metrangolo2008,Legon2008,Legon2010,Cavallo2016,Costa2017,Metrangolo2001,Meyer2013,Auffinger2004,Lu2009,Parisini2011,Metrangolo2006,Berger2015,Gilday2015,Carter2011,Lu2010,Guillier2019,Bulfield2016,Govindaraj2019,Peluso2020,Clark2007,Politzer2017,Desiraju2013,Wilcken2013,Dikundwar2014,Lin2017,Chopra2011,Politzer2013,Metrangolo2011a,Chopra2012,Hathwar2014,Berger2011,Panini2015,Cole2022,Bauza2016,Varadwaj2018,Pavan2013,Han2017,Neaton2017,BarceloOliver2010,Osuna2011,Sava2020,VJanjic2020,Rocha2021,Chopra2006,Wang2015,Sirohiwal2017,Elakkat2019,VGrudova2020,Rana2020,Shahi2014,McDowell2012,Raghavendra2007,Cincic2008,Scholfield2013,VBartashevich2014,Deepa2014,Shaw2016,Lisac2019,Riley2020,Stephens2012,Tsuzuki2016,Riley2017,Kanao2020,Portela2021,Bauza2013,Hill2013,Kozuch2013,OteroDeLaRoza2014,Wang2014,Scheiner2016,Wang2017,Oliveira2017,Anderson2018,Kim2019,Michalczyk2023,Alkorta1998a,Karpfen2000a,Karpfen2000b,Karpfen2003,Wu2007,Alkorta2008,Isegawa2011,Becke2018,Soydacs2013,Tentscher2013,Madajczyk2021,Zhou2010,Li2012,Wang2007,Li2008,Li2010,Yuan2012,Clark2019,Bandyopadhyay2020,Parajuli2013,Li2006,ZhiFeng2009,Li2011,Li2014a,Esrafili2016a,Esrafili2016b,Esrafili2015,Alkorta2014,Li2014b,Yu2016,Yu2020,Wu2023,Alkorta1998b,Tachikawa1998,Igarashi2002,Wang2003,Gil2004,Solimannejad2005,An2008,CrespoOtero2008,Czako2009,Li2009,Jacox1979,Merritt2006,Rudic2009,Politzer2007,Metrangolo2011b,Li2014c,Varadwaj2015,Zhao2005a,Ruiz1996,Zhang1997,Karpfen1999,Mardirossian2011,Koohi2017,Scheiner2020,Orangi2020,Latajka1996,Lu2007,Eskandari2015,Grabowski2015,Esrafili2017,McDowell2018,Alkorta2020,Oliveira2016,Kossmann2010,Kitsaras2021,Tran2022,Mehta2021,Jones2015,Toulouse2022,Cohen2012,Mardirossian2017,Becke2014,Su2017,Su2016a,Perdew1996,Perdew1981,Zhang1998,MoriSanchez2006,Cohen2008,Bryenton2023,Johnson2013,Goerigk2017,Goerigk2019,Mardirossian2016,Becke1997,Daas2021,Rezac2017,Rezac2012,Belair2002,Suma2005,Suma2006,Du2006,Kos2008,Han2008,Janssen2009,Bartolomei2010,Bertolus2012,Cybulski2012,V2012,Warehime2015,Dreux2015,VRyazantsev2017,Grein2017a,Grein2017b,Liu2019,Ajili2022,Khanal2015,Jurecka2006,Rezac2011a,Smith2016,Rezac2013,Zhao2005b,Morgante2019,Steinmann2012,Rezac2020a,Rezac2020b,Kriz2021,Kriz2022,Rezac2022,Donchev2021,Sparrow2021,Witte2015,Sirianni2018,Su2016b,Frisch2016,Werner2020,Apra2020,Woon1993,DunningJr1989,Werner2007,Knizia2008,Peterson2008,Tenno2004,Kong2012,Hattig2012,Kesharwani2018a,Hill2014,Lane2009,Boese2015,Boese2011,Manna2017,Alkorta2024,Legon2014,Burns2014,Brauer2016,Hill2015,Marchetti2009,Sirianni2017,Riley2012,Marchetti2008,Brauer2014,Santra2022,Mester2023,deLange2011,Kesharwani2018b,Wagle2021,Perkins2020,Peterson2015,Patkowski2013,Patkowski2012,Kodrycka2019,Schwenke2004,Hill2009,Yousaf2009,Rezac2011b,Halkier1998,Karton2012,Boys1970,Boese2007,Knizia2009,Hattig2010,Bartlett1978,Watts1993,Raghavachari1989,Kats2013,Kats2014,Werner2003,Hirata2003,Grimme2003,Zhao2008a,Zhao2008b,Doemer2013,Mikherdov2020,Ostras2020,Kryukova2019,VKashina2019,Kendall1992,Weigend2005,Hariharan1973,Krishnan1980,Friedrich2016,Santra2007,Burns2011,Vamhindi2017,OterodelaRoza2013}
\thispagestyle{fancy}
\centering
{\fontsize{17}{21}\selectfont Single-electron Fluorine Bonding. I. The SEFB-15 and CSFB-14 Benchmark Datasets, Reference Binding Energies and Intermolecular Bond Distances: Assessment of M06-2X density functional\par}
\vspace{12pt}
Govinda Prasad Khanal\textsuperscript{a}\,\orcidicon{0000-0002-0991-7005}, and Elangannan Arunan\textsuperscript{a}*\,\orcidicon{0000-0001-9669-4307}\par
{\itshape\textsuperscript{a}Inorganic and Physical Chemistry, Indian Institute of Science, Bangalore, India -- 560 012}\par
\textbf{*E-mail Id:} arunan@iisc.ac.in\par
\vspace{8pt}
\vspace{6pt}{\bfseries\small Abstract}\par\normalfont
\GPKfig{6.378in}{2.644in}{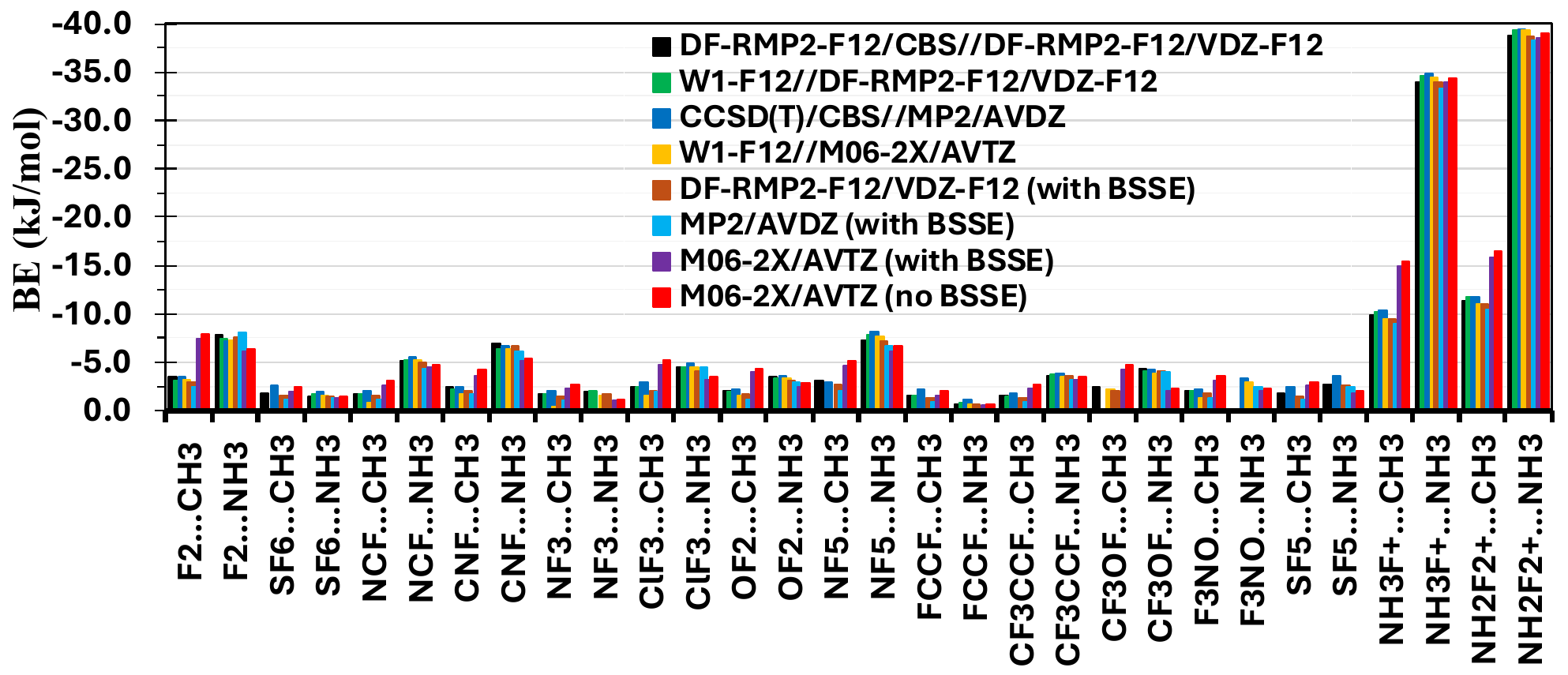}
{\small\justifying\setlength{\parindent}{0pt}\leftskip=0.4in\rightskip=0.4in
Halogen bonding in which a covalently bound fluorine acts as halogen-bond donor (F-bonding) is rare, and its open-shell variant with radical electron donors remains unexplored. We introduce SEFB-15 dataset (15 open-shell DF\textperiodcentered{}\textperiodcentered{}\textperiodcentered{}CH\textsubscript{3} complexes) exemplifying single-electron F-bonding (SEFB) and CSFB-14 dataset (14 closed-shell DF\textperiodcentered{}\textperiodcentered{}\textperiodcentered{}NH\textsubscript{3} complexes) representing closed-shell F-bonding (CSFB). Reference binding energies (BEs) at DF-RMP2-F12/CBS//DF-RMP2-F12/VDZ-F12 are validated against CCSD(T), CCSD(T)-F12b, W1-F12 and W2-F12 for selected systems (DF = F\textsubscript{2}, NCF, CNF, NH\textsubscript{3}F\textsuperscript{+} and NH\textsubscript{2}F\textsubscript{2}\textsuperscript{+}), with MA\%D of 4.5\% (DF\textperiodcentered{}\textperiodcentered{}\textperiodcentered{}CH\textsubscript{3}) and 3.4\% (DF\textperiodcentered{}\textperiodcentered{}\textperiodcentered{}NH\textsubscript{3}) against CCSD(T)-F12b/CBS//CCSD(T)-F12b/VDZ-F12. BEs from W1-F12 and W2-F12 are identical. Reference F-bond distances at DF-RMP2-F12/VDZ-F12 are validated against CCSD(T)-F12b, CCSD(T), CCSD, and DCSD for the same systems, with MA\%D of 0.9\% (SEFB) and 0.4\% (CSFB) against CCSD(T)-F12b/VDZ-F12. M06-2X is recommended for closed-shell halogen bonding with heavier halogens but has never been assessed for open-shell F-bonding. M06-2X/AVTZ systematically~overbinds~all SEFB-15 complexes and underestimates their F-bond distances, with MA\%Ds of 75.8\% on BEs and 8.9\% on F-bond distances.~M06-2X predicts comparable BEs for F\textsubscript{2}\textperiodcentered{}\textperiodcentered{}\textperiodcentered{}CH\textsubscript{3} and F\textsubscript{2}\textperiodcentered{}\textperiodcentered{}\textperiodcentered{}NH\textsubscript{3}, contrary to all ab initio results, an anomaly persisting across all 16 basis sets. Neither basis set nor BSSE correction resolves the overbinding. BSSE correction worsens BEs for DF\textperiodcentered{}\textperiodcentered{}\textperiodcentered{}NH\textsubscript{3} complexes. For CSFB-14, M06-2X/AVTZ yields MA\%Ds of 16.0\% for BEs and 3.3\% for F-bond distances. M06-2X is reliable for CSFB but unreliable for SEFB regardless of basis set. Companion paper II reports performance of other Minnesota functionals and \emph{ab initio} methods for these datasets.\par}
\vspace{6pt}
\justifying\setlength{\parindent}{1.5em}
\section{Introduction}\label{sec:introduction}

Noncovalent bonds (NCBs) are crucial to the structure and properties of water, proteins, DNA and RNA,\cite{MullerDethlefs2000,Hobza2010,Kolar2016,Rezac2016,Desiraju2001,Johnson2010,Stone2013,Neese2019} driving broad interest in their computational\cite{Rezac2016,AlHamdani2021,Villot2022,Ballesteros2021,AlHamdani2019} study. Halogen bonding (HalB) DX\textperiodcentered{}\textperiodcentered{}\textperiodcentered{}Y, with DX and Y as donor and acceptor was first recognized in the 19\textsuperscript{th} century\cite{Guthrie1863,Remsen1896,Benesi1949,Mulliken1950,Hassel1962,Bent1968,Hassel1970} and extensively studied since the late 1980s.\cite{Kolar2016,Blackstock1987,Andrews1988,Desiraju1995,Legon1999,Metrangolo2005,Metrangolo2008,Legon2008,Legon2010,Cavallo2016,Costa2017} It is one of well-studied NCBs with applications spanning materials science, drug design, and biology, among others.\cite{Metrangolo2001,Meyer2013,Auffinger2004,Lu2009,Parisini2011,Metrangolo2006,Berger2015,Gilday2015,Carter2011,Lu2010,Guillier2019,Bulfield2016,Govindaraj2019,Peluso2020} Its strength depends on the electron donor and the magnitude of the \ensuremath{\sigma}-hole on the halogen atom.\cite{Clark2007,Politzer2017} Most reported HalB interactions involve heavier halogens (Cl \textless{} Br \textless{} I).\cite{Cavallo2016,Desiraju2013} Due to its high electronegativity and low polarizability, fluorine was long considered incapable of forming a \ensuremath{\sigma}-hole and HalB.\cite{Wilcken2013,Dikundwar2014,Lin2017} If bonded to strongly electron-withdrawing groups (e.g., F\textsubscript{2}, F-CN, F-NC, F-CCF), however, fluorine can develop a \ensuremath{\sigma}-hole and form fluorine bonding (FB),\cite{Cavallo2016,Chopra2011,Politzer2013,Metrangolo2011a} confirmed experimentally and theoretically in crystals and functional materials.\cite{Metrangolo2011a,Chopra2012,Hathwar2014,Berger2011,Panini2015,Cole2022,Bauza2016,Varadwaj2018,Pavan2013,Han2017,Neaton2017,BarceloOliver2010,Osuna2011,Sava2020,VJanjic2020,Rocha2021,Chopra2006,Wang2015,Sirohiwal2017,Elakkat2019,VGrudova2020,Rana2020}

HalB acceptors Y in DX\textperiodcentered{}\textperiodcentered{}\textperiodcentered{}Y include lone pairs, \ensuremath{\pi}-systems, and anions.\cite{Cavallo2016,Shahi2014,McDowell2012,Raghavendra2007,Cincic2008,Scholfield2013,VBartashevich2014,Deepa2014,Shaw2016,Lisac2019,Riley2020,Stephens2012,Tsuzuki2016,Riley2017,Kanao2020,Portela2021} NH\textsubscript{3}, among the most widely considered electron donors for HalB,\cite{Cavallo2016,Shahi2014,Bauza2013,Hill2013,Kozuch2013,OteroDeLaRoza2014,Wang2014,Scheiner2016,Wang2017,Oliveira2017,Anderson2018,Kim2019,Michalczyk2023,Alkorta1998a,Karpfen2000a,Karpfen2000b,Karpfen2003,Wu2007,Alkorta2008,Isegawa2011,Becke2018} is adopted here to construct closed-shell FB (CSFB) complexes (DF\textperiodcentered{}\textperiodcentered{}\textperiodcentered{}NH\textsubscript{3}). CH\textsubscript{3} radical is also considered as an unconventional electron donor, forming open-shell FB complexes (DF\textperiodcentered{}\textperiodcentered{}\textperiodcentered{}CH\textsubscript{3}), referred to as single-electron FB (SEFB). Its single electron on the sp\textsuperscript{2}-hybridized carbon makes it a prototypical electron donor in radical-molecule complexes. The pivotal role of open-shell NCBs in chemistry cannot be exaggerated (refer to \cite{Soydacs2013,Tentscher2013,Madajczyk2021} and references therein). Quantum chemical studies on single-electron halogen bonding (SEHalB) with chlorine, bromine, and iodine have been reported,\cite{Shahi2014,Raghavendra2007,Zhou2010,Li2012,Wang2007,Li2008,Li2010,Yuan2012,Clark2019,Bandyopadhyay2020} but none has addressed SEFB. CH\textsubscript{3} radical is known to form single-electron NCB across the periodic table,\cite{Shahi2014,Raghavendra2007,Li2012,Parajuli2013,Li2006,ZhiFeng2009,Li2011,Li2014a,Esrafili2016a,Esrafili2016b,Esrafili2015,Alkorta2014,Li2014b,Yu2016,Yu2020,Wu2023} and Alkorta et al. established it as a single-electron donor via single-electron hydrogen bonding (SEHB)\cite{Alkorta1998b}\relax. Despite extensive computational work on SEHB,\cite{Shahi2014,Raghavendra2007,Li2012,Tachikawa1998,Igarashi2002,Wang2003,Gil2004,Solimannejad2005,An2008,CrespoOtero2008,Czako2009,Li2009} experimental reports remain scarce.\cite{Jacox1979,Merritt2006,Rudic2009}

Although several computational studies exist on single-electron NCBs, no benchmark assessment using diverse \emph{ab initio} methods and DFAs has, to our knowledge, been reported. Most studies rely on MP2, with limited use of higher-level methods or modern DFAs. CSFB complexes, including F\textsubscript{2}\textperiodcentered{}\textperiodcentered{}\textperiodcentered{}NH\textsubscript{3}, has similarly been studied with a limited set of donors and heavy reliance on MP2 or popular DFAs.\cite{Metrangolo2011a,Wang2015,Wang2014,Oliveira2017,Alkorta1998a,Karpfen2000a,Karpfen2000b,Karpfen2003,Politzer2007,Metrangolo2011b,Li2014c,Varadwaj2015,Zhao2005a,Ruiz1996,Zhang1997,Karpfen1999,Mardirossian2011,Koohi2017,Scheiner2020,Orangi2020,Latajka1996,Lu2007,Eskandari2015,Grabowski2015,Esrafili2017,McDowell2018,Alkorta2020,Oliveira2016} Concerns have also been raised about the reliability of MP2 for both open-shell\cite{Soydacs2013,Kossmann2010,Kitsaras2021,Tran2022} and closed-shell systems,\cite{Mehta2021} while higher-level correlated \emph{ab initio} methods, though more accurate, are often too costly for larger systems. DFAs\cite{Jones2015,Toulouse2022,Cohen2012,Mardirossian2017,Becke2014,Su2017,Su2016a,Perdew1996} are therefore used in practice, but conventional functionals suffer from self-interaction error (SIE)\cite{Perdew1981,Zhang1998,MoriSanchez2006} and delocalization error.\cite{Cohen2012,Cohen2008,Bryenton2023} Both of these deficiencies are more severe for open-shell systems such as radical-molecule complexes.\cite{Tentscher2013,Johnson2013} This makes selecting an appropriate DFA from the large pool of available functionals nontrivial.\cite{Mardirossian2017,Goerigk2017,Goerigk2019}

Among the available DFAs, Minnesota functionals are popular and extensively benchmarked,\cite{Mardirossian2016} but not for open-shell fluorine bonding or open-shell NCIs. Heavy parameterization can compromise the robustness and transferability of DFAs beyond their training data sets,\cite{Becke1997} raising questions about the reliability of highly parameterized Minnesota DFAs for DF\textperiodcentered{}\textperiodcentered{}\textperiodcentered{}CH\textsubscript{3} and DF\textperiodcentered{}\textperiodcentered{}\textperiodcentered{}NH\textsubscript{3}, which differ substantially from typical training sets. Hybrid DFAs with some Hartree-Fock exchange can partially mitigate delocalization error,\cite{Bryenton2023} raising further questions about the reliability of hybrid Minnesota DFAs, particularly M06-2X (54\% HF exchange), for the open-shell SEFB complex (DF\textperiodcentered{}\textperiodcentered{}\textperiodcentered{}CH\textsubscript{3}). Notably, prior benchmarks on closed-shell halogen bonding (CSHalB), without fluorine as the HalB donor, have reported M06-2X to be reliable for such interactions.\cite{Bauza2013,Kozuch2013,OteroDeLaRoza2014} Motivated by this, the present work evaluates M06-2X, alongside several correlated \emph{ab initio} methods, for BEs and F-bond distances of both SEFB (DF\textperiodcentered{}\textperiodcentered{}\textperiodcentered{}CH\textsubscript{3}) and CSFB (DF\textperiodcentered{}\textperiodcentered{}\textperiodcentered{}NH\textsubscript{3}) complexes.

Even within CSFB, F\textsubscript{2}\textperiodcentered{}\textperiodcentered{}\textperiodcentered{}NH\textsubscript{3} is known to pose challenges for conventional DFAs.\cite{Wang2014,Alkorta1998a,Karpfen2000a,Karpfen2003,Becke2018,Zhao2005a,Ruiz1996,Mardirossian2011,Daas2021,Rezac2017} Existing databases on HalB cover only closed-shell systems.\cite{Bauza2013,Kozuch2013,Rezac2012} TA13 database,\cite{Tentscher2013} the only comprehensive benchmark set for radical-molecule complexes, has no halogen-bonded systems and just one SEHB complex (HOH\textperiodcentered{}\textperiodcentered{}\textperiodcentered{}CH\textsubscript{3}). Benchmark studies on open-shell NCBs remain scarce though there exist several project-specific works, including our study on open-shell (O\textsubscript{2})\textsubscript{n}\textsuperscript{+} (n = 2-5).\cite{Soydacs2013,Madajczyk2021,Belair2002,Suma2005,Suma2006,Du2006,Kos2008,Han2008,Janssen2009,Bartolomei2010,Bertolus2012,Cybulski2012,V2012,Warehime2015,Dreux2015,VRyazantsev2017,Grein2017a,Grein2017b,Liu2019,Ajili2022,Khanal2015} Closed-shell NCBs, by contrast, have several benchmark sets including S22, S66, GMTKN55, and several others.\cite{Bauza2013,Kozuch2013,Zhao2005a,Mardirossian2017,Goerigk2017,Rezac2012,Jurecka2006,Rezac2011a,Smith2016,Rezac2013,Zhao2005b,Morgante2019,Steinmann2012,Rezac2020a,Rezac2020b,Kriz2021,Kriz2022,Rezac2022,Donchev2021,Sparrow2021} Among these, X40, XB18, XB51, and Bauza30 represent CSHalB complexes.\cite{Bauza2013,Kozuch2013,Rezac2012} Benchmarks for intermolecular distances are limited even for closed-shell systems\cite{Bauza2013,Kozuch2013,Witte2015,Sirianni2018,Su2016b} and essentially absent for open-shell NCIs. This combined lack of benchmark data for BEs and equilibrium geometries, particularly for open-shell SEFB systems, motivates the present work, which systematically assesses both quantities for SEFB and their CSFB counterparts.

This work reports the first study of single-electron fluorine bonding (SEFB). We introduce SEFB-15, a benchmark data set of DF\textperiodcentered{}\textperiodcentered{}\textperiodcentered{}CH\textsubscript{3} complexes, along with CSFB-14, a data set of closed-shell DF\textperiodcentered{}\textperiodcentered{}\textperiodcentered{}NH\textsubscript{3} complexes. Reference BEs and F-bond distances for both data sets are validated against high-level correlated \emph{ab initio} methods (CCSD, CCSD(T), CCSD(T)-F12b, DCSD, W1-F12, and W2-F12). Using these reference data, we assess the reliability of M06-2X for predicting BEs and F-bond distances of both SEFB and CSFB complexes. We also test M06-2X across 16 basis sets for selected systems from both databases. A systematic evaluation of the remaining Minnesota DFAs, additional \emph{ab initio} methods (including MP3, MP4, and QCISD), delocalization error, and Bader delocalization indices as descriptors of F-bonding strength are presented in the companion paper.

The remainder of this paper is organized as follows: Section \ref*{sec:computational-details} describes the computational details, Section \ref*{sec:statistical-error-analysis} the error metrics, Section \ref*{sec:results-and-discussions} the results and discussion, and Section \ref*{sec:conclusions-and-outlook} the conclusions and outlook.

\section{Computational details}\label{sec:computational-details}

All electronic structure calculations were performed using Gaussian 16 C.01\cite{Frisch2016} and MOLPRO 2015.1 program.\cite{Werner2020} For selected methods, additional calculations were carried out using NWChem 6.8 program\cite{Apra2020} as described below. Full geometry optimization and harmonic vibrational frequency calculations were performed for all DF\textperiodcentered{}\textperiodcentered{}\textperiodcentered{}CH\textsubscript{3}/NH\textsubscript{3} complexes and their corresponding monomers using every electronic structure method employed, unless otherwise stated, to confirm that the stationary points correspond to true minima.

The complete set of F-bonded complexes DF\textperiodcentered{}\textperiodcentered{}\textperiodcentered{}CH\textsubscript{3}/NH\textsubscript{3} (DF = F\textsubscript{2}, SF\textsubscript{6}, NCF, CNF, NF\textsubscript{3}, ClF\textsubscript{3}, OF\textsubscript{2}, NF\textsubscript{5}, FCCF, CF\textsubscript{3}CCF, CF\textsubscript{3}OF, F\textsubscript{3}NO, SF\textsubscript{5}, NH\textsubscript{3}F\textsuperscript{+}, and NH\textsubscript{2}F\textsubscript{2}\textsuperscript{+}) was investigated using a broad range of \emph{ab initio} electronic structure methods. Geometry optimization and frequency calculations were carried out using MP2 with aug-cc-pVDZ (AVDZ)\cite{Woon1993,DunningJr1989} basis set within Gaussian 16. The AVDZ basis set was adopted due to the size of several complexes, particularly open-shell SEFB systems such as SF\textsubscript{6}\textperiodcentered{}\textperiodcentered{}\textperiodcentered{}CH\textsubscript{3}, NF\textsubscript{5}\textperiodcentered{}\textperiodcentered{}\textperiodcentered{}CH\textsubscript{3}, CF\textsubscript{3}CCF\textperiodcentered{}\textperiodcentered{}\textperiodcentered{}CH\textsubscript{3}, and SF\textsubscript{5}\textperiodcentered{}\textperiodcentered{}\textperiodcentered{}CH\textsubscript{3}.

Explicitly correlated calculations were performed for the complete set of F-bonded complexes using density-fitting MP2 with explicit correlation, DF-RMP2-F12,\cite{Werner2007,Knizia2008} with the VDZ-F12\cite{Peterson2008} basis set and the diagonal fixed-amplitude 3C(FIX) ansatz,\cite{Tenno2004} specified as Ansatz=3c(fix) within MOLPRO program.\cite{Werner2020} F12 methods, including DF-RMP2-F12 converge substantially faster with basis set size than conventional CCSD(T).\cite{Kong2012,Hattig2012} Thus, they are routinely employed to generate high-quality reference data for NCBs, including halogen bonding,\cite{Shaw2016,Hill2013,Steinmann2012,Kesharwani2018a,Hill2014} hydrogen bonding,\cite{Lane2009,Boese2015,Boese2011,Manna2017} tetrel bonding,\cite{Alkorta2024} chalcogen bonding,\cite{Mehta2021} and other weak interactions.\cite{Madajczyk2021,Legon2014,Burns2014,Brauer2016,Hill2015,Marchetti2009,Sirianni2017,Riley2012,Marchetti2008,Brauer2014,Santra2022,Mester2023,deLange2011,Kesharwani2018b,Wagle2021,Perkins2020,Peterson2015,Patkowski2013,Patkowski2012,Kodrycka2019}

Complete basis set (CBS) limit correlation energies were evaluated at the DF-RMP2-F12 level using geometries optimized at DF-RMP2-F12/VDZ-F12. Specifically, DF-RMP2-F12/CBS{[}V(T,Q)Z-F12{]}//DF-RMP2-F12/VDZ-F12 correlation energies were obtained by using Schwenke-type two-point scheme,\cite{Schwenke2004,Hill2009} as given in Eq. (1).

\begin{equation}\label{eq:cbs-correlation}
\adjustbox{max width=0.9\linewidth}{$\displaystyle E_{\mathrm{CBS}}^{\mathrm{corr}}\mathrm{\lbrack(T,Q)Z\text{-}F12\rbrack} = (E_{4}^{\mathrm{corr}} - E_{3}^{\mathrm{corr}})F + E_{3}^{\mathrm{corr}}$}
\end{equation}\par\vspace{2pt}

Here, \(E_{n}^{\mathrm{corr}}\)(n = 3, 4) are the correlation energies obtained from single-point calculations using DF-RMP2-F12/VnZ-F12 level (n = 3 and 4) at DF-RMP2-F12/VDZ-F12 optimized geometries, and the CBS extrapolation coefficient (\(F\)) is taken from the literature.\cite{Hill2009} The CBS limit electronic energies were then obtained by combining these extrapolated correlation energies with Hartree--Fock (HF) reference energies, including the CABS singles correction,\cite{Yousaf2009} as expressed in Eq. (2).

\begin{equation}\label{eq:cbs-energy}
\adjustbox{max width=0.9\linewidth}{$\displaystyle E_{\mathrm{DF\text{-}RMP2\text{-}F12}}^{\mathrm{CBS\mathrm{\lbrack(T,Q)Z\text{-}F12\rbrack}}} = E_{\mathrm{CBS}}^{\mathrm{corr}}\mathrm{\lbrack(T,Q)Z\text{-}F12\rbrack} + E_{\mathrm{HF}}^{\mathrm{VQZ\text{-}F12}}$}
\end{equation}\par\vspace{2pt}

This protocol was applied to all F-bonded complexes and the corresponding monomers. Larger basis sets were not employed for geometry optimization due to computational limitations. The resulting CBS-limit binding energies (BEs), obtained from supermolecular approach, together with F-bond distances computed at the DF-RMP2-F12/VDZ-F12 level, were adopted as reference data in this work.

These reference BEs were cross-checked against two independent estimates of CCSD(T)/CBS. First, estimated CCSD(T)/CBS values were evaluated at MP2/AVDZ-optimized geometries using the composite scheme proposed by Hobza and co-workers\cite{Rezac2012,Rezac2011a,Rezac2011b}\relax:

\begin{equation}\label{eq:ccsdt-cbs}
\adjustbox{max width=0.9\linewidth}{$\displaystyle E_{\mathrm{CCSD(T)/CBS}} = E_{\mathrm{HF}}^{\mathrm{AVQZ}} + E_{\mathrm{MP2/CBS}}^{\mathrm{corr}} + \left( E_{\mathrm{CCSD(T)/AVTZ}}^{\mathrm{corr}}-E_{\mathrm{MP2/AVTZ}}^{\mathrm{corr}} \right)$}
\end{equation}\par\vspace{2pt}

where, \(E_{\mathrm{HF}}^{\mathrm{AVQZ}}\) is the HF energy computed with aug-cc-pVQZ (AVQZ) basis set, while \(E_{\mathrm{CCSD(T)/AVTZ}}^{\mathrm{corr}}\) and \(E_{\mathrm{MP2/AVTZ}}^{\mathrm{corr}}\) denote correlation energies obtained at CCSD(T)/aug-cc-pVTZ and MP2/aug-cc-pVTZ levels, respectively. The MP2/CBS correlation energy was obtained using Halkier's two-point extrapolation scheme,\cite{Halkier1998}

\begin{equation}\label{eq:mp2-avnz}
\adjustbox{max width=0.9\linewidth}{$\displaystyle E_{\mathrm{MP2/AVnZ}}^{\mathrm{corr}} = E_{\mathrm{MP2/CBS}}^{\mathrm{corr}} + \frac{A}{n^{3}},n = 3,4$}
\end{equation}\par\vspace{2pt}
leading to

\begin{equation}\label{eq:mp2-cbs}
\adjustbox{max width=0.9\linewidth}{$\displaystyle E_{\mathrm{MP2/CBS}}^{\mathrm{corr}} = \frac{64E_{4}^{\mathrm{corr}} - 27E_{3}^{\mathrm{corr}}}{37}$}
\end{equation}\par\vspace{2pt}

Second, CCSD(T)/CBS BEs for the full set were evaluated using the W1-F12\cite{Karton2012} protocol at geometries optimized at the DF-RMP2-F12/VDZ-F12 level, though W1-F12 calculations were not feasible for several larger complexes (SF\textsubscript{6}\textperiodcentered{}\textperiodcentered{}\textperiodcentered{}CH\textsubscript{3}, NF\textsubscript{5}\textperiodcentered{}\textperiodcentered{}\textperiodcentered{}CH\textsubscript{3}, CF\textsubscript{3}OF\textperiodcentered{}\textperiodcentered{}\textperiodcentered{}CH\textsubscript{3}, SF\textsubscript{5}\textperiodcentered{}\textperiodcentered{}\textperiodcentered{}CH\textsubscript{3}, and SF\textsubscript{5}\textperiodcentered{}\textperiodcentered{}\textperiodcentered{}NH\textsubscript{3}) due to their high computational cost.

BEs calculated using Gaussian 16 were corrected for basis-set superposition error (BSSE) using the built-in counterpoise scheme.\cite{Boys1970} For F12 calculations, BSSE corrections were applied manually using the standard counterpoise expression\cite{Boese2007}\relax:

\begin{equation}\label{eq:bsse}
\adjustbox{max width=0.9\linewidth}{$\displaystyle \Delta E_{\mathrm{BSSE}} = \lbrack E_{\mathrm{tot}}(\mathrm{DF}) - E_{\mathrm{tot}}(\mathrm{DF,ghost})\rbrack + \lbrack E_{\mathrm{tot}}(\mathrm{CH_{3}/NH_{3}}) - E_{\mathrm{tot}}(\mathrm{CH_{3}/NH_{3},ghost})\rbrack.$}
\end{equation}\par\vspace{2pt}

Monomer energies were evaluated both at their isolated equilibrium and dimer geometries. Both uncorrected and BSSE-corrected BEs are reported.

To further validate the reference BEs and F-bond distances, additional high-level calculations were performed for selected complexes DF\textperiodcentered{}\textperiodcentered{}\textperiodcentered{}CH\textsubscript{3}/NH\textsubscript{3} (DF = F\textsubscript{2}, NCF, CNF, NH\textsubscript{3}F\textsuperscript{+}, and NH\textsubscript{2}F\textsubscript{2}\textsuperscript{+}). The validation methods employed include CCSD(T)-F12b,\cite{Knizia2009,Hattig2010} CCSD,\cite{Bartlett1978} CCSD(T),\cite{Watts1993,Raghavachari1989} DCSD,\cite{Kats2013,Kats2014} DF-RMP2,\cite{Werner2003} MBPT(2),\cite{Hirata2003} and SCS-MP2\cite{Grimme2003} method. MBPT(2) calculations were performed using NWChem 6.8 program. CCSD(T)-F12b, CCSD(T), DCSD, DF-RMP2, and SCS-MP2 calculations were carried out using MOLPRO program. CCSD calculations were performed using Gaussian 16 suit of program.

In CCSD(T)-F12b, the perturbative triples correction (T) is not treated explicitly and does not fully benefit from explicit correlation.\cite{Knizia2009} To address this, the (T) contribution was scaled using the Werner--Marchetti approximation,\cite{Marchetti2009,Marchetti2008} \(\frac{E_{\mathrm{corr}}^{\mathrm{MP2\text{-}F12}}}{E_{\mathrm{corr}}^{\mathrm{MP2}}},\) implemented via the UCCSD(T)-F12b, SCALE\_TRIP = 1 option in MOLPRO program. CCSD(T)-F12b/CBS{[}V(T,Q)Z-F12{]}//CCSD(T)-F12b/VDZ-F12 BEs were obtained using the supermolecular approach, as stated above. The CCSD-F12b/3C(fix) correlation energies and the (T) contributions were extrapolated separately to the CBS limit using the VTZ-F12 and VQZ-F12 basis sets\cite{Hill2009} following Eqs. (\ref*{eq:cbs-correlation}) and (\ref*{eq:cbs-energy}), and combined with CABS-corrected HF/VQZ-F12 reference energies as described above. For all calculations with MOLPRO program, restricted Hartree--Fock (RHF) reference wave functions were employed, with DF-RHF reference orbitals for DF-RMP2 calculations. For Gaussian 16 calculations (MP2, CCSD, and M06-2X), the default set-up was used throughout. The geminal Slater exponent (\ensuremath{\beta}) was set to 0.9 for VDZ-F12 basis set and 1.0 for VTZ-F12 and VQZ-F12 basis sets, following the recommendations of Hill et al.\cite{Hill2009}

In addition to the \emph{ab initio} methods described above, density functional theory (DFT) calculations were performed using M06-2X\cite{Zhao2008a} functional to predict BEs and F-bond distances for all DF\textperiodcentered{}\textperiodcentered{}\textperiodcentered{}CH\textsubscript{3}/NH\textsubscript{3} complexes. M06-2X was selected based on its previously reported reliable performance for organic radicals,\cite{Zhao2008b} closed-shell halogen bonding,\cite{Bauza2013,Kozuch2013,OteroDeLaRoza2014} weak interactions involving halogen dimers including fluorine-containing systems such as (F\textsubscript{2})\textsubscript{2},\cite{Doemer2013} and halogen bonding in supramolecular\cite{Mikherdov2020,Ostras2020,Kryukova2019,VKashina2019} systems. M06-2X was employed with the aug-cc-pVTZ (AVTZ) basis set for all complexes. To assess the reliability of BEs from M06-2X/AVTZ, CCSD(T)/CBS reference values were computed using the W1-F12\cite{Karton2012} protocol at M06-2X/AVTZ-optimized geometries. To examine basis-set effects, M06-2X calculations were further performed for selected complexes (DF = F\textsubscript{2}, NCF, CNF, NH\textsubscript{3}F\textsuperscript{+}, and NH\textsubscript{2}F\textsubscript{2}\textsuperscript{+}) using 16 basis sets: cc-pVnZ (n = 2-5),\cite{DunningJr1989} aug-cc-pVnZ (n = 2-5),\cite{Kendall1992} def2-nZVP (n = 3, 4),\cite{Weigend2005} and several Pople-type split-valence basis sets with polarization and diffuse functions\cite{Hariharan1973,Krishnan1980} {[}6-311g(2d, 2p), 6-311++g(2d, 2p), 6-311g(2df, 2pd), 6-311++g(2df, 2pd), 6-311g(3df, 3pd) and 6-311++g(3df, 3pd){]}.

\section{Statistical error analysis}\label{sec:statistical-error-analysis}

To assess the performance of the ab initio methods and M06-2X functional employed in this work for predicting the binding energies (BEs) and F-bond distances (R) of the DF\textperiodcentered{}\textperiodcentered{}\textperiodcentered{}CH\textsubscript{3}/NH\textsubscript{3} complexes, a set of standard statistical measures was evaluated: mean signed error (MSE), mean absolute deviation (MAD), root-mean-square deviation (RMSD), maximum deviation (MAX), and mean absolute percentage deviation (MA\%D). These metrics are routinely employed in benchmark studies of intermolecular interactions, including halogen bonding and radical-molecule complexes.\cite{Bauza2013,Kozuch2013,OteroDeLaRoza2014,Anderson2018,Kim2019,Tentscher2013,Mardirossian2016,Witte2015,Su2016b,Boese2007,Doemer2013,Friedrich2016,Santra2007,Burns2011,Vamhindi2017}

The statistical quantities are defined as

\begin{equation}\label{eq:mse}
\adjustbox{max width=0.9\linewidth}{$\displaystyle \mathrm{MSE} = \frac{1}{n}\sum_{i = 1}^{n}\left\lbrack QTY_{\mathrm{method},i} - QTY_{\mathrm{ref},i} \right\rbrack$}
\end{equation}\par\vspace{2pt}

\begin{equation}\label{eq:mad}
\adjustbox{max width=0.9\linewidth}{$\displaystyle \mathrm{MAD} = \frac{1}{n}\sum_{i = 1}^{n}{\mid QTY_{\mathrm{method},i} - QTY_{\mathrm{ref},i} \mid}$}
\end{equation}\par\vspace{2pt}

\begin{equation}\label{eq:rmsd}
\adjustbox{max width=0.9\linewidth}{$\displaystyle \mathrm{RMSD} = \sqrt{\frac{1}{n}\sum_{i = 1}^{n}\left( QTY_{\mathrm{method},i}-QTY_{\mathrm{ref},i} \right)^{2}}$}
\end{equation}\par\vspace{2pt}

\begin{equation}\label{eq:max}
\adjustbox{max width=0.9\linewidth}{$\displaystyle \mathrm{MAX} = \underset{i}{\max} \mid QTY_{\mathrm{method},i} - QTY_{\mathrm{ref},i} \mid$}
\end{equation}\par\vspace{2pt}

\begin{equation}\label{eq:mapd}
\adjustbox{max width=0.9\linewidth}{$\displaystyle \mathrm{MA\%D} = 100 \times \frac{1}{n}\sum_{i = 1}^{n}\frac{\mid QTY_{\mathrm{method},i} - QTY_{\mathrm{ref},i} \mid}{\mid QTY_{\mathrm{ref},i} \mid}$}
\end{equation}\par\vspace{2pt}

where, QTY denotes either the binding energy (BE) or the intermolecular distance \(R(F\cdots C/N)\) of a given F-bonded complex (DF\textperiodcentered{}\textperiodcentered{}\textperiodcentered{}CH\textsubscript{3}/NH\textsubscript{3}), and the subscripts ``method'' and ``ref'' refer to the evaluated quantum-chemical method (ab initio or DFA) and the corresponding reference value, respectively.

For each complex i, the deviation \(\delta_{QTY}\) and the corresponding percentage deviation \(\%\delta_{QTY}\) relative to the reference value were likewise evaluated as

\begin{equation}\label{eq:deviation}
\adjustbox{max width=0.9\linewidth}{$\displaystyle \delta_{QTY} = QTY_{\mathrm{method},i} - QTY_{\mathrm{ref},i}$}
\end{equation}\par\vspace{2pt}

\begin{equation}\label{eq:percent-deviation}
\adjustbox{max width=0.9\linewidth}{$\displaystyle \%\delta_{QTY} = 100 \times \frac{QTY_{\mathrm{method},i} - QTY_{\mathrm{ref},i}}{\mid QTY_{\mathrm{ref},i} \mid}$}
\end{equation}\par\vspace{2pt}

\section{Results and Discussions}\label{sec:results-and-discussions}

\subsection{Geometries and energetics of complexes DF\textperiodcentered{}\textperiodcentered{}\textperiodcentered{}CH\textsubscript{3}/NH\textsubscript{3}}\label{sec:geometries-and-energetics-of}

\begin{figure}[!htbp]\centering
\GPKfig{6.57569in}{3.42986in}{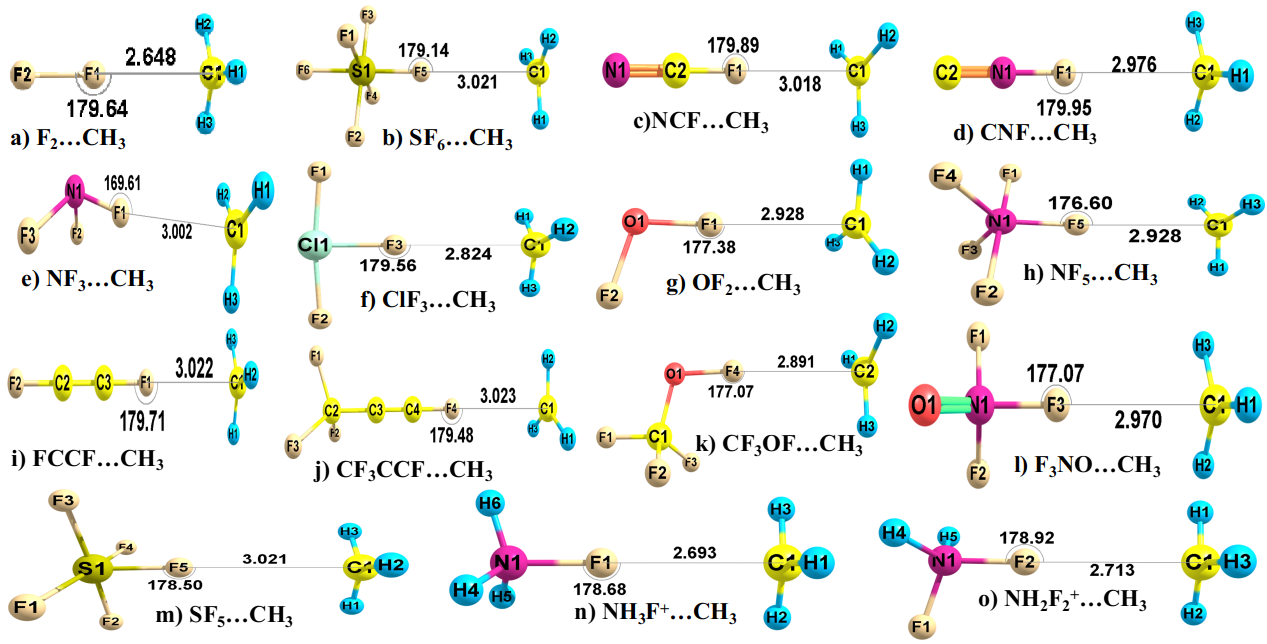}
\par\vspace{-1pt}\caption{Optimized structures for 15 single-electron F-bonded (SEFB) (\textbf{SEFB-15}) complexes DF\textperiodcentered{}\textperiodcentered{}\textperiodcentered{}CH\textsubscript{3} (SEFB-15 dataset) calculated at M06-2X/AVTZ level of theory. Structures for the corresponding closed-shell F-bonded (CSFB) complexes are depicted in \Fref{fig:the-optimized-structures-for} in Supporting Information.}\label{fig:optimized-structures-for-15}
\vspace{10pt}
\end{figure}

\Fref{fig:optimized-structures-for-15} shows the optimized structures obtained at M06-2X/AVTZ level for fifteen single-electron F-bonded (SEFB) complexes DF\textperiodcentered{}\textperiodcentered{}\textperiodcentered{}CH\textsubscript{3} (SEFB-15), formed with the CH\textsubscript{3} radical as electron donor.

The DF\textperiodcentered{}\textperiodcentered{}\textperiodcentered{}C angle and F\textperiodcentered{}\textperiodcentered{}\textperiodcentered{}C distance are shown for each complex in the figure. All complexes are linear, with DF\textperiodcentered{}\textperiodcentered{}\textperiodcentered{}C angles above 176\textdegree{}, except NF\textsubscript{3}\textperiodcentered{}\textperiodcentered{}\textperiodcentered{}CH\textsubscript{3}, (169.6\textdegree{}). All F\textperiodcentered{}\textperiodcentered{}\textperiodcentered{}C distances are significantly below the sum of their van der Waals radii (3.1 Å). The F\textsubscript{2}\textperiodcentered{}\textperiodcentered{}\textperiodcentered{}CH\textsubscript{3} complex has the shortest distance (2.648 Å) and, not surprisingly, has the highest binding energy among the neutral SEFB complexes, which is discussed next.

\FloatBarrier
\subsection{Reference binding energies and F-bond distances for DF\textperiodcentered{}\textperiodcentered{}\textperiodcentered{}CH\textsubscript{3}/NH\textsubscript{3}}\label{sec:reference-binding-energies-and}

\begin{figure}[!htbp]\centering
\GPKfig{7.05694in}{2.925in}{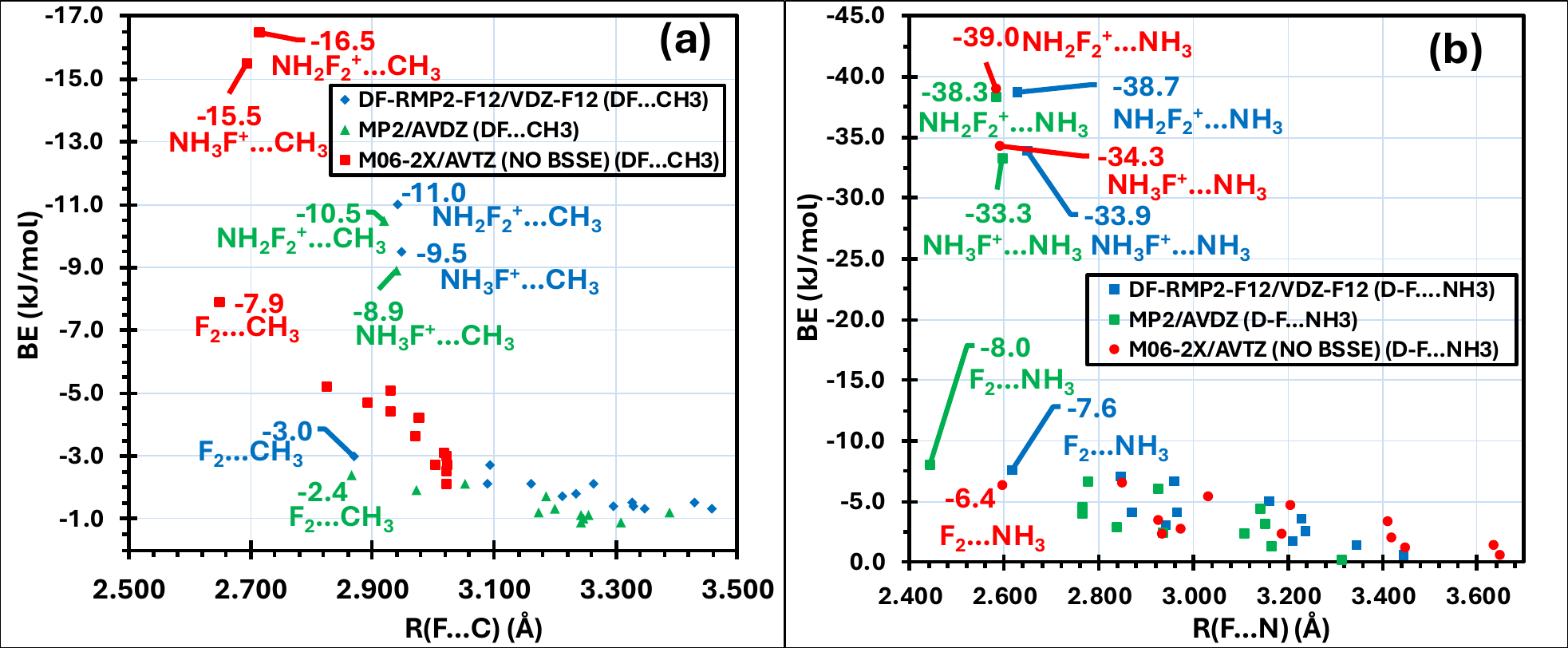}
\par\vspace{-1pt}\caption{Scatter plots of BEs versus F-bond distances R(F\textperiodcentered{}\textperiodcentered{}\textperiodcentered{}C/N) for F-bonded complexes DF\textperiodcentered{}\textperiodcentered{}\textperiodcentered{}CH\textsubscript{3} \textbf{(a)} and DF\textperiodcentered{}\textperiodcentered{}\textperiodcentered{}NH\textsubscript{3} \textbf{(b)} at DF-RMP2-F12/VDZ-F12, MP2/AVDZ and M06-2X/AVTZ. See \Tref{tab:bes-kj-mol-and} for the underlying data.}\label{fig:scatter-plots-of-bes}
\vspace{10pt}
\end{figure}

The CBS limit BEs for SEFB-15 (DF\textperiodcentered{}\textperiodcentered{}\textperiodcentered{}CH\textsubscript{3}) and CSFB-14 (DF\textperiodcentered{}\textperiodcentered{}\textperiodcentered{}NH\textsubscript{3}) complexes, calculated at DF-RMP2-F12/VDZ-F12, MP2/AVDZ and M06-2X/AVTZ (each at its own optimized geometry) are summarized in \mbox{\textbf{TABLE \ref{tab:cbs-limit-binding-energies}.}} The trend in BEs at M06-2X/AVTZ level is unaltered by BSSE correction and inconsistencies between M06-2X and high-level \emph{ab initio} methods persist even after correction for several SEFB complexes (DF\textperiodcentered{}\textperiodcentered{}\textperiodcentered{}CH\textsubscript{3}; DF = F\textsubscript{2}, NCF, CNF, ClF\textsubscript{3}, OF\textsubscript{2}, NF\textsubscript{5}, CF\textsubscript{3}CCF, CF\textsubscript{3}OF, F\textsubscript{3}NO, NH\textsubscript{3}F\textsuperscript{+} and NH\textsubscript{2}F\textsubscript{2}\textsuperscript{+}) (\Tref{tab:cbs-limit-binding-energies}, \Fref{fig:bar-graphs-showing-bes} and \Fref{fig:scatter-plots-of-bes}).

\par\Needspace*{200pt}\vspace{16pt}
\captionof{table}{CBS-limit binding energies (CBS B.E., kJ/mol) for F-bonded complexes DF\textperiodcentered{}\textperiodcentered{}\textperiodcentered{}CH\textsubscript{3}/NH\textsubscript{3}, computed at DF-RMP2-F12/{[}CBSV(T,Q)Z-F12{]}//DF-RMP2-F12/VDZ-F12. These serve as reference BEs throughout this work. Also, BEs from CCSD(T)/CBS//MP2/aug-cc-pVDZ via the Hobza method (see text and Eq. \ref*{eq:ccsdt-cbs}); W1-F12//DF-RMP2-F12/VDZ-F12; and W1-F12//M06-2X/aug-cc-pVTZ are included. BSSE-corrected BEs (\ensuremath{\Delta}\textsubscript{1+BSSE}) from DF-RMP2-F12/VDZ-F12, MP2/AVDZ, and M06-2X are given along with BSSE-uncorrected BEs (\ensuremath{\Delta}\textsubscript{1}) at M06-2X/AVTZ. Statistical errors on BEs for all complexes (MSE, MAD, RMSD, MAX, and MA\%D) are also reported.\textsuperscript{\#}}\label{tab:cbs-limit-binding-energies}\vspace{7pt}\nopagebreak
{\fontsize{9.5}{10.7}\fontfamily{ptm}\selectfont\renewcommand{\arraystretch}{1.0}\setlength{\tabcolsep}{1.5pt}
\setlength{\LTleft}{-7.3pt}\setlength{\LTright}{-7.3pt}\setlength{\LTpre}{0pt}\setlength{\LTpost}{0pt}
\begin{longtable}[x]{|>{\centering\arraybackslash}m{96.4pt}|>{\centering\arraybackslash}m{35.9pt}|>{\centering\arraybackslash}m{8.0pt}|>{\centering\arraybackslash}m{65.8pt}|>{\centering\arraybackslash}m{25.8pt}|>{\centering\arraybackslash}m{8.0pt}|>{\centering\arraybackslash}m{27.0pt}|>{\centering\arraybackslash}m{8.0pt}|>{\centering\arraybackslash}m{57.8pt}|>{\centering\arraybackslash}m{8.0pt}|>{\centering\arraybackslash}m{29.8pt}|>{\centering\arraybackslash}m{8.0pt}|>{\centering\arraybackslash}m{30.2pt}|>{\centering\arraybackslash}m{30.2pt}|}
\specialrule{0.9pt}{0pt}{0pt}
\multicolumn{14}{|>{\centering\arraybackslash}p{477.7pt}|}{\textbf{DF\textperiodcentered{}\textperiodcentered{}\textperiodcentered{}CH}\textbf{\textsubscript{3}}} \\* \hline\noalign{\penalty10000}
\multicolumn{5}{|>{\centering\arraybackslash}m{243.9pt}|}{Optimization @DF-RMP2-F12/\allowbreak{}VDZ-F12} & \multicolumn{5}{>{\centering\arraybackslash}m{120.7pt}|}{Optimization @MP2/\allowbreak{}AVDZ} & \multicolumn{4}{>{\centering\arraybackslash}m{107.1pt}|}{Optimization @M06-2X/\allowbreak{}AVTZ} \\* \hline\noalign{\penalty10000}
\multicolumn{1}{|>{\centering\arraybackslash}m{96.4pt}|}{\multirow{2}{96.4pt}{\centering DF\textperiodcentered{}\textperiodcentered{}\textperiodcentered{}CH\textsubscript{3}}} & \multicolumn{2}{>{\centering\arraybackslash}m{46.9pt}|}{VDZ- F12} & \multicolumn{2}{>{\centering\arraybackslash}m{94.6pt}|}{CBS BE} & \multicolumn{2}{>{\centering\arraybackslash}m{38.0pt}|}{AVDZ} & \multicolumn{3}{>{\centering\arraybackslash}m{79.8pt}|}{CCSD(T)/\allowbreak{}CBS BE} & \multirow{2}{29.8pt}{\centering \ensuremath{\Delta}\textsubscript{1}} & \multicolumn{2}{>{\centering\arraybackslash}m{41.2pt}|}{\multirow{2}{41.2pt}{\centering \ensuremath{\Delta}\textsubscript{1+} \textsubscript{BSSE}}} & \multirow{2}{30.2pt}{\centering W1- F12} \\* \cline{2-10}\noalign{\penalty10000}
\multicolumn{1}{|>{\centering\arraybackslash}m{96.4pt}|}{} & \multicolumn{2}{>{\centering\arraybackslash}m{46.9pt}|}{\ensuremath{\Delta}\textsubscript{1+} \textsubscript{BSSE}} & V(T,Q)Z- F12 (\textsuperscript{§}Reference) & W1-F12 & \multicolumn{2}{>{\centering\arraybackslash}m{38.0pt}|}{\ensuremath{\Delta}\textsubscript{1+} \textsubscript{BSSE}} & \multicolumn{3}{>{\centering\arraybackslash}m{79.8pt}|}{AV(T,Q)Z} &  & \multicolumn{2}{>{\centering\arraybackslash}m{41.2pt}|}{} &  \\* \hline\noalign{\penalty10000}
\endfirsthead
\multicolumn{14}{@{}l@{}}{\mbox{\textbf{\tablename~\ref*{tab:cbs-limit-binding-energies}.}} (Continued.)} \\[4pt]
\hline
\multicolumn{14}{|>{\centering\arraybackslash}p{477.7pt}|}{\textbf{DF\textperiodcentered{}\textperiodcentered{}\textperiodcentered{}CH}\textbf{\textsubscript{3}}} \\* \hline\noalign{\penalty10000}
\multicolumn{5}{|>{\centering\arraybackslash}m{243.9pt}|}{Optimization @DF-RMP2-F12/\allowbreak{}VDZ-F12} & \multicolumn{5}{>{\centering\arraybackslash}m{120.7pt}|}{Optimization @MP2/\allowbreak{}AVDZ} & \multicolumn{4}{>{\centering\arraybackslash}m{107.1pt}|}{Optimization @M06-2X/\allowbreak{}AVTZ} \\* \hline\noalign{\penalty10000}
\multicolumn{1}{|>{\centering\arraybackslash}m{96.4pt}|}{\multirow{2}{96.4pt}{\centering DF\textperiodcentered{}\textperiodcentered{}\textperiodcentered{}CH\textsubscript{3}}} & \multicolumn{2}{>{\centering\arraybackslash}m{46.9pt}|}{VDZ- F12} & \multicolumn{2}{>{\centering\arraybackslash}m{94.6pt}|}{CBS BE} & \multicolumn{2}{>{\centering\arraybackslash}m{38.0pt}|}{AVDZ} & \multicolumn{3}{>{\centering\arraybackslash}m{79.8pt}|}{CCSD(T)/\allowbreak{}CBS BE} & \multirow{2}{29.8pt}{\centering \ensuremath{\Delta}\textsubscript{1}} & \multicolumn{2}{>{\centering\arraybackslash}m{41.2pt}|}{\multirow{2}{41.2pt}{\centering \ensuremath{\Delta}\textsubscript{1+} \textsubscript{BSSE}}} & \multirow{2}{30.2pt}{\centering W1- F12} \\* \cline{2-10}\noalign{\penalty10000}
\multicolumn{1}{|>{\centering\arraybackslash}m{96.4pt}|}{} & \multicolumn{2}{>{\centering\arraybackslash}m{46.9pt}|}{\ensuremath{\Delta}\textsubscript{1+} \textsubscript{BSSE}} & V(T,Q)Z- F12 (\textsuperscript{§}Reference) & W1-F12 & \multicolumn{2}{>{\centering\arraybackslash}m{38.0pt}|}{\ensuremath{\Delta}\textsubscript{1+} \textsubscript{BSSE}} & \multicolumn{3}{>{\centering\arraybackslash}m{79.8pt}|}{AV(T,Q)Z} &  & \multicolumn{2}{>{\centering\arraybackslash}m{41.2pt}|}{} &  \\* \hline\noalign{\penalty10000}
\endhead
\multicolumn{1}{|>{\centering\arraybackslash}m{96.4pt}|}{1) F\textsubscript{2}\textperiodcentered{}\textperiodcentered{}\textperiodcentered{}CH\textsubscript{3}} & \multicolumn{2}{>{\centering\arraybackslash}m{46.9pt}|}{-3.0} & -3.4 & -3.1 & \multicolumn{2}{>{\centering\arraybackslash}m{38.0pt}|}{-2.4} & \multicolumn{3}{>{\centering\arraybackslash}m{79.8pt}|}{-3.5} & -7.9 & \multicolumn{2}{>{\centering\arraybackslash}m{41.2pt}|}{-7.4} & -3.2 \\* \hline\noalign{\penalty10000}
\multicolumn{1}{|>{\centering\arraybackslash}m{96.4pt}|}{2) SF\textsubscript{6}\textperiodcentered{}\textperiodcentered{}\textperiodcentered{}CH\textsubscript{3}} & \multicolumn{2}{>{\centering\arraybackslash}m{46.9pt}|}{-1.5} & -1.8 & N/\allowbreak{}A & \multicolumn{2}{>{\centering\arraybackslash}m{38.0pt}|}{-1.1} & \multicolumn{3}{>{\centering\arraybackslash}m{79.8pt}|}{-2.6} & -2.5 & \multicolumn{2}{>{\centering\arraybackslash}m{41.2pt}|}{-1.9} & N/\allowbreak{}A \\* \hline\noalign{\penalty10000}
\multicolumn{1}{|>{\centering\arraybackslash}m{96.4pt}|}{3) NCF\textperiodcentered{}\textperiodcentered{}\textperiodcentered{}CH\textsubscript{3}} & \multicolumn{2}{>{\centering\arraybackslash}m{46.9pt}|}{-1.5} & -1.7 & -1.7 & \multicolumn{2}{>{\centering\arraybackslash}m{38.0pt}|}{-1.2} & \multicolumn{3}{>{\centering\arraybackslash}m{79.8pt}|}{-2.0} & -3.1 & \multicolumn{2}{>{\centering\arraybackslash}m{41.2pt}|}{-2.6} & -0.8 \\* \hline\noalign{\penalty10000}
\multicolumn{1}{|>{\centering\arraybackslash}m{96.4pt}|}{4) CNF\textperiodcentered{}\textperiodcentered{}\textperiodcentered{}CH\textsubscript{3}} & \multicolumn{2}{>{\centering\arraybackslash}m{46.9pt}|}{-2.1} & -2.4 & -2.2 & \multicolumn{2}{>{\centering\arraybackslash}m{38.0pt}|}{-1.7} & \multicolumn{3}{>{\centering\arraybackslash}m{79.8pt}|}{-2.5} & -4.2 & \multicolumn{2}{>{\centering\arraybackslash}m{41.2pt}|}{-3.6} & -1.7 \\ \hline
\multicolumn{1}{|>{\centering\arraybackslash}m{96.4pt}|}{5) NF\textsubscript{3}\textperiodcentered{}\textperiodcentered{}\textperiodcentered{}CH\textsubscript{3}} & \multicolumn{2}{>{\centering\arraybackslash}m{46.9pt}|}{-1.4} & -1.7 & -1.7 & \multicolumn{2}{>{\centering\arraybackslash}m{38.0pt}|}{-1.0} & \multicolumn{3}{>{\centering\arraybackslash}m{79.8pt}|}{-2.1} & -2.7 & \multicolumn{2}{>{\centering\arraybackslash}m{41.2pt}|}{-2.3} & -0.4 \\ \hline
\multicolumn{1}{|>{\centering\arraybackslash}m{96.4pt}|}{6) ClF\textsubscript{3}\textperiodcentered{}\textperiodcentered{}\textperiodcentered{}CH\textsubscript{3}} & \multicolumn{2}{>{\centering\arraybackslash}m{46.9pt}|}{-2.1} & -2.4 & -2.4 & \multicolumn{2}{>{\centering\arraybackslash}m{38.0pt}|}{-1.9} & \multicolumn{3}{>{\centering\arraybackslash}m{79.8pt}|}{-2.9} & -5.2 & \multicolumn{2}{>{\centering\arraybackslash}m{41.2pt}|}{-4.7} & -1.5 \\ \hline
\multicolumn{1}{|>{\centering\arraybackslash}m{96.4pt}|}{7) OF\textsubscript{2}\textperiodcentered{}\textperiodcentered{}\textperiodcentered{}CH\textsubscript{3}} & \multicolumn{2}{>{\centering\arraybackslash}m{46.9pt}|}{-1.7} & -2.0 & -1.9 & \multicolumn{2}{>{\centering\arraybackslash}m{38.0pt}|}{-1.2} & \multicolumn{3}{>{\centering\arraybackslash}m{79.8pt}|}{-2.2} & -4.4 & \multicolumn{2}{>{\centering\arraybackslash}m{41.2pt}|}{-4.0} & -1.5 \\ \hline
\multicolumn{1}{|>{\centering\arraybackslash}m{96.4pt}|}{8) NF\textsubscript{5}\textperiodcentered{}\textperiodcentered{}\textperiodcentered{}CH\textsubscript{3}} & \multicolumn{2}{>{\centering\arraybackslash}m{46.9pt}|}{-2.7} & -3.1 & N/\allowbreak{}A & \multicolumn{2}{>{\centering\arraybackslash}m{38.0pt}|}{-2.1} & \multicolumn{3}{>{\centering\arraybackslash}m{79.8pt}|}{-3.0} & -5.1 & \multicolumn{2}{>{\centering\arraybackslash}m{41.2pt}|}{-4.6} & N/\allowbreak{}A \\ \hline
\multicolumn{1}{|>{\centering\arraybackslash}m{96.4pt}|}{9) FCCF\textperiodcentered{}\textperiodcentered{}\textperiodcentered{}CH\textsubscript{3}} & \multicolumn{2}{>{\centering\arraybackslash}m{46.9pt}|}{-1.3} & -1.5 & -1.5 & \multicolumn{2}{>{\centering\arraybackslash}m{38.0pt}|}{-0.9} & \multicolumn{3}{>{\centering\arraybackslash}m{79.8pt}|}{-2.2} & -2.1 & \multicolumn{2}{>{\centering\arraybackslash}m{41.2pt}|}{-1.6} & 0.2 \\ \hline
\multicolumn{1}{|>{\centering\arraybackslash}m{96.4pt}|}{10) CF\textsubscript{3}CCF\textperiodcentered{}\textperiodcentered{}\textperiodcentered{}CH\textsubscript{3}} & \multicolumn{2}{>{\centering\arraybackslash}m{46.9pt}|}{-1.3} & -1.5 & -1.4 & \multicolumn{2}{>{\centering\arraybackslash}m{38.0pt}|}{-0.9} & \multicolumn{3}{>{\centering\arraybackslash}m{79.8pt}|}{-1.8} & -2.7 & \multicolumn{2}{>{\centering\arraybackslash}m{41.2pt}|}{-2.3} & N/\allowbreak{}A \\ \hline
\multicolumn{1}{|>{\centering\arraybackslash}m{96.4pt}|}{11) CF\textsubscript{3}OF\textperiodcentered{}\textperiodcentered{}\textperiodcentered{}CH\textsubscript{3}} & \multicolumn{2}{>{\centering\arraybackslash}m{46.9pt}|}{-2.1} & -2.5 & N/\allowbreak{}A & \multicolumn{2}{>{\centering\arraybackslash}m{38.0pt}|}{b} & \multicolumn{3}{>{\centering\arraybackslash}m{79.8pt}|}{b} & -4.7 & \multicolumn{2}{>{\centering\arraybackslash}m{41.2pt}|}{-4.2} & -2.2 \\ \hline
\multicolumn{1}{|>{\centering\arraybackslash}m{96.4pt}|}{12) F\textsubscript{3}NO\textperiodcentered{}\textperiodcentered{}\textperiodcentered{}CH\textsubscript{3}} & \multicolumn{2}{>{\centering\arraybackslash}m{46.9pt}|}{-1.8} & -2.1 & -2.1 & \multicolumn{2}{>{\centering\arraybackslash}m{38.0pt}|}{-1.3} & \multicolumn{3}{>{\centering\arraybackslash}m{79.8pt}|}{-2.2} & -3.6 & \multicolumn{2}{>{\centering\arraybackslash}m{41.2pt}|}{-3.1} & -1.3 \\ \hline
\multicolumn{1}{|>{\centering\arraybackslash}m{96.4pt}|}{13) SF\textsubscript{5}\textperiodcentered{}\textperiodcentered{}\textperiodcentered{}CH\textsubscript{3}} & \multicolumn{2}{>{\centering\arraybackslash}m{46.9pt}|}{-1.4} & -1.8 & N/\allowbreak{}A & \multicolumn{2}{>{\centering\arraybackslash}m{38.0pt}|}{-1.1} & \multicolumn{3}{>{\centering\arraybackslash}m{79.8pt}|}{-2.5} & -3.0 & \multicolumn{2}{>{\centering\arraybackslash}m{41.2pt}|}{-2.6} & N/\allowbreak{}A \\ \hline
\multicolumn{1}{|>{\centering\arraybackslash}m{96.4pt}|}{14) NH\textsubscript{3}F+\textperiodcentered{}\textperiodcentered{}\textperiodcentered{}CH\textsubscript{3}} & \multicolumn{2}{>{\centering\arraybackslash}m{46.9pt}|}{-9.5} & -9.8 & -10.2 & \multicolumn{2}{>{\centering\arraybackslash}m{38.0pt}|}{-8.9} & \multicolumn{3}{>{\centering\arraybackslash}m{79.8pt}|}{-10.3} & -15.5 & \multicolumn{2}{>{\centering\arraybackslash}m{41.2pt}|}{-15.0} & -9.5 \\ \hline
\multicolumn{1}{|>{\centering\arraybackslash}m{96.4pt}|}{15) NH\textsubscript{2}F\textsubscript{2}\textsuperscript{+}\textperiodcentered{}\textperiodcentered{}\textperiodcentered{}CH\textsubscript{3}} & \multicolumn{2}{>{\centering\arraybackslash}m{46.9pt}|}{-11.0} & -11.4 & -11.7 & \multicolumn{2}{>{\centering\arraybackslash}m{38.0pt}|}{-10.5} & \multicolumn{3}{>{\centering\arraybackslash}m{79.8pt}|}{-11.8} & -16.5 & \multicolumn{2}{>{\centering\arraybackslash}m{41.2pt}|}{-15.9} & -11.0 \\ \hline
\multicolumn{1}{|>{\centering\arraybackslash}m{96.4pt}|}{\textbf{MSE}} & \multicolumn{2}{>{\centering\arraybackslash}m{46.9pt}|}{\textbf{0.3}} & \textbf{0.0} & \textbf{0.0} & \multicolumn{2}{>{\centering\arraybackslash}m{38.0pt}|}{\textbf{0.7}} & \multicolumn{3}{>{\centering\arraybackslash}m{79.8pt}|}{\textbf{-0.4}} & \textbf{-2.3} & \multicolumn{2}{>{\centering\arraybackslash}m{41.2pt}|}{\textbf{-1.8}} & \textbf{0.7} \\ \hline
\multicolumn{1}{|>{\centering\arraybackslash}m{96.4pt}|}{\textbf{MAD}} & \multicolumn{2}{>{\centering\arraybackslash}m{46.9pt}|}{\textbf{0.3}} & \textbf{0.0} & \textbf{0.1} & \multicolumn{2}{>{\centering\arraybackslash}m{38.0pt}|}{\textbf{0.7}} & \multicolumn{3}{>{\centering\arraybackslash}m{79.8pt}|}{\textbf{0.4}} & \textbf{2.3} & \multicolumn{2}{>{\centering\arraybackslash}m{41.2pt}|}{\textbf{1.8}} & \textbf{0.7} \\ \hline
\multicolumn{1}{|>{\centering\arraybackslash}m{96.4pt}|}{\textbf{RMSD}} & \multicolumn{2}{>{\centering\arraybackslash}m{46.9pt}|}{\textbf{0.3}} & \textbf{0.0} & \textbf{0.2} & \multicolumn{2}{>{\centering\arraybackslash}m{38.0pt}|}{\textbf{0.8}} & \multicolumn{3}{>{\centering\arraybackslash}m{79.8pt}|}{\textbf{0.4}} & \textbf{2.7} & \multicolumn{2}{>{\centering\arraybackslash}m{41.2pt}|}{\textbf{2.3}} & \textbf{0.9} \\ \hline
\multicolumn{1}{|>{\centering\arraybackslash}m{96.4pt}|}{\textbf{MAX}} & \multicolumn{2}{>{\centering\arraybackslash}m{46.9pt}|}{\textbf{0.4}} & \textbf{0.0} & \textbf{-0.4} & \multicolumn{2}{>{\centering\arraybackslash}m{38.0pt}|}{\textbf{1.0}} & \multicolumn{3}{>{\centering\arraybackslash}m{79.8pt}|}{\textbf{-0.8}} & \textbf{-5.7} & \multicolumn{2}{>{\centering\arraybackslash}m{41.2pt}|}{\textbf{-5.2}} & \textbf{1.3} \\ \hline
\multicolumn{1}{|>{\centering\arraybackslash}m{96.4pt}|}{\textbf{MA\%D}} & \multicolumn{2}{>{\centering\arraybackslash}m{46.9pt}|}{\textbf{13.1}} & \textbf{0.0} & \textbf{3.2} & \multicolumn{2}{>{\centering\arraybackslash}m{38.0pt}|}{\textbf{31.1}} & \multicolumn{3}{>{\centering\arraybackslash}m{79.8pt}|}{\textbf{17.6}} & \textbf{75.8} & \multicolumn{2}{>{\centering\arraybackslash}m{41.2pt}|}{\textbf{54.6}} & \textbf{36.1} \\ \specialrule{0.9pt}{0pt}{0pt}
\end{longtable}
\par\nopagebreak\Needspace*{110pt}
\begin{longtable}[x]{|>{\centering\arraybackslash}m{96.4pt}|>{\centering\arraybackslash}m{35.9pt}|>{\centering\arraybackslash}m{8.0pt}|>{\centering\arraybackslash}m{65.8pt}|>{\centering\arraybackslash}m{25.8pt}|>{\centering\arraybackslash}m{8.0pt}|>{\centering\arraybackslash}m{27.0pt}|>{\centering\arraybackslash}m{8.0pt}|>{\centering\arraybackslash}m{57.8pt}|>{\centering\arraybackslash}m{8.0pt}|>{\centering\arraybackslash}m{29.8pt}|>{\centering\arraybackslash}m{8.0pt}|>{\centering\arraybackslash}m{30.2pt}|>{\centering\arraybackslash}m{30.2pt}|}
\multicolumn{14}{|>{\centering\arraybackslash}m{477.7pt}|}{\textbf{DF\textperiodcentered{}\textperiodcentered{}\textperiodcentered{}NH}\textbf{\textsubscript{3}}} \\* \hline\noalign{\penalty10000}
\multicolumn{6}{|>{\centering\arraybackslash}m{254.9pt}|}{Optimization @DF-RMP2-F12/\allowbreak{}VDZ-F12} & \multicolumn{3}{>{\centering\arraybackslash}m{98.7pt}|}{Optimization @MP2/\allowbreak{}AVDZ} & \multicolumn{5}{>{\centering\arraybackslash}m{118.1pt}|}{Optimization @M06-2X/\allowbreak{}AVTZ} \\* \hline\noalign{\penalty10000}
\multicolumn{1}{|>{\centering\arraybackslash}m{96.4pt}|}{\multirow{2}{96.4pt}{\centering DF\textperiodcentered{}\textperiodcentered{}\textperiodcentered{}NH\textsubscript{3}}} & VDZ- F12 & \multicolumn{4}{>{\centering\arraybackslash}m{116.6pt}|}{CBS BE} & \multicolumn{2}{>{\centering\arraybackslash}m{38.0pt}|}{AVDZ} & CCSD(T)/ CBS BE & \multicolumn{3}{>{\centering\arraybackslash}m{51.8pt}|}{\multirow{2}{51.8pt}{\centering \ensuremath{\Delta}\textsubscript{1}}} & \multirow{2}{30.2pt}{\centering \ensuremath{\Delta}\textsubscript{1+} \textsubscript{BSSE}} & \multirow{2}{30.2pt}{\centering W1- F12} \\* \cline{2-9}\noalign{\penalty10000}
\multicolumn{1}{|>{\centering\arraybackslash}m{96.4pt}|}{} & \ensuremath{\Delta}\textsubscript{1+} \textsubscript{BSSE} & \multicolumn{2}{>{\centering\arraybackslash}m{76.8pt}|}{V(T,Q)Z- F12 (\textsuperscript{§}Reference)} & \multicolumn{2}{>{\centering\arraybackslash}m{36.8pt}|}{W1-F12} & \multicolumn{2}{>{\centering\arraybackslash}m{38.0pt}|}{\ensuremath{\Delta}\textsubscript{1+} \textsubscript{BSSE}} & AV(T,Q)Z & \multicolumn{3}{>{\centering\arraybackslash}m{51.8pt}|}{} &  &  \\* \hline\noalign{\penalty10000}
\endfirsthead
\multicolumn{14}{@{}l@{}}{\mbox{\textbf{\tablename~\ref*{tab:cbs-limit-binding-energies}.}} (Continued.)} \\[4pt]
\hline
\multicolumn{14}{|>{\centering\arraybackslash}m{477.7pt}|}{\textbf{DF\textperiodcentered{}\textperiodcentered{}\textperiodcentered{}NH}\textbf{\textsubscript{3}}} \\* \hline\noalign{\penalty10000}
\multicolumn{6}{|>{\centering\arraybackslash}m{254.9pt}|}{Optimization @DF-RMP2-F12/\allowbreak{}VDZ-F12} & \multicolumn{3}{>{\centering\arraybackslash}m{98.7pt}|}{Optimization @MP2/\allowbreak{}AVDZ} & \multicolumn{5}{>{\centering\arraybackslash}m{118.1pt}|}{Optimization @M06-2X/\allowbreak{}AVTZ} \\* \hline\noalign{\penalty10000}
\multicolumn{1}{|>{\centering\arraybackslash}m{96.4pt}|}{\multirow{2}{96.4pt}{\centering DF\textperiodcentered{}\textperiodcentered{}\textperiodcentered{}NH\textsubscript{3}}} & VDZ- F12 & \multicolumn{4}{>{\centering\arraybackslash}m{116.6pt}|}{CBS BE} & \multicolumn{2}{>{\centering\arraybackslash}m{38.0pt}|}{AVDZ} & CCSD(T)/ CBS BE & \multicolumn{3}{>{\centering\arraybackslash}m{51.8pt}|}{\multirow{2}{51.8pt}{\centering \ensuremath{\Delta}\textsubscript{1}}} & \multirow{2}{30.2pt}{\centering \ensuremath{\Delta}\textsubscript{1+} \textsubscript{BSSE}} & \multirow{2}{30.2pt}{\centering W1- F12} \\* \cline{2-9}\noalign{\penalty10000}
\multicolumn{1}{|>{\centering\arraybackslash}m{96.4pt}|}{} & \ensuremath{\Delta}\textsubscript{1+} \textsubscript{BSSE} & \multicolumn{2}{>{\centering\arraybackslash}m{76.8pt}|}{V(T,Q)Z- F12 (\textsuperscript{§}Reference)} & \multicolumn{2}{>{\centering\arraybackslash}m{36.8pt}|}{W1-F12} & \multicolumn{2}{>{\centering\arraybackslash}m{38.0pt}|}{\ensuremath{\Delta}\textsubscript{1+} \textsubscript{BSSE}} & AV(T,Q)Z & \multicolumn{3}{>{\centering\arraybackslash}m{51.8pt}|}{} &  &  \\* \hline\noalign{\penalty10000}
\endhead
\multicolumn{1}{|>{\centering\arraybackslash}m{96.4pt}|}{1) F\textsubscript{2}\textperiodcentered{}\textperiodcentered{}\textperiodcentered{}NH\textsubscript{3}} & -7.6 & \multicolumn{2}{>{\centering\arraybackslash}m{76.8pt}|}{-7.8} & \multicolumn{2}{>{\centering\arraybackslash}m{36.8pt}|}{-7.4} & \multicolumn{2}{>{\centering\arraybackslash}m{38.0pt}|}{-8.0} & -7.2 & \multicolumn{3}{>{\centering\arraybackslash}m{51.8pt}|}{-6.4} & -6.1 & -7.3 \\* \hline\noalign{\penalty10000}
\multicolumn{1}{|>{\centering\arraybackslash}m{96.4pt}|}{2) SF\textsubscript{6}\textperiodcentered{}\textperiodcentered{}\textperiodcentered{}NH\textsubscript{3}} & -1.4 & \multicolumn{2}{>{\centering\arraybackslash}m{76.8pt}|}{-1.4} & \multicolumn{2}{>{\centering\arraybackslash}m{36.8pt}|}{-1.7} & \multicolumn{2}{>{\centering\arraybackslash}m{38.0pt}|}{-1.3} & -1.9 & \multicolumn{3}{>{\centering\arraybackslash}m{51.8pt}|}{-1.4} & -1.3 & -1.6 \\* \hline\noalign{\penalty10000}
\multicolumn{1}{|>{\centering\arraybackslash}m{96.4pt}|}{3) NCF\textperiodcentered{}\textperiodcentered{}\textperiodcentered{}NH\textsubscript{3}} & -5.0 & \multicolumn{2}{>{\centering\arraybackslash}m{76.8pt}|}{-5.1} & \multicolumn{2}{>{\centering\arraybackslash}m{36.8pt}|}{-5.2} & \multicolumn{2}{>{\centering\arraybackslash}m{38.0pt}|}{-4.4} & -5.5 & \multicolumn{3}{>{\centering\arraybackslash}m{51.8pt}|}{-4.7} & -4.5 & -5.2 \\* \hline\noalign{\penalty10000}
\multicolumn{1}{|>{\centering\arraybackslash}m{96.4pt}|}{4) CNF\textperiodcentered{}\textperiodcentered{}\textperiodcentered{}NH\textsubscript{3}} & -6.7 & \multicolumn{2}{>{\centering\arraybackslash}m{76.8pt}|}{-6.9} & \multicolumn{2}{>{\centering\arraybackslash}m{36.8pt}|}{-6.4} & \multicolumn{2}{>{\centering\arraybackslash}m{38.0pt}|}{-6.1} & -6.7 & \multicolumn{3}{>{\centering\arraybackslash}m{51.8pt}|}{-5.4} & -5.1 & -6.4 \\ \hline
\multicolumn{1}{|>{\centering\arraybackslash}m{96.4pt}|}{5) NF\textsubscript{3}\textperiodcentered{}\textperiodcentered{}\textperiodcentered{}NH\textsubscript{3}} & -1.7 & \multicolumn{2}{>{\centering\arraybackslash}m{76.8pt}|}{-1.9} & \multicolumn{2}{>{\centering\arraybackslash}m{36.8pt}|}{-2.0} & \multicolumn{2}{>{\centering\arraybackslash}m{38.0pt}|}{b} & b & \multicolumn{3}{>{\centering\arraybackslash}m{51.8pt}|}{-1.2} & -1.0 & -1.6 \\ \hline
\multicolumn{1}{|>{\centering\arraybackslash}m{96.4pt}|}{6) ClF\textsubscript{3}\textperiodcentered{}\textperiodcentered{}\textperiodcentered{}NH\textsubscript{3}} & -4.1 & \multicolumn{2}{>{\centering\arraybackslash}m{76.8pt}|}{-4.5} & \multicolumn{2}{>{\centering\arraybackslash}m{36.8pt}|}{-4.5} & \multicolumn{2}{>{\centering\arraybackslash}m{38.0pt}|}{-4.5} & -4.9 & \multicolumn{3}{>{\centering\arraybackslash}m{51.8pt}|}{-3.5} & -3.2 & -4.5 \\ \hline
\multicolumn{1}{|>{\centering\arraybackslash}m{96.4pt}|}{7) OF\textsubscript{2}\textperiodcentered{}\textperiodcentered{}\textperiodcentered{}NH\textsubscript{3}} & -3.1 & \multicolumn{2}{>{\centering\arraybackslash}m{76.8pt}|}{-3.4} & \multicolumn{2}{>{\centering\arraybackslash}m{36.8pt}|}{-3.3} & \multicolumn{2}{>{\centering\arraybackslash}m{38.0pt}|}{-2.9} & -3.6 & \multicolumn{3}{>{\centering\arraybackslash}m{51.8pt}|}{-2.8} & -2.4 & -3.3 \\ \hline
\multicolumn{1}{|>{\centering\arraybackslash}m{96.4pt}|}{8) NF\textsubscript{5}\textperiodcentered{}\textperiodcentered{}\textperiodcentered{}NH\textsubscript{3}} & -7.1 & \multicolumn{2}{>{\centering\arraybackslash}m{76.8pt}|}{-7.3} & \multicolumn{2}{>{\centering\arraybackslash}m{36.8pt}|}{-7.8} & \multicolumn{2}{>{\centering\arraybackslash}m{38.0pt}|}{-6.7} & -8.2 & \multicolumn{3}{>{\centering\arraybackslash}m{51.8pt}|}{-6.6} & -6.2 & -7.7 \\ \hline
\multicolumn{1}{|>{\centering\arraybackslash}m{96.4pt}|}{9) FCCF\textperiodcentered{}\textperiodcentered{}\textperiodcentered{}NH\textsubscript{3}} & -0.6 & \multicolumn{2}{>{\centering\arraybackslash}m{76.8pt}|}{-0.6} & \multicolumn{2}{>{\centering\arraybackslash}m{36.8pt}|}{-0.8} & \multicolumn{2}{>{\centering\arraybackslash}m{38.0pt}|}{-0.2} & -1.2 & \multicolumn{3}{>{\centering\arraybackslash}m{51.8pt}|}{-0.6} & -0.5 & -0.7 \\ \hline
\multicolumn{1}{|>{\centering\arraybackslash}m{96.4pt}|}{10) CF\textsubscript{3}CCF\textperiodcentered{}\textperiodcentered{}\textperiodcentered{}NH\textsubscript{3}} & -3.6 & \multicolumn{2}{>{\centering\arraybackslash}m{76.8pt}|}{-3.6} & \multicolumn{2}{>{\centering\arraybackslash}m{36.8pt}|}{-3.7} & \multicolumn{2}{>{\centering\arraybackslash}m{38.0pt}|}{-3.2} & -3.9 & \multicolumn{3}{>{\centering\arraybackslash}m{51.8pt}|}{-3.4} & -3.2 & -3.5 \\ \hline
\multicolumn{1}{|>{\centering\arraybackslash}m{96.4pt}|}{11) CF\textsubscript{3}OF\textperiodcentered{}\textperiodcentered{}\textperiodcentered{}NH\textsubscript{3}} & -4.1 & \multicolumn{2}{>{\centering\arraybackslash}m{76.8pt}|}{-4.3} & \multicolumn{2}{>{\centering\arraybackslash}m{36.8pt}|}{-4.1} & \multicolumn{2}{>{\centering\arraybackslash}m{38.0pt}|}{-4.0} & -4.2 & \multicolumn{3}{>{\centering\arraybackslash}m{51.8pt}|}{-2.3} & -2.1 & -3.9 \\ \hline
\multicolumn{1}{|>{\centering\arraybackslash}m{96.4pt}|}{12) F\textsubscript{3}NO\textperiodcentered{}\textperiodcentered{}\textperiodcentered{}NH\textsubscript{3}} & b & \multicolumn{2}{>{\centering\arraybackslash}m{76.8pt}|}{b} & \multicolumn{2}{>{\centering\arraybackslash}m{36.8pt}|}{b} & \multicolumn{2}{>{\centering\arraybackslash}m{38.0pt}|}{-2.5} & -3.3 & \multicolumn{3}{>{\centering\arraybackslash}m{51.8pt}|}{-2.3} & -2.0 & -3.0 \\ \hline
\multicolumn{1}{|>{\centering\arraybackslash}m{96.4pt}|}{13) SF\textsubscript{5}\textperiodcentered{}\textperiodcentered{}\textperiodcentered{}NH\textsubscript{3}} & -2.6 & \multicolumn{2}{>{\centering\arraybackslash}m{76.8pt}|}{-2.7} & \multicolumn{2}{>{\centering\arraybackslash}m{36.8pt}|}{N/\allowbreak{}A} & \multicolumn{2}{>{\centering\arraybackslash}m{38.0pt}|}{-2.4} & -3.6 & \multicolumn{3}{>{\centering\arraybackslash}m{51.8pt}|}{-2.0} & -1.8 & N/\allowbreak{}A \\ \hline
\multicolumn{1}{|>{\centering\arraybackslash}m{96.4pt}|}{14) NH\textsubscript{3}F+\textperiodcentered{}\textperiodcentered{}\textperiodcentered{}NH\textsubscript{3}} & -33.9 & \multicolumn{2}{>{\centering\arraybackslash}m{76.8pt}|}{-33.9} & \multicolumn{2}{>{\centering\arraybackslash}m{36.8pt}|}{-34.6} & \multicolumn{2}{>{\centering\arraybackslash}m{38.0pt}|}{-33.3} & -34.8 & \multicolumn{3}{>{\centering\arraybackslash}m{51.8pt}|}{-34.3} & -34.0 & -34.5 \\ \hline
\multicolumn{1}{|>{\centering\arraybackslash}m{96.4pt}|}{15) NH\textsubscript{2}F\textsubscript{2}\textsuperscript{+}\textperiodcentered{}\textperiodcentered{}\textperiodcentered{}NH\textsubscript{3}} & -38.7 & \multicolumn{2}{>{\centering\arraybackslash}m{76.8pt}|}{-38.8} & \multicolumn{2}{>{\centering\arraybackslash}m{36.8pt}|}{-39.3} & \multicolumn{2}{>{\centering\arraybackslash}m{38.0pt}|}{-38.3} & -39.5 & \multicolumn{3}{>{\centering\arraybackslash}m{51.8pt}|}{-39.0} & -38.5 & -39.3 \\ \hline
\multicolumn{1}{|>{\centering\arraybackslash}m{96.4pt}|}{\textbf{MSE}} & \textbf{0.1} & \multicolumn{2}{>{\centering\arraybackslash}m{76.8pt}|}{\textbf{0.0}} & \multicolumn{2}{>{\centering\arraybackslash}m{36.8pt}|}{\textbf{-0.1}} & \multicolumn{2}{>{\centering\arraybackslash}m{38.0pt}|}{\textbf{0.4}} & \textbf{-0.4} & \multicolumn{3}{>{\centering\arraybackslash}m{51.8pt}|}{\textbf{0.6}} & \textbf{0.9} & \textbf{0.0} \\ \hline
\multicolumn{1}{|>{\centering\arraybackslash}m{96.4pt}|}{\textbf{MAD}} & \textbf{0.1} & \multicolumn{2}{>{\centering\arraybackslash}m{76.8pt}|}{\textbf{0.0}} & \multicolumn{2}{>{\centering\arraybackslash}m{36.8pt}|}{\textbf{0.3}} & \multicolumn{2}{>{\centering\arraybackslash}m{38.0pt}|}{\textbf{0.4}} & \textbf{0.5} & \multicolumn{3}{>{\centering\arraybackslash}m{51.8pt}|}{\textbf{0.7}} & \textbf{0.9} & \textbf{0.3} \\ \hline
\multicolumn{1}{|>{\centering\arraybackslash}m{96.4pt}|}{\textbf{RMSD}} & \textbf{0.2} & \multicolumn{2}{>{\centering\arraybackslash}m{76.8pt}|}{\textbf{0.0}} & \multicolumn{2}{>{\centering\arraybackslash}m{36.8pt}|}{\textbf{0.4}} & \multicolumn{2}{>{\centering\arraybackslash}m{38.0pt}|}{\textbf{0.5}} & \textbf{0.6} & \multicolumn{3}{>{\centering\arraybackslash}m{51.8pt}|}{\textbf{0.9}} & \textbf{1.1} & \textbf{0.4} \\* \hline\noalign{\penalty10000}
\multicolumn{1}{|>{\centering\arraybackslash}m{96.4pt}|}{\textbf{MAX}} & \textbf{0.4} & \multicolumn{2}{>{\centering\arraybackslash}m{76.8pt}|}{\textbf{0.0}} & \multicolumn{2}{>{\centering\arraybackslash}m{36.8pt}|}{\textbf{0.5}} & \multicolumn{2}{>{\centering\arraybackslash}m{38.0pt}|}{\textbf{0.8}} & \textbf{-0.9} & \multicolumn{3}{>{\centering\arraybackslash}m{51.8pt}|}{\textbf{2.0}} & \textbf{2.2} & \textbf{0.5} \\* \hline\noalign{\penalty10000}
\multicolumn{1}{|>{\centering\arraybackslash}m{96.4pt}|}{\textbf{MA\%D}} & \textbf{3.4} & \multicolumn{2}{>{\centering\arraybackslash}m{76.8pt}|}{\textbf{0.0}} & \multicolumn{2}{>{\centering\arraybackslash}m{36.8pt}|}{\textbf{7.3}} & \multicolumn{2}{>{\centering\arraybackslash}m{38.0pt}|}{\textbf{11.7}} & \textbf{17.7} & \multicolumn{3}{>{\centering\arraybackslash}m{51.8pt}|}{\textbf{16.0}} & \textbf{21.5} & \textbf{6.6} \\* \specialrule{0.9pt}{0pt}{0pt}\noalign{\penalty10000}
\end{longtable}}
\par\nopagebreak\vspace{1pt}{\setstretch{1}\setlength{\parskip}{1.5pt}\fontsize{9}{10.5}\selectfont\raggedright\noindent \textsuperscript{b}Convergence error during optimization of respective F-bonded complexes.\par \textsuperscript{N/A}W1-F12 calculations failed to converge.\par \textsuperscript{§}Obtained by the Schwenke-type extrapolation technique.\par \textsuperscript{\#} See \mbox{\textbf{Table \ref{tab:complete-basis-set-cbs}}} for BEs without BSSE correction at DF-RMP2-F12/VDZ-F12 and MP2/AVDZ. BEs with and without BSSE correction using DF-RMP2-F12 with various basis sets (VDZ-F12, VTZ-F12, VQZ-F12) are given in \mbox{\textbf{Table \ref{tab:bes-calculated-at-different}}}.\par}\vspace{12pt}

BSSE corrections calculated at M06-2X/AVTZ level are small for both sets (0.1-0.5 kJ/mol for DF\textperiodcentered{}\textperiodcentered{}\textperiodcentered{}NH\textsubscript{3} complexes; 0.4-0.6 kJ/mol for DF\textperiodcentered{}\textperiodcentered{}\textperiodcentered{}CH\textsubscript{3} complexes). Despite this, BSSE correction increases BEs of DF\textperiodcentered{}\textperiodcentered{}\textperiodcentered{}NH\textsubscript{3} complexes further from the reference (MA\%D: 16.0\% to 21.5\%), while reducing error on BEs of SEFB complexes (MA\%D: 75.8\% to 54.6\%) (\Tref{tab:cbs-limit-binding-energies}).

\begin{table}[!htbp]\centering\setlength{\parskip}{0pt}
\caption{Intermolecular distances {[}R(F\textperiodcentered{}\textperiodcentered{}\textperiodcentered{}C/N), Å{]} and F-bond angles (\ensuremath{\angle}DF\textperiodcentered{}\textperiodcentered{}\textperiodcentered{}C/N, \textdegree{}) for F-bonded complexes DF\textperiodcentered{}\textperiodcentered{}\textperiodcentered{}CH\textsubscript{3}/NH\textsubscript{3}, computed at DF-RMP2-F12/VDZ-F12, MP2/aug-cc-pVDZ and M06-2X/aug-cc-pVTZ. DF-RMP2-F12/VDZ-F12 values serve as reference.\textsuperscript{\#}}\label{tab:intermolecular-distances-r-f}\vspace{7pt}\nopagebreak
{\fontsize{9.0}{10.3}\fontfamily{ptm}\selectfont\renewcommand{\arraystretch}{1.17}\setlength{\tabcolsep}{1.5pt}
\centerline{\adjustbox{max width=518pt}{\begin{tabular}{|>{\centering\arraybackslash}m{67.9pt}|>{\centering\arraybackslash}m{31.6pt}|>{\centering\arraybackslash}m{31.6pt}|>{\centering\arraybackslash}m{31.8pt}|>{\centering\arraybackslash}m{27.9pt}|>{\centering\arraybackslash}m{30.9pt}|>{\centering\arraybackslash}m{33.3pt}|>{\centering\arraybackslash}m{67.9pt}|>{\centering\arraybackslash}m{32.8pt}|>{\centering\arraybackslash}m{30.7pt}|>{\centering\arraybackslash}m{30.2pt}|>{\centering\arraybackslash}m{29.3pt}|>{\centering\arraybackslash}m{30.7pt}|>{\centering\arraybackslash}m{29.3pt}|}
\specialrule{0.9pt}{0pt}{0pt}
\multicolumn{7}{|>{\centering\arraybackslash}m{273.0pt}|}{\textbf{DF\textperiodcentered{}\textperiodcentered{}\textperiodcentered{}CH}\textbf{\textsubscript{3}}} & \multicolumn{7}{>{\centering\arraybackslash}m{269.0pt}|}{\textbf{DF\textperiodcentered{}\textperiodcentered{}\textperiodcentered{}NH}\textbf{\textsubscript{3}}} \\* \hline\noalign{\penalty10000}
\multicolumn{1}{|>{\centering\arraybackslash}m{67.9pt}|}{\multirow{2}{67.9pt}{\centering \textbf{DF\textperiodcentered{}\textperiodcentered{}\textperiodcentered{}CH}\textbf{\textsubscript{3}}}} & \multicolumn{2}{>{\centering\arraybackslash}m{66.2pt}|}{\textbf{DF-RMP2-F12/} \textbf{VDZ-F12}} & \multicolumn{2}{>{\centering\arraybackslash}m{62.7pt}|}{\textbf{MP2/\allowbreak{}AVDZ}} & \multicolumn{2}{>{\centering\arraybackslash}m{67.2pt}|}{\textbf{M06-2X/\allowbreak{}AVTZ}} & \multirow{2}{67.9pt}{\centering \textbf{DF\textperiodcentered{}\textperiodcentered{}\textperiodcentered{}NH}\textbf{\textsubscript{3}}} & \multicolumn{2}{>{\centering\arraybackslash}m{66.5pt}|}{\textbf{DF-RMP2-F12/\allowbreak{}VDZ-F12}} & \multicolumn{2}{>{\centering\arraybackslash}m{62.5pt}|}{\textbf{MP2/\allowbreak{}AVDZ}} & \multicolumn{2}{>{\centering\arraybackslash}m{63.0pt}|}{\textbf{M06-2X/} \textbf{AVTZ}} \\* \cline{2-7}\cline{9-14}\noalign{\penalty10000}
\multicolumn{1}{|>{\centering\arraybackslash}m{67.9pt}|}{} & \textbf{R} & \textbf{\ensuremath{\angle}} & \textbf{R} & \textbf{\ensuremath{\angle}} & \textbf{R} & \textbf{\ensuremath{\angle}} &  & \textbf{R} & \textbf{\ensuremath{\angle}} & \textbf{R} & \textbf{\ensuremath{\angle}} & \textbf{R} & \textbf{\ensuremath{\angle}} \\* \hline\noalign{\penalty10000}
\multicolumn{1}{|>{\centering\arraybackslash}m{67.9pt}|}{F\textsubscript{2}\textperiodcentered{}\textperiodcentered{}\textperiodcentered{}CH\textsubscript{3}} & 2.869 & 178.8 & 2.865 & 180.0 & 2.648 & 179.6 & F\textsubscript{2}\textperiodcentered{}\textperiodcentered{}\textperiodcentered{}NH\textsubscript{3} & 2.616 & 180.0 & 2.442 & 180.0 & 2.596 & 178.1 \\* \hline\noalign{\penalty10000}
\multicolumn{1}{|>{\centering\arraybackslash}m{67.9pt}|}{SF\textsubscript{6}\textperiodcentered{}\textperiodcentered{}\textperiodcentered{}CH\textsubscript{3}} & 3.327 & 180.0 & 3.255 & 179.8 & 3.021 & 179.1 & SF\textsubscript{6}\textperiodcentered{}\textperiodcentered{}\textperiodcentered{}NH\textsubscript{3} & 3.345 & 180.0 & 3.166 & 177.2 & 3.634 & 177.0 \\* \hline\noalign{\penalty10000}
\multicolumn{1}{|>{\centering\arraybackslash}m{67.9pt}|}{NCF\textperiodcentered{}\textperiodcentered{}\textperiodcentered{}CH\textsubscript{3}} & 3.428 & 180.0 & 3.388 & 179.9 & 3.018 & 179.9 & NCF\textperiodcentered{}\textperiodcentered{}\textperiodcentered{}NH\textsubscript{3} & 3.160 & 180.0 & 3.141 & 180.0 & 3.205 & 180.0 \\* \hline\noalign{\penalty10000}
\multicolumn{1}{|>{\centering\arraybackslash}m{67.9pt}|}{CNF\textperiodcentered{}\textperiodcentered{}\textperiodcentered{}CH\textsubscript{3}} & 3.262 & 179.6 & 3.185 & 180.0 & 2.976 & 180.0 & CNF\textperiodcentered{}\textperiodcentered{}\textperiodcentered{}NH\textsubscript{3} & 2.960 & 178.7 & 2.924 & 180.0 & 3.031 & 180.0 \\ \hline
\multicolumn{1}{|>{\centering\arraybackslash}m{67.9pt}|}{NF\textsubscript{3}\textperiodcentered{}\textperiodcentered{}\textperiodcentered{}CH\textsubscript{3}} & 3.295 & 179.4 & 3.247 & 179.3 & 3.002 & 169.6 & NF\textsubscript{3}\textperiodcentered{}\textperiodcentered{}\textperiodcentered{}NH\textsubscript{3} & 3.209 & 179.1 & b & b & 3.448 & 168.2 \\ \hline
\multicolumn{1}{|>{\centering\arraybackslash}m{67.9pt}|}{ClF\textsubscript{3}\textperiodcentered{}\textperiodcentered{}\textperiodcentered{}CH\textsubscript{3}} & 3.160 & 169.8 & 2.972 & 178.5 & 2.824 & 179.6 & ClF\textsubscript{3}\textperiodcentered{}\textperiodcentered{}\textperiodcentered{}NH\textsubscript{3} & 2.965 & 179.4 & 2.763 & 179.4 & 2.924 & 179.7 \\ \hline
\multicolumn{1}{|>{\centering\arraybackslash}m{67.9pt}|}{OF\textsubscript{2}\textperiodcentered{}\textperiodcentered{}\textperiodcentered{}CH\textsubscript{3}} & 3.212 & 179.7 & 3.172 & 178.7 & 2.928 & 177.4 & OF\textsubscript{2}\textperiodcentered{}\textperiodcentered{}\textperiodcentered{}NH\textsubscript{3} & 2.942 & 176.7 & 2.838 & 176.2 & 2.973 & 157.1 \\ \hline
\multicolumn{1}{|>{\centering\arraybackslash}m{67.9pt}|}{NF\textsubscript{5}\textperiodcentered{}\textperiodcentered{}\textperiodcentered{}CH\textsubscript{3}} & 3.092 & 173.9 & 3.051 & 177.5 & 2.928 & 176.6 & NF\textsubscript{5}\textperiodcentered{}\textperiodcentered{}\textperiodcentered{}NH\textsubscript{3} & 2.846 & 178.8 & 2.776 & 178.9 & 2.849 & 173.3 \\ \hline
\multicolumn{1}{|>{\centering\arraybackslash}m{67.9pt}|}{FCCF\textperiodcentered{}\textperiodcentered{}\textperiodcentered{}CH\textsubscript{3}} & 3.346 & 178.0 & 3.243 & 179.9 & 3.022 & 179.7 & FCCF\textperiodcentered{}\textperiodcentered{}\textperiodcentered{}NH\textsubscript{3} & 3.444 & 179.6 & 3.312 & 180.0 & 3.648 & 180.0 \\ \hline
\multicolumn{1}{|>{\centering\arraybackslash}m{67.9pt}|}{CF\textsubscript{3}CCF\textperiodcentered{}\textperiodcentered{}\textperiodcentered{}CH\textsubscript{3}} & 3.457 & 180.0 & 3.308 & 179.6 & 3.023 & 179.5 & CF\textsubscript{3}CCF\textperiodcentered{}\textperiodcentered{}\textperiodcentered{}NH\textsubscript{3} & 3.227 & 180.0 & 3.153 & 180.0 & 3.409 & 180.0 \\ \hline
\multicolumn{1}{|>{\centering\arraybackslash}m{67.9pt}|}{CF\textsubscript{3}OF\textperiodcentered{}\textperiodcentered{}\textperiodcentered{}CH\textsubscript{3}} & 3.088 & 173.6 & b & b & 2.891 & 177.1 & CF\textsubscript{3}OF\textperiodcentered{}\textperiodcentered{}\textperiodcentered{}NH\textsubscript{3} & 2.869 & 178.8 & 2.763 & 177.9 & 2.932 & 176.0 \\ \hline
\multicolumn{1}{|>{\centering\arraybackslash}m{67.9pt}|}{F\textsubscript{3}NO\textperiodcentered{}\textperiodcentered{}\textperiodcentered{}CH\textsubscript{3}} & 3.234 & 162.7 & 3.200 & 175.1 & 2.970 & 177.1 & F\textsubscript{3}NO\textperiodcentered{}\textperiodcentered{}\textperiodcentered{}NH\textsubscript{3} & b & b & 2.935 & 159.9 & 3.186 & 175.3 \\ \hline
\multicolumn{1}{|>{\centering\arraybackslash}m{67.9pt}|}{SF\textsubscript{5}\textperiodcentered{}\textperiodcentered{}\textperiodcentered{}CH\textsubscript{3}} & 3.329 & 180.0 & 3.242 & 179.2 & 3.021 & 178.5 & SF\textsubscript{5}\textperiodcentered{}\textperiodcentered{}\textperiodcentered{}NH\textsubscript{3} & 3.236 & 179.7 & 3.106 & 179.9 & 3.419 & 172.4 \\ \hline
\multicolumn{1}{|>{\centering\arraybackslash}m{67.9pt}|}{NH\textsubscript{3}F+\textperiodcentered{}\textperiodcentered{}\textperiodcentered{}CH\textsubscript{3}} & 2.947 & 179.4 & 2.940 & 180.0 & 2.693 & 178.7 & NH\textsubscript{3}F+\textperiodcentered{}\textperiodcentered{}\textperiodcentered{}NH\textsubscript{3} & 2.647 & 180.0 & 2.596 & 180.0 & 2.590 & 180.0 \\ \hline
\multicolumn{1}{|>{\centering\arraybackslash}m{67.9pt}|}{NH\textsubscript{2}F\textsubscript{2}\textsuperscript{+}\textperiodcentered{}\textperiodcentered{}\textperiodcentered{}CH\textsubscript{3}} & 2.941 & 176.4 & 2.919 & 176.8 & 2.713 & 178.9 & NH\textsubscript{2}F\textsubscript{2}\textsuperscript{+}\textperiodcentered{}\textperiodcentered{}\textperiodcentered{}NH\textsubscript{3} & 2.627 & 177.3 & 2.581 & 178.2 & 2.581 & 178.1 \\ \hline
\multicolumn{1}{|>{\centering\arraybackslash}m{67.9pt}|}{\textbf{MSE}} & 0.0 & 0.0 & -0.065 & 1.9 & -0.287 & 1.3 & \textbf{MSE} & 0.0 & 0.0 & -0.102 & -0.1 & 0.082 & -3.4 \\ \hline
\multicolumn{1}{|>{\centering\arraybackslash}m{67.9pt}|}{\textbf{MAD}} & 0.0 & 0.0 & 0.065 & 2.3 & 0.287 & 3.3 & \textbf{MAD} & 0.0 & 0.0 & 0.102 & 0.5 & 0.105 & 3.8 \\ \hline
\multicolumn{1}{|>{\centering\arraybackslash}m{67.9pt}|}{\textbf{RMSD}} & 0.0 & 0.0 & 0.083 & 4.2 & 0.296 & 5.4 & \textbf{RMSD} & 0.0 & 0.0 & 0.116 & 0.9 & 0.138 & 6.6 \\* \hline\noalign{\penalty10000}
\multicolumn{1}{|>{\centering\arraybackslash}m{67.9pt}|}{\textbf{MAX}} & 0.0 & 0.0 & -0.188 & 12.4 & -0.434 & 14.4 & \textbf{MAX} & 0.0 & 0.0 & -0.202 & -2.8 & 0.289 & -19.6 \\* \hline\noalign{\penalty10000}
\multicolumn{1}{|>{\centering\arraybackslash}m{67.9pt}|}{\textbf{MA\%D}} & 0.0 & 0.0 & 2.0 & 1.3 & 8.9 & 1.2 & \textbf{MA\%D} & 0.0 & 0.0 & 3.4 & 0.3 & 3.3 & 2.2 \\* \specialrule{0.9pt}{0pt}{0pt}\noalign{\penalty10000}
\end{tabular}}}}
\par\nopagebreak\vspace{1pt}{\setstretch{1}\setlength{\parskip}{1.5pt}\fontsize{9}{10.5}\selectfont\raggedright\noindent \textsuperscript{b}Convergence error during optimization of respective F-bonded complexes.\par \textsuperscript{\#}Refer \Tref{tab:shifts-in-df-bond} for shifts in DF bonds (\ensuremath{\Delta}R(DF)) in DF\textperiodcentered{}\textperiodcentered{}\textperiodcentered{}CH\textsubscript{3}/NH\textsubscript{3} complexes calculated at DF-RMP2-F12/VDZ-F12, MP2/AVDZ and M06-2X/AVTZ level of theories.\par}
\end{table}

The F-F bond shift (\ensuremath{\Delta}R) for F\textsubscript{2}\textperiodcentered{}\textperiodcentered{}\textperiodcentered{}NH\textsubscript{3} is larger at MP2/AVDZ (0.0281 Å) than at DF-RMP2/VDZ-F12 (0.0141 Å) (\Tref{tab:shifts-in-df-bond}). The BSSE-corrected BE for this complex at MP2/AVDZ (-8.0 kJ/mol) is still comparable to the reference value (-7.8 kJ/mol) (\Tref{tab:cbs-limit-binding-energies}) even though the F-bond distance (2.442 Å) is underestimated by 6.7\% relative to the reference (2.616 Å) (\Tref{tab:intermolecular-distances-r-f}). Note that deviations in F-bond distances and BEs for selected complexes are explored further in a later section. Across the full dataset, most complexes are highly directional, with linear \ensuremath{\angle}DF\textperiodcentered{}\textperiodcentered{}\textperiodcentered{}C/N angles (\Tref{tab:intermolecular-distances-r-f}), except F\textsubscript{3}NO\textperiodcentered{}\textperiodcentered{}\textperiodcentered{}CH\textsubscript{3} at DF-RMP2/VDZ-F12 and F\textsubscript{3}NO\textperiodcentered{}\textperiodcentered{}\textperiodcentered{}NH\textsubscript{3} at MP2/AVDZ. No F-bonded structure could be obtained for F\textsubscript{3}NO\textperiodcentered{}\textperiodcentered{}\textperiodcentered{}NH\textsubscript{3} at DF-RMP2-F12/VDZ-F12. We, therefore, focus on 15 SEFB (DF\textperiodcentered{}\textperiodcentered{}\textperiodcentered{}CH\textsubscript{3}) and 14 (DF\textperiodcentered{}\textperiodcentered{}\textperiodcentered{}NH\textsubscript{3}) complexes throughout this work.

From \Tref{tab:intermolecular-distances-r-f}, one can see that MA\%D on F-bond distances for SEFB-15 complexes at MP2/AVDZ and M06-2X/AVTZ are 2.0\% and 8.9\%, respectively. This shows that MP2/AVDZ level of theory is better than M06-2X/AVTZ level for reliable prediction of F-bond distances for SEFB-15 complexes. The corresponding MA\%D values for bond angles are 1.3\% and 1.2\%, respectively, showing that both levels of theory are equally reliable for prediction of bond angles for SEFB-15 complexes.

For CSFB-14 complexes, MA\%D values on the F-bond distances are 3.4\% and 3.3\% for MP2/AVDZ and M06-2X/AVTZ, respectively, showing that both levels of theory are equally reliable for prediction of the F-bond distances. The corresponding MA\%D values for bond angles are 0.3\% and 2.2\%, respectively. This shows that MP2/AVDZ level of theory is better than M06-2X/AVTZ level of theory for reliable prediction of bond angles for CSFB-14 complexes.

\begin{figure}[!htbp]\centering
\GPKfig{7.021in}{2.636in}{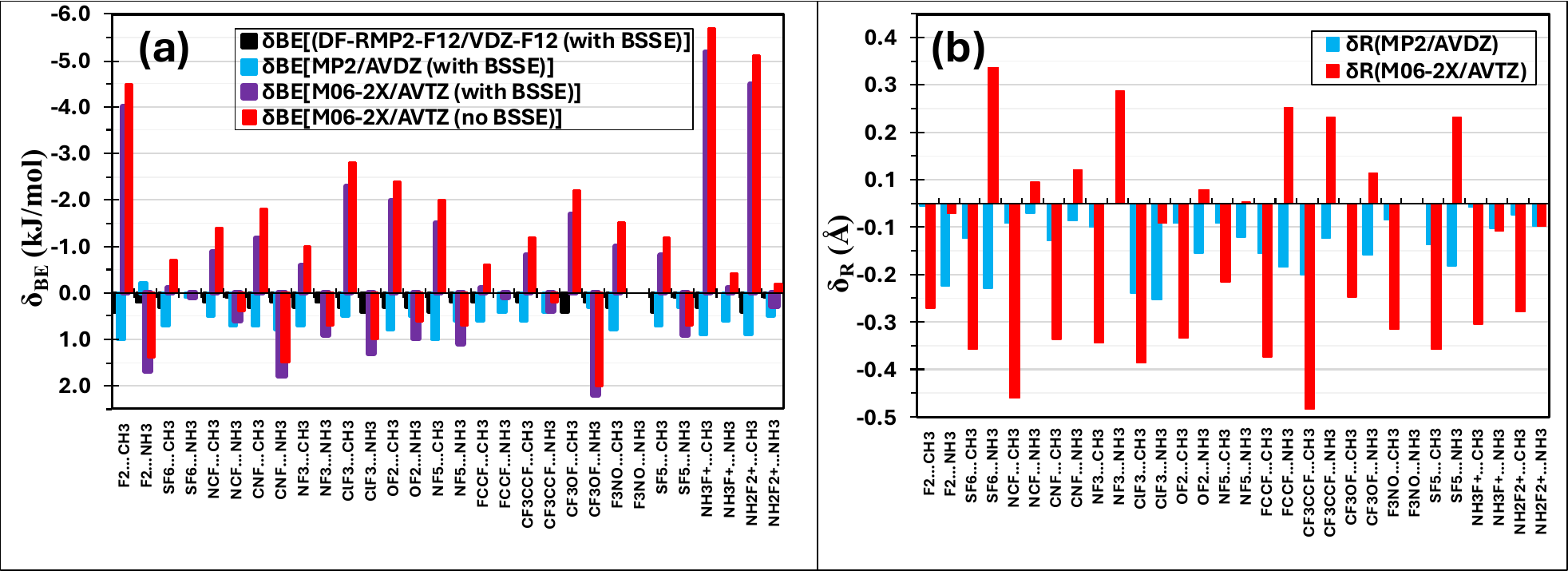}
\par\vspace{-1pt}\caption{Deviations on BEs in kJ/mol \textbf{(a)} and R(F\textperiodcentered{}\textperiodcentered{}\textperiodcentered{}C/N) in Å \textbf{(b)} for all DF\textperiodcentered{}\textperiodcentered{}\textperiodcentered{}CH\textsubscript{3}/NH\textsubscript{3} complexes, at DF-RMP2-F12/VDZ-F12, MP2/AVDZ and M06-2X/AVTZ, relative to the DF-RMP2-F12/CBS//DF-RMP2-F12/VDZ-F12 reference BEs and the DF-RMP2-F12/VDZ-F12 reference R values. The corresponding percentage deviations are provided in \Fref{fig:percentage-deviations-on-bes}.}\label{fig:deviations-on-bes-in}
\vspace{10pt}
\end{figure}

From \Fref{fig:bar-graphs-showing-bes}, BEs calculated at the different CBS limits (DF-RMP2-F12/CBS{[}V(T,Q)Z-F12{]}//DF-RMP2-F12/VDZ-F12,W1-F12//DF-RMP2-F12/VDZ-F12, CCSD(T)/CBS//MP2/AVDZ, and W1-F12//M06-2X/AVTZ) are comparable for most DF\textperiodcentered{}\textperiodcentered{}\textperiodcentered{}CH\textsubscript{3}/NH\textsubscript{3} complexes, as are BSSE-corrected BEs at DF-RMP2-F12/VDZ-F12. Trends in BEs at M06-2X/AVTZ with and without BSSE correction are unaltered (\Tref{tab:complete-basis-set-cbs}). BEs for SEFB complexes computed at M06-2X/AVTZ are overestimated relative to all four CBS-limit BEs, as shown by the purple (BSSE-corrected) and red (BSSE-uncorrected) bars in \mbox{\textbf{FIG. \ref{fig:bar-graphs-showing-bes}.}} To quantify these deviations, we calculated deviations in BEs for all SEFB-15 and CSFB-14 complexes relative to DF-RMP2-F12/CBS{[}V(T,Q)Z-F12{]}//DF-RMP2-F12/VDZ-F12 across three methods (DF-RMP2-F12/VDZ-F12, MP2/AVDZ, and M06-2X/AVTZ) (\Fref{fig:deviations-on-bes-in}). For percentage deviations on BEs and F-bond distances (R) for all complexes, one can refer to \Fref{fig:percentage-deviations-on-bes} and \textbf{TABLES \ref{tab:deviation-be-kj-mol}} and \textbf{S8}.

\begin{figure}[!htbp]\centering
\GPKfig{6.554in}{2.717in}{figures/fig02.pdf}
\par\vspace{-1pt}\caption{Bar graphs showing BEs (kJ/mol) for all complexes (DF\textperiodcentered{}\textperiodcentered{}\textperiodcentered{}CH\textsubscript{3}/NH\textsubscript{3}) calculated at different levels of theory, including the reference values (DF-RMP2-F12/CBS//DF-RMP2-F12/VDZ-F12, black bars). Green and orange bars represent BEs at W1-F12//DF-RMP2-F12/VDZ-F12 and W1-F12//M06-2X/AVTZ, respectively. Blue bars represent BEs at CCSD(T)/CBS//MP2/AVDZ using the Hobza method.\cite{Rezac2012,Rezac2011a,Rezac2011b} Purple and light-blue bars represent BSSE-corrected BEs at M06-2X/AVTZ and MP2/AVDZ, respectively, each using its own optimized geometry. Red bars represent BSSE-uncorrected BEs at M06-2X/AVTZ.}\label{fig:bar-graphs-showing-bes}
\vspace{10pt}
\end{figure}

To compare the performance of M06-2X/AVTZ against selected \emph{ab initio} methods (DF-RMP2-F12/VDZ-F12 and MP2/AVDZ) for reliable prediction of both BEs and F-bond distances, we plot BEs against F-bond distances R(F\textperiodcentered{}\textperiodcentered{}\textperiodcentered{}C/N) (\Fref{fig:scatter-plots-of-bes}; \Tref{tab:bes-kj-mol-and}). M06-2X/AVTZ overestimates BEs for several SEFB complexes (DF\textperiodcentered{}\textperiodcentered{}\textperiodcentered{}CH\textsubscript{3}), including F\textsubscript{2}\textperiodcentered{}\textperiodcentered{}\textperiodcentered{}CH\textsubscript{3}, NH\textsubscript{3}F\textsuperscript{+}\textperiodcentered{}\textperiodcentered{}\textperiodcentered{}CH\textsubscript{3} and NH\textsubscript{2}F\textsubscript{2}\textsuperscript{+}\textperiodcentered{}\textperiodcentered{}\textperiodcentered{}CH\textsubscript{3}(\textbf{\Fref{fig:scatter-plots-of-bes}(a)}). In the case of DF\textperiodcentered{}\textperiodcentered{}\textperiodcentered{}NH\textsubscript{3} complexes, the deviations are less prominent (\textbf{\Fref{fig:scatter-plots-of-bes}(b)}). Deviations on BEs (\ensuremath{\delta}\textsubscript{BE}) and R (\ensuremath{\delta}\textsubscript{R}) for all complexes are shown in \mbox{\textbf{FIG. \ref{fig:deviations-on-bes-in}.}}

Keeping the energy method fixed at W1-F12 and varying only the geometry isolates the error from geometry alone. BEs from W1-F12//M06-2X/AVTZ deviate far more from the reference values for SEFB complexes (MA\%D = 36.1\%) than W1-F12//DF-RMP2-F12/VDZ-F12 (3.2\%) (\Tref{tab:cbs-limit-binding-energies}). For CSFB, the two geometries yield comparable deviations (6.6\% vs. 7.3\%). This matches the MA\%Ds on F-bond-distances already reported above (8.9\% for SEFB vs. 3.3\% for CSFB, \Tref{tab:intermolecular-distances-r-f}). This indicates that not only the energies of SEFB complexes predicted by M06-2X/AVTZ but also their geometries are specifically less reliable.

In \textbf{Sections \ref*{sec:effect-of-basis-sets}-4.8}, we discuss basis-set effects on BEs and F-bond distances predicted by M06-2X. Below, we first characterize how this unreliability is distributed across the SEFB-15 donors.

Viewed as classes rather than as fifteen individual complexes, the SEFB-15 donors fall into two groups. Group 1 is the two cationic donors, NH\textsubscript{3}F\textsuperscript{+} and NH\textsubscript{2}F\textsubscript{2}\textsuperscript{+}, with the largest reference BEs (-9.8 and -11.4 kJ/mol at the DF-RMP2-F12/CBS limit). Group 2 is the thirteen neutral covalent donors (F\textsubscript{2}, SF\textsubscript{6}, NCF, CNF, NF\textsubscript{3}, ClF\textsubscript{3}, OF\textsubscript{2}, NF\textsubscript{5}, FCCF, CF\textsubscript{3}CCF, CF\textsubscript{3}OF, F\textsubscript{3}NO, SF\textsubscript{5}), with smaller, more tightly clustered reference BEs, from -1.5 kJ/mol (FCCF and CF\textsubscript{3}CCF) to -3.4 kJ/mol (F\textsubscript{2}, the strongest of the neutral donors) (\Tref{tab:cbs-limit-binding-energies}). For the cationic F-bond donors (NH\textsubscript{3}F\textsuperscript{+} and NH\textsubscript{2}F\textsubscript{2}\textsuperscript{+}), M06-2X/AVTZ overbinds both SEFB complexes NH\textsubscript{3}F\textsuperscript{+}\textperiodcentered{}\textperiodcentered{}\textperiodcentered{}CH\textsubscript{3} and NH\textsubscript{2}F\textsubscript{2}\textsuperscript{+}\textperiodcentered{}\textperiodcentered{}\textperiodcentered{}CH\textsubscript{3} (BSSE-corrected BEs of -15.0 and -15.9 kJ/mol vs. references of -9.8 and -11.4 kJ/mol), with the two largest absolute errors in the set (5.2 and 4.5 kJ/mol). All thirteen neutral donors are overbound as well, most prominently F\textsubscript{2}\textperiodcentered{}\textperiodcentered{}\textperiodcentered{}CH\textsubscript{3} (-7.4 kJ/mol BSSE-corrected vs. -3.4 kJ/mol reference), the largest percentage error of any SEFB complex. We attribute this to distinct electronic structure of F\textsubscript{2}: it is homonuclear, with no substituent to redistribute charge. We quantify this further in the companion paper, via monomer delocalization error and dimer delocalization index (DI) analysis.

\FloatBarrier
\subsection{Reliability of reference BEs and F-bond distances for DF\textperiodcentered{}\textperiodcentered{}\textperiodcentered{}CH\textsubscript{3}/NH\textsubscript{3}}\label{sec:reliability-of-reference-bes}

To confirm the reliability of geometries optimized at DF-RMP2-F12/VDZ-F12 level, we benchmarked them against several ab initio methods (CCSD(T)-F12b/VDZ-F12, DF-RMP2-F12/VDZ-F12, CCSD(T)/AVTZ, DCSD/AVTZ, DF-RMP2/AVTZ, MBPT(2)/VTZ, SCS-MP2/AVTZ, MP2/AVDZ-AV5Z) for five selected complexes DF\textperiodcentered{}\textperiodcentered{}\textperiodcentered{}CH\textsubscript{3}/NH\textsubscript{3} (DF = F\textsubscript{2}, NCF, CNF, NH\textsubscript{3}F\textsuperscript{+} and NH\textsubscript{2}F\textsubscript{2}\textsuperscript{+}). Two SEFB complexes (NH\textsubscript{3}F\textsuperscript{+}\textperiodcentered{}\textperiodcentered{}\textperiodcentered{}CH\textsubscript{3} and NH\textsubscript{2}F\textsubscript{2}\textsuperscript{+}\textperiodcentered{}\textperiodcentered{}\textperiodcentered{}CH\textsubscript{3}) could not be optimized at CCSD(T)/AVTZ or CCSD(T)-F12b/VDZ-F12 (\Tref{tab:f-bond-distances-r}). So, we used CCSD(T)-F12b/VTZ for these two complexes. For the rest, F-bond distances obtained from CCSD(T)-F12b/VDZ-F12 serve as reference. Various statistical measures (MSE, MAD, RMSD, MAX, MA\%D) for F-bond distances (R) of the selected complexes are summarized in \mbox{\textbf{TABLE \ref{tab:statistical-errors-mse-mad}.}}

\Tref{tab:f-bond-distances-r} summarizes the BEs and F-bond distances for the five selected complexes DF\textperiodcentered{}\textperiodcentered{}\textperiodcentered{}CH\textsubscript{3}/NH\textsubscript{3} (DF = F\textsubscript{2}, NCF, CNF, NH\textsubscript{3}F\textsuperscript{+} and NH\textsubscript{2}F\textsubscript{2}\textsuperscript{+}) across several \emph{ab initio} methods. For F\textsubscript{2}\textperiodcentered{}\textperiodcentered{}\textperiodcentered{}CH\textsubscript{3}, in particular, M06-2X/AVTZ overestimates BEs and underestimates F-bond distance relative to all \emph{ab initio} methods (\mbox{\textbf{TABLEs \ref{tab:cbs-limit-binding-energies}-\ref{tab:f-bond-distances-r}}}).

\begin{table}[!htbp]\centering\setlength{\parskip}{0pt}
\caption{F-bond distances [R(F\textperiodcentered{}\textperiodcentered{}\textperiodcentered{}C/\allowbreak{}N), Å] and BEs (kJ/\allowbreak{}mol) for selected DF\textperiodcentered{}\textperiodcentered{}\textperiodcentered{}CH\textsubscript{3}/NH\textsubscript{3} complexes (DF = F\textsubscript{2}, NCF, CNF, NH\textsubscript{3}F\textsuperscript{+} and NH\textsubscript{2}F\textsubscript{2}\textsuperscript{+}) at various ab initio levels of theory.\textsuperscript{\#}}\label{tab:f-bond-distances-r}\vspace{7pt}\nopagebreak
{\fontsize{9.0}{10.3}\fontfamily{ptm}\selectfont\renewcommand{\arraystretch}{1.17}\setlength{\tabcolsep}{1.5pt}
\centerline{\adjustbox{max width=518pt}{\begin{tabular}{|>{\centering\arraybackslash}m{69.1pt}|>{\centering\arraybackslash}m{28.1pt}|>{\centering\arraybackslash}m{30.2pt}|>{\centering\arraybackslash}m{28.1pt}|>{\centering\arraybackslash}m{28.1pt}|>{\centering\arraybackslash}m{28.1pt}|>{\centering\arraybackslash}m{28.1pt}|>{\centering\arraybackslash}m{28.1pt}|>{\centering\arraybackslash}m{28.1pt}|>{\centering\arraybackslash}m{28.1pt}|>{\centering\arraybackslash}m{28.1pt}|>{\centering\arraybackslash}m{28.1pt}|>{\centering\arraybackslash}m{28.1pt}|>{\centering\arraybackslash}m{28.1pt}|>{\centering\arraybackslash}m{28.1pt}|>{\centering\arraybackslash}m{28.1pt}|}
\specialrule{0.9pt}{0pt}{0pt}
\multicolumn{16}{|>{\centering\arraybackslash}p{537.6pt}|}{Fluorine bond distance (R(F\textperiodcentered{}\textperiodcentered{}\textperiodcentered{}C/\allowbreak{}N) (in Å) for DF\textperiodcentered{}\textperiodcentered{}\textperiodcentered{}CH\textsubscript{3}/NH\textsubscript{3} (DF = F\textsubscript{2}, NCF, CNF, NH\textsubscript{3}F\textsuperscript{+} and NH\textsubscript{2}F\textsubscript{2}\textsuperscript{+}) complexes.} \\* \hline\noalign{\penalty10000}
\multicolumn{1}{|>{\centering\arraybackslash}p{69.1pt}|}{\makebox[\linewidth][c]{\rotatebox{90}{\hspace{2pt}F-bonded complexes\hspace{2pt}}}} & \multicolumn{1}{>{\centering\arraybackslash}p{28.1pt}|}{\makebox[\linewidth][c]{\rotatebox{90}{\hspace{2pt}CCSD(T)-F12b/\allowbreak{}VDZ-F12\hspace{2pt}}}} & \multicolumn{1}{>{\centering\arraybackslash}p{30.2pt}|}{\makebox[\linewidth][c]{\rotatebox{90}{\hspace{2pt}CCSD(T)/ AVDZ\hspace{2pt}}}} & \multicolumn{1}{>{\centering\arraybackslash}p{28.1pt}|}{\makebox[\linewidth][c]{\rotatebox{90}{\hspace{2pt}CCSD(T)/ AVTZ\hspace{2pt}}}} & \multicolumn{1}{>{\centering\arraybackslash}p{28.1pt}|}{\makebox[\linewidth][c]{\rotatebox{90}{\hspace{2pt}CCSD/\allowbreak{}AVDZ\hspace{2pt}}}} & \multicolumn{1}{>{\centering\arraybackslash}p{28.1pt}|}{\makebox[\linewidth][c]{\rotatebox{90}{\hspace{2pt}CCSD/\allowbreak{}AVTZ\hspace{2pt}}}} & \multicolumn{1}{>{\centering\arraybackslash}p{28.1pt}|}{\makebox[\linewidth][c]{\rotatebox{90}{\hspace{2pt}DCSD/\allowbreak{}AVDZ\hspace{2pt}}}} & \multicolumn{1}{>{\centering\arraybackslash}p{28.1pt}|}{\makebox[\linewidth][c]{\rotatebox{90}{\hspace{2pt}DCSD/\allowbreak{}AVTZ\hspace{2pt}}}} & \multicolumn{1}{>{\centering\arraybackslash}p{28.1pt}|}{\makebox[\linewidth][c]{\rotatebox{90}{\hspace{2pt}DF-RMP2-F12/\allowbreak{}VDZ-F12\hspace{2pt}}}} & \multicolumn{1}{>{\centering\arraybackslash}p{28.1pt}|}{\makebox[\linewidth][c]{\rotatebox{90}{\hspace{2pt}DF-RMP2/\allowbreak{}AVTZ\hspace{2pt}}}} & \multicolumn{1}{>{\centering\arraybackslash}p{28.1pt}|}{\makebox[\linewidth][c]{\rotatebox{90}{\hspace{2pt}MBPT(2)/\allowbreak{}VTZ\hspace{2pt}}}} & \multicolumn{1}{>{\centering\arraybackslash}p{28.1pt}|}{\makebox[\linewidth][c]{\rotatebox{90}{\hspace{2pt}SCS-MP2/\allowbreak{}AVTZ\hspace{2pt}}}} & \multicolumn{1}{>{\centering\arraybackslash}p{28.1pt}|}{\makebox[\linewidth][c]{\rotatebox{90}{\hspace{2pt}MP2/\allowbreak{}AVDZ\hspace{2pt}}}} & \multicolumn{1}{>{\centering\arraybackslash}p{28.1pt}|}{\makebox[\linewidth][c]{\rotatebox{90}{\hspace{2pt}MP2/\allowbreak{}AVTZ\hspace{2pt}}}} & \multicolumn{1}{>{\centering\arraybackslash}p{28.1pt}|}{\makebox[\linewidth][c]{\rotatebox{90}{\hspace{2pt}MP2/\allowbreak{}AVQZ\hspace{2pt}}}} & \multicolumn{1}{>{\centering\arraybackslash}p{28.1pt}|}{\makebox[\linewidth][c]{\rotatebox{90}{\hspace{2pt}MP2/\allowbreak{}AV5Z\hspace{2pt}}}} \\* \hline\noalign{\penalty10000}
\multicolumn{1}{|>{\centering\arraybackslash}m{69.1pt}|}{\textbf{DF\textperiodcentered{}\textperiodcentered{}\textperiodcentered{}CH}\textbf{\textsubscript{3}}} & R & R & R & R & R & R & R & R & R & R & R & R & R & R & R \\* \hline\noalign{\penalty10000}
\multicolumn{1}{|>{\centering\arraybackslash}m{69.1pt}|}{F\textsubscript{2}\textperiodcentered{}\textperiodcentered{}\textperiodcentered{}CH\textsubscript{3}} & 2.850 & 2.616 & 2.829 & 2.938 & 3.031 & 2.759 & 2.900 & 2.869 & 2.841 & 2.979 & 2.838 & 2.865 & 2.957 & 2.958 & 2.959 \\* \hline\noalign{\penalty10000}
\multicolumn{1}{|>{\centering\arraybackslash}m{69.1pt}|}{NCF\textperiodcentered{}\textperiodcentered{}\textperiodcentered{}CH\textsubscript{3}} & 3.415 & 3.357 & 3.343 & 3.428 & 3.427 & 3.403 & 3.391 & 3.428 & 3.378 & 3.470 & 3.377 & 3.388 & 3.383 & 3.411 & 3.411 \\* \hline\noalign{\penalty10000}
\multicolumn{1}{|>{\centering\arraybackslash}m{69.1pt}|}{CNF\textperiodcentered{}\textperiodcentered{}\textperiodcentered{}CH\textsubscript{3}} & 3.254 & 3.159 & 3.171 & 3.235 & 3.258 & 3.199 & 3.216 & 3.262 & 3.187 & 3.302 & 3.186 & 3.185 & 3.201 & 3.241 & 3.242 \\ \hline
\multicolumn{1}{|>{\centering\arraybackslash}m{69.1pt}|}{\textsuperscript{\#}NH\textsubscript{3}F+\textperiodcentered{}\textperiodcentered{}\textperiodcentered{}CH\textsubscript{3}} & 2.897 & 2.882 & N/\allowbreak{}A & 2.969 & 2.967 & 2.935 & 2.975 & 2.947 & 2.970 & 2.970 & 2.970 & 2.940 & 2.941 & 2.957 & 2.959 \\ \hline
\multicolumn{1}{|>{\centering\arraybackslash}m{69.1pt}|}{\textsuperscript{\#}NH\textsubscript{2}F\textsubscript{2}\textsuperscript{+}\textperiodcentered{}\textperiodcentered{}\textperiodcentered{}CH\textsubscript{3}} & 2.893 & 2.876 & N/\allowbreak{}A & 2.948 & 2.954 & 2.924 & N/\allowbreak{}A & 2.941 & 2.952 & 2.953 & 2.951 & 2.919 & 2.926 & 2.945 & 2.946 \\ \hline
\multicolumn{1}{|>{\centering\arraybackslash}m{69.1pt}|}{\textbf{DF\textperiodcentered{}\textperiodcentered{}\textperiodcentered{}NH}\textbf{\textsubscript{3}}} &  &  &  &  &  &  &  &  &  &  &  &  &  &  &  \\ \hline
\multicolumn{1}{|>{\centering\arraybackslash}m{69.1pt}|}{F\textsubscript{2}\textperiodcentered{}\textperiodcentered{}\textperiodcentered{}NH\textsubscript{3}} & 2.626 & 2.452 & 2.614 & 2.649 & 2.771 & 2.521 & 2.671 & 2.616 & 2.596 & 2.640 & 2.593 & 2.442 & 2.595 & 2.606 & 2.607 \\ \hline
\multicolumn{1}{|>{\centering\arraybackslash}m{69.1pt}|}{NCF\textperiodcentered{}\textperiodcentered{}\textperiodcentered{}NH\textsubscript{3}} & 3.157 & 3.127 & 3.129 & 3.184 & 3.174 & 3.169 & 3.166 & 3.160 & 3.145 & 3.159 & 3.144 & 3.141 & 3.147 & 3.147 & 3.149 \\ \hline
\multicolumn{1}{|>{\centering\arraybackslash}m{69.1pt}|}{CNF\textperiodcentered{}\textperiodcentered{}\textperiodcentered{}NH\textsubscript{3}} & 2.981 & 2.934 & 2.952 & 2.992 & 3.013 & 2.964 & 2.981 & 2.960 & 2.960 & 2.958 & 2.950 & 2.924 & 2.955 & 2.957 & 2.958 \\ \hline
\multicolumn{1}{|>{\centering\arraybackslash}m{69.1pt}|}{NH\textsubscript{3}F+\textperiodcentered{}\textperiodcentered{}\textperiodcentered{}NH\textsubscript{3}} & 2.633 & 2.583 & 2.625 & 2.648 & 2.678 & 2.605 & 2.644 & 2.647 & 2.636 & 2.620 & 2.636 & 2.596 & 2.636 & 2.646 & 2.646 \\ \hline
\multicolumn{1}{|>{\centering\arraybackslash}m{69.1pt}|}{NH\textsubscript{2}F\textsubscript{2}\textsuperscript{+}\textperiodcentered{}\textperiodcentered{}\textperiodcentered{}NH\textsubscript{3}} & 2.616 & 2.569 & 2.606 & 2.622 & 2.655 & 2.587 & 2.611 & 2.627 & 2.613 & 2.601 & 2.612 & 2.581 & 2.615 & 2.625 & 2.627 \\ \specialrule{0.9pt}{0pt}{0pt}
\end{tabular}}}
\par\nointerlineskip
\centerline{\adjustbox{max width=518pt}{\begin{tabular}{|>{\centering\arraybackslash}m{69.1pt}|>{\centering\arraybackslash}m{28.1pt}|>{\centering\arraybackslash}m{30.2pt}|>{\centering\arraybackslash}m{28.1pt}|>{\centering\arraybackslash}m{28.1pt}|>{\centering\arraybackslash}m{28.1pt}|>{\centering\arraybackslash}m{28.1pt}|>{\centering\arraybackslash}m{28.1pt}|>{\centering\arraybackslash}m{28.1pt}|>{\centering\arraybackslash}m{28.1pt}|>{\centering\arraybackslash}m{28.1pt}|>{\centering\arraybackslash}m{28.1pt}|>{\centering\arraybackslash}m{28.1pt}|>{\centering\arraybackslash}m{28.1pt}|>{\centering\arraybackslash}m{28.1pt}|>{\centering\arraybackslash}m{28.1pt}|}
\multicolumn{16}{|>{\centering\arraybackslash}m{537.6pt}|}{BEs for DF\textperiodcentered{}\textperiodcentered{}\textperiodcentered{}CH\textsubscript{3}/NH\textsubscript{3} (DF = F\textsubscript{2}, NCF, CNF, NH\textsubscript{3}F\textsuperscript{+} and NH\textsubscript{2}F\textsubscript{2}\textsuperscript{+}) complexes} \\* \hline\noalign{\penalty10000}
\multicolumn{1}{|>{\centering\arraybackslash}m{69.1pt}|}{\multirow{2}{69.1pt}{\centering \textbf{DF\textperiodcentered{}\textperiodcentered{}\textperiodcentered{}CH}\textbf{\textsubscript{3}}}} & \multicolumn{1}{>{\centering\arraybackslash}p{28.1pt}|}{\makebox[\linewidth][c]{\rotatebox{90}{\hspace{2pt}CCSD(T)-F12b/ VDZ-F12\hspace{2pt}}}} & \multicolumn{1}{>{\centering\arraybackslash}p{30.2pt}|}{\makebox[\linewidth][c]{\rotatebox{90}{\hspace{2pt}CCSD(T)/ AVDZ\hspace{2pt}}}} & \multicolumn{1}{>{\centering\arraybackslash}p{28.1pt}|}{\makebox[\linewidth][c]{\rotatebox{90}{\hspace{2pt}CCSD(T)/ AVTZ\hspace{2pt}}}} & \multicolumn{1}{>{\centering\arraybackslash}p{28.1pt}|}{\makebox[\linewidth][c]{\rotatebox{90}{\hspace{2pt}CCSD/\allowbreak{}AVDZ\hspace{2pt}}}} & \multicolumn{1}{>{\centering\arraybackslash}p{28.1pt}|}{\makebox[\linewidth][c]{\rotatebox{90}{\hspace{2pt}CCSD/\allowbreak{}AVTZ\hspace{2pt}}}} & \multicolumn{1}{>{\centering\arraybackslash}p{28.1pt}|}{\makebox[\linewidth][c]{\rotatebox{90}{\hspace{2pt}DCSD/\allowbreak{}AVDZ\hspace{2pt}}}} & \multicolumn{1}{>{\centering\arraybackslash}p{28.1pt}|}{\makebox[\linewidth][c]{\rotatebox{90}{\hspace{2pt}DCSD/\allowbreak{}AVTZ\hspace{2pt}}}} & \multicolumn{1}{>{\centering\arraybackslash}p{28.1pt}|}{\makebox[\linewidth][c]{\rotatebox{90}{\hspace{2pt}DF-RMP2-F12/\allowbreak{}VDZ-F12\hspace{2pt}}}} & \multicolumn{1}{>{\centering\arraybackslash}p{28.1pt}|}{\makebox[\linewidth][c]{\rotatebox{90}{\hspace{2pt}DF-RMP2/\allowbreak{}AVTZ\hspace{2pt}}}} & \multicolumn{1}{>{\centering\arraybackslash}p{28.1pt}|}{\makebox[\linewidth][c]{\rotatebox{90}{\hspace{2pt}MBPT(2)/\allowbreak{}VTZ\hspace{2pt}}}} & \multicolumn{1}{>{\centering\arraybackslash}p{28.1pt}|}{\makebox[\linewidth][c]{\rotatebox{90}{\hspace{2pt}SCS-MP2/\allowbreak{}AVTZ\hspace{2pt}}}} & \multicolumn{1}{>{\centering\arraybackslash}p{28.1pt}|}{\makebox[\linewidth][c]{\rotatebox{90}{\hspace{2pt}MP2/\allowbreak{}AVDZ\hspace{2pt}}}} & \multicolumn{1}{>{\centering\arraybackslash}p{28.1pt}|}{\makebox[\linewidth][c]{\rotatebox{90}{\hspace{2pt}MP2/\allowbreak{}AVTZ\hspace{2pt}}}} & \multicolumn{1}{>{\centering\arraybackslash}p{28.1pt}|}{\makebox[\linewidth][c]{\rotatebox{90}{\hspace{2pt}MP2/\allowbreak{}AVQZ\hspace{2pt}}}} & \multicolumn{1}{>{\centering\arraybackslash}p{28.1pt}|}{\makebox[\linewidth][c]{\rotatebox{90}{\hspace{2pt}MP2/\allowbreak{}AV5Z\hspace{2pt}}}} \\* \cline{2-16}\noalign{\penalty10000}
\multicolumn{1}{|>{\centering\arraybackslash}m{69.1pt}|}{} & BE & BE & BE & BE & BE & BE & BE & BE & BE & BE & BE & BE & BE & BE & BE \\* \hline\noalign{\penalty10000}
\multicolumn{1}{|>{\centering\arraybackslash}m{69.1pt}|}{F\textsubscript{2}\textperiodcentered{}\textperiodcentered{}\textperiodcentered{}CH\textsubscript{3}} & -3.1 & -5.5 & -3.9 & -3.7 & -2.6 & -4.8 & -3.4 & -3.2 & -3.9 & -2.9 & -3.9 & -4.6 & -3.4 & -3.1 & -3.0 \\* \hline\noalign{\penalty10000}
\multicolumn{1}{|>{\centering\arraybackslash}m{69.1pt}|}{NCF\textperiodcentered{}\textperiodcentered{}\textperiodcentered{}CH\textsubscript{3}} & -1.6 & -2.5 & -2.2 & -2.1 & -1.8 & -2.2 & -1.9 & -1.7 & -2.1 & -1.7 & -2.1 & -2.3 & -2.0 & -1.9 & -1.7 \\* \hline\noalign{\penalty10000}
\multicolumn{1}{|>{\centering\arraybackslash}m{69.1pt}|}{CNF\textperiodcentered{}\textperiodcentered{}\textperiodcentered{}CH\textsubscript{3}} & -2.1 & -3.5 & -2.9 & -3.0 & -2.3 & -3.2 & -2.6 & -2.2 & -2.9 & -2.4 & -2.9 & -3.4 & -2.7 & -2.4 & -2.2 \\* \hline\noalign{\penalty10000}
\multicolumn{1}{|>{\centering\arraybackslash}m{69.1pt}|}{NH\textsubscript{3}F+\textperiodcentered{}\textperiodcentered{}\textperiodcentered{}CH\textsubscript{3}} & -10.8 & -11.4 & N/\allowbreak{}A & -10.2 & -9.6 & -8.0 & -8.6 & -9.7 & -8.8 & -9.0 & -8.8 & -10.7 & -10.2 & -9.8 & -9.4 \\ \hline
\multicolumn{1}{|>{\centering\arraybackslash}m{69.1pt}|}{NH\textsubscript{2}F\textsubscript{2}\textsuperscript{+}\textperiodcentered{}\textperiodcentered{}\textperiodcentered{}CH\textsubscript{3}} & -12.2 & -13.4 & N/\allowbreak{}A & -12.1 & -11.3 & -9.2 & N/\allowbreak{}A & -11.2 & -10.2 & -10.6 & -10.2 & -12.6 & -11.9 & -11.4 & -11.0 \\ \hline
\multicolumn{1}{|>{\centering\arraybackslash}m{69.1pt}|}{\textbf{DF\textperiodcentered{}\textperiodcentered{}\textperiodcentered{}NH}\textbf{\textsubscript{3}}} &  &  &  &  &  &  &  &  &  &  &  &  &  &  &  \\ \hline
\multicolumn{1}{|>{\centering\arraybackslash}m{69.1pt}|}{F\textsubscript{2}\textperiodcentered{}\textperiodcentered{}\textperiodcentered{}NH\textsubscript{3}} & -7.8 & -11.8 & -8.4 & -8.6 & -6.2 & -10.7 & -7.6 & -8.0 & -8.5 & -8.1 & -8.5 & -11.9 & -8.5 & -8.2 & -7.8 \\ \hline
\multicolumn{1}{|>{\centering\arraybackslash}m{69.1pt}|}{NCF\textperiodcentered{}\textperiodcentered{}\textperiodcentered{}NH\textsubscript{3}} & -5.5 & -6.0 & -5.7 & -5.4 & -5.1 & -5.5 & -5.2 & -5.5 & -5.6 & -6.1 & -5.6 & -6.0 & -5.6 & -5.4 & -5.2 \\ \hline
\multicolumn{1}{|>{\centering\arraybackslash}m{69.1pt}|}{CNF\textperiodcentered{}\textperiodcentered{}\textperiodcentered{}NH\textsubscript{3}} & -6.7 & -7.8 & -7.1 & -7.1 & -6.3 & -7.4 & -6.7 & -7.2 & -7.5 & -7.9 & -7.5 & -8.4 & -7.5 & -7.2 & -6.9 \\* \hline\noalign{\penalty10000}
\multicolumn{1}{|>{\centering\arraybackslash}m{69.1pt}|}{NH\textsubscript{3}F+\textperiodcentered{}\textperiodcentered{}\textperiodcentered{}NH\textsubscript{3}} & -35.1 & -37.0 & -35.3 & -35.0 & -33.4 & -36.3 & -34.6 & -34.4 & -34.6 & -35.5 & -34.6 & -36.3 & -34.6 & -34.2 & -34.0 \\* \hline\noalign{\penalty10000}
\multicolumn{1}{|>{\centering\arraybackslash}m{69.1pt}|}{NH\textsubscript{2}F\textsubscript{2}\textsuperscript{+}\textperiodcentered{}\textperiodcentered{}\textperiodcentered{}NH\textsubscript{3}} & -39.8 & -42.3 & -40.2 & -40.3 & -38.3 & -41.5 & -39.8 & -39.3 & -39.6 & -40.7 & -39.6 & -41.7 & -39.6 & -39.1 & -38.8 \\* \specialrule{0.9pt}{0pt}{0pt}\noalign{\penalty10000}
\end{tabular}}}}
\par\nopagebreak\vspace{1pt}{\setstretch{1}\setlength{\parskip}{1.5pt}\fontsize{9}{10.5}\selectfont\raggedright\noindent \textsuperscript{\#}Optimization at CCSD(T)-F12b/VTZ-F12 was computationally infeasible for NH\textsubscript{3}F\textsuperscript{+}\textperiodcentered{}\textperiodcentered{}\textperiodcentered{}CH\textsubscript{3} and NH\textsubscript{2}F\textsubscript{2}\textsuperscript{+}\textperiodcentered{}\textperiodcentered{}\textperiodcentered{}CH\textsubscript{3}; CCSD(T)-F12b/VDZ-F12 was used instead for all selected complexes DF\textperiodcentered{}\textperiodcentered{}\textperiodcentered{}CH\textsubscript{3}/NH\textsubscript{3} (DF = F\textsubscript{2}, NCF, CNF, NH\textsubscript{3}F\textsuperscript{+} and NH\textsubscript{2}F\textsubscript{2}\textsuperscript{+}). This also failed to converge forNH\textsubscript{3}F\textsuperscript{+}\textperiodcentered{}\textperiodcentered{}\textperiodcentered{}CH\textsubscript{3} and NH\textsubscript{2}F\textsubscript{2}\textsuperscript{+}\textperiodcentered{}\textperiodcentered{}\textperiodcentered{}CH\textsubscript{3}. So, these two complexes were optimized at CCSD(T)-F12b/VTZ.\par \textsuperscript{N/A}Convergence failure during optimization.\par}
\end{table}

\begin{table}[!htbp]\centering\setlength{\parskip}{0pt}
\caption{Statistical errors (MSE, MAD, RMSD, MAX, MA\%D) on F-bond distances for selected DF\textperiodcentered{}\textperiodcentered{}\textperiodcentered{}CH\textsubscript{3}/NH\textsubscript{3} complexes (DF = F\textsubscript{2}, NCF, CNF, NH\textsubscript{3}F\textsuperscript{+} and NH\textsubscript{2}F\textsubscript{2}\textsuperscript{+}), calculated at several ab initio levels relative to CCSD(T)-F12b/\allowbreak{}VDZ-F12. Distances are expressed in Å.\textsuperscript{\# }}\label{tab:statistical-errors-mse-mad}\vspace{7pt}\nopagebreak
{\fontsize{9.0}{10.3}\fontfamily{ptm}\selectfont\renewcommand{\arraystretch}{1.17}\setlength{\tabcolsep}{1.5pt}
\centerline{\adjustbox{max width=518pt}{\begin{tabular}{|>{\centering\arraybackslash}m{46.2pt}|>{\centering\arraybackslash}m{39.2pt}|>{\centering\arraybackslash}m{32.1pt}|>{\centering\arraybackslash}m{32.1pt}|>{\centering\arraybackslash}m{32.1pt}|>{\centering\arraybackslash}m{32.1pt}|>{\centering\arraybackslash}m{32.1pt}|>{\centering\arraybackslash}m{32.1pt}|>{\centering\arraybackslash}m{32.1pt}|>{\centering\arraybackslash}m{32.1pt}|>{\centering\arraybackslash}m{32.1pt}|>{\centering\arraybackslash}m{32.8pt}|>{\centering\arraybackslash}m{31.4pt}|>{\centering\arraybackslash}m{32.1pt}|>{\centering\arraybackslash}m{32.1pt}|}
\specialrule{0.9pt}{0pt}{0pt}
\multicolumn{1}{|>{\centering\arraybackslash}m{46.2pt}|}{\makebox[\linewidth][c]{\rotatebox{90}{\hspace{2pt}\textbf{Error statistics on R values}\hspace{2pt}}}} & \makebox[\linewidth][c]{\rotatebox{90}{\hspace{2pt}CCSD(T)-F12b/\allowbreak{}VDZ-F12\hspace{2pt}}} & \makebox[\linewidth][c]{\rotatebox{90}{\hspace{2pt}CCSD(T)/ AVDZ\hspace{2pt}}} & \makebox[\linewidth][c]{\rotatebox{90}{\hspace{2pt}CCSD/ AVDZ\hspace{2pt}}} & \makebox[\linewidth][c]{\rotatebox{90}{\hspace{2pt}CCSD/ AVTZ\hspace{2pt}}} & \makebox[\linewidth][c]{\rotatebox{90}{\hspace{2pt}DCSD/ AVDZ\hspace{2pt}}} & \makebox[\linewidth][c]{\rotatebox{90}{\hspace{2pt}DCSD/ AVTZ\hspace{2pt}}} & \makebox[\linewidth][c]{\rotatebox{90}{\hspace{2pt}DF-RMP2-F12/\allowbreak{}VDZ-F12\hspace{2pt}}} & \makebox[\linewidth][c]{\rotatebox{90}{\hspace{2pt}DF-RMP2/ AVTZ\hspace{2pt}}} & \makebox[\linewidth][c]{\rotatebox{90}{\hspace{2pt}MBPT(2)/ VTZ\hspace{2pt}}} & \makebox[\linewidth][c]{\rotatebox{90}{\hspace{2pt}SCS-MP2/\allowbreak{}AVTZ\hspace{2pt}}} & \makebox[\linewidth][c]{\rotatebox{90}{\hspace{2pt}MP2/\allowbreak{}AVDZ\hspace{2pt}}} & \makebox[\linewidth][c]{\rotatebox{90}{\hspace{2pt}MP2/\allowbreak{}AVTZ\hspace{2pt}}} & \makebox[\linewidth][c]{\rotatebox{90}{\hspace{2pt}MP2/\allowbreak{}AVQZ\hspace{2pt}}} & \makebox[\linewidth][c]{\rotatebox{90}{\hspace{2pt}MP2/\allowbreak{}AV5Z\hspace{2pt}}} \\* \hline\noalign{\penalty10000}
\multicolumn{15}{|>{\centering\arraybackslash}m{545.0pt}|}{\textbf{DF\textperiodcentered{}\textperiodcentered{}\textperiodcentered{}CH}\textbf{\textsubscript{3}}} \\* \hline\noalign{\penalty10000}
\multicolumn{1}{|>{\centering\arraybackslash}m{46.2pt}|}{MSE} & 0.0 & -0.084 & 0.042 & 0.066 & -0.018 & 0.017 & 0.028 & 0.004 & 0.073 & 0.003 & -0.002 & 0.020 & 0.041 & 0.042 \\* \hline\noalign{\penalty10000}
\multicolumn{1}{|>{\centering\arraybackslash}m{46.2pt}|}{MAD} & 0.0 & 0.084 & 0.049 & 0.066 & 0.045 & 0.048 & 0.028 & 0.049 & 0.073 & 0.050 & 0.036 & 0.054 & 0.047 & 0.048 \\* \hline\noalign{\penalty10000}
\multicolumn{1}{|>{\centering\arraybackslash}m{46.2pt}|}{RMSD} & 0.0 & 0.116 & 0.057 & 0.091 & 0.053 & 0.051 & 0.033 & 0.054 & 0.079 & 0.055 & 0.041 & 0.061 & 0.060 & 0.061 \\* \hline\noalign{\penalty10000}
\multicolumn{1}{|>{\centering\arraybackslash}m{46.2pt}|}{MAX} & 0.0 & -0.234 & 0.088 & 0.181 & -0.091 & 0.078 & 0.050 & 0.073 & 0.129 & 0.073 & -0.069 & 0.107 & 0.108 & 0.109 \\* \hline\noalign{\penalty10000}
\multicolumn{1}{|>{\centering\arraybackslash}m{46.2pt}|}{MA\%D} & 0.0 & 2.8 & 1.7 & 2.3 & 1.5 & 1.6 & \textbf{0.9} & 1.6 & 2.4 & 1.6 & 1.2 & 1.8 & 1.6 & 1.7 \\ \specialrule{0.9pt}{0pt}{0pt}
\end{tabular}}}
\par\nointerlineskip
\centerline{\adjustbox{max width=518pt}{\begin{tabular}{|>{\centering\arraybackslash}m{46.2pt}|>{\centering\arraybackslash}m{39.2pt}|>{\centering\arraybackslash}m{32.1pt}|>{\centering\arraybackslash}m{32.1pt}|>{\centering\arraybackslash}m{32.1pt}|>{\centering\arraybackslash}m{32.1pt}|>{\centering\arraybackslash}m{32.1pt}|>{\centering\arraybackslash}m{32.1pt}|>{\centering\arraybackslash}m{32.1pt}|>{\centering\arraybackslash}m{32.1pt}|>{\centering\arraybackslash}m{32.1pt}|>{\centering\arraybackslash}m{32.8pt}|>{\centering\arraybackslash}m{31.4pt}|>{\centering\arraybackslash}m{32.1pt}|>{\centering\arraybackslash}m{32.1pt}|}
\multicolumn{15}{|>{\centering\arraybackslash}m{545.0pt}|}{\textbf{DF\textperiodcentered{}\textperiodcentered{}\textperiodcentered{}NH}\textbf{\textsubscript{3}}} \\* \hline\noalign{\penalty10000}
\multicolumn{1}{|>{\centering\arraybackslash}m{46.2pt}|}{MSE} & 0.0 & -0.070 & 0.016 & 0.056 & -0.033 & 0.012 & -0.001 & -0.013 & -0.007 & -0.016 & -0.066 & -0.013 & -0.006 & -0.005 \\* \hline\noalign{\penalty10000}
\multicolumn{1}{|>{\centering\arraybackslash}m{46.2pt}|}{MAD} & 0.0 & 0.070 & 0.016 & 0.056 & 0.038 & 0.014 & 0.012 & 0.014 & 0.013 & 0.017 & 0.066 & 0.014 & 0.015 & 0.015 \\* \hline\noalign{\penalty10000}
\multicolumn{1}{|>{\centering\arraybackslash}m{46.2pt}|}{RMSD} & 0.0 & 0.087 & 0.018 & 0.072 & 0.051 & 0.021 & 0.013 & 0.017 & 0.015 & 0.021 & 0.089 & 0.019 & 0.016 & 0.016 \\* \hline\noalign{\penalty10000}
\multicolumn{1}{|>{\centering\arraybackslash}m{46.2pt}|}{MAX} & 0.0 & -0.174 & 0.027 & 0.145 & -0.105 & 0.045 & -0.021 & -0.030 & -0.023 & -0.033 & -0.184 & -0.031 & -0.024 & -0.023 \\* \hline\noalign{\penalty10000}
\multicolumn{1}{|>{\centering\arraybackslash}m{46.2pt}|}{MA\%D} & 0.0 & 2.6 & 0.6 & 2.1 & 1.4 & 0.5 & \textbf{0.4} & 0.5 & 0.5 & 0.6 & 2.4 & 0.5 & 0.5 & 0.5 \\* \specialrule{0.9pt}{0pt}{0pt}\noalign{\penalty10000}
\end{tabular}}}}
\par\nopagebreak\vspace{1pt}{\setstretch{1}\setlength{\parskip}{1.5pt}\fontsize{9}{10.5}\selectfont\raggedright\noindent \textsuperscript{\#} NH\textsubscript{3}F\textsuperscript{+}\textperiodcentered{}\textperiodcentered{}\textperiodcentered{}CH\textsubscript{3} and NH\textsubscript{2}F\textsubscript{2}\textsuperscript{+}\textperiodcentered{}\textperiodcentered{}\textperiodcentered{}CH\textsubscript{3} complexes were optimized at CCSD(T)-F12b/VTZ level.\par}
\end{table}

\begin{figure}[!htbp]\centering
\GPKfig{7.04722in}{3.07222in}{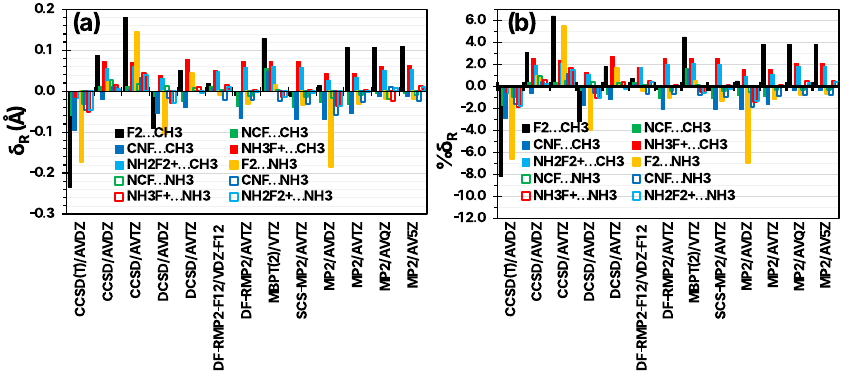}
\par\vspace{-1pt}\caption{Deviations {[}\ensuremath{\delta}\textsubscript{R} = R(ab initio) - R(CCSD(T)-F12b/VDZ-F12){]} \textbf{(a)} and percentage deviations on F-bond distances {[}\%\ensuremath{\delta}\textsubscript{R} = 100 \texttimes{} \ensuremath{\delta}\textsubscript{R}/R(CCSD(T)-F12b/VDZ-F12){]} \textbf{(b)} on F-bond distances for selected DF\textperiodcentered{}\textperiodcentered{}\textperiodcentered{}CH\textsubscript{3}/NH\textsubscript{3} complexes (DF = F\textsubscript{2}, NCF, CNF, NH\textsubscript{3}F\textsuperscript{+}, and NH\textsubscript{2}F\textsubscript{2}\textsuperscript{+}), calculated at several ab initio levels relative to CCSD(T)-F12b/VDZ-F12 (see text for details).}\label{fig:deviations-r-r-ab}
\vspace{10pt}
\end{figure}

\Tref{tab:statistical-errors-mse-mad} and \Fref{fig:deviations-r-r-ab} confirm that DF-RMP2-F12/VDZ-F12 is reliable for F-bond distances of the selected complexes (DF\textperiodcentered{}\textperiodcentered{}\textperiodcentered{}CH\textsubscript{3}: MSE/MAD/RMSD/MAX/MA\%D = 0.028/0.028/0.033/0.050 Å/0.9\%; DF\textperiodcentered{}\textperiodcentered{}\textperiodcentered{}NH\textsubscript{3}: -0.001/0.012/0.013/-0.021 Å/0.4\%), while CCSD(T)/AVDZ performs worst for both. For SEFB complexes, MP2/AVDZ and CCSD/AVDZ also perform well (MA\%D = 1.2\% and 1.7\%), with MP2/AVDZ slightly better than MP2/AVTZ (1.2\% vs. 1.8\%) (\Tref{tab:statistical-errors-mse-mad}). For CSFB complexes, however, MP2/AVDZ performs slightly inferior (MA\%D = 2.4\%), comparable to CCSD(T)/AVDZ (2.6\%), while CCSD/AVDZ and MP2/AVTZ remain reliable (0.6\% and 0.5\%) (\Tref{tab:statistical-errors-mse-mad}). F-bond distances from DF-RMP2-F12/VDZ-F12 are therefore adopted as reference for all 29 complexes, as a more affordable alternative to CCSD(T)-F12b/VDZ-F12.

\begin{figure}[!htbp]\centering
\GPKfig{7.01181in}{3.14792in}{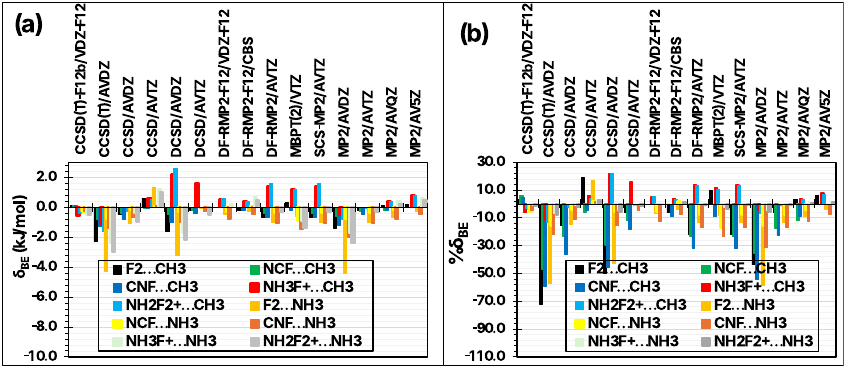}
\par\vspace{-1pt}\caption{Deviation {[}\ensuremath{\delta}\textsubscript{BE} = BE(ab initio) -- BE(CCSD(T)-F12b/CBS{[}V(T,Q)Z-F12{]}//CCSD(T)-F12b/VDZ-F12) \textbf{(a)} and percentage deviation {[}\%\ensuremath{\delta}BE = 100 \texttimes{} \ensuremath{\delta}\textsubscript{BE} / \textbar BE(CCSD(T)-F12b/CBS{[}V(T,Q)Z-F12{]}//CCSD(T)-F12b/VDZ-F12)\textbar{]} \textbf{(b)} on BEs for selected DF\textperiodcentered{}\textperiodcentered{}\textperiodcentered{}CH\textsubscript{3}/NH\textsubscript{3} complexes (DF = F\textsubscript{2}, NCF, CNF, NH\textsubscript{3}F\textsuperscript{+}, NH\textsubscript{2}F\textsubscript{2}\textsuperscript{+}), calculated at several ab initio methods relative to CCSD(T)-F12b/CBS{[}V(T,Q)Z-F12{]}//CCSD(T)-F12b/VDZ-F12 (see text for details).}\label{fig:deviation-be-be-ab}
\vspace{10pt}
\end{figure}

Deviations (\ensuremath{\delta}\textsubscript{R}) and percentage deviations (\%\ensuremath{\delta}\textsubscript{R}) on F-bond distances for the five selected complexes, relative to CCSD(T)-F12b/VDZ-F12, are shown in \Fref{fig:deviations-r-r-ab} (data in \Tref{tab:deviation-r-in-and}). F-bond distances from DF-RMP2-F12/VDZ-F12 are closest to the CCSD(T)-F12b/VDZ-F12 reference. Larger deviations occur for CCSD(T)/AVDZ, CCSD/AVTZ, DCSD/AVDZ and MP2/AVDZ (for F\textsubscript{2}\textperiodcentered{}\textperiodcentered{}\textperiodcentered{}NH\textsubscript{3}). \%\ensuremath{\delta}\textsubscript{R} exceeds 5.0\% for CCSD(T)/AVDZ and CCSD/AVTZ for both (F\textsubscript{2}\textperiodcentered{}\textperiodcentered{}\textperiodcentered{}CH\textsubscript{3}/NH\textsubscript{3}) complexes and MP2/AVDZ (F\textsubscript{2}\textperiodcentered{}\textperiodcentered{}\textperiodcentered{}NH\textsubscript{3}). CCSD/AVDZ outperforms CCSD/AVTZ for reproducing CCSD(T)-F12b bond lengths. Also, increasing the size of basis set from AVTZ to AVQZ, MP2 yields no further improvement, consistent with faster convergence of intermolecular distances with basis set size than BEs do, as noted by Riley et al.\cite{Riley2012} regarding suitability of triple-zeta basis sets in weakly bound NCIs. Deviations on BEs for the same set of selected complexes are discussed next (\Fref{fig:deviation-be-be-ab}).

Having established DF-RMP2-F12/VDZ-F12 as a reliable, affordable alternative to CCSD(T)-F12b for F-bond distances (\Fref{fig:deviations-r-r-ab}, \Tref{tab:statistical-errors-mse-mad}), we next check the reliability of the corresponding CBS-limit BEs. This is DF-RMP2-F12/CBS{[}V(T,Q)Z-F12{]}//DF-RMP2-F12/VDZ-F12, denoted DF-RMP2-F12/CBS* in \mbox{\textbf{TABLE \ref{tab:bes-kj-mol-of}.}} We compare it against CCSD(T)-F12b/CBS{[}V(T,Q)Z-F12{]}//CCSD(T)-F12b/VDZ-F12, denoted CCSD(T)-F12b/CBS. CBS BEs were computed at both geometries (using CCSD(T)-F12b/VTZ for NH\textsubscript{3}F\textsuperscript{+}\textperiodcentered{}\textperiodcentered{}\textperiodcentered{}CH\textsubscript{3} and NH\textsubscript{2}F\textsubscript{2}\textsuperscript{+}\textperiodcentered{}\textperiodcentered{}\textperiodcentered{}CH\textsubscript{3}, as in \mbox{\textbf{TABLEs \ref{tab:f-bond-distances-r}}} and \textbf{IV}). \Tref{tab:bes-kj-mol-of} shows the two CBS limit BEs are comparable. We then calculated statistical error analysis against CCSD(T)-F12b/CBS as reference, across CCSD(T)-F12b/VDZ-F12, CCSD(T)/AVDZ, CCSD/AVDZ, CCSD/AVTZ, DCSD/AVDZ, DCSD/AVTZ, DF-RMP2-F12/VDZ-F12, DF-RMP2-F12/CBS*, W1-F12 and W2-F12//DF-RMP2-F12/VDZ-F12,CCSD(T)-F12b/CBS{[}V(T,Q)Z-F12{]}//DF-RMP2-F12/VDZ-F12, DF-RMP2/AVTZ, MBPT(2)/VTZ, SCS-MP2/AVTZ, and MP2/AVDZ-AV5Z (\Tref{tab:mse-mad-rmsd-max}; deviations in \Fref{fig:deviation-be-be-ab}, \Tref{tab:percentage-deviation-be-and}).

\Fref{fig:deviation-be-be-ab} shows the largest deviations on BEs for F\textsubscript{2}\textperiodcentered{}\textperiodcentered{}\textperiodcentered{}CH\textsubscript{3} and F\textsubscript{2}\textperiodcentered{}\textperiodcentered{}\textperiodcentered{}NH\textsubscript{3} occur at CCSD(T)/AVDZ, DCSD/AVDZ, and MP2/AVDZ: \ensuremath{\delta}\textsubscript{BE}(\%\ensuremath{\delta}\textsubscript{BE}) = -2.3 kJ/mol (-71.9\%), -1.6 kJ/mol (-50.0\%) and -1.4 kJ/mol (-43.8\%) for F\textsubscript{2}\textperiodcentered{}\textperiodcentered{}\textperiodcentered{}CH\textsubscript{3}, versus -4.3 kJ/mol (-57.3\%), -3.2 kJ/mol (-42.7\%) and -4.4 kJ/mol (-58.7\%) for F\textsubscript{2}\textperiodcentered{}\textperiodcentered{}\textperiodcentered{}NH\textsubscript{3} (\mbox{\textbf{Table \ref{tab:percentage-deviation-be-and}}}). CCSD/AVDZ overestimates BEs of both complexes (by -15.6\% and -14.7\%) but CCSD/AVTZ instead under-binds both {[}0.6 kJ/mol (18.8\%) for F\textsubscript{2}\textperiodcentered{}\textperiodcentered{}\textperiodcentered{}CH\textsubscript{3} and 1.3 kJ/mol (17.3\%) for F\textsubscript{2}\textperiodcentered{}\textperiodcentered{}\textperiodcentered{}NH\textsubscript{3}{]}. For DCSD and MP2, increasing the basis set size from AVDZ to AVTZ results in sharp reduction in the error (\mbox{\textbf{Table \ref{tab:percentage-deviation-be-and}}}). Therefore, DCSD and MP2 with AVTZ basis set are the best affordable choices for BEs of these complexes.

\subsection{Cross-validation of reference BEs with CCSD(T)-F12b/CBS, W1-F12 and W2-F12}\label{sec:cross-validation-of-reference}

We have already discussed CCSD(T)/CBS BEs obtained via the W1-F12 composite method at DF-RMP2-F12/VDZ-F12 and M06-2X/AVTZ geometries in \Tref{tab:cbs-limit-binding-energies} and \mbox{\textbf{FIG. \ref{fig:bar-graphs-showing-bes}.}} To further confirm the reliability of BEs from DF-RMP2-F12/CBS{[}V(T,Q)-F12{]}//DF-RMP2-F12/VDZ-F12 level, we computed additional CBS BEs for five selected complexes DF\textperiodcentered{}\textperiodcentered{}\textperiodcentered{}CH\textsubscript{3}/NH\textsubscript{3} (DF = F\textsubscript{2}, NCF, CNF, NH\textsubscript{3}F\textsuperscript{+} and NH\textsubscript{2}F\textsubscript{2}\textsuperscript{+}) at both DF-RMP2-F12/VDZ-F12 and CCSD(T)-F12b/VDZ-F12 geometries (\Tref{tab:bes-kj-mol-of}). At the DF-RMP2-F12/VDZ-F12 geometry, we calculated BEs at DF-RMP2-F12/CBS*, W1-F12, W2-F12, and CCSD(T)-F12b/CBS** levels (refer footnotes in \Tref{tab:bes-kj-mol-of} for symbols * and **). At the CCSD(T)-F12b/VDZ-F12 geometry, we calculated BEs at CCSD(T)-F12b/CBS level. BEs obtained from CCSD(T)-F12b/CBS level are used as the reference values to benchmark several \emph{ab initio} methods for these complexes (\Tref{tab:mse-mad-rmsd-max}).

\begin{table}[!htbp]\centering\setlength{\parskip}{0pt}
\caption{BEs (kJ/\allowbreak{}mol) of selected DF\textperiodcentered{}\textperiodcentered{}\textperiodcentered{}CH\textsubscript{3}/NH\textsubscript{3} complexes (DF = F\textsubscript{2}, NCF, CNF, NH\textsubscript{3}F\textsuperscript{+} and NH\textsubscript{2}F\textsubscript{2}\textsuperscript{+}) at DF-RMP2-F12/\allowbreak{}VDZ-F12-optimized geometries (DF-RMP2-F12/\allowbreak{}VDZ-F12, DF-RMP2-F12/\allowbreak{}CBS, W1-F12, W2-F12, CCSD(T)-F12b/\allowbreak{}CBS) and at CCSD(T)-F12b/\allowbreak{}VDZ-F12-optimized geometries (CCSD(T)-F12b/\allowbreak{}VDZ-F12, CCSD(T)-F12b/\allowbreak{}CBS). CBS-limit BEs were obtained using the Schwenke-type two-point extrapolation scheme.}\label{tab:bes-kj-mol-of}\vspace{7pt}\nopagebreak
{\fontsize{10.0}{11.5}\fontfamily{ptm}\selectfont\renewcommand{\arraystretch}{1.17}\setlength{\tabcolsep}{1.5pt}
\centerline{\adjustbox{max width=518pt}{\begin{tabular}{|>{\centering\arraybackslash}m{92.7pt}|>{\centering\arraybackslash}m{73.5pt}|>{\centering\arraybackslash}m{55.5pt}|>{\centering\arraybackslash}m{33.0pt}|>{\centering\arraybackslash}m{33.0pt}|>{\centering\arraybackslash}m{64.5pt}|>{\centering\arraybackslash}m{73.4pt}|>{\centering\arraybackslash}m{53.7pt}|}
\specialrule{0.9pt}{0pt}{0pt}
\multicolumn{8}{|>{\centering\arraybackslash}m{500.2pt}|}{DF\textperiodcentered{}\textperiodcentered{}\textperiodcentered{}CH\textsubscript{3}} \\* \hline\noalign{\penalty10000}
\multicolumn{1}{|>{\centering\arraybackslash}m{92.7pt}|}{\multirow{2}{92.7pt}{\centering DF\textperiodcentered{}\textperiodcentered{}\textperiodcentered{}CH\textsubscript{3}}} & \multicolumn{5}{>{\centering\arraybackslash}m{271.5pt}|}{Optimization @DF-RMP2-F12/\allowbreak{}VDZ-F12} & \multicolumn{2}{>{\centering\arraybackslash}m{130.1pt}|}{Optimization@CCSD(T)-F12b/\allowbreak{}VDZ-F12} \\* \cline{2-8}\noalign{\penalty10000}
\multicolumn{1}{|>{\centering\arraybackslash}m{92.7pt}|}{} & DF-RMP2-F12/ VDZ-F12 & DF-RMP2-F12/\allowbreak{}CBS* & W1-F12 & W2-F12 & CCSD(T)-F12b/\allowbreak{}CBS** & CCSD(T)-F12b/ VDZ-F12 & CCSD(T)-F12b/\allowbreak{}CBS \\* \hline\noalign{\penalty10000}
\multicolumn{1}{|>{\centering\arraybackslash}m{92.7pt}|}{F\textsubscript{2}\textperiodcentered{}\textperiodcentered{}\textperiodcentered{}CH\textsubscript{3}} & -3.2 & -3.4 & -3.1 & -3.1 & -3.2 & -3.1 & \textbf{-3.2} \\* \hline\noalign{\penalty10000}
\multicolumn{1}{|>{\centering\arraybackslash}m{92.7pt}|}{NCF\textperiodcentered{}\textperiodcentered{}\textperiodcentered{}CH\textsubscript{3}} & -1.7 & -1.7 & -1.7 & -1.7 & -1.7 & -1.6 & \textbf{-1.7} \\* \hline\noalign{\penalty10000}
\multicolumn{1}{|>{\centering\arraybackslash}m{92.7pt}|}{CNF\textperiodcentered{}\textperiodcentered{}\textperiodcentered{}CH\textsubscript{3}} & -2.2 & -2.4 & -2.2 & -2.2 & -2.2 & -2.1 & \textbf{-2.2} \\* \hline\noalign{\penalty10000}
\multicolumn{1}{|>{\centering\arraybackslash}m{92.7pt}|}{NH\textsubscript{3}F+\textperiodcentered{}\textperiodcentered{}\textperiodcentered{}CH\textsubscript{3}} & -9.7 & -9.8 & -10.2 & -10.2 & -10.2 & -10.8 & \textbf{-10.2} \\ \hline
\multicolumn{1}{|>{\centering\arraybackslash}m{92.7pt}|}{NH\textsubscript{2}F\textsubscript{2}\textsuperscript{+}\textperiodcentered{}\textperiodcentered{}\textperiodcentered{}CH\textsubscript{3}} & -11.2 & -11.4 & -11.7 & -11.7 & -11.7 & -12.2 & \textbf{-11.8} \\ \specialrule{0.9pt}{0pt}{0pt}
\end{tabular}}}
\par\nointerlineskip
\centerline{\adjustbox{max width=518pt}{\begin{tabular}{|>{\centering\arraybackslash}m{92.7pt}|>{\centering\arraybackslash}m{73.5pt}|>{\centering\arraybackslash}m{55.5pt}|>{\centering\arraybackslash}m{33.0pt}|>{\centering\arraybackslash}m{33.0pt}|>{\centering\arraybackslash}m{64.5pt}|>{\centering\arraybackslash}m{73.4pt}|>{\centering\arraybackslash}m{53.7pt}|}
\multicolumn{1}{|>{\centering\arraybackslash}m{92.7pt}|}{\multirow{3}{92.7pt}{\centering DF\textperiodcentered{}\textperiodcentered{}\textperiodcentered{}NH\textsubscript{3}}} & \multicolumn{7}{>{\centering\arraybackslash}m{404.6pt}|}{DF\textperiodcentered{}\textperiodcentered{}\textperiodcentered{}NH\textsubscript{3}} \\* \cline{2-8}\noalign{\penalty10000}
\multicolumn{1}{|>{\centering\arraybackslash}m{92.7pt}|}{} & \multicolumn{5}{>{\centering\arraybackslash}m{271.5pt}|}{Optimization@DF-RMP2-F12/\allowbreak{}VDZ-F12} & \multicolumn{2}{>{\centering\arraybackslash}m{130.1pt}|}{Optimization@CCSD(T)-F12b/\allowbreak{}VDZ-F12} \\* \cline{2-8}\noalign{\penalty10000}
\multicolumn{1}{|>{\centering\arraybackslash}m{92.7pt}|}{} & DF-RMP2-F12/ VDZ-F12 & DF-RMP2-F12/\allowbreak{}CBS* & W1-F12 & W2-F12 & CCSD(T)-F12b/\allowbreak{}CBS** & CCSD(T)-F12b/ VDZ-F12 & CCSD(T)-F12b/\allowbreak{}CBS \\* \hline\noalign{\penalty10000}
\multicolumn{1}{|>{\centering\arraybackslash}m{92.7pt}|}{F\textsubscript{2}\textperiodcentered{}\textperiodcentered{}\textperiodcentered{}NH\textsubscript{3}} & -8.0 & -7.8 & -7.4 & -7.4 & -7.5 & -7.8 & \textbf{-7.5} \\* \hline\noalign{\penalty10000}
\multicolumn{1}{|>{\centering\arraybackslash}m{92.7pt}|}{NCF\textperiodcentered{}\textperiodcentered{}\textperiodcentered{}NH\textsubscript{3}} & -5.5 & -5.1 & -5.2 & -5.2 & -5.2 & -5.5 & \textbf{-5.2} \\* \hline\noalign{\penalty10000}
\multicolumn{1}{|>{\centering\arraybackslash}m{92.7pt}|}{CNF\textperiodcentered{}\textperiodcentered{}\textperiodcentered{}NH\textsubscript{3}} & -7.2 & -6.9 & -6.4 & -6.3 & -6.4 & -6.7 & \textbf{-6.4} \\* \hline\noalign{\penalty10000}
\multicolumn{1}{|>{\centering\arraybackslash}m{92.7pt}|}{NH\textsubscript{3}F+\textperiodcentered{}\textperiodcentered{}\textperiodcentered{}NH\textsubscript{3}} & -34.4 & -33.9 & -34.6 & -34.5 & -34.6 & -35.1 & \textbf{-34.6} \\* \hline\noalign{\penalty10000}
\multicolumn{1}{|>{\centering\arraybackslash}m{92.7pt}|}{NH\textsubscript{2}F\textsubscript{2}\textsuperscript{+}\textperiodcentered{}\textperiodcentered{}\textperiodcentered{}NH\textsubscript{3}} & -39.3 & -38.8 & -39.3 & -39.3 & -39.3 & -39.8 & \textbf{-39.3} \\* \specialrule{0.9pt}{0pt}{0pt}\noalign{\penalty10000}
\end{tabular}}}}
\par\nopagebreak\vspace{1pt}{\setstretch{1}\setlength{\parskip}{1.5pt}\fontsize{9}{10.5}\selectfont\raggedright\noindent * Complete basis set limit BEs calculated at DF-RMP2-F12/CBS{[}V(T,Q)Z-F12{]}//DF-RMP2-F12/VDZ-F12 level of theory, using Schwenke-type two-point extrapolation scheme.\par **Complete basis set limit BEs calculated at CCSD(T)-F12b/CBS{[}V(T,Q)Z-F12{]}//DF-RMP2-F12/VDZ-F12 level of theory, using Schwenke-type extrapolation.\par}
\end{table}

W1-F12 and W2-F12 (non-relativistic, valence-electron) levels produce identical BEs for the selected DF\textperiodcentered{}\textperiodcentered{}\textperiodcentered{}CH\textsubscript{3} and DF\textperiodcentered{}\textperiodcentered{}\textperiodcentered{}NH\textsubscript{3} complexes. DF-RMP2-F12/CBS* is likewise close to both. This agreement across independent CBS-extrapolation and composite methods supports treating DF-RMP2-F12/CBS as a trustworthy reference BE. We will test this conclusion more rigorously against 15 other \emph{ab initio} methods next (\Tref{tab:mse-mad-rmsd-max}).

\begin{table}[!htbp]\centering\setlength{\parskip}{0pt}
\caption{MSE, MAD, RMSD, MAX, and MA\%D on BEs (kJ/\allowbreak{}mol) for selected DF\textperiodcentered{}\textperiodcentered{}\textperiodcentered{}CH\textsubscript{3}/NH\textsubscript{3} complexes (DF = F\textsubscript{2}, NCF, CNF, NH\textsubscript{3}F\textsuperscript{+} and NH\textsubscript{2}F\textsubscript{2}\textsuperscript{+}), calculated using several ab initio methods relative to CCSD(T)-F12b/\allowbreak{}CBS[V(T,Q)Z-F12]//CCSD(T)-F12b/\allowbreak{}VDZ-F12 (denoted CCSD(T)-F12b/\allowbreak{}CBS). DF-RMP2-F12/\allowbreak{}CBS and CCSD(T)-F12b/\allowbreak{}CBS values were obtained using the Schwenke-type extrapolation scheme. Deviations and percentage deviations on BEs relative to CCSD(T)-F12b/\allowbreak{}CBS//CCSD(T)-F12b/\allowbreak{}VDZ-F12 and to CCSD(T)-F12b/\allowbreak{}VDZ-F12 are given in \Tref{tab:percentage-deviation-be-and} and \Tref{tab:percentage-deviation-be-and2}, respectively.\textsuperscript{\#}}\label{tab:mse-mad-rmsd-max}\vspace{7pt}\nopagebreak
{\fontsize{9.0}{10.3}\fontfamily{ptm}\selectfont\renewcommand{\arraystretch}{1.17}\setlength{\tabcolsep}{1.5pt}
\centerline{\adjustbox{max width=518pt}{\begin{tabular}{|>{\centering\arraybackslash}p{39.5pt}|>{\centering\arraybackslash}p{25.4pt}|>{\centering\arraybackslash}p{32.5pt}|>{\centering\arraybackslash}p{32.5pt}|>{\centering\arraybackslash}p{25.4pt}|>{\centering\arraybackslash}p{39.5pt}|>{\centering\arraybackslash}p{25.4pt}|>{\centering\arraybackslash}p{25.4pt}|>{\centering\arraybackslash}p{28.4pt}|>{\centering\arraybackslash}p{29.4pt}|>{\centering\arraybackslash}p{25.4pt}|>{\centering\arraybackslash}p{25.4pt}|>{\centering\arraybackslash}p{25.4pt}|>{\centering\arraybackslash}p{35.0pt}|>{\centering\arraybackslash}p{22.5pt}|>{\centering\arraybackslash}p{25.4pt}|>{\centering\arraybackslash}p{25.6pt}|}
\specialrule{0.9pt}{0pt}{0pt}
\multicolumn{1}{|>{\centering\arraybackslash}m{39.5pt}|}{\multirow{2}{39.5pt}{\centering \makebox[\linewidth][c]{\rotatebox{90}{\hspace{2pt}\textbf{Statistical Errors on BEs}\hspace{2pt}}}}} & \multicolumn{1}{>{\centering\arraybackslash}m{25.4pt}|}{\multirow{2}{25.4pt}{\centering \makebox[\linewidth][c]{\rotatebox{90}{\hspace{2pt}CCSD(T)-F12b/\allowbreak{}VDZ-F12\hspace{2pt}}}}} & \multicolumn{1}{>{\centering\arraybackslash}m{32.5pt}|}{\multirow{2}{32.5pt}{\centering \makebox[\linewidth][c]{\rotatebox{90}{\hspace{2pt}CCSD(T)/\allowbreak{}AVDZ\hspace{2pt}}}}} & \multicolumn{1}{>{\centering\arraybackslash}m{32.5pt}|}{\multirow{2}{32.5pt}{\centering \makebox[\linewidth][c]{\rotatebox{90}{\hspace{2pt}CCSD/\allowbreak{}AVDZ\hspace{2pt}}}}} & \multicolumn{1}{>{\centering\arraybackslash}m{25.4pt}|}{\multirow{2}{25.4pt}{\centering \makebox[\linewidth][c]{\rotatebox{90}{\hspace{2pt}CCSD/\allowbreak{}AVTZ\hspace{2pt}}}}} & \multicolumn{1}{>{\centering\arraybackslash}m{39.5pt}|}{\multirow{2}{39.5pt}{\centering \makebox[\linewidth][c]{\rotatebox{90}{\hspace{2pt}DCSD/\allowbreak{}AVDZ\hspace{2pt}}}}} & \multicolumn{1}{>{\centering\arraybackslash}m{25.4pt}|}{\multirow{2}{25.4pt}{\centering \makebox[\linewidth][c]{\rotatebox{90}{\hspace{2pt}DCSD/\allowbreak{}AVTZ\hspace{2pt}}}}} & \multicolumn{3}{>{\centering\arraybackslash}m{89.2pt}|}{opt@DF-RMP2- F12/\allowbreak{}VDZ-F12} & \multicolumn{1}{>{\centering\arraybackslash}m{25.4pt}|}{\multirow{2}{25.4pt}{\centering \makebox[\linewidth][c]{\rotatebox{90}{\hspace{2pt}DF-RMP2/\allowbreak{}AVTZ\hspace{2pt}}}}} & \multicolumn{1}{>{\centering\arraybackslash}m{25.4pt}|}{\multirow{2}{25.4pt}{\centering \makebox[\linewidth][c]{\rotatebox{90}{\hspace{2pt}MBPT(2)/\allowbreak{}VTZ\hspace{2pt}}}}} & \multicolumn{1}{>{\centering\arraybackslash}m{25.4pt}|}{\multirow{2}{25.4pt}{\centering \makebox[\linewidth][c]{\rotatebox{90}{\hspace{2pt}SCS-MP2/\allowbreak{}AVTZ\hspace{2pt}}}}} & \multicolumn{1}{>{\centering\arraybackslash}m{35.0pt}|}{\multirow{2}{35.0pt}{\centering \makebox[\linewidth][c]{\rotatebox{90}{\hspace{2pt}MP2/\allowbreak{}AVDZ\hspace{2pt}}}}} & \multicolumn{1}{>{\centering\arraybackslash}m{22.5pt}|}{\multirow{2}{22.5pt}{\centering \makebox[\linewidth][c]{\rotatebox{90}{\hspace{2pt}MP2/\allowbreak{}AVTZ\hspace{2pt}}}}} & \multicolumn{1}{>{\centering\arraybackslash}m{25.4pt}|}{\multirow{2}{25.4pt}{\centering \makebox[\linewidth][c]{\rotatebox{90}{\hspace{2pt}MP2/\allowbreak{}AVQZ\hspace{2pt}}}}} & \multicolumn{1}{>{\centering\arraybackslash}m{25.6pt}|}{\multirow{2}{25.6pt}{\centering \makebox[\linewidth][c]{\rotatebox{90}{\hspace{2pt}MP2/\allowbreak{}AV5Z\hspace{2pt}}}}} \\* \cline{8-10}\noalign{\penalty10000}
\multicolumn{1}{|>{\centering\arraybackslash}p{39.5pt}|}{} &  &  &  &  &  &  & \multicolumn{1}{>{\centering\arraybackslash}p{25.4pt}|}{\makebox[\linewidth][c]{\rotatebox{90}{\hspace{2pt}\textbf{DF-RMP2F12/} \textbf{CBS}\textbf{*}\hspace{2pt}}}} & \multicolumn{1}{>{\centering\arraybackslash}p{28.4pt}|}{\makebox[\linewidth][c]{\rotatebox{90}{\hspace{2pt}W1-F12\textsuperscript{\$}\hspace{2pt}}}} & \multicolumn{1}{>{\centering\arraybackslash}p{29.4pt}|}{\makebox[\linewidth][c]{\rotatebox{90}{\hspace{2pt}CCSD(T)-F12b/ CBS**\hspace{2pt}}}} &  &  &  &  &  &  &  \\* \hline\noalign{\penalty10000}
\multicolumn{17}{|>{\centering\arraybackslash}p{535.7pt}|}{\textbf{DF\textperiodcentered{}\textperiodcentered{}\textperiodcentered{}CH}\textbf{\textsubscript{3}}} \\* \hline\noalign{\penalty10000}
\multicolumn{1}{|>{\centering\arraybackslash}p{39.5pt}|}{MSE} & -0.1 & -1.4 & -0.4 & 0.3 & 0.3 & 0.2 & \textbf{0.1} & 0.0 & 0.0 & 0.2 & 0.5 & 0.2 & -0.9 & -0.2 & 0.1 & 0.4 \\* \hline\noalign{\penalty10000}
\multicolumn{1}{|>{\centering\arraybackslash}p{39.5pt}|}{MAD} & 0.3 & 1.4 & 0.4 & 0.4 & 1.6 & 0.6 & \textbf{0.2} & 0.0 & 0.0 & 1.0 & 0.6 & 1.0 & 0.9 & 0.2 & 0.3 & 0.4 \\* \hline\noalign{\penalty10000}
\multicolumn{1}{|>{\centering\arraybackslash}p{39.5pt}|}{RMSD} & 0.3 & 1.5 & 0.5 & 0.4 & 1.8 & 0.8 & \textbf{0.3} & 0.1 & 0.0 & 1.1 & 0.8 & 1.1 & 1.0 & 0.3 & 0.3 & 0.5 \\* \hline\noalign{\penalty10000}
\multicolumn{1}{|>{\centering\arraybackslash}p{39.5pt}|}{MAX} & -0.6 & -2.3 & -0.8 & 0.6 & 2.6 & 1.6 & \textbf{0.4} & 0.1 & 0.1 & 1.6 & 1.2 & 1.6 & -1.4 & -0.5 & -0.2 & 0.8 \\* \hline\noalign{\penalty10000}
\multicolumn{1}{|>{\centering\arraybackslash}p{39.5pt}|}{MA\%D} & 4.6 & 40.7 & 15.6 & 7.9 & -33.7 & 13.0 & \textbf{4.5} & 0.8 & 0.2 & 20.9 & 8.1 & 20.9 & 29.1 & 9.5 & 6.3 & 4.2 \\ \specialrule{0.9pt}{0pt}{0pt}
\end{tabular}}}
\par\nointerlineskip
\centerline{\adjustbox{max width=518pt}{\begin{tabular}{|>{\centering\arraybackslash}p{39.5pt}|>{\centering\arraybackslash}p{25.4pt}|>{\centering\arraybackslash}p{32.5pt}|>{\centering\arraybackslash}p{32.5pt}|>{\centering\arraybackslash}p{25.4pt}|>{\centering\arraybackslash}p{39.5pt}|>{\centering\arraybackslash}p{25.4pt}|>{\centering\arraybackslash}p{25.4pt}|>{\centering\arraybackslash}p{28.4pt}|>{\centering\arraybackslash}p{29.4pt}|>{\centering\arraybackslash}p{25.4pt}|>{\centering\arraybackslash}p{25.4pt}|>{\centering\arraybackslash}p{25.4pt}|>{\centering\arraybackslash}p{35.0pt}|>{\centering\arraybackslash}p{22.5pt}|>{\centering\arraybackslash}p{25.4pt}|>{\centering\arraybackslash}p{25.6pt}|}
\multicolumn{17}{|>{\centering\arraybackslash}m{535.7pt}|}{\textbf{DF\textperiodcentered{}\textperiodcentered{}\textperiodcentered{}NH}\textbf{\textsubscript{3}}} \\* \hline\noalign{\penalty10000}
\multicolumn{1}{|>{\centering\arraybackslash}p{39.5pt}|}{MSE} & -0.4 & -2.4 & -0.7 & 0.7 & -1.7 & -0.2 & \textbf{0.1} & 0.0 & 0.0 & -0.6 & -1.1 & -0.6 & -2.3 & -0.6 & -0.2 & 0.1 \\* \hline\noalign{\penalty10000}
\multicolumn{1}{|>{\centering\arraybackslash}p{39.5pt}|}{MAD} & 0.4 & 2.4 & 0.7 & 0.7 & 1.7 & 0.2 & \textbf{0.4} & 0.0 & 0.0 & 0.6 & 1.1 & 0.6 & 2.3 & 0.6 & 0.5 & 0.4 \\* \hline\noalign{\penalty10000}
\multicolumn{1}{|>{\centering\arraybackslash}p{39.5pt}|}{RMSD} & 0.4 & 2.7 & 0.8 & 0.9 & 2.0 & 0.3 & \textbf{0.5} & 0.0 & 0.0 & 0.7 & 1.1 & 0.7 & 2.6 & 0.7 & 0.5 & 0.4 \\* \hline\noalign{\penalty10000}
\multicolumn{1}{|>{\centering\arraybackslash}p{39.5pt}|}{MAX} & -0.5 & -4.3 & -1.1 & 1.3 & -3.2 & -0.5 & \textbf{0.7} & 0.1 & 0.0 & -1.1 & -1.5 & -1.1 & -4.4 & -1.1 & -0.8 & 0.6 \\* \hline\noalign{\penalty10000}
\multicolumn{1}{|>{\centering\arraybackslash}p{39.5pt}|}{MA\%D} & 3.4 & 21.8 & 6.6 & 5.4 & 14.9 & 1.5 & \textbf{3.4} & 0.3 & 0.0 & 7.8 & 11.0 & 7.8 & 23.3 & 7.8 & 5.5 & 3.0 \\* \specialrule{0.9pt}{0pt}{0pt}\noalign{\penalty10000}
\end{tabular}}}}
\par\nopagebreak\vspace{1pt}{\setstretch{1}\setlength{\parskip}{1.5pt}\fontsize{9}{10.5}\selectfont\raggedright\noindent \textsuperscript{\#}Optimization of NH\textsubscript{3}F\textsuperscript{+}\textperiodcentered{}\textperiodcentered{}\textperiodcentered{}CH\textsubscript{3} and NH\textsubscript{2}F\textsubscript{2}\textsuperscript{+}\textperiodcentered{}\textperiodcentered{}\textperiodcentered{}CH\textsubscript{3} was carried out at CCSD(T)-F12b/VTZ level of theory.\par \textsuperscript{*,**}Refer to footnote in \Tref{tab:bes-kj-mol-of}.\par \textsuperscript{\$}Non-relativistic, valence-electron CCSD(T)/CBS BEs calculated at DF-RMP2-F12/VDZ-F12 optimized geometries.\par}
\end{table}

MAD (MA\%D) on BEs for the five selected DF\textperiodcentered{}\textperiodcentered{}\textperiodcentered{}CH\textsubscript{3} and DF\textperiodcentered{}\textperiodcentered{}\textperiodcentered{}NH\textsubscript{3} complexes at DF-RMP2-F12/CBS* relative to CCSD(T)-F12b/CBS are 0.2 kJ/mol (4.5\%) and 0.4 kJ/mol (3.4\%), respectively (\Tref{tab:mse-mad-rmsd-max}). Also, from \mbox{\textbf{TABLEs \ref{tab:bes-kj-mol-of}}} and \textbf{VI}, one can see that CCSD(T)-F12b/VDZ-F12, DF-RMP2-F12/CBS* and MP2/AV5Z are comparable for reproducing CCSD(T)-F12b/CBS BEs for the selected SEFB and CSFB complexes. The remaining methods in \Tref{tab:mse-mad-rmsd-max}, calculated with fixed conventional basis sets, show larger deviations, reaching MA\%D of 40.7\% (DF\textperiodcentered{}\textperiodcentered{}\textperiodcentered{}CH\textsubscript{3}) and 21.8\% (DF\textperiodcentered{}\textperiodcentered{}\textperiodcentered{}NH\textsubscript{3}) for CCSD(T)/AVDZ, which is also the less reliable method for F-bond distances (\Sref{sec:reliability-of-reference-bes}, \Tref{tab:statistical-errors-mse-mad}).

Thus, one can safely consider BEs obtained at DF-RMP2-F12/{[}CBS(T,Q)Z-F12{]}//DF-RMP2-F12/VDZ-F12 level as reliable values for both types of complexes DF\textperiodcentered{}\textperiodcentered{}\textperiodcentered{}CH\textsubscript{3} and DF\textperiodcentered{}\textperiodcentered{}\textperiodcentered{}NH\textsubscript{3}. This conclusion is consistent with the comparison across the full data set (\Tref{tab:cbs-limit-binding-energies} and \Fref{fig:bar-graphs-showing-bes}), where DF-RMP2-F12/CBS agrees with three independent CBS-limit BE estimates {[}W1-F12//DF-RMP2-F12/VDZ-F12, CCSD(T)/CBS//MP2/AVDZ (Hobza method) and W1-F12//M06-2X/AVTZ{]} for most of the SEFB-15 and CSFB-14 complexes, not only the five selected F-bond donors here (F\textsubscript{2}, NCF, CNF, NH\textsubscript{3}F\textsuperscript{+} and NH\textsubscript{2}F\textsubscript{2}\textsuperscript{+}). Due to its affordability for all the complexes DF\textperiodcentered{}\textperiodcentered{}\textperiodcentered{}CH\textsubscript{3}/NH\textsubscript{3} in this paper, we have therefore considered BEs calculated at DF-RMP2-F12/{[}CBS(T,Q)Z-F12{]}//DF-RMP2-F12/VDZ-F12 (denoted as DF-RMP2-F12/CBS in \Fref{fig:deviation-be-be-ab}) throughout. This is consistent with the established use of DF-RMP2-F12 to generate high-quality reference data for various noncovalent bonding, as noted in \Sref{sec:computational-details}.

\FloatBarrier
\subsection{All-electron correlation and relativistic correction in binding energy of DF\textperiodcentered{}\textperiodcentered{}\textperiodcentered{}CH\textsubscript{3}/NH\textsubscript{3} within W1-F12 protocol}\label{sec:all-electron-correlation-and}

Having validated the non-relativistic, valence-electron CCSD(T)/CBS BEs from W1-F12/W2-F12 (TABLE \textbf{V}), we now check whether all-electron correlation or relativistic effects change these BEs. In \Tref{tab:ccsd-t-cbs-bes}, we compare BEs with relativistic and non-relativistic treatments for the SEFB-15 and CSFB-14 complexes where W1-F12 converged (W2-F12 was not feasible for most of the complexes, except for the five selected complexes reported in \Tref{tab:bes-kj-mol-of}).

\begin{table}[!htbp]\centering\setlength{\parskip}{0pt}
\caption{CCSD(T)/CBS BEs (kJ/mol) for DF\textperiodcentered{}\textperiodcentered{}\textperiodcentered{}CH\textsubscript{3}/NH\textsubscript{3} complexes at three levels of treatment: non-relativistic, valence-electron (N-R, val-e); non-relativistic, all-electron (N-R, all-e); and relativistic, all-electron (Rel, all-e). All values were obtained via W1-F12 theory using structures optimized at DF-RMP2-F12/VDZ-F12 and M06-2X/AVTZ.}\label{tab:ccsd-t-cbs-bes}\vspace{7pt}\nopagebreak
{\fontsize{9.0}{10.3}\fontfamily{ptm}\selectfont\renewcommand{\arraystretch}{1.17}\setlength{\tabcolsep}{1.5pt}
\centerline{\adjustbox{max width=518pt}{\begin{tabular}{|>{\centering\arraybackslash}m{67.9pt}|>{\centering\arraybackslash}m{26.9pt}|>{\centering\arraybackslash}m{28.2pt}|>{\centering\arraybackslash}m{25.6pt}|>{\centering\arraybackslash}m{27.9pt}|>{\centering\arraybackslash}m{27.9pt}|>{\centering\arraybackslash}m{26.4pt}|>{\centering\arraybackslash}m{67.9pt}|>{\centering\arraybackslash}m{28.9pt}|>{\centering\arraybackslash}b{27.9pt}|>{\centering\arraybackslash}m{29.3pt}|>{\centering\arraybackslash}m{27.9pt}|>{\centering\arraybackslash}m{28.9pt}|>{\centering\arraybackslash}m{26.9pt}|}
\specialrule{0.9pt}{0pt}{0pt}
\multicolumn{7}{|>{\centering\arraybackslash}p{248.8pt}|}{DF\textperiodcentered{}\textperiodcentered{}\textperiodcentered{}CH\textsubscript{3}} & \multicolumn{7}{>{\centering\arraybackslash}p{255.8pt}|}{DF\textperiodcentered{}\textperiodcentered{}\textperiodcentered{}NH\textsubscript{3}} \\* \hline\noalign{\penalty10000}
\multicolumn{1}{|>{\centering\arraybackslash}m{67.9pt}|}{\multirow{2}{67.9pt}{\centering DF\textperiodcentered{}\textperiodcentered{}\textperiodcentered{}CH\textsubscript{3}}} & \multicolumn{3}{>{\centering\arraybackslash}m{86.7pt}|}{W1-F12//DF-RMP2-F12/\allowbreak{}VDZ-F12} & \multicolumn{3}{>{\centering\arraybackslash}m{88.2pt}|}{W1-F12//M06-2X/ AVTZ} & \multirow{2}{67.9pt}{\centering DF\textperiodcentered{}\textperiodcentered{}\textperiodcentered{}NH\textsubscript{3}} & \multicolumn{3}{>{\centering\arraybackslash}m{92.1pt}|}{W1-F12//DF-RMP2-F12/\allowbreak{}VDZ-F12} & \multicolumn{3}{>{\centering\arraybackslash}m{89.7pt}|}{W1-F12//M06-2X/ AVTZ} \\* \cline{2-7}\cline{9-14}\noalign{\penalty10000}
\multicolumn{1}{|>{\centering\arraybackslash}m{67.9pt}|}{} & N-R, val-e & N-R, all-e & Rel, all-e & N-R, val-e & N-R, all-e & Rel, all-e &  & N-R, val-e & N-R, all-e & Rel, all-e & N-R, val-e & N-R, all-e & Rel, all-e \\* \hline\noalign{\penalty10000}
\multicolumn{1}{|>{\centering\arraybackslash}m{67.9pt}|}{F\textsubscript{2}\textperiodcentered{}\textperiodcentered{}\textperiodcentered{}CH\textsubscript{3}} & -3.1 & -3.1 & -3.1 & -3.2 & -3.1 & -3.2 & F\textsubscript{2}\textperiodcentered{}\textperiodcentered{}\textperiodcentered{}NH\textsubscript{3} & -7.4 & -7.4 & -7.5 & -7.3 & -7.3 & -7.4 \\* \hline\noalign{\penalty10000}
\multicolumn{1}{|>{\centering\arraybackslash}m{67.9pt}|}{SF\textsubscript{6}\textperiodcentered{}\textperiodcentered{}\textperiodcentered{}CH\textsubscript{3}} & N/\allowbreak{}A & N/\allowbreak{}A & N/\allowbreak{}A & N/\allowbreak{}A & N/\allowbreak{}A & N/\allowbreak{}A & SF\textsubscript{6}\textperiodcentered{}\textperiodcentered{}\textperiodcentered{}NH\textsubscript{3} & -1.7 & -1.6 & -1.7 & -1.6 & -1.6 & -1.6 \\* \hline\noalign{\penalty10000}
\multicolumn{1}{|>{\centering\arraybackslash}m{67.9pt}|}{NCF\textperiodcentered{}\textperiodcentered{}\textperiodcentered{}CH\textsubscript{3}} & -1.7 & -1.7 & -1.7 & -0.8 & -0.8 & -0.9 & NCF\textperiodcentered{}\textperiodcentered{}\textperiodcentered{}NH\textsubscript{3} & -5.2 & -5.2 & -5.2 & -5.2 & -5.2 & -5.2 \\* \hline\noalign{\penalty10000}
\multicolumn{1}{|>{\centering\arraybackslash}m{67.9pt}|}{CNF\textperiodcentered{}\textperiodcentered{}\textperiodcentered{}CH\textsubscript{3}} & -2.2 & -2.2 & -2.2 & -1.7 & -1.7 & -1.8 & CNF\textperiodcentered{}\textperiodcentered{}\textperiodcentered{}NH\textsubscript{3} & -6.4 & -6.4 & -6.5 & -6.4 & -6.4 & -6.5 \\ \hline
\multicolumn{1}{|>{\centering\arraybackslash}m{67.9pt}|}{NF\textsubscript{3}\textperiodcentered{}\textperiodcentered{}\textperiodcentered{}CH\textsubscript{3}} & -1.7 & -1.7 & -1.7 & -0.4 & -0.4 & -0.5 & NF\textsubscript{3}\textperiodcentered{}\textperiodcentered{}\textperiodcentered{}NH\textsubscript{3} & -2.0 & -2.0 & -2.0 & -1.6 & -1.6 & -1.6 \\ \hline
\multicolumn{1}{|>{\centering\arraybackslash}m{67.9pt}|}{ClF\textsubscript{3}\textperiodcentered{}\textperiodcentered{}\textperiodcentered{}CH\textsubscript{3}} & -2.4 & -2.4 & -2.5 & -1.5 & -1.5 & -1.5 & ClF\textsubscript{3}\textperiodcentered{}\textperiodcentered{}\textperiodcentered{}NH\textsubscript{3} & -4.5 & -4.5 & -4.5 & -4.5 & -4.5 & -4.5 \\ \hline
\multicolumn{1}{|>{\centering\arraybackslash}m{67.9pt}|}{OF\textsubscript{2}\textperiodcentered{}\textperiodcentered{}\textperiodcentered{}CH\textsubscript{3}} & -1.9 & -1.9 & -1.9 & -1.5 & -1.5 & -1.5 & OF\textsubscript{2}\textperiodcentered{}\textperiodcentered{}\textperiodcentered{}NH\textsubscript{3} & -3.3 & -3.3 & -3.3 & -3.3 & -3.2 & -3.3 \\ \hline
\multicolumn{1}{|>{\centering\arraybackslash}m{67.9pt}|}{NF\textsubscript{5}\textperiodcentered{}\textperiodcentered{}\textperiodcentered{}CH\textsubscript{3}} & N/\allowbreak{}A & N/\allowbreak{}A & N/\allowbreak{}A & N/\allowbreak{}A & N/\allowbreak{}A & N/\allowbreak{}A & NF\textsubscript{5}\textperiodcentered{}\textperiodcentered{}\textperiodcentered{}NH\textsubscript{3} & -7.8 & -7.8 & -7.9 & -7.7 & -7.7 & -7.8 \\ \hline
\multicolumn{1}{|>{\centering\arraybackslash}m{67.9pt}|}{FCCF\textperiodcentered{}\textperiodcentered{}\textperiodcentered{}CH\textsubscript{3}} & -1.5 & -1.5 & -1.5 & 0.2 & 0.2 & 0.1 & FCCF\textperiodcentered{}\textperiodcentered{}\textperiodcentered{}NH\textsubscript{3} & -0.8 & -0.8 & -0.8 & -0.7 & -0.7 & -0.7 \\ \hline
\multicolumn{1}{|>{\centering\arraybackslash}m{67.9pt}|}{CF\textsubscript{3}CCF\textperiodcentered{}\textperiodcentered{}\textperiodcentered{}CH\textsubscript{3}} & -1.4 & -1.4 & -1.5 & N/\allowbreak{}A & N/\allowbreak{}A & N/\allowbreak{}A & CF\textsubscript{3}CCF\textperiodcentered{}\textperiodcentered{}\textperiodcentered{}NH\textsubscript{3} & -3.7 & -3.6 & -3.7 & -3.5 & -3.5 & -3.5 \\ \hline
\multicolumn{1}{|>{\centering\arraybackslash}m{67.9pt}|}{CF\textsubscript{3}OF\textperiodcentered{}\textperiodcentered{}\textperiodcentered{}CH\textsubscript{3}} & N/\allowbreak{}A & N/\allowbreak{}A & N/\allowbreak{}A & -2.2 & -2.2 & -2.2 & CF\textsubscript{3}OF\textperiodcentered{}\textperiodcentered{}\textperiodcentered{}NH\textsubscript{3} & -4.1 & -4.1 & -4.1 & -3.9 & -3.9 & -4.0 \\ \hline
\multicolumn{1}{|>{\centering\arraybackslash}m{67.9pt}|}{F\textsubscript{3}NO\textperiodcentered{}\textperiodcentered{}\textperiodcentered{}CH\textsubscript{3}} & -2.1 & -2.0 & -2.1 & -1.3 & -1.3 & -1.3 & F\textsubscript{3}NO\textperiodcentered{}\textperiodcentered{}\textperiodcentered{}NH\textsubscript{3} & b & b & b & -3.0 & -2.9 & -3.0 \\ \hline
\multicolumn{1}{|>{\centering\arraybackslash}m{67.9pt}|}{SF\textsubscript{5}\textperiodcentered{}\textperiodcentered{}\textperiodcentered{}CH\textsubscript{3}} & N/\allowbreak{}A & N/\allowbreak{}A & N/\allowbreak{}A & N/\allowbreak{}A & N/\allowbreak{}A & N/\allowbreak{}A & SF\textsubscript{5}\textperiodcentered{}\textperiodcentered{}\textperiodcentered{}NH\textsubscript{3} & N/\allowbreak{}A & N/\allowbreak{}A & N/\allowbreak{}A & N/\allowbreak{}A & N/\allowbreak{}A & N/\allowbreak{}A \\* \hline\noalign{\penalty10000}
\multicolumn{1}{|>{\centering\arraybackslash}m{67.9pt}|}{NH\textsubscript{3}F+\textperiodcentered{}\textperiodcentered{}\textperiodcentered{}CH\textsubscript{3}} & -10.2 & -10.1 & -10.2 & -9.5 & -9.3 & -9.5 & NH\textsubscript{3}F+\textperiodcentered{}\textperiodcentered{}\textperiodcentered{}NH\textsubscript{3} & -34.6 & -34.5 & -34.6 & -34.5 & -34.4 & -34.6 \\* \hline\noalign{\penalty10000}
\multicolumn{1}{|>{\centering\arraybackslash}m{67.9pt}|}{NH\textsubscript{2}F\textsubscript{2}\textsuperscript{+}\textperiodcentered{}\textperiodcentered{}\textperiodcentered{}CH\textsubscript{3}} & -11.7 & -11.6 & -11.7 & -11.0 & -10.9 & -11.0 & NH\textsubscript{2}F\textsubscript{2}\textsuperscript{+}\textperiodcentered{}\textperiodcentered{}\textperiodcentered{}NH\textsubscript{3} & -39.3 & -39.2 & -39.4 & -39.3 & -39.2 & -39.4 \\* \specialrule{0.9pt}{0pt}{0pt}\noalign{\penalty10000}
\end{tabular}}}}
\par\nopagebreak\vspace{1pt}{\setstretch{1}\setlength{\parskip}{1.5pt}\fontsize{9}{10.5}\selectfont\raggedright\noindent \textsuperscript{b}Convergence error during optimization of respective F-bonded complexes.\par \textsuperscript{N/A}W1-F12 calculations failed to converge\par}
\end{table}

From \Tref{tab:ccsd-t-cbs-bes}, it can be observed that all-electron consideration has almost no effect on the CCSD(T)/CBS BEs within the W1-F12 protocol for F-bonded complexes DF\textperiodcentered{}\textperiodcentered{}\textperiodcentered{}CH\textsubscript{3}/NH\textsubscript{3}. The maximum difference between valence-electron, non-relativistic CCSD(T)/CBS BEs (denoted as ``NR, val-e'' in \Tref{tab:ccsd-t-cbs-bes}) and those obtained from all-electron, non-relativistic CCSD(T)/CBS level (denoted as ``N-R, all-e'' in \Tref{tab:ccsd-t-cbs-bes}) of theory is 0.1 kJ/mol (for complexes F\textsubscript{3}NO\textperiodcentered{}\textperiodcentered{}\textperiodcentered{}CH\textsubscript{3}, NH\textsubscript{3}F\textsuperscript{+}\textperiodcentered{}\textperiodcentered{}\textperiodcentered{}CH\textsubscript{3}, NH\textsubscript{2}F\textsubscript{2}\textsuperscript{+}\textperiodcentered{}\textperiodcentered{}\textperiodcentered{}CH\textsubscript{3}, SF\textsubscript{6}\textperiodcentered{}\textperiodcentered{}\textperiodcentered{}NH\textsubscript{3}, CF\textsubscript{3}CCF\textperiodcentered{}\textperiodcentered{}\textperiodcentered{}NH\textsubscript{3}, NH\textsubscript{3}F\textsuperscript{+}\textperiodcentered{}\textperiodcentered{}\textperiodcentered{}NH\textsubscript{3} and NH\textsubscript{2}F\textsubscript{2}\textsuperscript{+}\textperiodcentered{}\textperiodcentered{}\textperiodcentered{}NH\textsubscript{3}) at DF-RMP2-F12/VDZ-F12 optimized geometries, and 0.2 kJ/mol (for NH\textsubscript{3}F\textsuperscript{+}\textperiodcentered{}\textperiodcentered{}\textperiodcentered{}CH\textsubscript{3}) at M06-2X/AVTZ geometries.

Relativistic treatment is similarly insignificant. The maximum difference between all-electron, non-relativistic CCSD(T)/CBS BEs (denoted as ``N-R, all-e'' in \Tref{tab:ccsd-t-cbs-bes}) and all-electron, relativistic CCSD(T)/CBS BEs (denoted as ``Rel, all-e'' in \Tref{tab:ccsd-t-cbs-bes}) is 0.2 kJ/mol at DF-RMP2-F12/VDZ-F12 geometries (NH\textsubscript{2}F\textsubscript{2}\textsuperscript{+}\textperiodcentered{}\textperiodcentered{}\textperiodcentered{}NH\textsubscript{3}). At M06-2X/AVTZ geometries, the maximum difference is also 0.2 kJ/mol (NH\textsubscript{3}F\textsuperscript{+}\textperiodcentered{}\textperiodcentered{}\textperiodcentered{}CH\textsubscript{3}, NH\textsubscript{3}F\textsuperscript{+}\textperiodcentered{}\textperiodcentered{}\textperiodcentered{}NH\textsubscript{3}, NH\textsubscript{2}F\textsubscript{2}\textsuperscript{+}\textperiodcentered{}\textperiodcentered{}\textperiodcentered{}NH\textsubscript{3}). These deviations remain small relative to their BEs. To sum up, reference BEs from non-relativistic, valence-electron treatment are sufficient to check the reliability of M06-2X and popular ab initio methods for reliable prediction of BEs for the F-bonded complexes DF\textperiodcentered{}\textperiodcentered{}\textperiodcentered{}CH\textsubscript{3}/NH\textsubscript{3}.

Small relativistic corrections are expected here. The donors and acceptor are built almost entirely from first- and second-row atoms (H, C, N, O, F). Sulfur and chlorine (SF\textsubscript{6}, ClF\textsubscript{3}, SF\textsubscript{5}) are the only third-row elements present. Relativistic effects scale with atomic number and are small for light elements. This agrees with prior benchmarks on radical-molecule complexes, where relativistic corrections were largest for the heaviest donors tested, including Cl and Br, but below 1 kJ/mol in every case.\cite{Tentscher2013} It also agrees with a closed-shell halogen-bonding benchmark study more broadly. Relativistic corrections were small for chlorine but more significant for bromine.\cite{Kozuch2013} This supports our use of non-relativistic, valence-electron CCSD(T)/CBS BEs from the W1-F12 protocol to check the reliability of M06-2X in this work.

\subsection{Effect of basis sets for BEs for DF\textperiodcentered{}\textperiodcentered{}\textperiodcentered{}CH\textsubscript{3} from M06-2X}\label{sec:effect-of-basis-sets}

We have already discussed that M06-2X/AVTZ predicts overestimated BEs for several DF\textperiodcentered{}\textperiodcentered{}\textperiodcentered{}CH\textsubscript{3}/NH\textsubscript{3} complexes, including F\textsubscript{2}\textperiodcentered{}\textperiodcentered{}\textperiodcentered{}CH\textsubscript{3} and F\textsubscript{2}\textperiodcentered{}\textperiodcentered{}\textperiodcentered{}NH\textsubscript{3}. This is not consistent with results from DF-RMP2-F12, MP2, W1-F12, W2-F12 (\mbox{\textbf{TABLEs \ref{tab:cbs-limit-binding-energies}}}, \textbf{V}, \textbf{VII} and \mbox{\textbf{FIGs. \ref{fig:bar-graphs-showing-bes}}}-\textbf{4}) and other high-level methods, CCSD(T)-F12b, CCSD(T), CCSD, DCSD, DF-RMP2, MBPT(2), MP2, and SCS-MP2 (\textbf{TABLEs} \textbf{III}, \textbf{VI} and \Fref{fig:deviation-be-be-ab}). This raises the question of whether the discrepancy of M06-2X/AVTZ is a basis-set artifact or not. We test this below and return to the method itself in later sections.

\begin{figure}[!htbp]\centering
\GPKfig{7.0324in}{3.28387in}{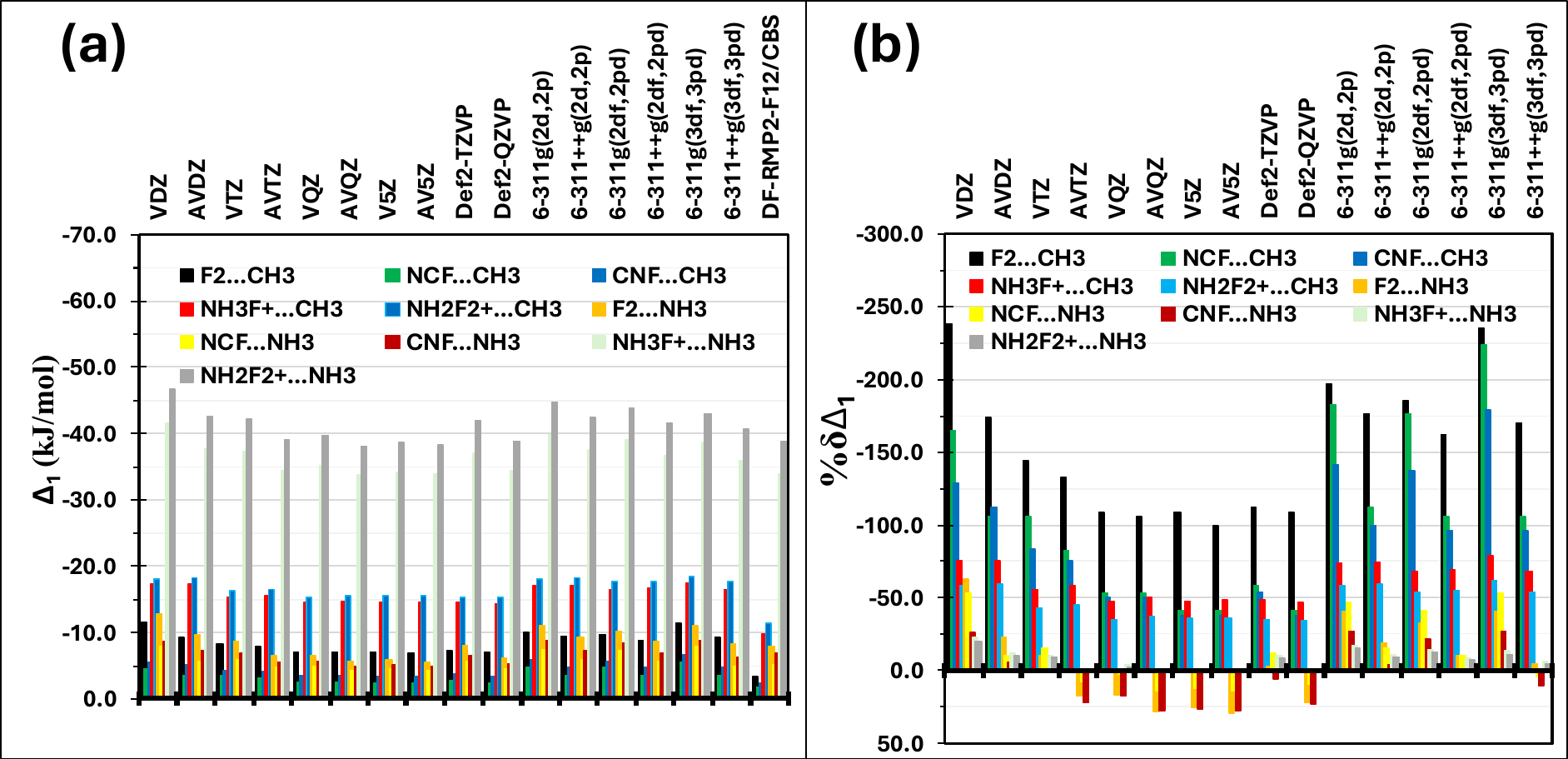}
\GPKfig{7.00758in}{3.09695in}{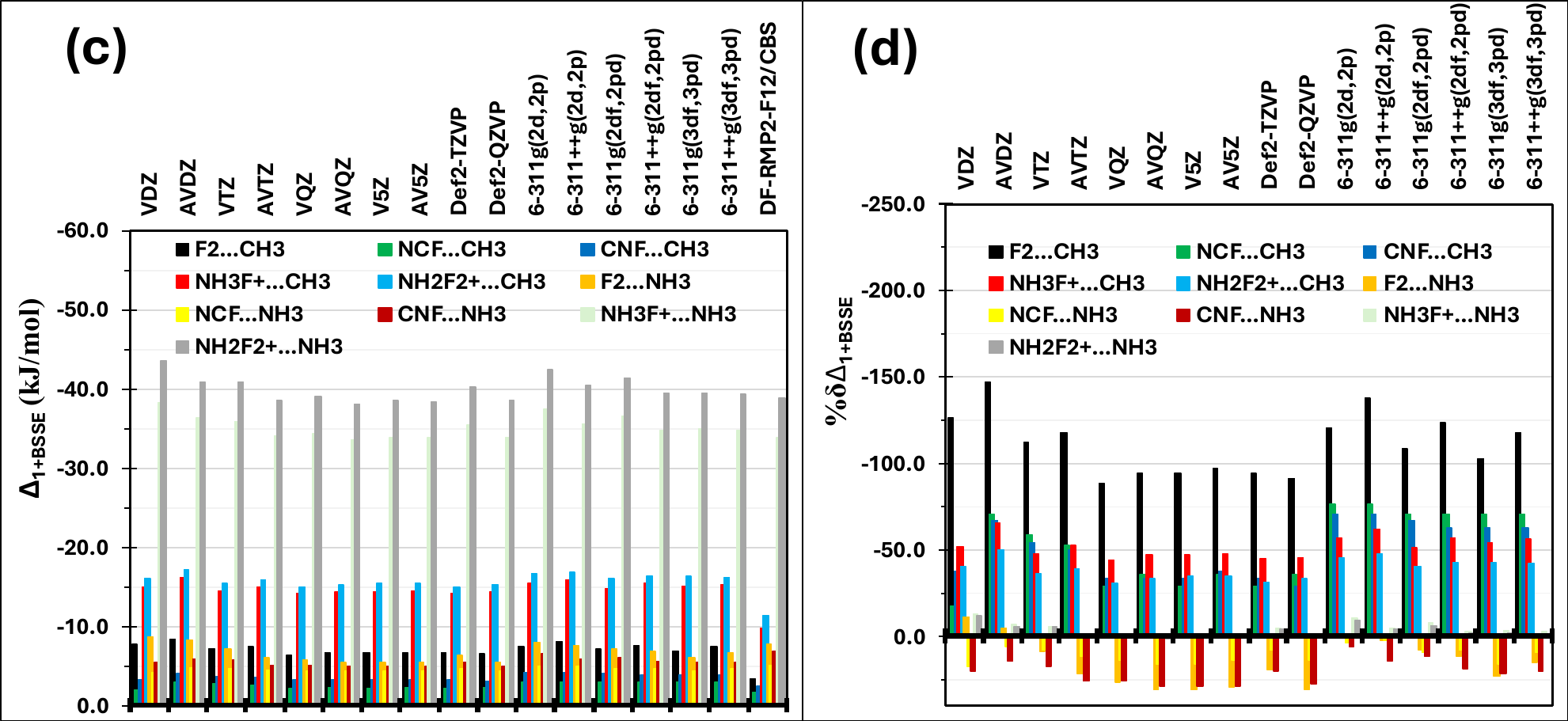}
\par\vspace{-1pt}\caption{BSSE-uncorrected BEs (\ensuremath{\Delta}\textsubscript{1}, kJ/mol) \textbf{(a)} and percentage deviation (\%\ensuremath{\delta}\ensuremath{\Delta}\textsubscript{1}) \textbf{(b)} for selected DF\textperiodcentered{}\textperiodcentered{}\textperiodcentered{}CH\textsubscript{3}/NH\textsubscript{3} complexes (DF = F\textsubscript{2}, NCF, CNF, NH\textsubscript{3}F\textsuperscript{+} and NH\textsubscript{2}F\textsubscript{2}\textsuperscript{+}), predicted by M06-2X with 16 different basis sets. Corresponding BSSE-corrected BEs (\ensuremath{\Delta}\textsubscript{1+BSSE} and \%\ensuremath{\delta}\ensuremath{\Delta}\textsubscript{1+BSSE}) are shown in \textbf{(c)} and \textbf{(d)}. Deviations and percentage deviations are relative to reference BEs at DF-RMP2-F12/CBS{[}V(T,Q)Z-F12{]}//DF-RMP2-F12/VDZ-F12.}\label{fig:bsse-uncorrected-bes-1}
\vspace{10pt}
\end{figure}

To test this, we recalculated BEs (\Tref{tab:bes-kj-mol-with}) using M06-2X functional with 16 basis sets for five representative complexes DF\textperiodcentered{}\textperiodcentered{}\textperiodcentered{}CH\textsubscript{3}/NH\textsubscript{3} (DF = F\textsubscript{2}, NCF, CNF, NH\textsubscript{3}F\textsuperscript{+}, NH\textsubscript{2}F\textsubscript{2}\textsuperscript{+}) already benchmarked against several \emph{ab initio} methods. Full geometry optimization was carried out at each of the 16 basis sets for all complexes. Percentage deviations on BEs with respect to DF-RMP2-F12/CBS level are shown in \Fref{fig:bsse-uncorrected-bes-1} and \mbox{\textbf{TABLE \ref{tab:percentage-deviation-on-bes}.}} The over-binding persists across all 16 basis sets with and without BSSE treatments. BEs for F\textsubscript{2}\textperiodcentered{}\textperiodcentered{}\textperiodcentered{}CH\textsubscript{3} remain almost comparable to F\textsubscript{2}\textperiodcentered{}\textperiodcentered{}\textperiodcentered{}NH\textsubscript{3} at every basis set used (\Fref{fig:bsse-uncorrected-bes-1}). All five SEFB complexes are over-bound at every basis set (as reflected by negative percentage deviations on BEs in \Tref{tab:percentage-deviation-on-bes}), while their CSFB counterparts are mostly under-bound (as indicated by positive values of percentage deviation on BEs) (\Tref{tab:percentage-deviation-on-bes}, \Fref{fig:bsse-uncorrected-bes-1}). Without BSSE correction, the smallest deviation on BEs for each SEFB complex ranges from -34.2\% (NH\textsubscript{2}F\textsubscript{2}\textsuperscript{+}\textperiodcentered{}\textperiodcentered{}\textperiodcentered{}CH\textsubscript{3}, def2-QZVP) to -100.0\% (F\textsubscript{2}\textperiodcentered{}\textperiodcentered{}\textperiodcentered{}CH\textsubscript{3}, AV5Z) (\Tref{tab:percentage-deviation-on-bes}). With BSSE correction, this range is from -17.6\% (NCF\textperiodcentered{}\textperiodcentered{}\textperiodcentered{}CH\textsubscript{3}, VDZ) to -88.2\% (F\textsubscript{2}\textperiodcentered{}\textperiodcentered{}\textperiodcentered{}CH\textsubscript{3}, VQZ). For CSFB complexes (DF\textperiodcentered{}\textperiodcentered{}\textperiodcentered{}NH\textsubscript{3}), without BSSE correction, the smallest deviation ranges from 0.0\% (CNF\textperiodcentered{}\textperiodcentered{}\textperiodcentered{}NH\textsubscript{3}, 6-311++g(2df,2pd) and NH\textsubscript{3}F\textsuperscript{+}\textperiodcentered{}\textperiodcentered{}\textperiodcentered{}NH\textsubscript{3}, AV5Z) to -2.6\% (F\textsubscript{2}\textperiodcentered{}\textperiodcentered{}\textperiodcentered{}NH\textsubscript{3}, def2-TZVP). With BSSE correction, this range is 0.0\% (NH\textsubscript{3}F\textsuperscript{+}\textperiodcentered{}\textperiodcentered{}\textperiodcentered{}NH\textsubscript{3}, V5Z and def2-QZVP) to 5.8\% (CNF\textperiodcentered{}\textperiodcentered{}\textperiodcentered{}NH\textsubscript{3}, 6-311g(2d,2p)). Here, we note that AVTZ is generally considered a practical CBS-limit basis set for DFT calculations.\cite{OterodelaRoza2013}

\begin{table}[!htbp]\centering\setlength{\parskip}{0pt}
\caption{BEs (kJ/\allowbreak{}mol), with (\ensuremath{\Delta}\textsubscript{1+BSSE}) and without (\ensuremath{\Delta}\textsubscript{1}) BSSE correction, for selected DF\textperiodcentered{}\textperiodcentered{}\textperiodcentered{}CH\textsubscript{3}/NH\textsubscript{3} complexes (DF = F\textsubscript{2}, NCF, CNF, NH\textsubscript{3}F\textsuperscript{+} and NH\textsubscript{2}F\textsubscript{2}\textsuperscript{+}) obtained from M06-2X functional using 16 basis sets. Following basis sets are denoted as shorthand notation: cc-pVDZ\ensuremath{\rightarrow}VDZ, Aug-cc-pVDZ\ensuremath{\rightarrow}AVDZ, cc-pVTZ\ensuremath{\rightarrow}VTZ, Aug-cc-pVTZ\ensuremath{\rightarrow}AVTZ, cc-pVQZ\ensuremath{\rightarrow}VQZ, Aug-cc-pVQZ\ensuremath{\rightarrow}AVQZ, cc-pV5Z\ensuremath{\rightarrow}V5Z, Aug-cc-pV5Z\ensuremath{\rightarrow}AV5Z. These BEs may be compared with the DF-RMP2-F12/\allowbreak{}CBS[V(T,Q)Z-F12]//DF-RMP2-F12/\allowbreak{}VDZ-F12 CBS-limit values (\mbox{\textbf{Table \ref{tab:bes-kj-mol-of}}}); see \mbox{\textbf{Table \ref{tab:mse-mad-rmsd-max2}}} for the corresponding statistical analysis.}\label{tab:bes-kj-mol-with}\vspace{7pt}\nopagebreak
{\fontsize{8.0}{9.2}\fontfamily{ptm}\selectfont\renewcommand{\arraystretch}{1.17}\setlength{\tabcolsep}{1.5pt}
\centerline{\adjustbox{max width=518pt}{\begin{tabular}{|>{\centering\arraybackslash}m{56.1pt}|>{\centering\arraybackslash}m{33.6pt}|>{\centering\arraybackslash}m{22.6pt}|>{\centering\arraybackslash}m{25.0pt}|>{\centering\arraybackslash}m{25.0pt}|>{\centering\arraybackslash}m{25.1pt}|>{\centering\arraybackslash}m{25.0pt}|>{\centering\arraybackslash}m{28.8pt}|>{\centering\arraybackslash}m{25.0pt}|>{\centering\arraybackslash}m{25.0pt}|>{\centering\arraybackslash}m{25.0pt}|>{\centering\arraybackslash}m{25.0pt}|>{\centering\arraybackslash}m{25.0pt}|>{\centering\arraybackslash}m{25.0pt}|>{\centering\arraybackslash}m{8.0pt}|>{\centering\arraybackslash}p{20.7pt}|>{\centering\arraybackslash}m{23.6pt}|>{\centering\arraybackslash}m{22.3pt}|>{\centering\arraybackslash}m{25.3pt}|}
\specialrule{0.9pt}{0pt}{0pt}
\multicolumn{2}{|>{\centering\arraybackslash}m{92.7pt}|}{DF\textperiodcentered{}\textperiodcentered{}\textperiodcentered{}CH\textsubscript{3}/NH\textsubscript{3}} & \multicolumn{1}{>{\centering\arraybackslash}p{22.6pt}|}{\makebox[\linewidth][c]{\rotatebox{90}{\hspace{2pt}VDZ\hspace{2pt}}}} & \multicolumn{1}{>{\centering\arraybackslash}p{25.0pt}|}{\makebox[\linewidth][c]{\rotatebox{90}{\hspace{2pt}AVDZ\hspace{2pt}}}} & \multicolumn{1}{>{\centering\arraybackslash}p{25.0pt}|}{\makebox[\linewidth][c]{\rotatebox{90}{\hspace{2pt}VTZ\hspace{2pt}}}} & \multicolumn{1}{>{\centering\arraybackslash}p{25.1pt}|}{\makebox[\linewidth][c]{\rotatebox{90}{\hspace{2pt}\textbf{AVTZ}\hspace{2pt}}}} & \multicolumn{1}{>{\centering\arraybackslash}p{25.0pt}|}{\makebox[\linewidth][c]{\rotatebox{90}{\hspace{2pt}VQZ\hspace{2pt}}}} & \multicolumn{1}{>{\centering\arraybackslash}p{28.8pt}|}{\makebox[\linewidth][c]{\rotatebox{90}{\hspace{2pt}AVQZ\hspace{2pt}}}} & \multicolumn{1}{>{\centering\arraybackslash}p{25.0pt}|}{\makebox[\linewidth][c]{\rotatebox{90}{\hspace{2pt}V5Z\hspace{2pt}}}} & \multicolumn{1}{>{\centering\arraybackslash}p{25.0pt}|}{\makebox[\linewidth][c]{\rotatebox{90}{\hspace{2pt}AV5Z\hspace{2pt}}}} & \multicolumn{1}{>{\centering\arraybackslash}p{25.0pt}|}{\makebox[\linewidth][c]{\rotatebox{90}{\hspace{2pt}Def2-TZVP\hspace{2pt}}}} & \multicolumn{1}{>{\centering\arraybackslash}p{25.0pt}|}{\makebox[\linewidth][c]{\rotatebox{90}{\hspace{2pt}Def2-QZVP\hspace{2pt}}}} & \multicolumn{1}{>{\centering\arraybackslash}p{25.0pt}|}{\makebox[\linewidth][c]{\rotatebox{90}{\hspace{2pt}6-311g(2d,2p)\hspace{2pt}}}} & \multicolumn{2}{>{\centering\arraybackslash}p{36.0pt}|}{\makebox[\linewidth][c]{\rotatebox{90}{\hspace{2pt}6-311++g(2d,2p)\hspace{2pt}}}} & \makebox[\linewidth][c]{\rotatebox{90}{\hspace{2pt}6-311g(2df,2pd)\hspace{2pt}}} & \multicolumn{1}{>{\centering\arraybackslash}p{23.6pt}|}{\makebox[\linewidth][c]{\rotatebox{90}{\hspace{2pt}6-311++g(2df,2pd)\hspace{2pt}}}} & \multicolumn{1}{>{\centering\arraybackslash}p{22.3pt}|}{\makebox[\linewidth][c]{\rotatebox{90}{\hspace{2pt}6-311g(3df,3pd)\hspace{2pt}}}} & \multicolumn{1}{>{\centering\arraybackslash}p{25.3pt}|}{\makebox[\linewidth][c]{\rotatebox{90}{\hspace{2pt}6-311++g(3df,3pd)\hspace{2pt}}}} \\* \hline\noalign{\penalty10000}
\multicolumn{1}{|>{\centering\arraybackslash}m{56.1pt}|}{\multirow{2}{56.1pt}{\centering F\textsubscript{2}\textperiodcentered{}\textperiodcentered{}\textperiodcentered{}CH\textsubscript{3}}} & \textbf{\ensuremath{\Delta}}\textbf{\textsubscript{1}} & -11.5 & -9.3 & -8.3 & -7.9 & -7.1 & -7.0 & -7.1 & -6.8 & -7.2 & -7.1 & -10.1 & -9.4 & \multicolumn{2}{>{\centering\arraybackslash}m{31.7pt}|}{-9.7} & -8.9 & -11.4 & -9.2 \\* \cline{2-19}\noalign{\penalty10000}
\multicolumn{1}{|>{\centering\arraybackslash}m{56.1pt}|}{} & \textbf{\ensuremath{\Delta}}\textbf{\textsubscript{1+BSSE}} & -7.7 & -8.4 & -7.2 & -7.4 & -6.4 & -6.6 & -6.6 & -6.7 & -6.6 & -6.5 & -7.5 & -8.1 & \multicolumn{2}{>{\centering\arraybackslash}m{31.7pt}|}{-7.1} & -7.6 & -6.9 & -7.4 \\* \hline\noalign{\penalty10000}
\multicolumn{1}{|>{\centering\arraybackslash}m{56.1pt}|}{\multirow{2}{56.1pt}{\centering NCF\textperiodcentered{}\textperiodcentered{}\textperiodcentered{}CH\textsubscript{3}}} & \textbf{\ensuremath{\Delta}}\textbf{\textsubscript{1}} & -4.5 & -3.5 & -3.5 & -3.1 & -2.6 & -2.6 & -2.4 & -2.4 & -2.7 & -2.4 & -4.8 & -3.6 & \multicolumn{2}{>{\centering\arraybackslash}m{31.7pt}|}{-4.7} & -3.5 & -5.5 & -3.5 \\* \cline{2-19}\noalign{\penalty10000}
\multicolumn{1}{|>{\centering\arraybackslash}m{56.1pt}|}{} & \textbf{\ensuremath{\Delta}}\textbf{\textsubscript{1+BSSE}} & -2.0 & -2.9 & -2.7 & -2.6 & -2.2 & -2.3 & -2.2 & -2.3 & -2.2 & -2.3 & -3.0 & -3.0 & \multicolumn{2}{>{\centering\arraybackslash}m{31.7pt}|}{-2.9} & -2.9 & -2.9 & -2.9 \\* \hline\noalign{\penalty10000}
\multicolumn{1}{|>{\centering\arraybackslash}m{56.1pt}|}{\multirow{2}{56.1pt}{\centering CNF\textperiodcentered{}\textperiodcentered{}\textperiodcentered{}CH\textsubscript{3}}} & \textbf{\ensuremath{\Delta}}\textbf{\textsubscript{1}} & -5.5 & -5.1 & -4.4 & -4.2 & -3.6 & -3.6 & -3.3 & -3.3 & -3.7 & -3.3 & -5.8 & -4.8 & \multicolumn{2}{>{\centering\arraybackslash}m{31.7pt}|}{-5.7} & -4.7 & -6.7 & -4.7 \\* \cline{2-19}\noalign{\penalty10000}
\multicolumn{1}{|>{\centering\arraybackslash}m{56.1pt}|}{} & \textbf{\ensuremath{\Delta}}\textbf{\textsubscript{1+BSSE}} & -3.3 & -4.0 & -3.7 & -3.6 & -3.2 & -3.2 & -3.2 & -3.3 & -3.2 & -3.1 & -4.1 & -4.1 & \multicolumn{2}{>{\centering\arraybackslash}m{31.7pt}|}{-4.0} & -3.9 & -3.9 & -3.9 \\* \hline\noalign{\penalty10000}
\multicolumn{1}{|>{\centering\arraybackslash}m{56.1pt}|}{\multirow{2}{56.1pt}{\centering NH\textsubscript{3}F\textsuperscript{+}\textperiodcentered{}\textperiodcentered{}\textperiodcentered{}CH\textsubscript{3}}} & \textbf{\ensuremath{\Delta}}\textbf{\textsubscript{1}} & -17.2 & -17.2 & -15.3 & -15.5 & -14.5 & -14.7 & -14.5 & -14.6 & -14.6 & -14.4 & -17.0 & -17.1 & \multicolumn{2}{>{\centering\arraybackslash}m{31.7pt}|}{-16.5} & -16.6 & -17.5 & -16.5 \\* \cline{2-19}\noalign{\penalty10000}
\multicolumn{1}{|>{\centering\arraybackslash}m{56.1pt}|}{} & \textbf{\ensuremath{\Delta}}\textbf{\textsubscript{1+BSSE}} & -14.9 & -16.2 & -14.5 & -15.0 & -14.1 & -14.4 & -14.4 & -14.5 & -14.2 & -14.3 & -15.4 & -15.9 & \multicolumn{2}{>{\centering\arraybackslash}m{31.7pt}|}{-14.8} & -15.4 & -15.1 & -15.3 \\* \hline\noalign{\penalty10000}
\multicolumn{1}{|>{\centering\arraybackslash}m{56.1pt}|}{\multirow{2}{56.1pt}{\centering NH\textsubscript{2}F\textsubscript{2}\textsuperscript{+}\textperiodcentered{}\textperiodcentered{}\textperiodcentered{}CH\textsubscript{3}}} & \textbf{\ensuremath{\Delta}}\textbf{\textsubscript{1}} & -18.1 & -18.2 & -16.3 & -16.5 & -15.4 & -15.6 & -15.5 & -15.5 & -15.4 & -15.3 & -18.1 & -18.2 & \multicolumn{2}{>{\centering\arraybackslash}m{31.7pt}|}{-17.6} & -17.7 & -18.5 & -17.6 \\* \cline{2-19}\noalign{\penalty10000}
\multicolumn{1}{|>{\centering\arraybackslash}m{56.1pt}|}{} & \textbf{\ensuremath{\Delta}}\textbf{\textsubscript{1+BSSE}} & -16.0 & -17.1 & -15.5 & -15.9 & -14.9 & -15.2 & -15.4 & -15.4 & -15.0 & -15.2 & -16.6 & -16.8 & \multicolumn{2}{>{\centering\arraybackslash}m{31.7pt}|}{-16.0} & -16.3 & -16.3 & -16.2 \\* \hline\noalign{\penalty10000}
\multicolumn{1}{|>{\centering\arraybackslash}m{56.1pt}|}{\multirow{2}{56.1pt}{\centering {\leavevmode\color[HTML]{FF0000}F}{\leavevmode\color[HTML]{FF0000}\textsubscript{2}}{\leavevmode\color[HTML]{FF0000}\textperiodcentered{}\textperiodcentered{}\textperiodcentered{}NH}{\leavevmode\color[HTML]{FF0000}\textsubscript{3}}}} & \textbf{\ensuremath{\Delta}}\textbf{\textsubscript{1}} & -12.7 & -9.6 & -8.6 & {\leavevmode\color[HTML]{FF0000}-6.4} & -6.5 & -5.6 & -5.8 & -5.5 & -8.0 & -6.1 & -11.0 & -9.3 & \multicolumn{2}{>{\centering\arraybackslash}m{31.7pt}|}{-10.3} & -8.6 & -11.0 & -8.2 \\* \cline{2-19}\noalign{\penalty10000}
\multicolumn{1}{|>{\centering\arraybackslash}m{56.1pt}|}{} & \textbf{\ensuremath{\Delta}}\textbf{\textsubscript{1+BSSE}} & -8.7 & -8.2 & -7.1 & {\leavevmode\color[HTML]{FF0000}-6.1} & -5.7 & -5.4 & -5.4 & -5.5 & -6.3 & -5.4 & -7.9 & -7.6 & \multicolumn{2}{>{\centering\arraybackslash}m{31.7pt}|}{-7.2} & -6.9 & -6.0 & -6.6 \\* \hline\noalign{\penalty10000}
\multicolumn{1}{|>{\centering\arraybackslash}m{56.1pt}|}{\multirow{2}{56.1pt}{\centering {\leavevmode\color[HTML]{FF0000}NCF\textperiodcentered{}\textperiodcentered{}\textperiodcentered{}NH}{\leavevmode\color[HTML]{FF0000}\textsubscript{3}}}} & \textbf{\ensuremath{\Delta}}\textbf{\textsubscript{1}} & -7.8 & -5.6 & -5.9 & {\leavevmode\color[HTML]{FF0000}-4.7} & -5.0 & -4.4 & -4.5 & -4.4 & -5.7 & -4.6 & -7.5 & -5.9 & \multicolumn{2}{>{\centering\arraybackslash}m{31.7pt}|}{-7.2} & -5.6 & -7.8 & -4.9 \\* \cline{2-19}\noalign{\penalty10000}
\multicolumn{1}{|>{\centering\arraybackslash}m{56.1pt}|}{} & \textbf{\ensuremath{\Delta}}\textbf{\textsubscript{1+BSSE}} & -4.2 & -4.8 & -4.7 & {\leavevmode\color[HTML]{FF0000}-4.5} & -4.4 & -4.3 & -4.3 & -4.4 & -4.7 & -4.4 & -4.9 & -5.0 & \multicolumn{2}{>{\centering\arraybackslash}m{31.7pt}|}{-4.6} & -4.7 & -4.3 & -4.6 \\* \hline\noalign{\penalty10000}
\multicolumn{1}{|>{\centering\arraybackslash}m{56.1pt}|}{\multirow{2}{56.1pt}{\centering {\leavevmode\color[HTML]{FF0000}CNF\textperiodcentered{}\textperiodcentered{}\textperiodcentered{}NH}{\leavevmode\color[HTML]{FF0000}\textsubscript{3}}}} & \textbf{\ensuremath{\Delta}}\textbf{\textsubscript{1}} & -8.7 & -7.3 & -6.8 & {\leavevmode\color[HTML]{FF0000}-5.4} & -5.7 & -5.0 & -5.1 & -5.0 & -6.5 & -5.3 & -8.8 & -7.2 & \multicolumn{2}{>{\centering\arraybackslash}m{31.7pt}|}{-8.4} & -6.9 & -8.8 & -6.2 \\* \cline{2-19}\noalign{\penalty10000}
\multicolumn{1}{|>{\centering\arraybackslash}m{56.1pt}|}{} & \textbf{\ensuremath{\Delta}}\textbf{\textsubscript{1+BSSE}} & -5.5 & -5.9 & -5.7 & {\leavevmode\color[HTML]{FF0000}-5.1} & -5.1 & -4.9 & -4.9 & -4.9 & -5.5 & -5.0 & -6.5 & -5.9 & \multicolumn{2}{>{\centering\arraybackslash}m{31.7pt}|}{-6.1} & -5.6 & -5.4 & -5.5 \\* \hline\noalign{\penalty10000}
\multicolumn{1}{|>{\centering\arraybackslash}m{56.1pt}|}{\multirow{2}{56.1pt}{\centering {\leavevmode\color[HTML]{FF0000}NH}{\leavevmode\color[HTML]{FF0000}\textsubscript{3}}{\leavevmode\color[HTML]{FF0000}F}{\leavevmode\color[HTML]{FF0000}\textsuperscript{+}}{\leavevmode\color[HTML]{FF0000}\textperiodcentered{}\textperiodcentered{}\textperiodcentered{}NH}{\leavevmode\color[HTML]{FF0000}\textsubscript{3}}}} & \textbf{\ensuremath{\Delta}}\textbf{\textsubscript{1}} & -41.6 & -37.8 & -37.3 & {\leavevmode\color[HTML]{FF0000}-34.3} & -35.2 & -33.7 & -34.1 & -33.9 & -37.2 & -34.3 & -39.8 & -37.5 & \multicolumn{2}{>{\centering\arraybackslash}m{31.7pt}|}{-39.0} & -36.7 & -38.6 & -36.0 \\* \cline{2-19}\noalign{\penalty10000}
\multicolumn{1}{|>{\centering\arraybackslash}m{56.1pt}|}{} & \textbf{\ensuremath{\Delta}}\textbf{\textsubscript{1+BSSE}} & -38.3 & -36.3 & -35.9 & {\leavevmode\color[HTML]{FF0000}-34.0} & -34.4 & -33.6 & -33.9 & -33.8 & -35.5 & -33.9 & -37.4 & -35.6 & \multicolumn{2}{>{\centering\arraybackslash}m{31.7pt}|}{-36.5} & -34.8 & -35.0 & -34.8 \\* \hline\noalign{\penalty10000}
\multicolumn{1}{|>{\centering\arraybackslash}m{56.1pt}|}{\multirow{2}{56.1pt}{\centering {\leavevmode\color[HTML]{FF0000}NH}{\leavevmode\color[HTML]{FF0000}\textsubscript{2}}{\leavevmode\color[HTML]{FF0000}F}{\leavevmode\color[HTML]{FF0000}\textsubscript{2}}{\leavevmode\color[HTML]{FF0000}\textsuperscript{+}}{\leavevmode\color[HTML]{FF0000}\textperiodcentered{}\textperiodcentered{}\textperiodcentered{}NH}{\leavevmode\color[HTML]{FF0000}\textsubscript{3}}}} & \textbf{\ensuremath{\Delta}}\textbf{\textsubscript{1}} & -46.7 & -42.6 & -42.3 & {\leavevmode\color[HTML]{FF0000}-39.0} & -39.7 & -38.2 & -38.7 & -38.4 & -42.0 & -38.9 & -44.8 & -42.5 & \multicolumn{2}{>{\centering\arraybackslash}m{31.7pt}|}{-43.8} & -41.6 & -43.1 & -40.7 \\* \cline{2-19}\noalign{\penalty10000}
\multicolumn{1}{|>{\centering\arraybackslash}m{56.1pt}|}{} & \textbf{\ensuremath{\Delta}}\textbf{\textsubscript{1+BSSE}} & -43.5 & -40.9 & -40.9 & {\leavevmode\color[HTML]{FF0000}-38.5} & -39.0 & -38.1 & -38.5 & -38.4 & -40.3 & -38.5 & -42.4 & -40.4 & \multicolumn{2}{>{\centering\arraybackslash}m{31.7pt}|}{-41.3} & -39.5 & -39.5 & -39.4 \\* \specialrule{0.9pt}{0pt}{0pt}\noalign{\penalty10000}
\end{tabular}}}}
\end{table}

At M06-2X/AVTZ level, \%\ensuremath{\delta}\textsubscript{BE} without BSSE correction ranges from -44.7\% (NH\textsubscript{2}F\textsubscript{2}\textsuperscript{+}\textperiodcentered{}\textperiodcentered{}\textperiodcentered{}CH\textsubscript{3}) to -132.4\% (F\textsubscript{2}\textperiodcentered{}\textperiodcentered{}\textperiodcentered{}CH\textsubscript{3}) for the same set of selected SEFB complexes, whereas BSSE correction improves percentage deviations by 5 to 30\% (\Tref{tab:percentage-deviation-on-bes}). For the DF\textperiodcentered{}\textperiodcentered{}\textperiodcentered{}NH\textsubscript{3} complexes at M06-2X/AVTZ, \%\ensuremath{\delta}\textsubscript{BE} without BSSE correction ranges from -0.5\% (NH\textsubscript{2}F\textsubscript{2}\textsuperscript{+}\textperiodcentered{}\textperiodcentered{}\textperiodcentered{}NH\textsubscript{3}) to 21.7\% (CNF\textperiodcentered{}\textperiodcentered{}\textperiodcentered{}NH\textsubscript{3}). It is worthwhile to mention that BSSE correction typically increases the deviations on BEs for DF\textperiodcentered{}\textperiodcentered{}\textperiodcentered{}NH\textsubscript{3} complexes with NH\textsubscript{3}F\textsuperscript{+}\textperiodcentered{}\textperiodcentered{}\textperiodcentered{}NH\textsubscript{3} as one exception where BSSE correction slightly reduces deviation on BE from reference value considered (\Tref{tab:percentage-deviation-on-bes}).

\begin{table}[!htbp]\centering\setlength{\parskip}{0pt}
\caption{Percentage deviation on BEs without (\%\ensuremath{\delta}\ensuremath{\Delta}\textsubscript{1}) and with (\%\ensuremath{\delta}\ensuremath{\Delta}\textsubscript{1+BSSE}) BSSE correction, for selected DF\textperiodcentered{}\textperiodcentered{}\textperiodcentered{}CH\textsubscript{3}/NH\textsubscript{3} complexes (DF = F\textsubscript{2}, NCF, CNF, NH\textsubscript{3}F\textsuperscript{+} and NH\textsubscript{2}F\textsubscript{2}\textsuperscript{+}) using M06-2X with 16 basis sets, relative to DF-RMP2-F12/\allowbreak{}CBS[V(T,Q)Z-F12]//DF-RMP2-F12/\allowbreak{}VDZ-F12 (denoted as DF-RMP2-F12/\allowbreak{}CBS). The smallest deviation for each complex is highlighted in bold; M06-2X/\allowbreak{}AVTZ results are italicized; values in red indicate an increase in magnitude of percentage deviations on BEs after BSSE correction. See \mbox{\textbf{Tables \ref{tab:bsse-uncorrected-bes-1}}} and \textbf{S12} for the underlying BEs (without and with BSSE correction, respectively), and \mbox{\textbf{Tables \ref{tab:deviations-on-bsse-uncorrected}}} and \textbf{S14} for the corresponding deviations.}\label{tab:percentage-deviation-on-bes}\vspace{7pt}\nopagebreak
{\fontsize{9.0}{10.3}\fontfamily{ptm}\selectfont\renewcommand{\arraystretch}{1.05}\setlength{\tabcolsep}{1.5pt}
\centerline{\adjustbox{max width=518pt}{\begin{tabular}{|>{\centering\arraybackslash}m{88.0pt}|>{\centering\arraybackslash}m{43.7pt}|>{\centering\arraybackslash}m{31.3pt}|>{\centering\arraybackslash}m{30.6pt}|>{\centering\arraybackslash}m{30.2pt}|>{\centering\arraybackslash}m{30.5pt}|>{\centering\arraybackslash}m{80.7pt}|>{\centering\arraybackslash}m{34.3pt}|>{\centering\arraybackslash}m{34.0pt}|>{\centering\arraybackslash}m{35.7pt}|>{\centering\arraybackslash}m{35.7pt}|>{\centering\arraybackslash}m{35.7pt}|}
\specialrule{0.9pt}{0pt}{0pt}
\multicolumn{6}{|>{\centering\arraybackslash}m{269.5pt}|}{\textbf{DF\textperiodcentered{}\textperiodcentered{}\textperiodcentered{}CH}\textbf{\textsubscript{3}}\textbf{ (without BSSE correction)}} & \multicolumn{6}{>{\centering\arraybackslash}m{271.0pt}|}{\textbf{DF\textperiodcentered{}\textperiodcentered{}\textperiodcentered{}CH}\textbf{\textsubscript{3}}\textbf{ (with BSSE correction)}} \\* \hline\noalign{\penalty10000}
\multicolumn{1}{|>{\centering\arraybackslash}m{88.0pt}|}{\multirow{2}{88.0pt}{\centering Basis sets}} & \multicolumn{1}{>{\centering\arraybackslash}m{43.7pt}|}{\makebox[\linewidth][c]{\rotatebox{90}{\hspace{2pt}\textbf{F}\textbf{\textsubscript{2}}\textbf{\textperiodcentered{}\textperiodcentered{}\textperiodcentered{}CH}\textbf{\textsubscript{3}}\hspace{2pt}}}} & \multicolumn{1}{>{\centering\arraybackslash}m{31.3pt}|}{\makebox[\linewidth][c]{\rotatebox{90}{\hspace{2pt}\textbf{NCF\textperiodcentered{}\textperiodcentered{}\textperiodcentered{}CH}\textbf{\textsubscript{3}}\hspace{2pt}}}} & \multicolumn{1}{>{\centering\arraybackslash}m{30.6pt}|}{\makebox[\linewidth][c]{\rotatebox{90}{\hspace{2pt}\textbf{CNF\textperiodcentered{}\textperiodcentered{}\textperiodcentered{}CH}\textbf{\textsubscript{3}}\hspace{2pt}}}} & \multicolumn{1}{>{\centering\arraybackslash}m{30.2pt}|}{\makebox[\linewidth][c]{\rotatebox{90}{\hspace{2pt}\textbf{NH}\textbf{\textsubscript{3}}\textbf{F}\textbf{\textsuperscript{+}}\textbf{\textperiodcentered{}\textperiodcentered{}\textperiodcentered{}CH}\textbf{\textsubscript{3}}\hspace{2pt}}}} & \multicolumn{1}{>{\centering\arraybackslash}m{30.5pt}|}{\makebox[\linewidth][c]{\rotatebox{90}{\hspace{2pt}\textbf{NH}\textbf{\textsubscript{2}}\textbf{F}\textbf{\textsubscript{2}}\textbf{\textsuperscript{+}}\textbf{\textperiodcentered{}\textperiodcentered{}\textperiodcentered{}CH}\textbf{\textsubscript{3}}\hspace{2pt}}}} & \multicolumn{1}{>{\centering\arraybackslash}m{80.7pt}|}{\multirow{2}{80.7pt}{\centering Basis sets}} & \multicolumn{1}{>{\centering\arraybackslash}m{34.3pt}|}{\makebox[\linewidth][c]{\rotatebox{90}{\hspace{2pt}\textbf{F}\textbf{\textsubscript{2}}\textbf{\textperiodcentered{}\textperiodcentered{}\textperiodcentered{}CH}\textbf{\textsubscript{3}}\hspace{2pt}}}} & \multicolumn{1}{>{\centering\arraybackslash}m{34.0pt}|}{\makebox[\linewidth][c]{\rotatebox{90}{\hspace{2pt}\textbf{NCF\textperiodcentered{}\textperiodcentered{}\textperiodcentered{}CH}\textbf{\textsubscript{3}}\hspace{2pt}}}} & \multicolumn{1}{>{\centering\arraybackslash}m{35.7pt}|}{\makebox[\linewidth][c]{\rotatebox{90}{\hspace{2pt}\textbf{CNF\textperiodcentered{}\textperiodcentered{}\textperiodcentered{}CH}\textbf{\textsubscript{3}}\hspace{2pt}}}} & \multicolumn{1}{>{\centering\arraybackslash}m{35.7pt}|}{\makebox[\linewidth][c]{\rotatebox{90}{\hspace{2pt}\textbf{NH}\textbf{\textsubscript{3}}\textbf{F}\textbf{\textsuperscript{+}}\textbf{\textperiodcentered{}\textperiodcentered{}\textperiodcentered{}CH}\textbf{\textsubscript{3}}\hspace{2pt}}}} & \multicolumn{1}{>{\centering\arraybackslash}m{35.7pt}|}{\makebox[\linewidth][c]{\rotatebox{90}{\hspace{2pt}\textbf{NH}\textbf{\textsubscript{2}}\textbf{F}\textbf{\textsubscript{2}}\textbf{\textsuperscript{+}}\textbf{\textperiodcentered{}\textperiodcentered{}\textperiodcentered{}CH}\textbf{\textsubscript{3}}\hspace{2pt}}}} \\* \cline{2-6}\cline{8-12}\noalign{\penalty10000}
\multicolumn{1}{|>{\centering\arraybackslash}m{88.0pt}|}{} & \%\ensuremath{\delta}\ensuremath{\Delta}\textbf{\textsubscript{1}} & \%\ensuremath{\delta}\ensuremath{\Delta}\textbf{\textsubscript{1}} & \%\ensuremath{\delta}\ensuremath{\Delta}\textbf{\textsubscript{1}} & \%\ensuremath{\delta}\ensuremath{\Delta}\textbf{\textsubscript{1}} & \%\ensuremath{\delta}\ensuremath{\Delta}\textbf{\textsubscript{1}} &  & \%\ensuremath{\delta}\ensuremath{\Delta}\textsubscript{1+} \textsubscript{BSSE} & \%\ensuremath{\delta}\ensuremath{\Delta}\textsubscript{1+} \textsubscript{BSSE} & \%\ensuremath{\delta}\ensuremath{\Delta}\textsubscript{1+}\allowbreak\textsubscript{BSSE} & \%\ensuremath{\delta}\ensuremath{\Delta}\textsubscript{1+}\allowbreak\textsubscript{BSSE} & \%\ensuremath{\delta}\ensuremath{\Delta}\textsubscript{1+}\allowbreak\textsubscript{BSSE} \\* \hline\noalign{\penalty10000}
\multicolumn{1}{|>{\centering\arraybackslash}m{88.0pt}|}{VDZ} & -238.2 & -164.7 & -129.2 & -75.5 & -58.8 & VDZ & -126.5 & \textbf{-17.6} & -37.5 & -52.0 & -40.4 \\* \hline\noalign{\penalty10000}
\multicolumn{1}{|>{\centering\arraybackslash}m{88.0pt}|}{AVDZ} & -173.5 & -105.9 & -112.5 & -75.5 & -59.6 & AVDZ & -147.1 & -70.6 & -66.7 & -65.3 & -50.0 \\* \hline\noalign{\penalty10000}
\multicolumn{1}{|>{\centering\arraybackslash}m{88.0pt}|}{VTZ} & -144.1 & -105.9 & -83.3 & -56.1 & -43.0 & VTZ & -111.8 & -58.8 & -54.2 & -48.0 & -36.0 \\* \hline\noalign{\penalty10000}
\multicolumn{1}{|>{\centering\arraybackslash}m{88.0pt}|}{\emph{AVTZ}} & \emph{-132.4} & \emph{-82.4} & \emph{-75.0} & \emph{-58.2} & \emph{-44.7} & \emph{AVTZ} & \emph{-117.6} & \emph{-52.9} & \emph{-50.0} & \emph{-53.1} & \emph{-39.5} \\ \hline
\multicolumn{1}{|>{\centering\arraybackslash}m{88.0pt}|}{VQZ} & -108.8 & -52.9 & -50.0 & -48.0 & -35.1 & VQZ & -88.2 & -29.4 & -33.3 & \textbf{-43.9} & \textbf{-30.7} \\ \hline
\multicolumn{1}{|>{\centering\arraybackslash}m{88.0pt}|}{AVQZ} & -105.9 & -52.9 & -50.0 & -50.0 & -36.8 & AVQZ & -94.1 & -35.3 & -33.3 & -46.9 & -33.3 \\ \hline
\multicolumn{1}{|>{\centering\arraybackslash}m{88.0pt}|}{V5Z} & -108.8 & \textbf{-41.2} & \textbf{-37.5} & -48.0 & -36.0 & V5Z & -94.1 & -29.4 & -33.3 & -46.9 & -35.1 \\ \hline
\multicolumn{1}{|>{\centering\arraybackslash}m{88.0pt}|}{AV5Z} & \textbf{-100.0} & \textbf{-41.2} & \textbf{-37.5} & -49.0 & -36.0 & AV5Z & -97.1 & -35.3 & -37.5 & -48.0 & -35.1 \\ \hline
\multicolumn{1}{|>{\centering\arraybackslash}m{88.0pt}|}{def2-TZVP} & -111.8 & -58.8 & -54.2 & -49.0 & -35.1 & def2-TZVP & -94.1 & -29.4 & -33.3 & -44.9 & -31.6 \\ \hline
\multicolumn{1}{|>{\centering\arraybackslash}m{88.0pt}|}{def2-QZVP} & -108.8 & \textbf{-41.2} & \textbf{-37.5} & \textbf{-46.9} & \textbf{-34.2} & def2-QZVP & \textbf{-91.2} & -35.3 & \textbf{-29.2} & -45.9 & -33.3 \\ \hline
\multicolumn{1}{|>{\centering\arraybackslash}m{88.0pt}|}{6-311g(2d,2p)} & -197.1 & -182.4 & -141.7 & -73.5 & -58.8 & 6-311g(2d,2p) & -120.6 & -76.5 & -70.8 & -57.1 & -45.6 \\ \hline
\multicolumn{1}{|>{\centering\arraybackslash}m{88.0pt}|}{6-311++g(2d,2p)} & -176.5 & -111.8 & -100.0 & -74.5 & -59.6 & 6-311++g(2d,2p) & -138.2 & -76.5 & -70.8 & -62.2 & -47.4 \\ \hline
\multicolumn{1}{|>{\centering\arraybackslash}m{88.0pt}|}{6-311g(2df,2pd)} & -185.3 & -176.5 & -137.5 & -68.4 & -54.4 & 6-311g(2df,2pd) & -108.8 & -70.6 & -66.7 & -51.0 & -40.4 \\ \hline
\multicolumn{1}{|>{\centering\arraybackslash}m{88.0pt}|}{6-311++g(2df,2pd)} & -161.8 & -105.9 & -95.8 & -69.4 & -55.3 & 6-311++g(2df,2pd) & -123.5 & -70.6 & -62.5 & -57.1 & -43.0 \\ \hline
\multicolumn{1}{|>{\centering\arraybackslash}m{88.0pt}|}{6-311g(3df,3pd)} & -235.3 & -223.5 & -179.2 & -78.6 & -62.3 & 6-311g(3df,3pd) & -102.9 & -70.6 & -62.5 & -54.1 & -43.0 \\ \hline
\multicolumn{1}{|>{\centering\arraybackslash}m{88.0pt}|}{6-311++g(3df,3pd)} & -170.6 & -105.9 & -95.8 & -68.4 & -54.4 & 6-311++g(3df,3pd) & -117.6 & -70.6 & -62.5 & -56.1 & -42.1 \\ \hline
\multicolumn{6}{|>{\centering\arraybackslash}m{269.5pt}|}{\rule[-4.5pt]{0pt}{13.5pt}\textbf{DF\textperiodcentered{}\textperiodcentered{}\textperiodcentered{}NH}\textbf{\textsubscript{3}}\textbf{ (without BSSE correction)}} & \multicolumn{6}{>{\centering\arraybackslash}m{271.0pt}|}{\textbf{DF\textperiodcentered{}\textperiodcentered{}\textperiodcentered{}NH}\textbf{\textsubscript{3}}\textbf{ (with BSSE correction)}} \\ \hline
\multicolumn{1}{|>{\centering\arraybackslash}m{88.0pt}|}{\multirow{2}{88.0pt}{\centering Basis sets}} & \multicolumn{1}{>{\centering\arraybackslash}m{43.7pt}|}{\makebox[\linewidth][c]{\rotatebox{90}{\hspace{2pt}F\textsubscript{2}\textperiodcentered{}\textperiodcentered{}\textperiodcentered{}NH\textsubscript{3}\hspace{2pt}}}} & \multicolumn{1}{>{\centering\arraybackslash}m{31.3pt}|}{\makebox[\linewidth][c]{\rotatebox{90}{\hspace{2pt}NCF\textperiodcentered{}\textperiodcentered{}\textperiodcentered{}NH\textsubscript{3}\hspace{2pt}}}} & \multicolumn{1}{>{\centering\arraybackslash}m{30.6pt}|}{\makebox[\linewidth][c]{\rotatebox{90}{\hspace{2pt}CNF\textperiodcentered{}\textperiodcentered{}\textperiodcentered{}NH\textsubscript{3}\hspace{2pt}}}} & \multicolumn{1}{>{\centering\arraybackslash}m{30.2pt}|}{\makebox[\linewidth][c]{\rotatebox{90}{\hspace{2pt}NH\textsubscript{3}F\textsuperscript{+}\textperiodcentered{}\textperiodcentered{}\textperiodcentered{}NH\textsubscript{3}\hspace{2pt}}}} & \multicolumn{1}{>{\centering\arraybackslash}m{30.5pt}|}{\makebox[\linewidth][c]{\rotatebox{90}{\hspace{2pt}NH\textsubscript{2}F\textsubscript{2}\textsuperscript{+}\textperiodcentered{}\textperiodcentered{}\textperiodcentered{}NH\textsubscript{3}\hspace{2pt}}}} & \multicolumn{1}{>{\centering\arraybackslash}m{80.7pt}|}{\multirow{2}{80.7pt}{\centering Basis sets}} & \multicolumn{1}{>{\centering\arraybackslash}m{34.3pt}|}{\makebox[\linewidth][c]{\rotatebox{90}{\hspace{2pt}F\textsubscript{2}\textperiodcentered{}\textperiodcentered{}\textperiodcentered{}NH\textsubscript{3}\hspace{2pt}}}} & \multicolumn{1}{>{\centering\arraybackslash}m{34.0pt}|}{\makebox[\linewidth][c]{\rotatebox{90}{\hspace{2pt}NCF\textperiodcentered{}\textperiodcentered{}\textperiodcentered{}NH\textsubscript{3}\hspace{2pt}}}} & \multicolumn{1}{>{\centering\arraybackslash}m{35.7pt}|}{\makebox[\linewidth][c]{\rotatebox{90}{\hspace{2pt}CNF\textperiodcentered{}\textperiodcentered{}\textperiodcentered{}NH\textsubscript{3}\hspace{2pt}}}} & \multicolumn{1}{>{\centering\arraybackslash}m{35.7pt}|}{\makebox[\linewidth][c]{\rotatebox{90}{\hspace{2pt}NH\textsubscript{3}F\textsuperscript{+}\textperiodcentered{}\textperiodcentered{}\textperiodcentered{}NH\textsubscript{3}\hspace{2pt}}}} & \multicolumn{1}{>{\centering\arraybackslash}m{35.7pt}|}{\makebox[\linewidth][c]{\rotatebox{90}{\hspace{2pt}NH\textsubscript{2}F\textsubscript{2}\textsuperscript{+}\textperiodcentered{}\textperiodcentered{}\textperiodcentered{}NH\textsubscript{3}\hspace{2pt}}}} \\* \cline{2-6}\cline{8-12}\noalign{\penalty10000}
\multicolumn{1}{|>{\centering\arraybackslash}m{88.0pt}|}{} & \%\ensuremath{\delta}\ensuremath{\Delta}\textbf{\textsubscript{1}} & \%\ensuremath{\delta}\ensuremath{\Delta}\textbf{\textsubscript{1}} & \%\ensuremath{\delta}\ensuremath{\Delta}\textbf{\textsubscript{1}} & \%\ensuremath{\delta}\ensuremath{\Delta}\textbf{\textsubscript{1}} & \%\ensuremath{\delta}\ensuremath{\Delta}\textbf{\textsubscript{1}} &  & \%\ensuremath{\delta}\ensuremath{\Delta}\textsubscript{1+} \textsubscript{BSSE} & \%\ensuremath{\delta}\ensuremath{\Delta}\textsubscript{1+} \textsubscript{BSSE} & \%\ensuremath{\delta}\ensuremath{\Delta}\textsubscript{1+}\allowbreak\textsubscript{BSSE} & \%\ensuremath{\delta}\ensuremath{\Delta}\textsubscript{1+}\allowbreak\textsubscript{BSSE} & \%\ensuremath{\delta}\ensuremath{\Delta}\textsubscript{1+}\allowbreak\textsubscript{BSSE} \\* \hline\noalign{\penalty10000}
\multicolumn{1}{|>{\centering\arraybackslash}m{88.0pt}|}{VDZ} & -62.8 & -52.9 & -26.1 & -22.7 & -20.4 & VDZ & -11.5 & 17.6 & 20.3 & -13.0 & -12.1 \\* \hline\noalign{\penalty10000}
\multicolumn{1}{|>{\centering\arraybackslash}m{88.0pt}|}{AVDZ} & -23.1 & -9.8 & {\leavevmode\color[HTML]{FF0000}-5.8} & -11.5 & -9.8 & AVDZ & -5.1 & 5.9 & {\leavevmode\color[HTML]{FF0000}14.5} & -7.1 & -5.4 \\* \hline\noalign{\penalty10000}
\multicolumn{1}{|>{\centering\arraybackslash}m{88.0pt}|}{VTZ} & -10.3 & -15.7 & {\leavevmode\color[HTML]{FF0000}1.4} & -10.0 & -9.0 & VTZ & 9.0 & 7.8 & {\leavevmode\color[HTML]{FF0000}17.4} & -5.9 & -5.4 \\* \hline\noalign{\penalty10000}
\multicolumn{1}{|>{\centering\arraybackslash}m{88.0pt}|}{\emph{AVTZ}} & {\leavevmode\color[HTML]{FF0000}\emph{17.9}} & {\leavevmode\color[HTML]{FF0000}\emph{7.8}} & {\leavevmode\color[HTML]{FF0000}\emph{21.7}} & \emph{-1.2} & {\leavevmode\color[HTML]{FF0000}\emph{-0.5}} & \emph{AVTZ} & {\leavevmode\color[HTML]{FF0000}\emph{21.8}} & {\leavevmode\color[HTML]{FF0000}\emph{11.8}} & {\leavevmode\color[HTML]{FF0000}\emph{26.1}} & \emph{-0.3} & {\leavevmode\color[HTML]{FF0000}\emph{\textbf{0.8}}} \\ \hline
\multicolumn{1}{|>{\centering\arraybackslash}m{88.0pt}|}{VQZ} & {\leavevmode\color[HTML]{FF0000}16.7} & {\leavevmode\color[HTML]{FF0000}\textbf{2.0}} & {\leavevmode\color[HTML]{FF0000}17.4} & -3.8 & -2.3 & VQZ & {\leavevmode\color[HTML]{FF0000}26.9} & {\leavevmode\color[HTML]{FF0000}13.7} & {\leavevmode\color[HTML]{FF0000}26.1} & -1.5 & -0.5 \\ \hline
\multicolumn{1}{|>{\centering\arraybackslash}m{88.0pt}|}{AVQZ} & {\leavevmode\color[HTML]{FF0000}28.2} & {\leavevmode\color[HTML]{FF0000}13.7} & {\leavevmode\color[HTML]{FF0000}27.5} & {\leavevmode\color[HTML]{FF0000}0.6} & {\leavevmode\color[HTML]{FF0000}1.5} & AVQZ & {\leavevmode\color[HTML]{FF0000}30.8} & {\leavevmode\color[HTML]{FF0000}15.7} & {\leavevmode\color[HTML]{FF0000}29.0} & {\leavevmode\color[HTML]{FF0000}0.9} & {\leavevmode\color[HTML]{FF0000}1.8} \\ \hline
\multicolumn{1}{|>{\centering\arraybackslash}m{88.0pt}|}{V5Z} & {\leavevmode\color[HTML]{FF0000}25.6} & {\leavevmode\color[HTML]{FF0000}11.8} & {\leavevmode\color[HTML]{FF0000}26.1} & -0.6 & {\leavevmode\color[HTML]{FF0000}\textbf{0.3}} & V5Z & {\leavevmode\color[HTML]{FF0000}30.8} & {\leavevmode\color[HTML]{FF0000}15.7} & {\leavevmode\color[HTML]{FF0000}29.0} & \textbf{0.0} & {\leavevmode\color[HTML]{FF0000}\textbf{0.8}} \\ \hline
\multicolumn{1}{|>{\centering\arraybackslash}m{88.0pt}|}{AV5Z} & 29.5 & 13.7 & {\leavevmode\color[HTML]{FF0000}27.5} & {\leavevmode\color[HTML]{FF0000}\textbf{0.0}} & 1.0 & AV5Z & 29.5 & 13.7 & {\leavevmode\color[HTML]{FF0000}29.0} & {\leavevmode\color[HTML]{FF0000}0.3} & 1.0 \\ \hline
\multicolumn{1}{|>{\centering\arraybackslash}m{88.0pt}|}{Def2-TZVP} & {\leavevmode\color[HTML]{FF0000}\textbf{-2.6}} & -11.8 & {\leavevmode\color[HTML]{FF0000}5.8} & -9.7 & -8.2 & Def2-TZVP & {\leavevmode\color[HTML]{FF0000}19.2} & 7.8 & {\leavevmode\color[HTML]{FF0000}20.3} & -4.7 & -3.9 \\ \hline
\multicolumn{1}{|>{\centering\arraybackslash}m{88.0pt}|}{Def2-QZVP} & {\leavevmode\color[HTML]{FF0000}21.8} & {\leavevmode\color[HTML]{FF0000}9.8} & {\leavevmode\color[HTML]{FF0000}23.2} & -1.2 & {\leavevmode\color[HTML]{FF0000}-0.3} & Def2-QZVP & {\leavevmode\color[HTML]{FF0000}30.8} & {\leavevmode\color[HTML]{FF0000}13.7} & {\leavevmode\color[HTML]{FF0000}27.5} & \textbf{0.0} & {\leavevmode\color[HTML]{FF0000}\textbf{0.8}} \\ \hline
\multicolumn{1}{|>{\centering\arraybackslash}m{88.0pt}|}{6-311g(2d,2p)} & -41.0 & -47.1 & -27.5 & -17.4 & -15.5 & 6-311g(2d,2p) & \textbf{-1.3} & 3.9 & \textbf{5.8} & -10.3 & -9.3 \\ \hline
\multicolumn{1}{|>{\centering\arraybackslash}m{88.0pt}|}{6-311++g(2d,2p)} & -19.2 & -15.7 & -4.3 & -10.6 & -9.5 & 6-311++g(2d,2p) & 2.6 & \textbf{2.0} & 14.5 & -5.0 & -4.1 \\ \hline
\multicolumn{1}{|>{\centering\arraybackslash}m{88.0pt}|}{6-311g(2df,2pd)} & -32.1 & -41.2 & -21.7 & -15.0 & -12.9 & 6-311g(2df,2pd) & 7.7 & 9.8 & 11.6 & -7.7 & -6.4 \\ \hline
\multicolumn{1}{|>{\centering\arraybackslash}m{88.0pt}|}{6-311++g(2df,2pd)} & {\leavevmode\color[HTML]{FF0000}-10.3} & -9.8 & {\leavevmode\color[HTML]{FF0000}\textbf{0.0}} & -8.3 & -7.2 & 6-311++g(2df,2pd) & {\leavevmode\color[HTML]{FF0000}11.5} & 7.8 & {\leavevmode\color[HTML]{FF0000}18.8} & -2.7 & -1.8 \\* \hline\noalign{\penalty10000}
\multicolumn{1}{|>{\centering\arraybackslash}m{88.0pt}|}{6-311g(3df,3pd)} & -41.0 & -52.9 & -27.5 & -13.9 & -11.1 & 6-311g(3df,3pd) & 23.1 & 15.7 & 21.7 & -3.2 & -1.8 \\* \hline\noalign{\penalty10000}
\multicolumn{1}{|>{\centering\arraybackslash}m{88.0pt}|}{6-311++g(3df,3pd)} & {\leavevmode\color[HTML]{FF0000}-5.1} & {\leavevmode\color[HTML]{FF0000}3.9} & {\leavevmode\color[HTML]{FF0000}10.1} & -6.2 & -4.9 & 6-311++g(3df,3pd) & {\leavevmode\color[HTML]{FF0000}15.4} & {\leavevmode\color[HTML]{FF0000}9.8} & {\leavevmode\color[HTML]{FF0000}20.3} & -2.7 & -1.5 \\* \specialrule{0.9pt}{0pt}{0pt}\noalign{\penalty10000}
\end{tabular}}}}
\end{table}

\Tref{tab:cbs-limit-binding-energies} already reports BEs with and without BSSE correction for all 29 complexes at M06-2X/AVTZ level. We justify that choice of basis set and BSSE treatment here, based on the results above, for two reasons. First, AVTZ is a practical CBS-limit basis set for DFT calculations.\cite{OterodelaRoza2013} Even after BSSE correction, M06-2X/AVTZ still shows large deviations on BEs from the reference values (\Tref{tab:percentage-deviation-on-bes}, \Fref{fig:bsse-uncorrected-bes-1}). Also, the anomaly regarding BEs of F\textsubscript{2}\textperiodcentered{}\textperiodcentered{}\textperiodcentered{}CH\textsubscript{3}/F\textsubscript{2}\textperiodcentered{}\textperiodcentered{}\textperiodcentered{}NH\textsubscript{3} persists, contrary to all \emph{ab initio} methods. Second, BSSE correction mostly increases the deviation on BEs for CSFB complexes, including F\textsubscript{2}\textperiodcentered{}\textperiodcentered{}\textperiodcentered{}NH\textsubscript{3}. We therefore use BSSE-uncorrected BEs for all 29 complexes at M06-2X/AVTZ throughout this work. Our main conclusions remain unchanged if BSSE-corrected BEs are used instead.

Kozuch and Martin\cite{Kozuch2013} tested basis-set convergence for dissociation energies of closed-shell halogen-bonded complexes with other DFAs. The authors found that AVQZ basis set is close to the complete-basis-set limit. For practical purpose, they recommended AVTZ basis set with BSSE correction for higher accuracy. Double-zeta basis sets performed poorly, with strong overbinding from basis-set superposition error. In the present study, this is not the case for BEs of the selected SEFB complexes predicted by M06-2X with several basis sets. Even with the largest basis set tested (AV5Z) and after BSSE correction, \%\ensuremath{\delta}\textsubscript{BE} for SEFB complexes remains large, for example -97.1\% for F\textsubscript{2}\textperiodcentered{}\textperiodcentered{}\textperiodcentered{}CH\textsubscript{3} (\Tref{tab:percentage-deviation-on-bes}). Increasing basis set alone therefore does not resolve the inaccurate BEs of SEFB complexes found here.

\subsection{Effect of basis sets on F-bond distances by M06-2X}\label{sec:effect-of-basis-sets2}

We assess the performance of M06-2X functional across the 16 basis sets discussed in the previous section to reproduce reference F-bond distances from DF-RMP2-F12/VDZ-F12 (\Tref{tab:intermolecular-distances-r-f}). The F-bond distances are collected in \Tref{tab:f-bond-distances-r2} and percentage deviations in \mbox{\textbf{TABLE \ref{tab:percentage-deviations-on-f}.}}

\begin{table}[!htbp]\centering\setlength{\parskip}{0pt}
\caption{F-bond distances [R(F\textperiodcentered{}\textperiodcentered{}\textperiodcentered{}C/\allowbreak{}N), Å] for selected DF\textperiodcentered{}\textperiodcentered{}\textperiodcentered{}CH\textsubscript{3}/NH\textsubscript{3} complexes (DF = F\textsubscript{2}, NCF, CNF, NH\textsubscript{3}F\textsuperscript{+} and NH\textsubscript{2}F\textsubscript{2}\textsuperscript{+}) obtained from M06-2X with 16 basis sets. Shorthand notation for selected Dunning basis sets are: VDZ\ensuremath{\rightarrow} cc-pVDZ, AVDZ\ensuremath{\rightarrow} aug-cc-pVDZ, VTZ\ensuremath{\rightarrow} cc-pVTZ, AVTZ\ensuremath{\rightarrow} aug-cc-pVTZ, VQZ\ensuremath{\rightarrow} cc-pVQZ, AVQZ\ensuremath{\rightarrow} aug-cc-pVQZ, V5Z\ensuremath{\rightarrow} cc-pV5Z, AV5Z\ensuremath{\rightarrow} aug-cc-pV5Z. See \textbf{T}\textbf{ABLE }\textbf{S15} for the corresponding F-bond angles.}\label{tab:f-bond-distances-r2}\vspace{7pt}\nopagebreak
{\fontsize{8.0}{9.2}\fontfamily{ptm}\selectfont\renewcommand{\arraystretch}{1.17}\setlength{\tabcolsep}{1.5pt}
\centerline{\adjustbox{max width=518pt}{\begin{tabular}{|>{\centering\arraybackslash}m{60.6pt}|>{\centering\arraybackslash}m{26.7pt}|>{\centering\arraybackslash}m{26.7pt}|>{\centering\arraybackslash}m{26.7pt}|>{\centering\arraybackslash}m{26.7pt}|>{\centering\arraybackslash}m{30.9pt}|>{\centering\arraybackslash}m{26.7pt}|>{\centering\arraybackslash}m{26.7pt}|>{\centering\arraybackslash}m{30.9pt}|>{\centering\arraybackslash}m{27.2pt}|>{\centering\arraybackslash}m{26.7pt}|>{\centering\arraybackslash}m{26.2pt}|>{\centering\arraybackslash}m{27.0pt}|>{\centering\arraybackslash}m{26.7pt}|>{\centering\arraybackslash}m{26.6pt}|>{\centering\arraybackslash}m{25.6pt}|>{\centering\arraybackslash}m{28.6pt}|}
\specialrule{0.9pt}{0pt}{0pt}
\multicolumn{1}{|>{\centering\arraybackslash}m{60.6pt}|}{DF\textperiodcentered{}\textperiodcentered{}\textperiodcentered{}CH\textsubscript{3}/NH\textsubscript{3}} & \multicolumn{1}{>{\centering\arraybackslash}p{26.7pt}|}{\makebox[\linewidth][c]{\rotatebox{90}{\hspace{2pt}VDZ\hspace{2pt}}}} & \multicolumn{1}{>{\centering\arraybackslash}p{26.7pt}|}{\makebox[\linewidth][c]{\rotatebox{90}{\hspace{2pt}AVDZ\hspace{2pt}}}} & \multicolumn{1}{>{\centering\arraybackslash}p{26.7pt}|}{\makebox[\linewidth][c]{\rotatebox{90}{\hspace{2pt}VTZ\hspace{2pt}}}} & \multicolumn{1}{>{\centering\arraybackslash}p{26.7pt}|}{\makebox[\linewidth][c]{\rotatebox{90}{\hspace{2pt}\textbf{AVTZ}\hspace{2pt}}}} & \multicolumn{1}{>{\centering\arraybackslash}p{30.9pt}|}{\makebox[\linewidth][c]{\rotatebox{90}{\hspace{2pt}VQZ\hspace{2pt}}}} & \multicolumn{1}{>{\centering\arraybackslash}p{26.7pt}|}{\makebox[\linewidth][c]{\rotatebox{90}{\hspace{2pt}AVQZ\hspace{2pt}}}} & \multicolumn{1}{>{\centering\arraybackslash}p{26.7pt}|}{\makebox[\linewidth][c]{\rotatebox{90}{\hspace{2pt}V5Z\hspace{2pt}}}} & \multicolumn{1}{>{\centering\arraybackslash}p{30.9pt}|}{\makebox[\linewidth][c]{\rotatebox{90}{\hspace{2pt}AV5Z\hspace{2pt}}}} & \multicolumn{1}{>{\centering\arraybackslash}p{27.2pt}|}{\makebox[\linewidth][c]{\rotatebox{90}{\hspace{2pt}Def2-TZVP\hspace{2pt}}}} & \multicolumn{1}{>{\centering\arraybackslash}p{26.7pt}|}{\makebox[\linewidth][c]{\rotatebox{90}{\hspace{2pt}Def2-QZVP\hspace{2pt}}}} & \multicolumn{1}{>{\centering\arraybackslash}p{26.2pt}|}{\makebox[\linewidth][c]{\rotatebox{90}{\hspace{2pt}6-311g(2d,2p)\hspace{2pt}}}} & \multicolumn{1}{>{\centering\arraybackslash}p{27.0pt}|}{\makebox[\linewidth][c]{\rotatebox{90}{\hspace{2pt}6-311++g(2d,2p)\hspace{2pt}}}} & \multicolumn{1}{>{\centering\arraybackslash}p{26.7pt}|}{\makebox[\linewidth][c]{\rotatebox{90}{\hspace{2pt}6-311g(2df,2pd)\hspace{2pt}}}} & \multicolumn{1}{>{\centering\arraybackslash}p{26.6pt}|}{\makebox[\linewidth][c]{\rotatebox{90}{\hspace{2pt}6-311++g(2df,2pd)\hspace{2pt}}}} & \multicolumn{1}{>{\centering\arraybackslash}p{25.6pt}|}{\makebox[\linewidth][c]{\rotatebox{90}{\hspace{2pt}6-311g(3df,3pd)\hspace{2pt}}}} & \multicolumn{1}{>{\centering\arraybackslash}p{28.6pt}|}{\makebox[\linewidth][c]{\rotatebox{90}{\hspace{2pt}6-311++g(3df,3pd)\hspace{2pt}}}} \\* \hline\noalign{\penalty10000}
\multicolumn{1}{|>{\centering\arraybackslash}m{60.6pt}|}{F\textsubscript{2}\textperiodcentered{}\textperiodcentered{}\textperiodcentered{}CH\textsubscript{3}} & 2.514 & 2.614 & 2.648 & 2.648 & 2.660 & 2.661 & 2.661 & 2.663 & 2.664 & 2.658 & 2.629 & 2.633 & 2.644 & 2.646 & 2.640 & 2.645 \\* \hline\noalign{\penalty10000}
\multicolumn{1}{|>{\centering\arraybackslash}m{60.6pt}|}{\textbf{NCF\textperiodcentered{}\textperiodcentered{}\textperiodcentered{}CH}\textbf{\textsubscript{3}}} & 2.932 & 3.004 & 3.011 & 3.018 & 3.036 & 3.037 & 3.031 & 3.033 & 3.025 & 3.033 & 2.975 & 3.002 & 2.974 & 3.000 & 2.980 & 3.005 \\ \hline
\multicolumn{1}{|>{\centering\arraybackslash}m{60.6pt}|}{\textbf{CNF\textperiodcentered{}\textperiodcentered{}\textperiodcentered{}CH}\textbf{\textsubscript{3}}} & 2.868 & 2.935 & 2.965 & 2.976 & 2.989 & 2.987 & 2.988 & 2.987 & 2.979 & 2.988 & 2.911 & 2.946 & 2.913 & 2.948 & 2.919 & 2.959 \\ \hline
\multicolumn{1}{|>{\centering\arraybackslash}m{60.6pt}|}{NH\textsubscript{3}F\textsuperscript{+}\textperiodcentered{}\textperiodcentered{}\textperiodcentered{}CH\textsubscript{3}} & 2.607 & 2.689 & 2.686 & 2.693 & 2.703 & 2.707 & 2.707 & 2.711 & 2.699 & 2.708 & 2.662 & 2.684 & 2.676 & 2.701 & 2.686 & 2.714 \\ \hline
\multicolumn{1}{|>{\centering\arraybackslash}m{60.6pt}|}{NH\textsubscript{2}F\textsubscript{2}\textsuperscript{+}\textperiodcentered{}\textperiodcentered{}\textperiodcentered{}CH\textsubscript{3}} & 2.621 & 2.705 & 2.703 & 2.713 & 2.721 & 2.727 & 2.727 & 2.733 & 2.718 & 2.728 & 2.678 & 2.706 & 2.689 & 2.716 & 2.702 & 2.728 \\ \hline
\multicolumn{1}{|>{\centering\arraybackslash}m{60.6pt}|}{{\leavevmode\color[HTML]{FF0000}F}{\leavevmode\color[HTML]{FF0000}\textsubscript{2}}{\leavevmode\color[HTML]{FF0000}\textperiodcentered{}\textperiodcentered{}\textperiodcentered{}NH}{\leavevmode\color[HTML]{FF0000}\textsubscript{3}}} & 2.490 & 2.554 & 2.563 & 2.596 & 2.590 & 2.602 & 2.606 & 2.612 & 2.572 & 2.603 & 2.553 & 2.578 & 2.564 & 2.595 & 2.570 & 2.588 \\ \hline
\multicolumn{1}{|>{\centering\arraybackslash}m{60.6pt}|}{{\leavevmode\color[HTML]{FF0000}NCF\textperiodcentered{}\textperiodcentered{}\textperiodcentered{}NH}{\leavevmode\color[HTML]{FF0000}\textsubscript{3}}} & 2.904 & 3.088 & 3.119 & 3.205 & 3.183 & 3.235 & 3.225 & 3.231 & 3.073 & 3.211 & 2.952 & 3.067 & 2.961 & 3.091 & 2.969 & 3.159 \\ \hline
\multicolumn{1}{|>{\centering\arraybackslash}m{60.6pt}|}{{\leavevmode\color[HTML]{FF0000}CNF\textperiodcentered{}\textperiodcentered{}\textperiodcentered{}NH}{\leavevmode\color[HTML]{FF0000}\textsubscript{3}}} & 2.804 & 2.881 & 2.903 & 3.031 & 3.008 & 3.072 & 3.066 & 3.072 & 2.898 & 3.050 & 2.839 & 2.893 & 2.852 & 2.909 & 2.856 & 2.940 \\* \hline\noalign{\penalty10000}
\multicolumn{1}{|>{\centering\arraybackslash}m{60.6pt}|}{{\leavevmode\color[HTML]{FF0000}NH}{\leavevmode\color[HTML]{FF0000}\textsubscript{3}}{\leavevmode\color[HTML]{FF0000}F}{\leavevmode\color[HTML]{FF0000}\textsuperscript{+}}{\leavevmode\color[HTML]{FF0000}\textperiodcentered{}\textperiodcentered{}\textperiodcentered{}NH}{\leavevmode\color[HTML]{FF0000}\textsubscript{3}}} & 2.489 & 2.563 & 2.556 & 2.590 & 2.579 & 2.594 & 2.594 & 2.596 & 2.570 & 2.592 & 2.529 & 2.574 & 2.538 & 2.585 & 2.555 & 2.588 \\* \hline\noalign{\penalty10000}
\multicolumn{1}{|>{\centering\arraybackslash}m{60.6pt}|}{{\leavevmode\color[HTML]{FF0000}NH}{\leavevmode\color[HTML]{FF0000}\textsubscript{2}}{\leavevmode\color[HTML]{FF0000}F}{\leavevmode\color[HTML]{FF0000}\textsubscript{2}}{\leavevmode\color[HTML]{FF0000}\textsuperscript{+}}{\leavevmode\color[HTML]{FF0000}\textperiodcentered{}\textperiodcentered{}\textperiodcentered{}NH}{\leavevmode\color[HTML]{FF0000}\textsubscript{3}}} & 2.478 & 2.554 & 2.546 & 2.581 & 2.574 & 2.585 & 2.586 & 2.586 & 2.567 & 2.582 & 2.514 & 2.566 & 2.523 & 2.577 & 2.542 & 2.578 \\* \specialrule{0.9pt}{0pt}{0pt}\noalign{\penalty10000}
\end{tabular}}}}
\end{table}

\Tref{tab:f-bond-distances-r2} shows that F-bond distances (R) for the selected SEFB complexes predicted by M06-2X generally increase with basis set size, with little further change beyond VQZ basis set. These distances remain shorter than the reference values from DF-RMP2-F12/VDZ-F12 (\Tref{tab:intermolecular-distances-r-f}) at every basis set tested. For F\textsubscript{2}\textperiodcentered{}\textperiodcentered{}\textperiodcentered{}CH\textsubscript{3}, R reaches 2.663 Å at M06-2X/AV5Z, against a reference of 2.869 Å (\Tref{tab:intermolecular-distances-r-f}), a difference of 0.206 Å. For NCF\textperiodcentered{}\textperiodcentered{}\textperiodcentered{}CH\textsubscript{3}, the difference is larger, 0.395 Å (3.033 Å versus 3.428 Å reference). The other three SEFB complexes fall closer to F\textsubscript{2}..CH\textsubscript{3}. This deviation persists not only at AV5Z, but also across all 16 basis sets. The three basis-set families used, Dunning, Alrich and Pople, do not converge to the same value. For F\textsubscript{2}\textperiodcentered{}\textperiodcentered{}\textperiodcentered{}CH\textsubscript{3}, the Dunning and Alrich basis sets converge near 2.6 Å, while the Pople 6-311g basis sets remain lower, around 2.63 to 2.65 Å. For the CSFB complexes, F-bond distances approach the reference values (\Tref{tab:intermolecular-distances-r-f}) closely across all three basis-set families (\Tref{tab:f-bond-distances-r2}). For F\textsubscript{2}\textperiodcentered{}\textperiodcentered{}\textperiodcentered{}NH\textsubscript{3}, R is 2.612 Å at M06-2X/AV5Z against a reference of 2.616 Å, a difference of only 0.004 Å.

\begin{table}[!htbp]\centering\setlength{\parskip}{0pt}
\caption{Percentage deviations on\textbf{ }F-bond distances [\%\ensuremath{\delta}R, Å] for selected DF\textperiodcentered{}\textperiodcentered{}\textperiodcentered{}CH\textsubscript{3}/NH\textsubscript{3} complexes (DF = F\textsubscript{2}, NCF, CNF, NH\textsubscript{3}F\textsuperscript{+}\textsuperscript{ }and NH\textsubscript{2}F\textsubscript{2}\textsuperscript{+}) using M06-2X functional with 16 basis sets with respect to F-bond distances calculated at DF-RMP2-F12/\allowbreak{}VDZ-F12 level of theory. Following are shorthand notation for selected Dunning basis sets: VDZ\ensuremath{\rightarrow} cc-pVDZ, AVDZ\ensuremath{\rightarrow} aug-cc-pVDZ, VTZ\ensuremath{\rightarrow} cc-pVTZ, AVTZ\ensuremath{\rightarrow} aug-cc-pVTZ, VQZ\ensuremath{\rightarrow} cc-pVQZ, AVQZ\ensuremath{\rightarrow} aug-cc-pVQZ, V5Z\ensuremath{\rightarrow} cc-pV5Z, AV5Z\ensuremath{\rightarrow} aug-cc-pV5Z.}\label{tab:percentage-deviations-on-f}\vspace{7pt}\nopagebreak
{\fontsize{9.0}{10.3}\fontfamily{ptm}\selectfont\renewcommand{\arraystretch}{1.17}\setlength{\tabcolsep}{1.5pt}
\centerline{\adjustbox{max width=518pt}{\begin{tabular}{|>{\centering\arraybackslash}m{84.8pt}|>{\centering\arraybackslash}p{31.3pt}|>{\centering\arraybackslash}p{33.0pt}|>{\centering\arraybackslash}p{32.1pt}|>{\centering\arraybackslash}p{32.1pt}|>{\centering\arraybackslash}p{32.4pt}|>{\centering\arraybackslash}m{80.7pt}|>{\centering\arraybackslash}p{36.5pt}|>{\centering\arraybackslash}p{36.5pt}|>{\centering\arraybackslash}p{32.1pt}|>{\centering\arraybackslash}p{32.1pt}|>{\centering\arraybackslash}p{32.2pt}|}
\specialrule{0.9pt}{0pt}{0pt}
\multicolumn{6}{|>{\centering\arraybackslash}m{260.7pt}|}{\textbf{DF\textperiodcentered{}\textperiodcentered{}\textperiodcentered{}CH}\textbf{\textsubscript{3}}} & \multicolumn{6}{>{\centering\arraybackslash}m{265.3pt}|}{\textbf{DF\textperiodcentered{}\textperiodcentered{}\textperiodcentered{}NH}\textbf{\textsubscript{3}}} \\* \hline\noalign{\penalty10000}
\multicolumn{1}{|>{\centering\arraybackslash}m{84.8pt}|}{\multirow{2}{84.8pt}{\centering Basis sets}} & \multicolumn{1}{>{\centering\arraybackslash}m{31.3pt}|}{\makebox[\linewidth][c]{\rotatebox{90}{\hspace{2pt}\textbf{F}\textbf{\textsubscript{2}}\textbf{\textperiodcentered{}\textperiodcentered{}\textperiodcentered{}CH}\textbf{\textsubscript{3}}\hspace{2pt}}}} & \multicolumn{1}{>{\centering\arraybackslash}m{33.0pt}|}{\makebox[\linewidth][c]{\rotatebox{90}{\hspace{2pt}\textbf{NCF\textperiodcentered{}\textperiodcentered{}\textperiodcentered{}CH}\textbf{\textsubscript{3}}\hspace{2pt}}}} & \multicolumn{1}{>{\centering\arraybackslash}m{32.1pt}|}{\makebox[\linewidth][c]{\rotatebox{90}{\hspace{2pt}\textbf{CNF\textperiodcentered{}\textperiodcentered{}\textperiodcentered{}CH}\textbf{\textsubscript{3}}\hspace{2pt}}}} & \multicolumn{1}{>{\centering\arraybackslash}m{32.1pt}|}{\makebox[\linewidth][c]{\rotatebox{90}{\hspace{2pt}\textbf{NH}\textbf{\textsubscript{3}}\textbf{F}\textbf{\textsuperscript{+}}\textbf{\textperiodcentered{}\textperiodcentered{}\textperiodcentered{}CH}\textbf{\textsubscript{3}}\hspace{2pt}}}} & \multicolumn{1}{>{\centering\arraybackslash}m{32.4pt}|}{\makebox[\linewidth][c]{\rotatebox{90}{\hspace{2pt}\textbf{NH}\textbf{\textsubscript{2}}\textbf{F}\textbf{\textsubscript{2}}\textbf{\textsuperscript{+}}\textbf{\textperiodcentered{}\textperiodcentered{}\textperiodcentered{}CH}\textbf{\textsubscript{3}}\hspace{2pt}}}} & \multirow{2}{80.7pt}{\centering Basis sets} & \multicolumn{1}{>{\centering\arraybackslash}m{36.5pt}|}{\makebox[\linewidth][c]{\rotatebox{90}{\hspace{2pt}\textbf{F}\textbf{\textsubscript{2}}\textbf{\textperiodcentered{}\textperiodcentered{}\textperiodcentered{}NH}\textbf{\textsubscript{3}}\hspace{2pt}}}} & \multicolumn{1}{>{\centering\arraybackslash}m{36.5pt}|}{\makebox[\linewidth][c]{\rotatebox{90}{\hspace{2pt}\textbf{NCF\textperiodcentered{}\textperiodcentered{}\textperiodcentered{}NH}\textbf{\textsubscript{3}}\hspace{2pt}}}} & \multicolumn{1}{>{\centering\arraybackslash}m{32.1pt}|}{\makebox[\linewidth][c]{\rotatebox{90}{\hspace{2pt}\textbf{CNF\textperiodcentered{}\textperiodcentered{}\textperiodcentered{}NH}\textbf{\textsubscript{3}}\hspace{2pt}}}} & \multicolumn{1}{>{\centering\arraybackslash}m{32.1pt}|}{\makebox[\linewidth][c]{\rotatebox{90}{\hspace{2pt}\textbf{NH}\textbf{\textsubscript{3}}\textbf{F}\textbf{\textsuperscript{+}}\textbf{\textperiodcentered{}\textperiodcentered{}\textperiodcentered{}NH}\textbf{\textsubscript{3}}\hspace{2pt}}}} & \multicolumn{1}{>{\centering\arraybackslash}m{32.2pt}|}{\makebox[\linewidth][c]{\rotatebox{90}{\hspace{2pt}\textbf{NH}\textbf{\textsubscript{2}}\textbf{F}\textbf{\textsubscript{2}}\textbf{\textsuperscript{+}}\textbf{\textperiodcentered{}\textperiodcentered{}\textperiodcentered{}NH}\textbf{\textsubscript{3}}\hspace{2pt}}}} \\* \cline{2-6}\cline{8-12}\noalign{\penalty10000}
\multicolumn{1}{|>{\centering\arraybackslash}m{84.8pt}|}{} & \%\ensuremath{\delta}R & \%\ensuremath{\delta}R & \%\ensuremath{\delta}R & \%\ensuremath{\delta}R & \%\ensuremath{\delta}R &  & \%\ensuremath{\delta}R & \%\ensuremath{\delta}R & \%\ensuremath{\delta}R & \%\ensuremath{\delta}R & \%\ensuremath{\delta}R \\* \hline\noalign{\penalty10000}
\multicolumn{1}{|>{\centering\arraybackslash}m{84.8pt}|}{VDZ} & -12.4 & -14.5 & -12.1 & -11.5 & -10.9 & VDZ & -4.8 & -8.1 & -5.3 & -6.0 & -5.7 \\* \hline\noalign{\penalty10000}
\multicolumn{1}{|>{\centering\arraybackslash}m{84.8pt}|}{AVDZ} & -8.9 & -12.4 & -10.0 & -8.8 & -8.0 & AVDZ & -2.4 & -2.3 & -2.7 & -3.2 & -2.8 \\* \hline\noalign{\penalty10000}
\multicolumn{1}{|>{\centering\arraybackslash}m{84.8pt}|}{VTZ} & -7.7 & -12.2 & -9.1 & -8.9 & -8.1 & VTZ & -2.0 & -1.3 & -1.9 & -3.4 & -3.1 \\* \hline\noalign{\penalty10000}
\multicolumn{1}{|>{\centering\arraybackslash}m{84.8pt}|}{\emph{AVTZ}} & \emph{-7.7} & \emph{-12.0} & \emph{-8.8} & \emph{-8.6} & \emph{-7.8} & \emph{AVTZ} & \emph{-0.8} & \emph{1.4} & \emph{2.4} & \emph{-2.2} & \emph{-1.8} \\ \hline
\multicolumn{1}{|>{\centering\arraybackslash}m{84.8pt}|}{VQZ} & -7.3 & \textbf{-11.4} & \textbf{-8.4} & -8.3 & -7.5 & VQZ & -1.0 & 0.7 & 1.6 & -2.6 & -2.0 \\ \hline
\multicolumn{1}{|>{\centering\arraybackslash}m{84.8pt}|}{AVQZ} & -7.2 & \textbf{-11.4} & \textbf{-8.4} & -8.1 & -7.3 & AVQZ & -0.5 & 2.4 & 3.8 & -2.0 & \textbf{-1.6} \\ \hline
\multicolumn{1}{|>{\centering\arraybackslash}m{84.8pt}|}{V5Z} & -7.2 & -11.6 & \textbf{-8.4} & -8.1 & -7.3 & V5Z & -0.4 & 2.1 & 3.6 & -2.0 & \textbf{-1.6} \\ \hline
\multicolumn{1}{|>{\centering\arraybackslash}m{84.8pt}|}{AV5Z} & -7.2 & -11.5 & \textbf{-8.4} & \textbf{-8.0} & \textbf{-7.1} & AV5Z & \textbf{-0.2} & 2.2 & 3.8 & \textbf{-1.9} & \textbf{-1.6} \\ \hline
\multicolumn{1}{|>{\centering\arraybackslash}m{84.8pt}|}{Def2-TZVP} & \textbf{-7.1} & -11.8 & -8.7 & -8.4 & -7.6 & Def2-TZVP & -1.7 & -2.8 & -2.1 & -2.9 & -2.3 \\ \hline
\multicolumn{1}{|>{\centering\arraybackslash}m{84.8pt}|}{Def2-QZVP} & -7.4 & -11.5 & \textbf{-8.4} & -8.1 & -7.2 & Def2-QZVP & -0.5 & 1.6 & 3.0 & -2.1 & -1.7 \\ \hline
\multicolumn{1}{|>{\centering\arraybackslash}m{84.8pt}|}{6-311g(2d,2p)} & -8.4 & -13.2 & -10.8 & -9.7 & -8.9 & 6-311g(2d,2p) & -2.4 & -6.6 & -4.1 & -4.5 & -4.3 \\ \hline
\multicolumn{1}{|>{\centering\arraybackslash}m{84.8pt}|}{6-311++g(2d,2p)} & -8.2 & -12.4 & -9.7 & -8.9 & -8.0 & 6-311++g(2d,2p) & -1.5 & -2.9 & -2.3 & -2.8 & -2.3 \\ \hline
\multicolumn{1}{|>{\centering\arraybackslash}m{84.8pt}|}{6-311g(2df,2pd)} & -7.8 & -13.2 & -10.7 & -9.2 & -8.6 & 6-311g(2df,2pd) & -2.0 & -6.3 & -3.6 & -4.1 & -4.0 \\ \hline
\multicolumn{1}{|>{\centering\arraybackslash}m{84.8pt}|}{6-311++g(2df,2pd)} & -7.8 & -12.5 & -9.6 & -8.3 & -7.7 & 6-311++g(2df,2pd) & -0.8 & -2.2 & -1.7 & -2.3 & -1.9 \\* \hline\noalign{\penalty10000}
\multicolumn{1}{|>{\centering\arraybackslash}m{84.8pt}|}{6-311g(3df,3pd)} & -8.0 & -13.1 & -10.5 & -8.9 & -8.1 & 6-311g(3df,3pd) & -1.8 & -6.0 & -3.5 & -3.5 & -3.2 \\* \hline\noalign{\penalty10000}
\multicolumn{1}{|>{\centering\arraybackslash}m{84.8pt}|}{6-311++g(3df,3pd)} & -7.8 & -12.3 & -9.3 & -7.9 & -7.2 & 6-311++g(3df,3pd) & -1.1 & \textbf{0.0} & \textbf{-0.7} & -2.2 & -1.9 \\* \specialrule{0.9pt}{0pt}{0pt}\noalign{\penalty10000}
\end{tabular}}}}
\end{table}

In \Tref{tab:percentage-deviations-on-f}, negative percentage deviations on F-bond distances (\%\ensuremath{\delta}R) indicate that M06-2X underestimates the F-bond distances relative to reference values, while positive \%\ensuremath{\delta}R indicates overestimation. For the selected SEFB complexes (DF\textperiodcentered{}\textperiodcentered{}\textperiodcentered{}CH\textsubscript{3}), \%\ensuremath{\delta}R values are negative and above 5.0\% in magnitude at every basis set used, from -7.1\% at best to -14.5\% at worst (VDZ). So, none of the 16 basis sets used, in conjunction with M06-2X, reproduces the reference F-bond distances of the selected SEFB complexes obtained at DF-RMP2-F12/VDZ-F12 level (\Tref{tab:intermolecular-distances-r-f}).

For CSFB complexes (DF\textperiodcentered{}\textperiodcentered{}\textperiodcentered{}NH\textsubscript{3}), two of the five, NCF\textperiodcentered{}\textperiodcentered{}\textperiodcentered{}NH\textsubscript{3} and CNF\textperiodcentered{}\textperiodcentered{}\textperiodcentered{}NH\textsubscript{3}, change sign of \%\ensuremath{\delta}R from negative to positive with basis sets AVTZ or VQZ onward, indicating that M06-2X shifts from underestimating to overestimating these distances as the basis set grows. F-bond distances for F\textsubscript{2}\textperiodcentered{}\textperiodcentered{}\textperiodcentered{}NH\textsubscript{3}, NH\textsubscript{3}F\textsuperscript{+}\textperiodcentered{}\textperiodcentered{}\textperiodcentered{}NH\textsubscript{3} and NH\textsubscript{2}F\textsubscript{2}\textsuperscript{+}\textperiodcentered{}\textperiodcentered{}\textperiodcentered{}NH\textsubscript{3} complexes predicted by M06-2X with all 16 basis sets remain negative (\Tref{tab:percentage-deviations-on-f}). The magnitude of \%\ensuremath{\delta}R for the selected DF\textperiodcentered{}\textperiodcentered{}\textperiodcentered{}NH\textsubscript{3} complexes in general falls within 5.0\%, with clear exception observed at M06-2X/VDZ level (Refer \Tref{tab:percentage-deviations-on-f} for detail).

In fact, VDZ is the worst basis set for every complex in both data sets (SEFB and CSFB) alike. It should be avoided regardless of complex type. Increasing basis sets from AVTZ to AVQZ and AV5Z shows that AVTZ is already close to the basis set limit for F-bond distances. For all five SEFB complexes, \%\ensuremath{\delta}R changes by no more than 0.7\% from AVTZ to AV5Z. For the CSFB complexes, the change is similarly small, no more than 1.4\% (\Tref{tab:percentage-deviations-on-f}). Therefore, AVTZ is a reasonable practical choice for F-bond distances in both data sets, close to the basis-set limit rather than merely below the threshold of 5.0\%. For SEFB complexes, however, M06-2X with no basis set, including AVQZ and AV5Z, reproduces reference F-bond distances obtained from DF-RMP2-F12/VDZ-F12 level.

Overall, increasing the size of basis set helps to reproduce reference F-bond distances of the selected CSFB complexes but not in the case of selected SEFB complexes. Combined with the basis-set independence of the over-binding of SEFB complexes already discussed in \Sref{sec:effect-of-basis-sets}, M06-2X combined with any of the 16 basis sets used is not suitable for predicting both BEs and F-bond distances of SEFB complexes.

We have mentioned in \Sref{sec:effect-of-basis-sets} that Kozuch and Martin\cite{Kozuch2013} tested basis-set convergence for dissociation energies of closed-shell halogen-bonded complexes. They did not examine the effects of size of basis set on halogen-bond distances. The present work examines the effect of basis sets on the F-bond distances of hitherto unexplored open-shell SEFB and CSFB complexes.

\FloatBarrier
\subsection{Overall effect of basis sets for BEs of DF\textperiodcentered{}\textperiodcentered{}\textperiodcentered{}CH\textsubscript{3}/NH\textsubscript{3} by M06-2X}\label{sec:overall-effect-of-basis}

\begin{figure}[!htbp]\centering
\GPKfig{6.46781in}{2.21637in}{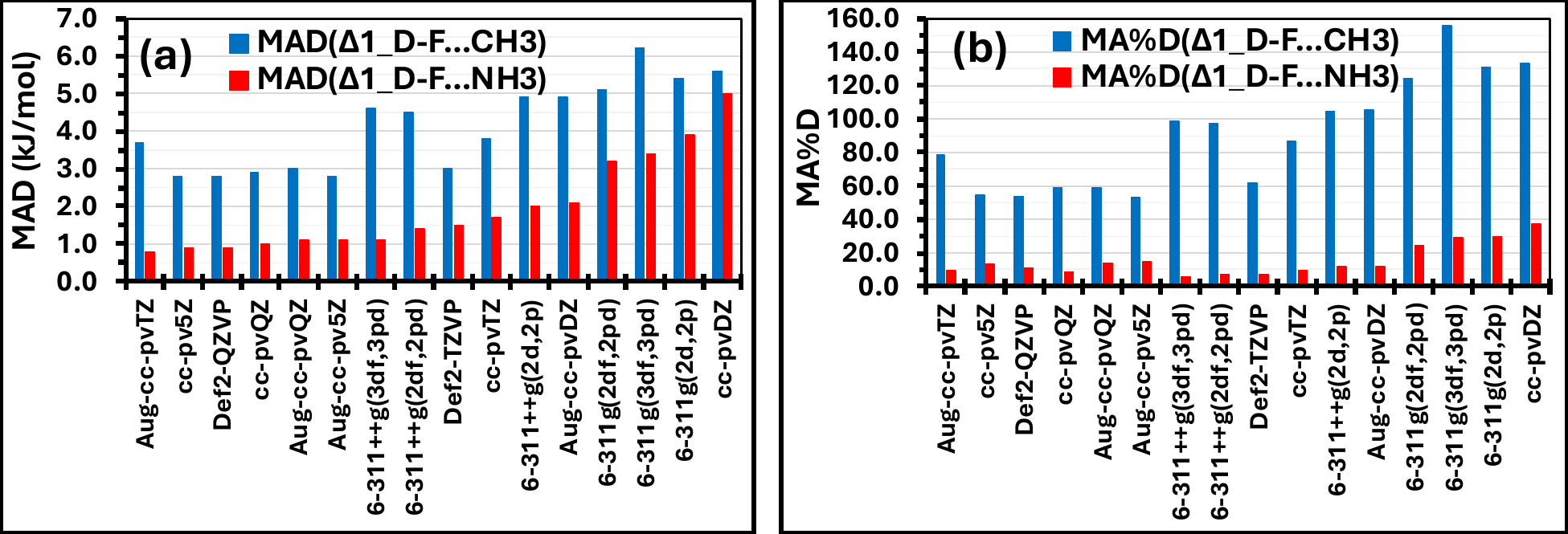}
\GPKfig{6.48147in}{2.51187in}{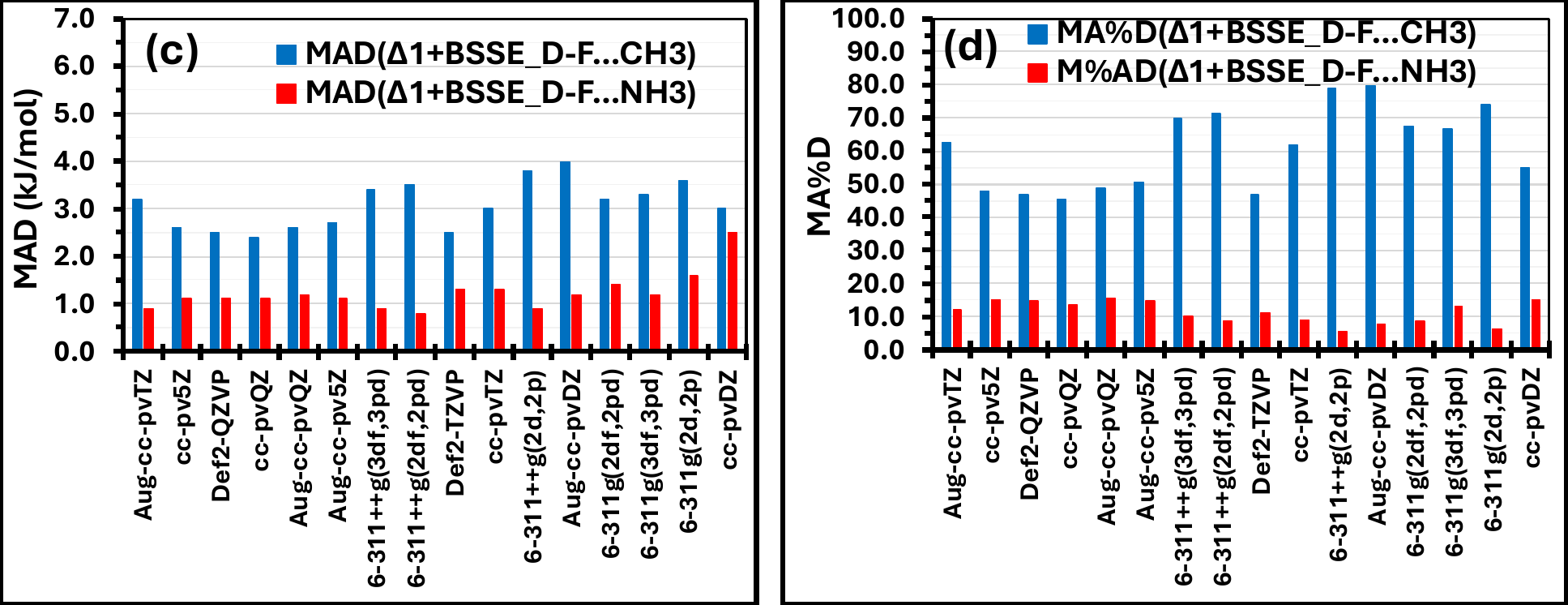}
\par\vspace{-1pt}\caption{MAD \textbf{(a)} and MA\%D \textbf{(b)} on BSSE-uncorrected BEs (\ensuremath{\Delta}\textsubscript{1}) for selected DF\textperiodcentered{}\textperiodcentered{}\textperiodcentered{}CH\textsubscript{3}/NH\textsubscript{3} complexes (DF = F\textsubscript{2}, NCF, CNF, NH\textsubscript{3}F\textsuperscript{+} and NH\textsubscript{2}F\textsubscript{2}\textsuperscript{+}), obtained from M06-2X with 16 basis sets. Corresponding MAD and MA\%D for BSSE-corrected BEs are shown in \textbf{(c)} and \textbf{(d)}. Basis sets are ordered by increasing MADs on BSSE-uncorrected BEs for DF\textperiodcentered{}\textperiodcentered{}\textperiodcentered{}NH\textsubscript{3} complexes. Reference BEs are from DF-RMP2-F12/CBS{[}V(T,Q)Z-F12{]}//DF-RMP2-F12/VDZ-F12 level of theory. MSE and RMSD values on BEs are shown in \Fref{fig:mean-signed-error-mse}.}\label{fig:mad-a-and-ma}
\vspace{10pt}
\end{figure}

In \Tref{tab:mse-mad-rmsd-max2}, we summarize error statistics (MSE, MAD, RMSD, MAX, MA\%D) on BEs for the five selected complexes predicted by M06-2X across the 16 basis sets. Also, for graphical visualization, MAD and MA\%D values are shown in \mbox{\textbf{FIG. \ref{fig:mad-a-and-ma}.}} MSEs on both BSSE-corrected and uncorrected BEs are negative for SEFB complexes at every basis set, confirming over-binding regardless of the basis sets used. For DF\textperiodcentered{}\textperiodcentered{}\textperiodcentered{}NH\textsubscript{3} complexes, MSEs are negative at smaller basis sets but become positive from AVTZ or VQZ onward.

\begin{table}[!htbp]\centering\setlength{\parskip}{0pt}
\caption{MSE, MAD, RMSD, MAX, and MA\%D on BEs (kJ/\allowbreak{}mol) for selected DF\textperiodcentered{}\textperiodcentered{}\textperiodcentered{}CH\textsubscript{3}/NH\textsubscript{3} complexes (DF = F\textsubscript{2}, NCF, CNF, NH\textsubscript{3}F\textsuperscript{+}\textsuperscript{ }and NH\textsubscript{2}F\textsubscript{2}\textsuperscript{+}) obtained from M06-2X with 16 different basis sets, relative to reference BEs at DF-RMP2-F12/\allowbreak{}CBS[V(T,Q)Z-F12]//DF-RMP2-F12/\allowbreak{}VDZ-F12. BEs without (\ensuremath{\Delta}\textsubscript{1}) and with (\ensuremath{\Delta}\textsubscript{1+BSSE}) BSSE correction are analyzed separately. M06-2X/\allowbreak{}AVTZ results are shown in italics; the smallest MAD and MA\%D values for each complex type (SEFB and CSFB) are shown in boldface.}\label{tab:mse-mad-rmsd-max2}\vspace{7pt}\nopagebreak
{\fontsize{9.0}{10.3}\fontfamily{ptm}\selectfont\renewcommand{\arraystretch}{1.18}\setlength{\tabcolsep}{1.5pt}
\centerline{\adjustbox{max width=518pt}{\begin{tabular}{|>{\centering\arraybackslash}m{80.7pt}|>{\centering\arraybackslash}m{27.9pt}|>{\centering\arraybackslash}m{32.4pt}|>{\centering\arraybackslash}m{38.5pt}|>{\centering\arraybackslash}m{31.5pt}|>{\centering\arraybackslash}m{39.5pt}|>{\centering\arraybackslash}m{80.7pt}|>{\centering\arraybackslash}m{28.3pt}|>{\centering\arraybackslash}m{31.5pt}|>{\centering\arraybackslash}m{32.2pt}|>{\centering\arraybackslash}m{31.5pt}|>{\centering\arraybackslash}m{33.9pt}|}
\specialrule{0.9pt}{0pt}{0pt}
\multicolumn{6}{|>{\centering\arraybackslash}m{265.6pt}|}{\textbf{DF\textperiodcentered{}\textperiodcentered{}\textperiodcentered{}CH}\textbf{\textsubscript{3}}\textbf{ (without BSSE correction)}} & \multicolumn{6}{>{\centering\arraybackslash}m{253.0pt}|}{\textbf{DF\textperiodcentered{}\textperiodcentered{}\textperiodcentered{}CH}\textbf{\textsubscript{3}}\textbf{ (with BSSE correction)}} \\* \hline\noalign{\penalty10000}
\multicolumn{1}{|>{\centering\arraybackslash}m{80.7pt}|}{Basis sets} & MSE & MAD & RMSD & MAX & MA\%D & Basis sets & MSE & MAD & RMSD & MAX & MA\%D \\* \hline\noalign{\penalty10000}
\multicolumn{1}{|>{\centering\arraybackslash}m{80.7pt}|}{VDZ} & -5.6 & 5.6 & 6.0 & -8.1 & 133.3 & VDZ & -3.0 & 3.0 & 3.6 & -5.1 & 54.8 \\ \hline
\multicolumn{1}{|>{\centering\arraybackslash}m{80.7pt}|}{AVDZ} & -4.9 & 4.9 & 5.4 & -7.4 & 105.4 & AVDZ & -4.0 & 4.0 & 4.5 & -6.4 & 79.9 \\ \hline
\multicolumn{1}{|>{\centering\arraybackslash}m{80.7pt}|}{VTZ} & -3.8 & 3.8 & 4.1 & -5.5 & 86.5 & VTZ & -3.0 & 3.0 & 3.3 & -4.7 & 61.7 \\ \hline
\multicolumn{1}{|>{\centering\arraybackslash}m{80.7pt}|}{\emph{AVTZ}} & \emph{-3.7} & \emph{3.7} & \emph{4.1} & \emph{-5.7} & \emph{78.5} & \emph{AVTZ} & \emph{-3.2} & \emph{3.2} & \emph{3.6} & \emph{-5.2} & \emph{62.6} \\ \hline
\multicolumn{1}{|>{\centering\arraybackslash}m{80.7pt}|}{VQZ} & -2.9 & 2.9 & 3.3 & -4.7 & 59.0 & VQZ & -2.4 & \textbf{2.4} & 2.9 & -4.3 & \textbf{45.1} \\ \hline
\multicolumn{1}{|>{\centering\arraybackslash}m{80.7pt}|}{AVQZ} & -3.0 & 3.0 & 3.4 & -4.9 & 59.1 & AVQZ & -2.6 & 2.6 & 3.1 & -4.6 & 48.6 \\ \hline
\multicolumn{1}{|>{\centering\arraybackslash}m{80.7pt}|}{V5Z} & -2.8 & \textbf{2.8} & 3.3 & -4.7 & 54.3 & V5Z & -2.6 & 2.6 & 3.1 & -4.6 & 47.8 \\ \hline
\multicolumn{1}{|>{\centering\arraybackslash}m{80.7pt}|}{AV5Z} & -2.8 & \textbf{2.8} & 3.2 & -4.8 & 52.7 & AV5Z & -2.7 & 2.7 & 3.2 & -4.7 & 50.6 \\ \hline
\multicolumn{1}{|>{\centering\arraybackslash}m{80.7pt}|}{Def2-TZVP} & -3.0 & 3.0 & 3.4 & -4.8 & 61.8 & Def2-TZVP & -2.5 & 2.5 & 2.9 & -4.4 & 46.7 \\ \hline
\multicolumn{1}{|>{\centering\arraybackslash}m{80.7pt}|}{Def2-QZVP} & -2.8 & \textbf{2.8} & 3.2 & -4.6 & \textbf{53.7} & Def2-QZVP & -2.5 & 2.5 & 3.0 & -4.5 & 47.0 \\ \hline
\multicolumn{1}{|>{\centering\arraybackslash}m{80.7pt}|}{6-311g(2d,2p)} & -5.4 & 5.4 & 5.7 & -7.2 & 130.7 & 6-311g(2d,2p) & -3.6 & 3.6 & 4.0 & -5.6 & 74.1 \\ \hline
\multicolumn{1}{|>{\centering\arraybackslash}m{80.7pt}|}{6-311++g(2d,2p)} & -4.9 & 4.9 & 5.4 & -7.3 & 104.5 & 6-311++g(2d,2p) & -3.8 & 3.8 & 4.3 & -6.1 & 79.0 \\ \hline
\multicolumn{1}{|>{\centering\arraybackslash}m{80.7pt}|}{6-311g(2df,2pd)} & -5.1 & 5.1 & 5.3 & -6.7 & 124.4 & 6-311g(2df,2pd) & -3.2 & 3.2 & 3.6 & -5.0 & 67.5 \\ \hline
\multicolumn{1}{|>{\centering\arraybackslash}m{80.7pt}|}{6-311++g(2df,2pd)} & -4.5 & 4.5 & 5.0 & -6.8 & 97.6 & 6-311++g(2df,2pd) & -3.5 & 3.5 & 3.9 & -5.6 & 71.3 \\ \hline
\multicolumn{1}{|>{\centering\arraybackslash}m{80.7pt}|}{6-311g(3df,3pd)} & -6.2 & 6.2 & 6.4 & -7.7 & 155.8 & 6-311g(3df,3pd) & -3.3 & 3.3 & 3.7 & -5.3 & 66.6 \\ \hline
\multicolumn{1}{|>{\centering\arraybackslash}m{80.7pt}|}{6-311++g(3df,3pd)} & -4.6 & 4.6 & 5.0 & -6.7 & 99.0 & 6-311++g(3df,3pd) & -3.4 & 3.4 & 3.8 & -5.5 & 69.8 \\ \hline
\multicolumn{6}{|>{\centering\arraybackslash}m{265.6pt}|}{\rule[-4.5pt]{0pt}{13.5pt}\textbf{DF\textperiodcentered{}\textperiodcentered{}\textperiodcentered{}NH\textsubscript{3} (without BSSE correction)}} & \multicolumn{6}{>{\centering\arraybackslash}m{253.0pt}|}{\textbf{DF\textperiodcentered{}\textperiodcentered{}\textperiodcentered{}NH\textsubscript{3} (with BSSE correction)}} \\ \hline
\multicolumn{1}{|>{\centering\arraybackslash}m{80.7pt}|}{Basis sets} & MSE & MAD & RMSD & MAX & MA\%D & Basis sets & MSE & MAD & RMSD & MAX & MA\%D \\ \hline
\multicolumn{1}{|>{\centering\arraybackslash}m{80.7pt}|}{VDZ} & -5.0 & 5.0 & 5.6 & -7.9 & 37.0 & VDZ & -1.5 & 2.5 & 3.0 & -4.7 & 14.9 \\ \hline
\multicolumn{1}{|>{\centering\arraybackslash}m{80.7pt}|}{AVDZ} & -2.1 & 2.1 & 2.6 & -3.9 & 12.0 & AVDZ & -0.7 & 1.2 & 1.5 & -2.4 & 7.6 \\ \hline
\multicolumn{1}{|>{\centering\arraybackslash}m{80.7pt}|}{VTZ} & -1.7 & 1.7 & 2.2 & -3.5 & 9.3 & VTZ & -0.4 & 1.3 & 1.4 & -2.1 & 9.1 \\ \hline
\multicolumn{1}{|>{\centering\arraybackslash}m{80.7pt}|}{\emph{AVTZ}} & \emph{0.5} & \emph{\textbf{0.8}} & \emph{1.0} & \emph{1.5} & \emph{9.8} & \emph{AVTZ} & \emph{0.9} & \emph{0.9} & \emph{1.1} & \emph{1.8} & \emph{12.1} \\ \hline
\multicolumn{1}{|>{\centering\arraybackslash}m{80.7pt}|}{VQZ} & 0.1 & 1.0 & 1.1 & 1.3 & 8.4 & VQZ & 0.8 & 1.1 & 1.3 & 2.1 & 13.7 \\ \hline
\multicolumn{1}{|>{\centering\arraybackslash}m{80.7pt}|}{AVQZ} & 1.1 & 1.1 & 1.4 & 2.2 & 14.3 & AVQZ & 1.2 & 1.2 & 1.5 & 2.4 & 15.6 \\ \hline
\multicolumn{1}{|>{\centering\arraybackslash}m{80.7pt}|}{V5Z} & 0.9 & 0.9 & 1.2 & 2.0 & 12.9 & V5Z & 1.1 & 1.1 & 1.4 & 2.4 & 15.2 \\ \hline
\multicolumn{1}{|>{\centering\arraybackslash}m{80.7pt}|}{AV5Z} & 1.1 & 1.1 & 1.4 & 2.3 & 14.4 & AV5Z & 1.1 & 1.1 & 1.4 & 2.3 & 14.7 \\ \hline
\multicolumn{1}{|>{\centering\arraybackslash}m{80.7pt}|}{Def2-TZVP} & -1.4 & 1.5 & 2.1 & -3.3 & 7.6 & Def2-TZVP & 0.0 & 1.3 & 1.4 & -1.6 & 11.2 \\ \hline
\multicolumn{1}{|>{\centering\arraybackslash}m{80.7pt}|}{Def2-QZVP} & 0.7 & 0.9 & 1.1 & 1.7 & 11.2 & Def2-QZVP & 1.1 & 1.1 & 1.4 & 2.4 & 14.6 \\ \hline
\multicolumn{1}{|>{\centering\arraybackslash}m{80.7pt}|}{6-311g(2d,2p)} & -3.9 & 3.9 & 4.3 & -6.0 & 29.7 & 6-311g(2d,2p) & -1.3 & 1.6 & 2.3 & -3.6 & 6.1 \\ \hline
\multicolumn{1}{|>{\centering\arraybackslash}m{80.7pt}|}{6-311++g(2d,2p)} & -2.0 & 2.0 & 2.4 & -3.7 & 11.9 & 6-311++g(2d,2p) & -0.4 & 0.9 & 1.1 & -1.7 & \textbf{5.6} \\ \hline
\multicolumn{1}{|>{\centering\arraybackslash}m{80.7pt}|}{6-311g(2df,2pd)} & -3.2 & 3.2 & 3.6 & -5.1 & 24.6 & 6-311g(2df,2pd) & -0.6 & 1.4 & 1.7 & -2.6 & 8.6 \\ \hline
\multicolumn{1}{|>{\centering\arraybackslash}m{80.7pt}|}{6-311++g(2df,2pd)} & -1.4 & 1.4 & 1.8 & -2.8 & 7.1 & 6-311++g(2df,2pd) & 0.2 & \textbf{0.8} & 0.9 & 1.3 & 8.5 \\* \hline\noalign{\penalty10000}
\multicolumn{1}{|>{\centering\arraybackslash}m{80.7pt}|}{6-311g(3df,3pd)} & -3.4 & 3.4 & 3.5 & -4.7 & 29.3 & 6-311g(3df,3pd) & 0.5 & 1.2 & 1.3 & 1.8 & 13.1 \\* \hline\noalign{\penalty10000}
\multicolumn{1}{|>{\centering\arraybackslash}m{80.7pt}|}{6-311++g(3df,3pd)} & -0.7 & 1.1 & 1.3 & -2.1 & \textbf{6.1} & 6-311++g(3df,3pd) & 0.3 & 0.9 & 1.0 & 1.4 & 9.9 \\* \specialrule{0.9pt}{0pt}{0pt}\noalign{\penalty10000}
\end{tabular}}}}
\end{table}

From \Tref{tab:mse-mad-rmsd-max2}, one can see that MAD(MA\%D) values for BSSE-uncorrected BEs for the selected DF\textperiodcentered{}\textperiodcentered{}\textperiodcentered{}CH\textsubscript{3} complexes calculated from M06-2X functional combined with several basis sets range from 2.8 kJ/mol (52.7 \%) to 6.2 kJ/mol (155.8 \%), whereas the respective range for BSSE-corrected BEs for these complexes is 2.4 kJ/mol (45.1 \%) to 4.0 kJ/mol (79.9 \%). On the other hand, the respective range in the case of the selected DF\textperiodcentered{}\textperiodcentered{}\textperiodcentered{}NH\textsubscript{3} complexes are 0.8 kJ/mol (6.1 \%) to 5.0 kJ/mol (37.0 \%), without BSSE-correction and 0.8 kJ/mol (5.6 \%) to 2.5 kJ/mol (15.6 \%), with BSSE-correction. \textbf{From \Tref{tab:mse-mad-rmsd-max2}}, \textbf{\Fref{fig:mad-a-and-ma} (a)-(d)}, one can consider that BEs without BSSE correction calculated at M06-2X /AVTZ level of theory for the selected DF\textperiodcentered{}\textperiodcentered{}\textperiodcentered{}NH\textsubscript{3} complexes (MAD (MA\%D) = 0.8 kJ/mol (9.8 \%)) {[}note that with BSSE-correction, MAD (MA\%D) = 0.9 kJ/mol (12.1 \%)){]} are reliable with respect to the reference BEs for the respective complexes. The reliability for F-bond distances at the same level of theory is discussed in \Sref{sec:overall-assessment-of-basis}. Note that readers can refer to \Tref{tab:deviations-and-percentage-deviations} and \Fref{fig:deviation-1-1-bsse} to get an overview of deviations on BEs for all the 15 SEFB complexes (DF\textperiodcentered{}\textperiodcentered{}\textperiodcentered{}CH\textsubscript{3}) and 14 CSFB complexes with NH\textsubscript{3} as electron donor (DF\textperiodcentered{}\textperiodcentered{}\textperiodcentered{}NH\textsubscript{3}).

\FloatBarrier
\subsection{Overall assessment of basis sets on M06-2X for F-bond distances of DF\textperiodcentered{}\textperiodcentered{}\textperiodcentered{}CH\textsubscript{3}/NH\textsubscript{3}}\label{sec:overall-assessment-of-basis}

\begin{figure}[!htbp]\centering
\GPKfig{6.51042in}{2.47431in}{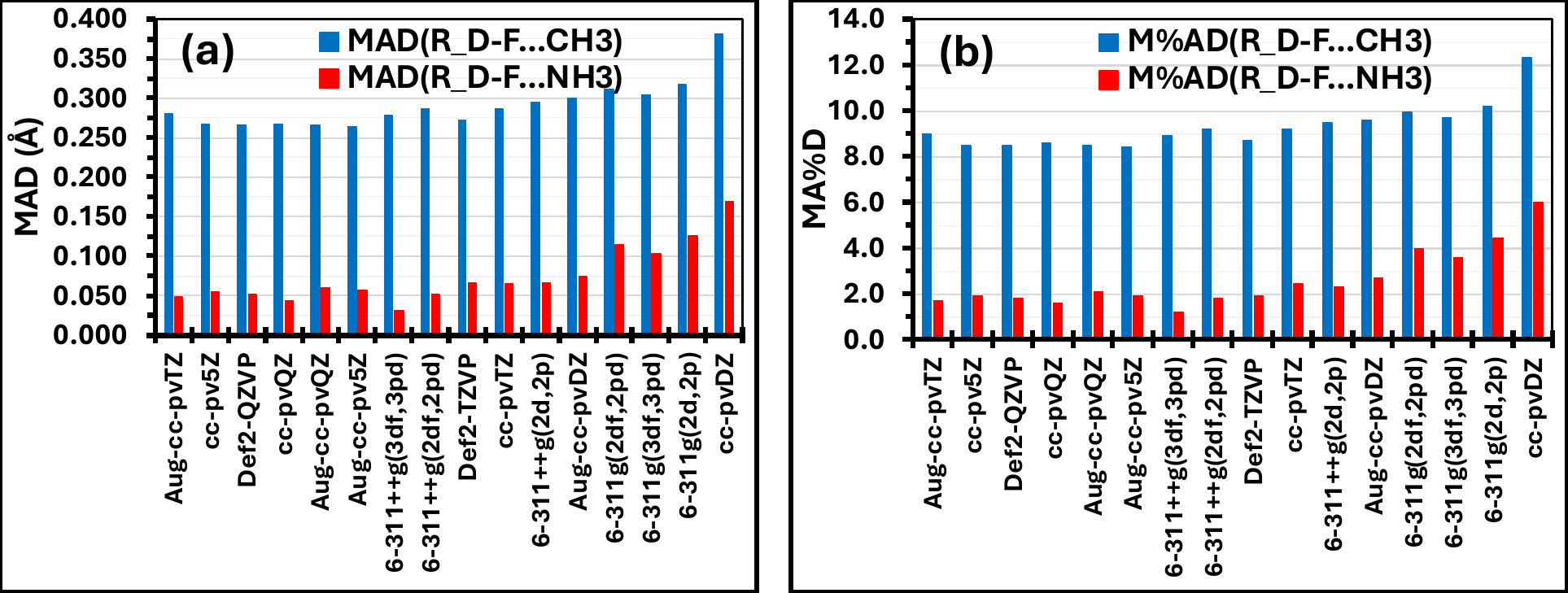}
\par\vspace{-1pt}\caption{MAD \textbf{(a)} and MA\%D \textbf{(b)} on F-bond distances for selected DF\textperiodcentered{}\textperiodcentered{}\textperiodcentered{}CH\textsubscript{3}/NH\textsubscript{3} complexes (DF = F\textsubscript{2}, NCF, CNF, NH\textsubscript{3}F\textsuperscript{+} and NH\textsubscript{2}F\textsubscript{2}\textsuperscript{+}), obtained from M06-2X with 16 basis sets. Basis sets are ordered by increasing MADs on BSSE-uncorrected BEs for DF\textperiodcentered{}\textperiodcentered{}\textperiodcentered{}NH\textsubscript{3} complexes. Reference F-bond distances are from DF-RMP2-F12/VDZ-F12 level of theory. MSE and RMSD values on F-bond distances are shown in \Fref{fig:mean-signed-error-mse}.}\label{fig:mad-a-and-ma2}
\vspace{10pt}
\end{figure}

Various statistical errors on F-bond distances of the selected F-bonded complexes predicted by M06-2X functional using several basis sets are summarized in \mbox{\textbf{TABLE \ref{tab:mse-mad-rmsd-max3}.}} Also, the corresponding MAD and MA\%D values are graphically shown in \mbox{\textbf{FIG. \ref{fig:mad-a-and-ma2}.}}

\begin{table}[!htbp]\centering\setlength{\parskip}{0pt}
\caption{MSE, MAD, RMSD, MAX, and MA\%D on F-bond distances (R, Å) for selected DF\textperiodcentered{}\textperiodcentered{}\textperiodcentered{}CH\textsubscript{3}/NH\textsubscript{3} complexes (DF = F\textsubscript{2}, NCF, CNF, NH\textsubscript{3}F\textsuperscript{+}\textsuperscript{ }and NH\textsubscript{2}F\textsubscript{2}\textsuperscript{+}), obtained from M06-2X with 16 different basis sets, relative to reference distances at DF-RMP2-F12/\allowbreak{}VDZ-F12. M06-2X/\allowbreak{}AVTZ results are shown in italics; the smallest MAD and MA\%D values for SEFB (DF\textperiodcentered{}\textperiodcentered{}\textperiodcentered{}CH\textsubscript{3}) and CSFB (DF\textperiodcentered{}\textperiodcentered{}\textperiodcentered{}NH\textsubscript{3}) are shown separately in boldface.}\label{tab:mse-mad-rmsd-max3}\vspace{7pt}\nopagebreak
{\fontsize{9.0}{10.3}\fontfamily{ptm}\selectfont\renewcommand{\arraystretch}{1.17}\setlength{\tabcolsep}{1.5pt}
\centerline{\adjustbox{max width=518pt}{\begin{tabular}{|>{\centering\arraybackslash}m{82.0pt}|>{\centering\arraybackslash}b{32.1pt}|>{\centering\arraybackslash}m{32.0pt}|>{\centering\arraybackslash}m{32.3pt}|>{\centering\arraybackslash}b{32.1pt}|>{\centering\arraybackslash}m{33.9pt}|>{\centering\arraybackslash}m{82.0pt}|>{\centering\arraybackslash}m{32.0pt}|>{\centering\arraybackslash}m{32.1pt}|>{\centering\arraybackslash}b{39.1pt}|>{\centering\arraybackslash}b{32.0pt}|>{\centering\arraybackslash}m{33.9pt}|}
\specialrule{0.9pt}{0pt}{0pt}
\multicolumn{6}{|>{\centering\arraybackslash}m{259.4pt}|}{\textbf{DF\textperiodcentered{}\textperiodcentered{}\textperiodcentered{}CH}\textbf{\textsubscript{3}}} & \multicolumn{6}{>{\centering\arraybackslash}m{266.2pt}|}{\textbf{DF\textperiodcentered{}\textperiodcentered{}\textperiodcentered{}NH}\textbf{\textsubscript{3}}} \\* \hline\noalign{\penalty10000}
\multicolumn{1}{|>{\centering\arraybackslash}m{82.0pt}|}{Basis sets} & MSE & MAD & RMSD & MAX & MA\%D & Basis sets & MSE & MAD & RMSD & MAX & MA\%D \\* \hline\noalign{\penalty10000}
\multicolumn{1}{|>{\centering\arraybackslash}m{82.0pt}|}{VDZ} & -0.381 & 0.381 & 0.386 & -0.496 & 12.3 & VDZ & -0.169 & 0.169 & 0.175 & -0.256 & 6.0 \\ \hline
\multicolumn{1}{|>{\centering\arraybackslash}m{82.0pt}|}{AVDZ} & -0.300 & 0.300 & 0.308 & -0.424 & 9.6 & AVDZ & -0.074 & 0.074 & 0.074 & -0.084 & 2.7 \\ \hline
\multicolumn{1}{|>{\centering\arraybackslash}m{82.0pt}|}{VTZ} & -0.287 & 0.287 & 0.295 & -0.417 & 9.2 & VTZ & -0.065 & 0.065 & 0.067 & -0.091 & 2.4 \\ \hline
\multicolumn{1}{|>{\centering\arraybackslash}m{82.0pt}|}{\emph{AVTZ}} & \emph{-0.280} & \emph{0.280} & \emph{0.288} & \emph{-0.410} & \emph{9.0} & \emph{AVTZ} & \emph{-0.001} & \emph{0.048} & \emph{0.051} & \emph{0.071} & \emph{1.7} \\ \hline
\multicolumn{1}{|>{\centering\arraybackslash}m{82.0pt}|}{VQZ} & -0.268 & 0.268 & 0.276 & -0.392 & 8.6 & VQZ & -0.015 & 0.044 & 0.047 & -0.068 & 1.6 \\ \hline
\multicolumn{1}{|>{\centering\arraybackslash}m{82.0pt}|}{AVQZ} & -0.266 & 0.266 & 0.274 & -0.391 & 8.5 & AVQZ & 0.016 & 0.059 & 0.068 & 0.112 & 2.1 \\ \hline
\multicolumn{1}{|>{\centering\arraybackslash}m{82.0pt}|}{V5Z} & -0.267 & 0.267 & 0.275 & -0.397 & 8.5 & V5Z & 0.013 & 0.055 & 0.063 & 0.106 & 1.9 \\ \hline
\multicolumn{1}{|>{\centering\arraybackslash}m{82.0pt}|}{AV5Z} & -0.264 & \textbf{0.264} & 0.273 & -0.395 & \textbf{8.4} & AV5Z & 0.017 & 0.056 & 0.066 & 0.112 & 1.9 \\ \hline
\multicolumn{1}{|>{\centering\arraybackslash}m{82.0pt}|}{Def2-TZVP} & -0.272 & 0.272 & 0.281 & -0.403 & 8.7 & Def2-TZVP & -0.066 & 0.066 & 0.068 & -0.087 & 1.9 \\ \hline
\multicolumn{1}{|>{\centering\arraybackslash}m{82.0pt}|}{Def2-QZVP} & -0.266 & 0.266 & 0.275 & -0.395 & 8.5 & Def2-QZVP & 0.006 & 0.051 & 0.056 & 0.090 & 1.8 \\ \hline
\multicolumn{1}{|>{\centering\arraybackslash}m{82.0pt}|}{6-311g(2d,2p)} & -0.318 & 0.318 & 0.328 & -0.453 & 10.2 & 6-311g(2d,2p) & -0.125 & 0.125 & 0.133 & -0.208 & 4.4 \\ \hline
\multicolumn{1}{|>{\centering\arraybackslash}m{82.0pt}|}{6-311++g(2d,2p)} & -0.295 & 0.295 & 0.304 & -0.426 & 9.5 & 6-311++g(2d,2p) & -0.066 & 0.066 & 0.069 & -0.093 & 2.3 \\ \hline
\multicolumn{1}{|>{\centering\arraybackslash}m{82.0pt}|}{6-311g(2df,2pd)} & -0.310 & 0.310 & 0.321 & -0.454 & 9.9 & 6-311g(2df,2pd) & -0.114 & 0.114 & 0.124 & -0.199 & 4.0 \\ \hline
\multicolumn{1}{|>{\centering\arraybackslash}m{82.0pt}|}{6-311++g(2df,2pd)} & -0.287 & 0.287 & 0.298 & -0.428 & 9.2 & 6-311++g(2df,2pd) & -0.051 & 0.051 & 0.053 & -0.069 & 1.8 \\* \hline\noalign{\penalty10000}
\multicolumn{1}{|>{\centering\arraybackslash}m{82.0pt}|}{6-311g(3df,3pd)} & -0.304 & 0.304 & 0.315 & -0.448 & 9.7 & 6-311g(3df,3pd) & -0.104 & 0.104 & 0.114 & -0.191 & 3.6 \\* \hline\noalign{\penalty10000}
\multicolumn{1}{|>{\centering\arraybackslash}m{82.0pt}|}{6-311++g(3df,3pd)} & -0.279 & 0.279 & 0.290 & -0.423 & 8.9 & 6-311++g(3df,3pd) & -0.031 & \textbf{0.031} & 0.038 & -0.059 & \textbf{1.2} \\* \specialrule{0.9pt}{0pt}{0pt}\noalign{\penalty10000}
\end{tabular}}}}
\end{table}

From \Tref{tab:mse-mad-rmsd-max3}, one can see that F-bond distances for DF\textperiodcentered{}\textperiodcentered{}\textperiodcentered{}CH\textsubscript{3} complexes predicted by M06-2X with any of the 16 basis sets never reach MA\%D below 5.0\%. Also, MADs (MA\%Ds) on F-bond distances for the selected SEFB complexes range from 0.264 to 0.381 Å (8.4 to 12.3\%). For the respective CSFB complexes, MA\%Ds remain below 5.0\% at every basis set except VDZ basis set. Range of MADs (MA\%Ds) on F-bond distances of these complexes range from 0.031 to 0.169 Å (1.2 to 6.0\%) (\Tref{tab:mse-mad-rmsd-max3}). At AVTZ specifically, F-bond distances for the selected CSFB complexes are reliable, with MAD (MA\%D) of 0.048 Å (1.7\%) (\Tref{tab:mse-mad-rmsd-max3}, \Fref{fig:mad-a-and-ma2}), consistent with the reliability of BEs at AVTZ already noted in \Sref{sec:overall-effect-of-basis}. Deviations for all 29 complexes (15 DF\textperiodcentered{}\textperiodcentered{}\textperiodcentered{}CH\textsubscript{3} and 14 DF\textperiodcentered{}\textperiodcentered{}\textperiodcentered{}NH\textsubscript{3}) are depicted in \Tref{tab:deviations-and-percentage-deviations} and \mbox{\textbf{FIG. \ref{fig:deviation-1-1-bsse}.}}

Overall, we found that deviations on BEs and F-bond distances for the selected SEFB complexes predicted by M06-2X across the 16 basis sets used remain appreciably larger than reference values (\mbox{\textbf{TABLEs \ref{tab:percentage-deviation-on-bes}}}, \textbf{XI}, \textbf{S13}, \textbf{S14}, \textbf{FIG.} \textbf{7}-\textbf{9}). Also, it has been already mentioned that the anomaly in BEs for F\textsubscript{2}\textperiodcentered{}\textperiodcentered{}\textperiodcentered{}CH\textsubscript{3}/F\textsubscript{2}\textperiodcentered{}\textperiodcentered{}\textperiodcentered{}NH\textsubscript{3} persists relative to several \emph{ab initio} benchmarks. On the other hand, the results for DF\textperiodcentered{}\textperiodcentered{}\textperiodcentered{}NH\textsubscript{3} complexes obtained from M06-2X functional with a reasonable basis set remain close to reference values throughout. AVTZ remains a suitable basis set on the grounds of cost and accuracy for assessing M06-2X functional on these complexes. Note that BEs and F-bond distances for all 29 complexes from M06-2X/AVTZ have already been listed in \textbf{TABLES \ref{tab:cbs-limit-binding-energies}} and \textbf{II}. We briefly discuss statistical analysis for this full dataset in \Sref{sec:effect-of-bsse-correction}. The possible underlying cause for failure of M06-2X functional for modelling SEFB complexes will be discussed in \Sref{sec:m06-2x-avtz-for}.

\subsection{Effect of BSSE correction on binding energies for all DF\textperiodcentered{}\textperiodcentered{}\textperiodcentered{}CH\textsubscript{3}/NH\textsubscript{3} with M06-2X/AVTZ}\label{sec:effect-of-bsse-correction}

\begin{figure}[!htbp]\centering
\GPKfig{6.86237in}{3.05853in}{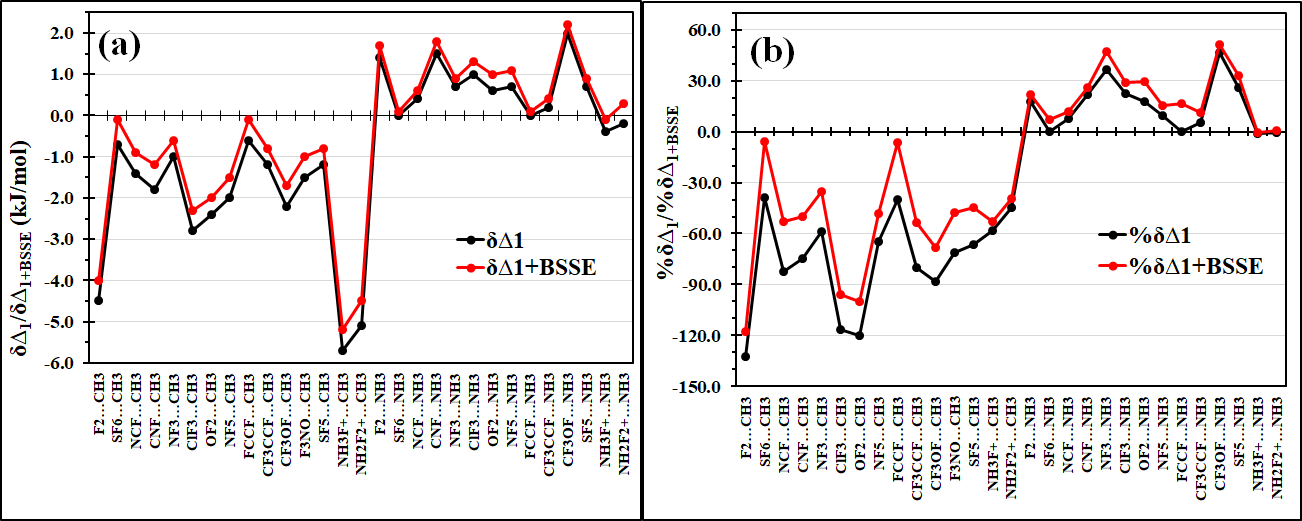}
\GPKfig{6.90992in}{3.0863in}{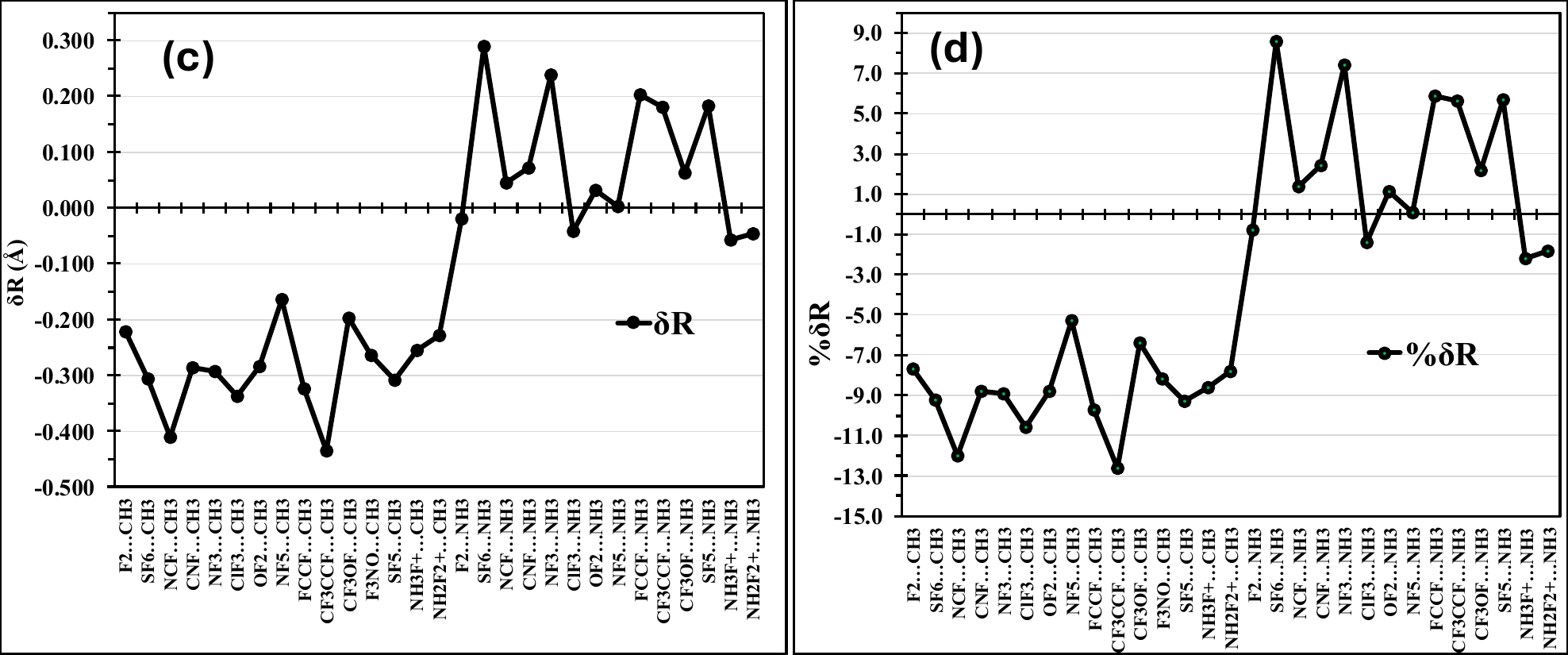}
\par\vspace{-1pt}\caption{Deviation (\ensuremath{\delta}\ensuremath{\Delta}\textsubscript{1}, \ensuremath{\delta}\ensuremath{\Delta}\textsubscript{1+BSSE}; kJ/mol) \textbf{(a)} and percentage deviation (\%\ensuremath{\delta}\ensuremath{\Delta}\textsubscript{1}, \%\ensuremath{\delta}\ensuremath{\Delta}\textsubscript{1+BSSE}) \textbf{(b)} on BSSE-uncorrected and BSSE-corrected BEs, and deviation (\ensuremath{\delta}R; Å) \textbf{(c)} and percentage deviation (\%\ensuremath{\delta}R) \textbf{(d)} on F-bond distances, at M06-2X/AVTZ for all DF\textperiodcentered{}\textperiodcentered{}\textperiodcentered{}CH\textsubscript{3}/NH\textsubscript{3} complexes (15 DF\textperiodcentered{}\textperiodcentered{}\textperiodcentered{}CH\textsubscript{3}, 14 DF\textperiodcentered{}\textperiodcentered{}\textperiodcentered{}NH\textsubscript{3}). Reference BEs are from DF-RMP2-F12/CBS{[}V(T,Q)Z-F12{]}//DF-RMP2-F12/VDZ-F12 (denoted DF-RMP2-F12/CBS); reference F-bond distances are from DF-RMP2-F12/VDZ-F12.}\label{fig:deviation-1-1-bsse}
\vspace{10pt}
\end{figure}

\Sref{sec:reference-binding-energies-and} already considers deviations on BEs predicted by M06-2X/AVTZ level for all 29 complexes graphically (\textbf{\Fref{fig:deviations-on-bes-in}(a)}). Also, \Tref{tab:cbs-limit-binding-energies} reveals the underlying BEs and their error statistics (MSE, MAD, RMSD, MAX, MA\%D) with and without BSSE correction. Neither \Tref{tab:cbs-limit-binding-energies} nor \Fref{fig:deviations-on-bes-in} reflects the percentage deviation on BE for each complex. \Tref{tab:deviations-and-percentage-deviations} provides this directly (\ensuremath{\delta}\ensuremath{\Delta}\textsubscript{1}, \%\ensuremath{\delta}\ensuremath{\Delta}\textsubscript{1}, \ensuremath{\delta}\ensuremath{\Delta}\textsubscript{1+BSSE}, \%\ensuremath{\delta}\ensuremath{\Delta}\textsubscript{1+BSSE}). \textbf{\Fref{fig:deviation-1-1-bsse}(a-b)} reflects it directly, enabling assessment of the effect of BSSE correction complex by complex. Also, for comparison, in \Tref{tab:deviations-and-percentage-deviations}, we have shown error statistics determined for all the 29 complexes (already tabulated in \Tref{tab:cbs-limit-binding-energies}) along with those values obtained for the sub-set of the selected SEFB and CSFB complexes considered (bracketed values).

\begin{table}[!htbp]\centering\setlength{\parskip}{0pt}
\caption{Deviations and percentage deviations on BEs, without (\ensuremath{\delta}\ensuremath{\Delta}\textsubscript{1}, \%\ensuremath{\delta}\ensuremath{\Delta}\textsubscript{1}) and with (\ensuremath{\delta}\ensuremath{\Delta}\textsubscript{1+BSSE}, \%\ensuremath{\delta}\ensuremath{\Delta}\textsubscript{1+BSSE}) BSSE correction, and on F-bond distances (\ensuremath{\delta}R, \%\ensuremath{\delta}R), for all DF\textperiodcentered{}\textperiodcentered{}\textperiodcentered{}CH\textsubscript{3}/NH\textsubscript{3} complexes at M06-2X/\allowbreak{}AVTZ. \ensuremath{\delta}\ensuremath{\Delta}\textsubscript{1} and \ensuremath{\delta}\ensuremath{\Delta}\textsubscript{1+BSSE} are in kJ/\allowbreak{}mol; \ensuremath{\delta}R is in Å; \%\ensuremath{\delta}\ensuremath{\Delta}\textsubscript{1}, \%\ensuremath{\delta}\ensuremath{\Delta}\textsubscript{1+BSSE}, and \%\ensuremath{\delta}R are in percentages. Reference BEs are from DF-RMP2-F12/\allowbreak{}CBS[V(T,Q)Z-F12]//DF-RMP2-F12/\allowbreak{}VDZ-F12 (denoted as DF-RMP2-F12/\allowbreak{}CBS); reference F-bond distances are from DF-RMP2-F12/\allowbreak{}VDZ-F12. MSE, MAD, RMSD, MAX, and MA\%D are given for all SEFB-15 (DF\textperiodcentered{}\textperiodcentered{}\textperiodcentered{}CH\textsubscript{3}) and CSFB-14 (DF\textperiodcentered{}\textperiodcentered{}\textperiodcentered{}NH\textsubscript{3}) complexes, with corresponding values for the selected subset (DF = F\textsubscript{2}, NCF, CNF, NH\textsubscript{3}F\textsuperscript{+} and NH\textsubscript{2}F\textsubscript{2}\textsuperscript{+}) shown in brackets. The smallest values of each deviation are shown in boldface, the largest are italicized, and \%\ensuremath{\delta}R values \ensuremath{\leq}5.0\% (in magnitude) are shown in red. See \Tref{tab:cbs-limit-binding-energies} for the corresponding BEs.}\label{tab:deviations-and-percentage-deviations}\vspace{7pt}\nopagebreak
{\fontsize{9.0}{10.3}\fontfamily{ptm}\selectfont\renewcommand{\arraystretch}{1.17}\setlength{\tabcolsep}{1.5pt}
\centerline{\adjustbox{max width=518pt}{\begin{tabular}{|>{\centering\arraybackslash}p{69.5pt}|>{\centering\arraybackslash}p{28.1pt}|>{\centering\arraybackslash}p{31.3pt}|>{\centering\arraybackslash}p{26.3pt}|>{\centering\arraybackslash}p{35.7pt}|>{\centering\arraybackslash}p{34.4pt}|>{\centering\arraybackslash}p{31.8pt}|>{\centering\arraybackslash}p{68.9pt}|>{\centering\arraybackslash}p{23.1pt}|>{\centering\arraybackslash}p{25.0pt}|>{\centering\arraybackslash}p{29.7pt}|>{\centering\arraybackslash}p{32.3pt}|>{\centering\arraybackslash}p{39.3pt}|>{\centering\arraybackslash}p{29.6pt}|}
\specialrule{0.9pt}{0pt}{0pt}
\multicolumn{7}{|>{\centering\arraybackslash}p{274.9pt}|}{\textbf{DF\textperiodcentered{}\textperiodcentered{}\textperiodcentered{}CH}\textbf{\textsubscript{3}}\textbf{ @ M06-2X/\allowbreak{}AVTZ}} & \multicolumn{7}{>{\centering\arraybackslash}p{265.8pt}|}{\textbf{DF\textperiodcentered{}\textperiodcentered{}\textperiodcentered{}NH}\textbf{\textsubscript{3}}\textbf{ @ M06-2X/\allowbreak{}AVTZ}} \\* \hline\noalign{\penalty10000}
\multicolumn{1}{|>{\centering\arraybackslash}p{69.5pt}|}{\multirow{2}{69.5pt}{\centering DF\textperiodcentered{}\textperiodcentered{}\textperiodcentered{}CH\textsubscript{3}}} & \multicolumn{2}{>{\centering\arraybackslash}p{62.3pt}|}{BE without BSSE} & \multicolumn{2}{>{\centering\arraybackslash}p{65.0pt}|}{BE with BSSE} & \multicolumn{2}{>{\centering\arraybackslash}p{69.2pt}|}{F-bond distance} & \multirow{2}{68.9pt}{\centering DF\textperiodcentered{}\textperiodcentered{}\textperiodcentered{}NH\textsubscript{3}} & \multicolumn{2}{>{\centering\arraybackslash}p{51.1pt}|}{BE without BSSE} & \multicolumn{2}{>{\centering\arraybackslash}p{65.0pt}|}{BE with BSSE} & \multicolumn{2}{>{\centering\arraybackslash}p{71.9pt}|}{F-bond distance} \\* \cline{2-7}\cline{9-14}\noalign{\penalty10000}
\multicolumn{1}{|>{\centering\arraybackslash}p{69.5pt}|}{} & \ensuremath{\delta}\ensuremath{\Delta}\textsubscript{1} & \%\ensuremath{\delta}\ensuremath{\Delta}\textsubscript{1} & \ensuremath{\delta}\ensuremath{\Delta}\textsubscript{1} \textsubscript{+BSSE} & \%\ensuremath{\delta}\ensuremath{\Delta}\textsubscript{1+BSSE} & \ensuremath{\delta}R & \%\ensuremath{\delta}R &  & \ensuremath{\delta}\ensuremath{\Delta}\textsubscript{1} & \%\ensuremath{\delta}\ensuremath{\Delta}\textsubscript{1} & \ensuremath{\delta}\ensuremath{\Delta}\textsubscript{1} \textsubscript{+BSSE} & \%\ensuremath{\delta}\ensuremath{\Delta}\textsubscript{1} \textsubscript{+BSSE} & \ensuremath{\delta}R & \%\ensuremath{\delta}R \\* \hline\noalign{\penalty10000}
\multicolumn{1}{|>{\centering\arraybackslash}p{69.5pt}|}{F\textsubscript{2}\textperiodcentered{}\textperiodcentered{}\textperiodcentered{}CH\textsubscript{3}} & -4.5 & \emph{\textbf{-132.4}} & -4.0 & \emph{\textbf{-117.6}} & -0.221 & -7.7 & F\textsubscript{2}\textperiodcentered{}\textperiodcentered{}\textperiodcentered{}NH\textsubscript{3} & 1.4 & 17.9 & 1.7 & 21.8 & -0.020 & {\leavevmode\color[HTML]{FF0000}-0.8} \\* \hline\noalign{\penalty10000}
\multicolumn{1}{|>{\centering\arraybackslash}p{69.5pt}|}{SF\textsubscript{6}\textperiodcentered{}\textperiodcentered{}\textperiodcentered{}CH\textsubscript{3}} & -0.7 & -38.9 & \textbf{-0.1} & \textbf{-5.6} & -0.306 & -9.2 & SF\textsubscript{6}\textperiodcentered{}\textperiodcentered{}\textperiodcentered{}NH\textsubscript{3} & \textbf{0.0} & \textbf{0.0} & \textbf{0.1} & 7.1 & \emph{\textbf{0.289}} & \emph{\textbf{8.6}} \\* \hline\noalign{\penalty10000}
\multicolumn{1}{|>{\centering\arraybackslash}p{69.5pt}|}{NCF\textperiodcentered{}\textperiodcentered{}\textperiodcentered{}CH\textsubscript{3}} & -1.4 & -82.4 & -0.9 & -52.9 & -0.410 & -12.0 & NCF\textperiodcentered{}\textperiodcentered{}\textperiodcentered{}NH\textsubscript{3} & 0.4 & 7.8 & 0.6 & 11.8 & 0.045 & {\leavevmode\color[HTML]{FF0000}1.4} \\* \hline\noalign{\penalty10000}
\multicolumn{1}{|>{\centering\arraybackslash}p{69.5pt}|}{CNF\textperiodcentered{}\textperiodcentered{}\textperiodcentered{}CH\textsubscript{3}} & -1.8 & -75.0 & -1.2 & -50.0 & -0.286 & -8.8 & CNF\textperiodcentered{}\textperiodcentered{}\textperiodcentered{}NH\textsubscript{3} & 1.5 & 21.7 & 1.8 & 26.1 & 0.071 & {\leavevmode\color[HTML]{FF0000}2.4} \\ \hline
\multicolumn{1}{|>{\centering\arraybackslash}p{69.5pt}|}{NF\textsubscript{3}\textperiodcentered{}\textperiodcentered{}\textperiodcentered{}CH\textsubscript{3}} & -1.0 & -58.8 & -0.6 & -35.3 & -0.293 & -8.9 & NF\textsubscript{3}\textperiodcentered{}\textperiodcentered{}\textperiodcentered{}NH\textsubscript{3} & 0.7 & 36.8 & 0.9 & 47.4 & 0.239 & 7.4 \\ \hline
\multicolumn{1}{|>{\centering\arraybackslash}p{69.5pt}|}{ClF\textsubscript{3}\textperiodcentered{}\textperiodcentered{}\textperiodcentered{}CH\textsubscript{3}} & -2.8 & -116.7 & -2.3 & -95.8 & -0.336 & -10.6 & ClF\textsubscript{3}\textperiodcentered{}\textperiodcentered{}\textperiodcentered{}NH\textsubscript{3} & 1.0 & 22.2 & 1.3 & 28.9 & -0.041 & {\leavevmode\color[HTML]{FF0000}-1.4} \\ \hline
\multicolumn{1}{|>{\centering\arraybackslash}p{69.5pt}|}{OF\textsubscript{2}\textperiodcentered{}\textperiodcentered{}\textperiodcentered{}CH\textsubscript{3}} & -2.4 & -120.0 & -2.0 & -100.0 & -0.284 & -8.8 & OF\textsubscript{2}\textperiodcentered{}\textperiodcentered{}\textperiodcentered{}NH\textsubscript{3} & 0.6 & 17.6 & 1.0 & 29.4 & 0.031 & {\leavevmode\color[HTML]{FF0000}1.1} \\ \hline
\multicolumn{1}{|>{\centering\arraybackslash}p{69.5pt}|}{NF\textsubscript{5}\textperiodcentered{}\textperiodcentered{}\textperiodcentered{}CH\textsubscript{3}} & -2.0 & -64.5 & -1.5 & -48.4 & \textbf{-0.164} & {\leavevmode\color[HTML]{FF0000}\textbf{-5.3}} & NF\textsubscript{5}\textperiodcentered{}\textperiodcentered{}\textperiodcentered{}NH\textsubscript{3} & 0.7 & 9.6 & 1.1 & 15.1 & \textbf{0.003} & {\leavevmode\color[HTML]{FF0000}\textbf{0.1}} \\ \hline
\multicolumn{1}{|>{\centering\arraybackslash}p{69.5pt}|}{FCCF\textperiodcentered{}\textperiodcentered{}\textperiodcentered{}CH\textsubscript{3}} & \textbf{-0.6} & \textbf{-40.0} & \textbf{-0.1} & -6.7 & -0.324 & -9.7 & FCCF\textperiodcentered{}\textperiodcentered{}\textperiodcentered{}NH\textsubscript{3} & 0.0 & 0.0 & \textbf{0.1} & 16.7 & 0.204 & 5.9 \\ \hline
\multicolumn{1}{|>{\centering\arraybackslash}p{69.5pt}|}{CF\textsubscript{3}CCF\textperiodcentered{}\textperiodcentered{}\textperiodcentered{}CH\textsubscript{3}} & -1.2 & -80.0 & -0.8 & -53.3 & \emph{\textbf{-0.434}} & \emph{\textbf{-12.6}} & CF\textsubscript{3}CCF\textperiodcentered{}\textperiodcentered{}\textperiodcentered{}NH\textsubscript{3} & 0.2 & 5.6 & 0.4 & 11.1 & 0.182 & 5.6 \\ \hline
\multicolumn{1}{|>{\centering\arraybackslash}p{69.5pt}|}{CF\textsubscript{3}OF\textperiodcentered{}\textperiodcentered{}\textperiodcentered{}CH\textsubscript{3}} & -2.2 & -88.0 & -1.7 & -68.0 & -0.197 & -6.4 & CF\textsubscript{3}OF\textperiodcentered{}\textperiodcentered{}\textperiodcentered{}NH\textsubscript{3} & \emph{\textbf{2.0}} & \emph{\textbf{46.5}} & \emph{\textbf{2.2}} & \emph{\textbf{51.2}} & 0.063 & {\leavevmode\color[HTML]{FF0000}2.2} \\ \hline
\multicolumn{1}{|>{\centering\arraybackslash}p{69.5pt}|}{F\textsubscript{3}NO\textperiodcentered{}\textperiodcentered{}\textperiodcentered{}CH\textsubscript{3}} & -1.5 & -71.4 & -1.0 & -47.6 & -0.264 & -8.2 & F\textsubscript{3}NO\textperiodcentered{}\textperiodcentered{}\textperiodcentered{}NH\textsubscript{3} & b & b & b & b & b & b \\ \hline
\multicolumn{1}{|>{\centering\arraybackslash}p{69.5pt}|}{SF\textsubscript{5}\textperiodcentered{}\textperiodcentered{}\textperiodcentered{}CH\textsubscript{3}} & -1.2 & -66.7 & -0.8 & -44.4 & -0.308 & -9.3 & SF\textsubscript{5}\textperiodcentered{}\textperiodcentered{}\textperiodcentered{}NH\textsubscript{3} & 0.7 & 25.9 & 0.9 & 33.3 & 0.183 & 5.7 \\ \hline
\multicolumn{1}{|>{\centering\arraybackslash}p{69.5pt}|}{NH\textsubscript{3}F+\textperiodcentered{}\textperiodcentered{}\textperiodcentered{}CH\textsubscript{3}} & \emph{\textbf{-5.7}} & -58.2 & \emph{\textbf{-5.2}} & -53.1 & -0.254 & -8.6 & NH\textsubscript{3}F+\textperiodcentered{}\textperiodcentered{}\textperiodcentered{}NH\textsubscript{3} & -0.4 & -1.2 & -0.1 & \textbf{-0.3} & -0.057 & {\leavevmode\color[HTML]{FF0000}-2.2} \\ \hline
\multicolumn{1}{|>{\centering\arraybackslash}p{69.5pt}|}{NH\textsubscript{2}F\textsubscript{2}\textsuperscript{+}\textperiodcentered{}\textperiodcentered{}\textperiodcentered{}CH\textsubscript{3}} & -5.1 & -44.7 & -4.5 & -39.5 & -0.228 & -7.8 & NH\textsubscript{2}F\textsubscript{2}\textsuperscript{+}\textperiodcentered{}\textperiodcentered{}\textperiodcentered{}NH\textsubscript{3} & -0.2 & -0.5 & 0.3 & 0.8 & -0.046 & {\leavevmode\color[HTML]{FF0000}-1.8} \\ \hline
\multicolumn{1}{|>{\centering\arraybackslash}p{69.5pt}|}{MSE} & \multicolumn{2}{>{\centering\arraybackslash}p{62.3pt}|}{-2.3 (\emph{-3.7})} & \multicolumn{2}{>{\centering\arraybackslash}p{65.0pt}|}{-1.8 (\emph{-3.2})} & \multicolumn{2}{>{\centering\arraybackslash}p{69.2pt}|}{-0.287 (\emph{-0.280})} & MSE & \multicolumn{2}{>{\centering\arraybackslash}p{51.1pt}|}{0.6 (\emph{0.5})} & \multicolumn{2}{>{\centering\arraybackslash}p{65.0pt}|}{0.9 (\emph{0.9})} & \multicolumn{2}{>{\centering\arraybackslash}p{71.9pt}|}{0.082 (\emph{-0.001})} \\ \hline
\multicolumn{1}{|>{\centering\arraybackslash}p{69.5pt}|}{MAD} & \multicolumn{2}{>{\centering\arraybackslash}p{62.3pt}|}{2.3 (\emph{3.7})} & \multicolumn{2}{>{\centering\arraybackslash}p{65.0pt}|}{1.8 (\emph{3.2})} & \multicolumn{2}{>{\centering\arraybackslash}p{69.2pt}|}{0.287 (\emph{0.280})} & MAD & \multicolumn{2}{>{\centering\arraybackslash}p{51.1pt}|}{0.7 (\emph{0.8})} & \multicolumn{2}{>{\centering\arraybackslash}p{65.0pt}|}{0.9 (\emph{0.9})} & \multicolumn{2}{>{\centering\arraybackslash}p{71.9pt}|}{0.105 (\emph{0.048})} \\ \hline
\multicolumn{1}{|>{\centering\arraybackslash}p{69.5pt}|}{RMSD} & \multicolumn{2}{>{\centering\arraybackslash}p{62.3pt}|}{2.7 (\emph{4.1})} & \multicolumn{2}{>{\centering\arraybackslash}p{65.0pt}|}{2.3 (\emph{3.6})} & \multicolumn{2}{>{\centering\arraybackslash}p{69.2pt}|}{0.296 (\emph{0.288})} & RMSD & \multicolumn{2}{>{\centering\arraybackslash}p{51.1pt}|}{0.9 (\emph{1.0})} & \multicolumn{2}{>{\centering\arraybackslash}p{65.0pt}|}{1.1 (\emph{1.1})} & \multicolumn{2}{>{\centering\arraybackslash}p{71.9pt}|}{0.138 (\emph{0.051})} \\* \hline\noalign{\penalty10000}
\multicolumn{1}{|>{\centering\arraybackslash}p{69.5pt}|}{MAX} & \multicolumn{2}{>{\centering\arraybackslash}p{62.3pt}|}{-5.7 (\emph{-5.7})} & \multicolumn{2}{>{\centering\arraybackslash}p{65.0pt}|}{-5.2 (\emph{-5.2})} & \multicolumn{2}{>{\centering\arraybackslash}p{69.2pt}|}{-0.434 (\emph{-0.410})} & MAX & \multicolumn{2}{>{\centering\arraybackslash}p{51.1pt}|}{2.0 (\emph{1.5})} & \multicolumn{2}{>{\centering\arraybackslash}p{65.0pt}|}{2.2 (\emph{1.8})} & \multicolumn{2}{>{\centering\arraybackslash}p{71.9pt}|}{0.289 (\emph{0.071})} \\* \hline\noalign{\penalty10000}
\multicolumn{1}{|>{\centering\arraybackslash}p{69.5pt}|}{MA\%D} & \multicolumn{2}{>{\centering\arraybackslash}p{62.3pt}|}{75.8 (\emph{78.5})} & \multicolumn{2}{>{\centering\arraybackslash}p{65.0pt}|}{54.6 (\emph{62.6})} & \multicolumn{2}{>{\centering\arraybackslash}p{69.2pt}|}{8.9 (\emph{9.0})} & MA\%D & \multicolumn{2}{>{\centering\arraybackslash}p{51.1pt}|}{16.0 (\emph{9.8})} & \multicolumn{2}{>{\centering\arraybackslash}p{65.0pt}|}{21.5 (\emph{12.1})} & \multicolumn{2}{>{\centering\arraybackslash}p{71.9pt}|}{3.3 (\emph{1.7})} \\* \specialrule{0.9pt}{0pt}{0pt}\noalign{\penalty10000}
\end{tabular}}}}
\par\nopagebreak\vspace{1pt}{\setstretch{1}\setlength{\parskip}{1.5pt}\fontsize{9}{10.5}\selectfont\raggedright\noindent \textsuperscript{b}Convergence error during optimization of respective F-bonded complexes at DF-RMP2-F12/VDZ-F12 level of theory.\par}
\end{table}

In \textbf{Sections \ref*{sec:reference-binding-energies-and}} and \textbf{4.6}, we have already shown that BSSE correction improves BEs of all as well as selected SEFB complexes, though it worsens them for the corresponding CSFB complexes. \Tref{tab:deviations-and-percentage-deviations} reports deviations and percentage deviations on BEs and F-bond distances predicted by M06-2X/AVTZ for all SEFB and CSFB complexes considered. BSSE correction reduces the deviation on BE for each of the 15 SEFB complexes without exception (\Tref{tab:deviations-and-percentage-deviations}, \textbf{\Fref{fig:deviation-1-1-bsse}(a-b)}). It increases the deviation for each one of the 14 CSFB complexes except NH\textsubscript{3}F\textsuperscript{+}\textperiodcentered{}\textperiodcentered{}\textperiodcentered{}NH\textsubscript{3} (\Tref{tab:deviations-and-percentage-deviations}, \textbf{\Fref{fig:deviation-1-1-bsse}(a-b)}). This asymmetric trend follows from SEFB being uniformly over-bound (indicated by negative deviations on BEs in \Tref{tab:deviations-and-percentage-deviations} and \textbf{\Fref{fig:deviation-1-1-bsse}(a-b)}) and CSFB being mostly under-bound (indicated by positive deviations on BEs in \Tref{tab:deviations-and-percentage-deviations} and \textbf{\Fref{fig:deviation-1-1-bsse}(a-b)}). It therefore corrects the over-bound case for SEFB complexes but worsens the already reliable BEs of CSFB complexes predicted by the M06-2X/AVTZ level.

Among individual complexes, F\textsubscript{2}\textperiodcentered{}\textperiodcentered{}\textperiodcentered{}CH\textsubscript{3} has the largest \%\ensuremath{\delta}\textsubscript{BE} among SEFB complexes (-132.4\% without BSSE correction and -117.6\% with BSSE-correction). It does not, however, have the largest deviation in kJ/mol (-4.5 and -4.0 kJ/mol). The largest deviation occurs for NH\textsubscript{3}F\textsuperscript{+}\textperiodcentered{}\textperiodcentered{}\textperiodcentered{}CH\textsubscript{3} (-5.7 and -5.2 kJ/mol without and with BSSE-correction, respectively). For CSFB complexes, CF\textsubscript{3}OF\textperiodcentered{}\textperiodcentered{}\textperiodcentered{}NH\textsubscript{3} has the largest deviation without and with BSSE correction (2.0 and 2.2 kJ/mol, 46.5\% and 51.2\%) (\Tref{tab:deviations-and-percentage-deviations}). By contrast, F\textsubscript{2}\textperiodcentered{}\textperiodcentered{}\textperiodcentered{}NH\textsubscript{3} bears a smaller \%\ensuremath{\delta}\textsubscript{BE} (17.9\% and 21.8\%, respectively without and with BSSE correction), with its BE obtained from M06-2X/AVTZ (-6.4 and -6.1 kJ/mol, respectively without and with BSSE correction) below the reference (-7.8 kJ/mol) (\Tref{tab:cbs-limit-binding-energies}).

For the SEFB complexes, BSSE correction reduces the MA\%D error (from 78.5\% to 62.6\% for the subset, and from 75.8\% to 54.6\% for the full set), though the error remains large even after the correction. For the CSFB complexes, BSSE correction instead increases the MA\%D error (from 9.8\% to 12.1\% for the subset, and from 16.0\% to 21.5\% for the full set). Thus, we have found that deviations on BEs with and without BSSE correction are smaller in the subset of CSFB complexes than that in the full-set. Also, BEs of SEFB complexes predicted by M06-2X/AVTZ remain far less reliable than the corresponding CSFB complexes, regardless of BSSE correction or dataset size.

\FloatBarrier
\subsection{Deviations on F-bond distances of all DF\textperiodcentered{}\textperiodcentered{}\textperiodcentered{}CH\textsubscript{3}/NH\textsubscript{3} from M06-2X/AVTZ}\label{sec:deviations-on-f-bond}

In \Tref{tab:deviations-and-percentage-deviations}, we show the pattern of deviations and percentage deviations for F-bond distances across all 29 complexes from the SEFB-15 and CSFB-14 data sets. \Tref{tab:intermolecular-distances-r-f} and \textbf{\Fref{fig:deviations-on-bes-in}(b)} already give the F-bond distances and their overall error statistics, but not the deviation for each complex. \Tref{tab:deviations-and-percentage-deviations} and \textbf{\Fref{fig:deviation-1-1-bsse}(c-d)} reveal the deviation and percentage deviation for each complex. Also, values in parenthesis in \Tref{tab:deviations-and-percentage-deviations} for the selected SEFB and CSFB complexes are included for comparison. M06-2X/AVTZ underestimates F-bond distances for each of the 15 SEFB complexes (\Tref{tab:deviations-and-percentage-deviations}, \textbf{\Fref{fig:deviation-1-1-bsse}(c-d)}). Contrary to it, the same method shows a mixed pattern for CSFB complexes. F-bond distances of ten of the 14 CSFB complexes are overestimated, and those for the remaining four are underestimated (\Tref{tab:deviations-and-percentage-deviations}, \textbf{\Fref{fig:deviation-1-1-bsse}(c-d)}). In \Sref{sec:effect-of-basis-sets2}, we have already seen that there is basis set dependence for predicting F-bond distances of the selected five CSFB complexes using M06-2X functional (\Tref{tab:percentage-deviations-on-f}).

Among all SEFB complexes, NF\textsubscript{5}\textperiodcentered{}\textperiodcentered{}\textperiodcentered{}CH\textsubscript{3} has the smallest deviation (-0.164 Å, -5.3\%) and CF\textsubscript{3}CCF\textperiodcentered{}\textperiodcentered{}\textperiodcentered{}CH\textsubscript{3} the largest (-0.434 Å, -12.6\%). In the case of CSFB complexes, NF\textsubscript{5}\textperiodcentered{}\textperiodcentered{}\textperiodcentered{}NH\textsubscript{3} has the smallest deviation (0.003 Å, 0.1\%). The error statistics for the selected SEFB complexes are close to that obtained for the whole SEFB-15 database (MA\%D = 9.0\% versus 8.9\%) (\Tref{tab:deviations-and-percentage-deviations}). For CSFB complexes, the error statistics for the selected complexes are smaller than those of the CSFB-14 database (MA\%D = 1.7\% versus 3.3\%, nearly half), consistent with the same pattern found for BEs, as summarized in \mbox{\textbf{TABLE \ref{tab:deviations-and-percentage-deviations}.}}

Except for NF\textsubscript{5}\textperiodcentered{}\textperiodcentered{}\textperiodcentered{}CH\textsubscript{3}, deviations on F-bond distances of the remaining 14 SEFB complexes exceed the 5.0\% threshold. For CSFB complexes, the corresponding values for 9 out of 14 complexes remain within 5.0\%, and the remaining 5 exceed it (\Tref{tab:deviations-and-percentage-deviations}). Therefore, M06-2X/AVTZ is not suitable for predicting F-bond distances of SEFB complexes though it remains overall reliable for CSFB complexes.

\subsection{M06-2X/AVTZ for SEFB-15 and CSFB-14: context of literature}\label{sec:m06-2x-avtz-for}

In \textbf{Sections \ref*{sec:effect-of-basis-sets}-4.11}, we have shown that M06-2X/AVTZ over-binds SEFB complexes and underestimates their F-bond distances. This holds regardless of the basis sets or BSSE correction. With reasonable basis sets, such as AVTZ, BEs and F-bond distances for the selected CSFB complexes remain close to reference values throughout. Also, \Sref{sec:reference-binding-energies-and} already reveals that BEs without BSSE correction for CSFB complexes predicted by M06-2X/AVTZ mostly approach reference BEs obtained at DF-RMP2-F12/CBS (\Tref{tab:cbs-limit-binding-energies}). Also, deviations of F-bond distances obtained at the same level of theory are mostly within threshold of 5.0\% (\Tref{tab:deviations-and-percentage-deviations}). Keeping the method for energy fixed at W1-F12 and varying only the geometry, BEs deviate far more from reference when the geometry is obtained from M06-2X/AVTZ than from DF-RMP2-F12/VDZ-F12 for SEFB, while CSFB shows comparable deviations either way (\Tref{tab:cbs-limit-binding-energies}). Thus, M06-2X/AVTZ fails to predict not only reliable BEs for SEFB complexes, but also the geometries for them (\textbf{TABLES \ref{tab:cbs-limit-binding-energies}, \ref{tab:intermolecular-distances-r-f}} and \mbox{\textbf{FIGs. \ref{fig:bar-graphs-showing-bes}-\ref{fig:deviations-on-bes-in}}}).

Good performance of M06-2X/AVTZ for modelling CSFB-14 database is consistent with the established reputation of M06-2X for closed-shell halogen bonded complexes.\cite{Bauza2013,Kozuch2013,OteroDeLaRoza2014,Kim2019} Kozuch and Martin\cite{Kozuch2013} recommended M06-2X for geometries and dissociation energies of halogen-bonded complexes though they caution it may be over-fitted to its training set. Otero-de-la-Roza et al.\cite{OteroDeLaRoza2014}reported MAD of 0.020 Å for halogen-bond distances and 1.0 kJ/mol for BEs of closed-shell halogen bonded dataset XB18. Bauzá et al.\cite{Bauza2013} found M06-2X/AVTZ the most accurate method tested for halogen-bond distances. CSFB-14 database fits this established pattern, while SEFB-15 falls outside every dataset these studies tested.

TA13 database,\cite{Tentscher2013} the only comprehensive benchmark for radical-molecule complexes, contains one single-electron hydrogen-bonded (SEHB) complex, HOH\textperiodcentered{}\textperiodcentered{}\textperiodcentered{}CH\textsubscript{3}. The single electron in CH\textsubscript{3} radical acts as the hydrogen-bond acceptor. Similarly, CH\textsubscript{3} is the single electron donor in SEFB complexes. The deviation on BE of OH\textsubscript{2}\textperiodcentered{}\textperiodcentered{}\textperiodcentered{}CH\textsubscript{3} predicted by M06-2X is -3.6 kJ/mol against a -7.07 kJ/mol reference, a deviation of about 51\%. The absolute error looks small because the reference BE itself is weak. This is comparable in scale to percentage deviations on BSSE-corrected BES for several SEFB complexes, reported in \mbox{\textbf{TABLE \ref{tab:deviations-and-percentage-deviations}.}} The accuracy of M06-2X for CSFB complexes may relate to its own training data, which includes F\textsubscript{2}\textperiodcentered{}\textperiodcentered{}\textperiodcentered{}NH\textsubscript{3}, a charge-transfer complex in the CT7/04 database used in the parametrization of M06-type\cite{Zhao2008a} functionals.

Also, due to near 50\% exact exchange, M06-2X could be expected to reduce delocalization error.\cite{Bryenton2023} Due to its high exact-exchange fraction, Kim et al.\cite{Kim2019} attribute less density-driven error for M06-2X for predicting BEs of halogen bonded complexes. Whether this reasoning extends to the open-shell SEFB-15 dataset has not been tested.

In summary, good performance of M06-2X/AVTZ for CSFB-14 dataset is therefore consistent with its documented reliability for closed-shell halogen bonded systems. The failure for SEFB-15, in both energy and geometry, is already reflected in the one genuine SEHB analog (OH\textsubscript{2}\textperiodcentered{}\textperiodcentered{}\textperiodcentered{}CH\textsubscript{3}) in the literature.\cite{Tentscher2013} In the present study, we have not distinguished between a training-domain explanation, a density-driven explanation, and other possible causes. A broader assessment of these datasets across other Minnesota DFAs and popular \emph{ab initio} methods is reported in the companion paper \textbf{II}.

\section{Conclusions and outlook}\label{sec:conclusions-and-outlook}

We introduce two benchmark datasets for fluorine bonding in which a covalently bound fluorine atom acts as the halogen-bond donor: SEFB-15, consisting of 15 open-shell radical-molecule complexes DF\textperiodcentered{}\textperiodcentered{}\textperiodcentered{}CH\textsubscript{3} (DF = F\textsubscript{2}, SF\textsubscript{6}, NCF, CNF, NF\textsubscript{3}, ClF\textsubscript{3}, OF\textsubscript{2}, NF\textsubscript{5}, FCCF, CF\textsubscript{3}CCF, CF\textsubscript{3}OF, F\textsubscript{3}NO, SF\textsubscript{5}, NH\textsubscript{3}F\textsuperscript{+}, and NH\textsubscript{2}F\textsubscript{2}\textsuperscript{+}) exemplifying single-electron F-bonding (SEFB), and CSFB-14, consisting of 14 closed-shell complexes DF\textperiodcentered{}\textperiodcentered{}\textperiodcentered{}NH\textsubscript{3} (DF = F\textsubscript{2}, SF\textsubscript{6}, NCF, CNF, NF\textsubscript{3}, ClF\textsubscript{3}, OF\textsubscript{2}, NF\textsubscript{5}, FCCF, CF\textsubscript{3}CCF, CF\textsubscript{3}OF, SF\textsubscript{5}, NH\textsubscript{3}F\textsuperscript{+}, and NH\textsubscript{2}F\textsubscript{2}\textsuperscript{+}) representing closed-shell F-bonding (CSFB). Among the neutral SEFB complexes, F\textsubscript{2}\textperiodcentered{}\textperiodcentered{}\textperiodcentered{}CH\textsubscript{3} has the shortest F-bond distance and the highest binding energy.

For both datasets, the reference BEs at DF-RMP2-F12/CBS//DF-RMP2-F12/VDZ-F12 are confirmed by W1-F12, W2-F12, and CCSD(T)-F12b/CBS//CCSD(T)-F12b/VDZ-F12 for selected systems (DF = F\textsubscript{2}, NCF, CNF, NH\textsubscript{3}F\textsuperscript{+}, and NH\textsubscript{2}F\textsubscript{2}\textsuperscript{+}), establishing their reliability for the full datasets. W1-F12 and W2-F12 produce identical or near-identical BEs. Non-relativistic valence-electron, non-relativistic all-electron, and relativistic all-electron CCSD(T)/CBS BEs within the W1-F12 protocol differ by at most 0.1-0.2 kJ/mol, confirming that relativistic and all-electron corrections are negligible.

Reference F-bond distances at DF-RMP2-F12/VDZ-F12 are confirmed against CCSD(T)-F12b, CCSD(T), CCSD, and DCSD for the same selected systems, establishing DF-RMP2-F12/VDZ-F12 as a reliable and affordable alternative to CCSD(T)-F12b for F-bond distances. Among the ab initio methods assessed, CCSD(T)/AVDZ performs worst for F-bond distances in both datasets, and CCSD performs better with AVDZ than AVTZ for both SEFB and CSFB complexes. DCSD/AVTZ yields improved BEs over DCSD/AVDZ for both datasets. Moreover, MP2/AVDZ reproduces F-bond distances well for SEFB (MA\%D = 1.2\%) and CSFB (MA\%D = 2.4\%). Also, increasing the basis set beyond AVTZ does not further improve F-bond distances from MP2.

M06-2X has been widely recommended for closed-shell halogen bonding with heavier halogens,\cite{Bauza2013,Kozuch2013,OteroDeLaRoza2014} although previous benchmark studies that led to this recommendation considered neither radical-molecule F-bonded complexes nor closed-shell F-bonded complexes such as F\textsubscript{2}\textperiodcentered{}\textperiodcentered{}\textperiodcentered{}NH\textsubscript{3}. For SEFB-15, M06-2X/AVTZ gives large errors on both BEs and F-bond distances for all complexes, with F\textsubscript{2}\textperiodcentered{}\textperiodcentered{}\textperiodcentered{}CH\textsubscript{3} being particularly challenging. This failure is not a basis set artifact; testing 16 different basis sets {[}VDZ, AVDZ, VTZ, AVTZ, VQZ, AVQZ, V5Z, AV5Z, def2-TZVP, def2-QZVP, 6-311g(2d,2p), 6-311++g(2d,2p), 6-311(2df,2pd), 6-311++(2df,2pd), 6-311g(3df,3pd) and 6-311++g(3df,3pd){]} confirms this. Furthermore, BSSE correction, while reducing MA\%D on BEs of SEFB complexes from 75.8\% to 54.6\%, does not resolve the inconsistency with BEs from ab initio methods and increases MA\%D for DF\textperiodcentered{}\textperiodcentered{}\textperiodcentered{}NH\textsubscript{3} from 16.0\% to 21.5\%. The unreliability of M06-2X geometries for SEFB is further confirmed by W1-F12//M06-2X/AVTZ yielding MA\%D of 36.1\% on BEs against 3.2\% for W1-F12//DF-RMP2-F12/VDZ-F12. In contrast, M06-2X/AVTZ performs reliably for CSFB-14. These results call for caution in extending DFA recommendations based on closed-shell halogen bonding to open-shell systems. The SEFB-15 and CSFB-14 datasets introduced here provide reference data for benchmarking DFAs for F-bonding, and their broader assessment across Minnesota DFAs and popular ab initio methods is reported in the companion paper published in this issue.

\section*{Supporting Information:}

Raw data supporting various results in this manuscript are included in \mbox{\textbf{Figures \ref{fig:the-optimized-structures-for}-\ref{fig:mean-signed-error-mse}}} and \mbox{\textbf{Tables \ref{tab:complete-basis-set-cbs}-\ref{tab:f-bond-angles-df}.}} Also, optimized geometries obtained at several level of theories used in the manuscript are included in \mbox{\textbf{Tables S16-S46.}}

\par\vspace{4pt}\noindent{\bfseries\fontsize{13}{16}\selectfont ORCID}\par\nopagebreak\vspace{2pt}

Govinda Prasad Khanal: \href{https://orcid.org/0000-0002-0991-7005}{\textcolor[HTML]{00B0F0}{https://orcid.org/0000-0002-0991-7005}}

Elangannan Arunan: \href{https://orcid.org/0000-0001-9669-4307}{\textcolor[HTML]{00B0F0}{https://orcid.org/0000-0001-9669-4307}}

\par\vspace{4pt}\noindent{\bfseries\fontsize{13}{16}\selectfont Conflicts of interest}\par\nopagebreak\vspace{2pt}

There are no conflicts to declare.

\par\vspace{4pt}\noindent{\bfseries\fontsize{13}{16}\selectfont Acknowledgements}\par\nopagebreak\vspace{2pt}

G.P.K. and E.A. thank Prof. Sai G. Ramesh for maintaining the computational facilities at the Department of Inorganic and Physical Chemistry. G. P. K. thanks Indian Institute of Science (IISC) for scholarship for his PhD. Also, both authors thank Prof. Debashree Ghosh for fruitful suggestions. The authors also acknowledge the Supercomputer Education and Research Centre, IISC, for providing computational resources.

\section*{References}
{\setstretch{1}\fontsize{8}{9.5}\selectfont\setlength{\parskip}{1pt}\justifying\setlength{\parindent}{0pt}
\bibliographystyle{gpk}\bibliography{refs}
}
\clearpage
\setcounter{figure}{0}\setcounter{table}{0}\renewcommand{\thefigure}{S\arabic{figure}}\renewcommand{\thetable}{S\arabic{table}}\renewcommand{\theHfigure}{S\arabic{figure}}\renewcommand{\theHtable}{S\arabic{table}}\renewcommand{\figurename}{Figure}\renewcommand{\tablename}{Table}
\section*{Supporting Information for:}

\textbf{Single-electron Fluorine Bonding. I. The SEFB-15 and CSFB-14 Benchmark Datasets, Reference Binding Energies and Intermolecular Bond Distances: Assessment of M06-2X density functional}

Govinda Prasad Khanal\textsuperscript{a}, and Elangannan Arunan\textsuperscript{a}*

\textsuperscript{a}Inorganic and Physical Chemistry, Indian Institute of Science, Bangalore, India -- 560 012.

\textbf{*E-mail Id:} \href{mailto:arunan@iisc.ac.in}{\nolinkurl{arunan@iisc.ac.in}}

\begin{figure}[!htbp]\centering
\GPKfig{6.90278in}{3.58194in}{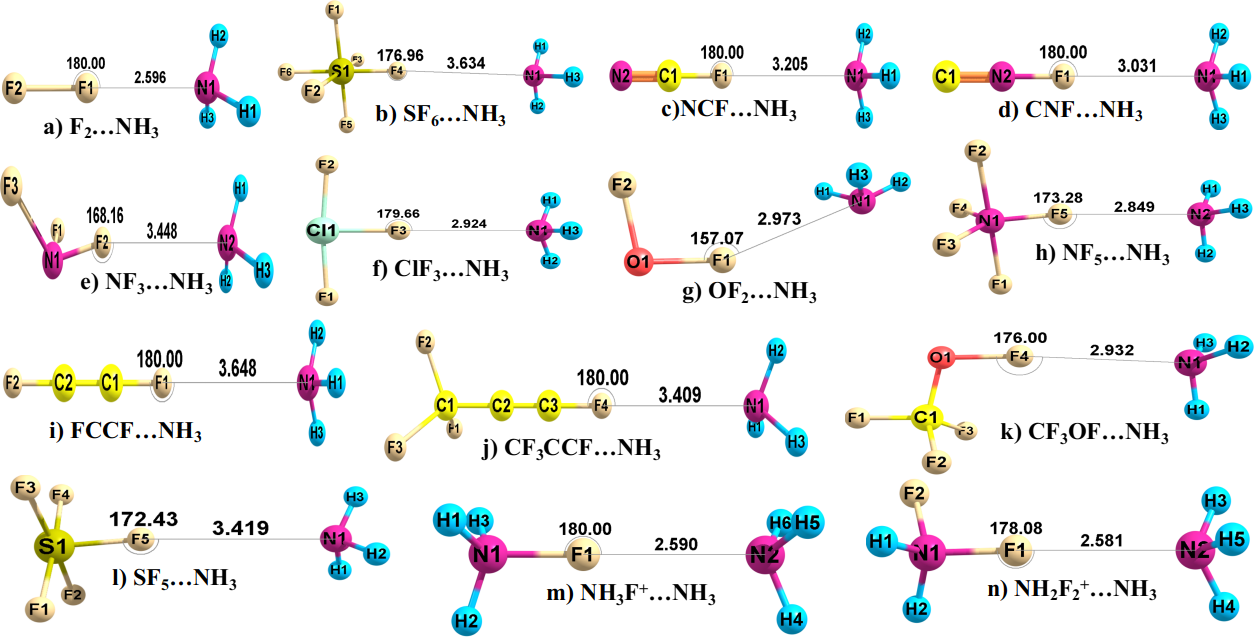}
\par\vspace{-1pt}\caption{The optimized structures for 14 closed-shell F-bonded (CSFB) complexes DF\textperiodcentered{}\textperiodcentered{}\textperiodcentered{}NH\textsubscript{3} calculated at M06-2X/AVTZ level of theory.}\label{fig:the-optimized-structures-for}
\vspace{10pt}
\end{figure}

\begin{figure}[!htbp]\centering
\GPKfig{6.5in}{2.79236in}{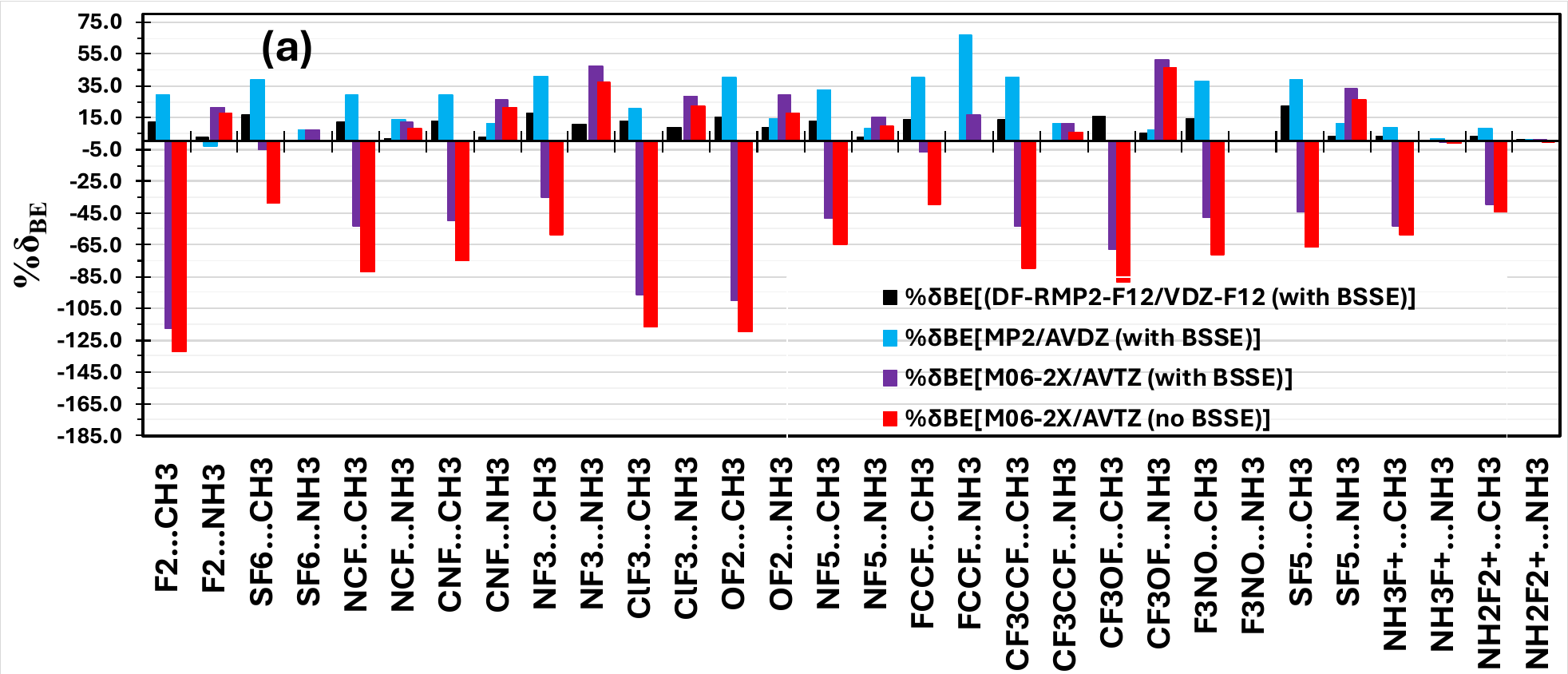}\GPKfig{6.5in}{3.33403in}{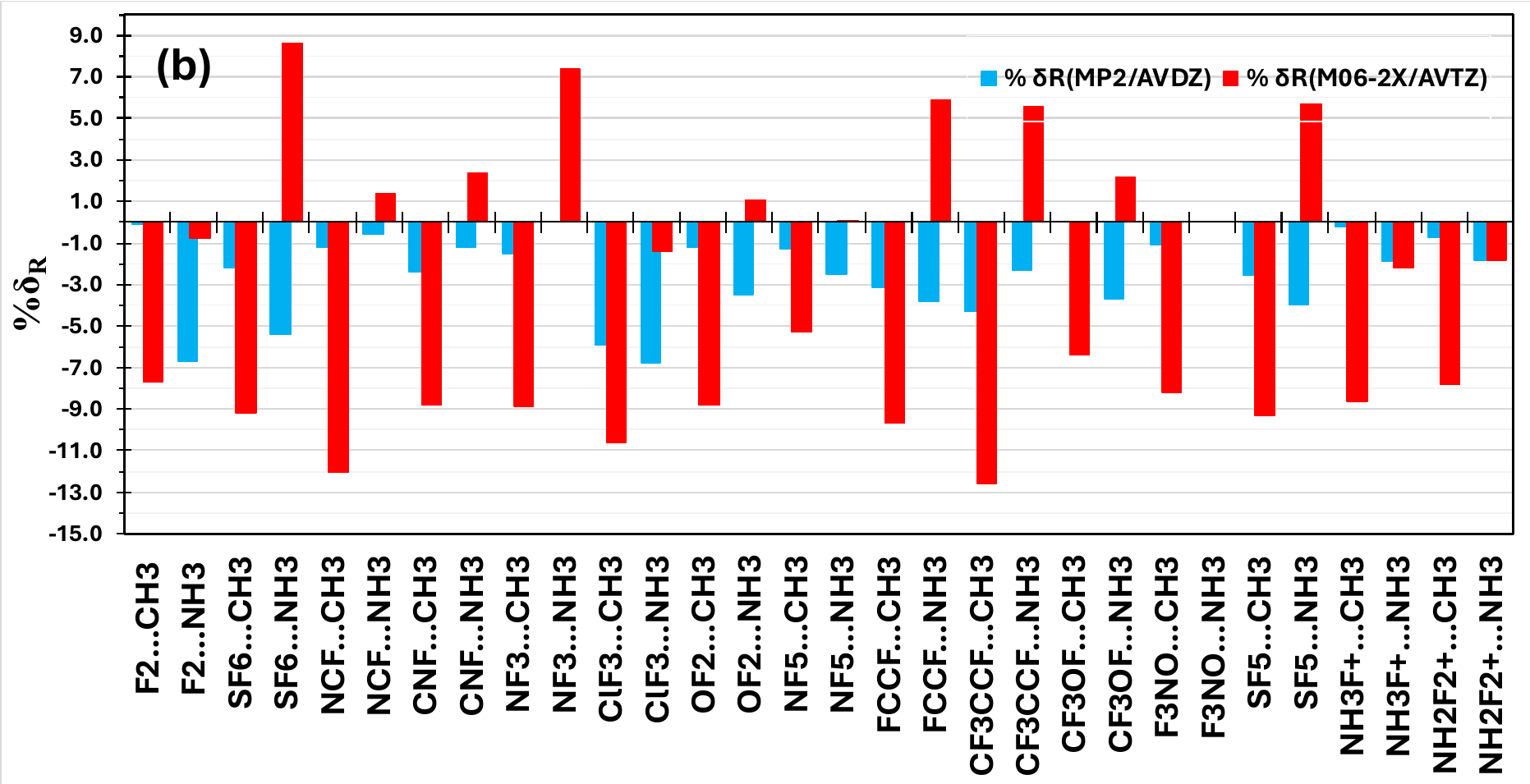}
\par\vspace{-1pt}{\renewcommand\figurename{FIG.}\caption{Percentage deviations on BEs \textbf{(a)} and that on F-bond distances \textbf{(b)} for all DF\textperiodcentered{}\textperiodcentered{}\textperiodcentered{}CH\textsubscript{3}/NH\textsubscript{3} complexes at DF-RMP2-F12/VDZ-F12, MP2/AVDZ and M06-2X/AVTZ level of theories. BEs calculated at DF-RMP2-F12/CBS//DF-RMP2-F12/VDZ-F12 and F-bond distances obtained at DF-RMP2-F12/VDZ-F12 are used as reference values.}\label{fig:percentage-deviations-on-bes}}
\vspace{10pt}
\end{figure}

\begin{figure}[!htbp]\centering
\GPKfig{6.5in}{3.00139in}{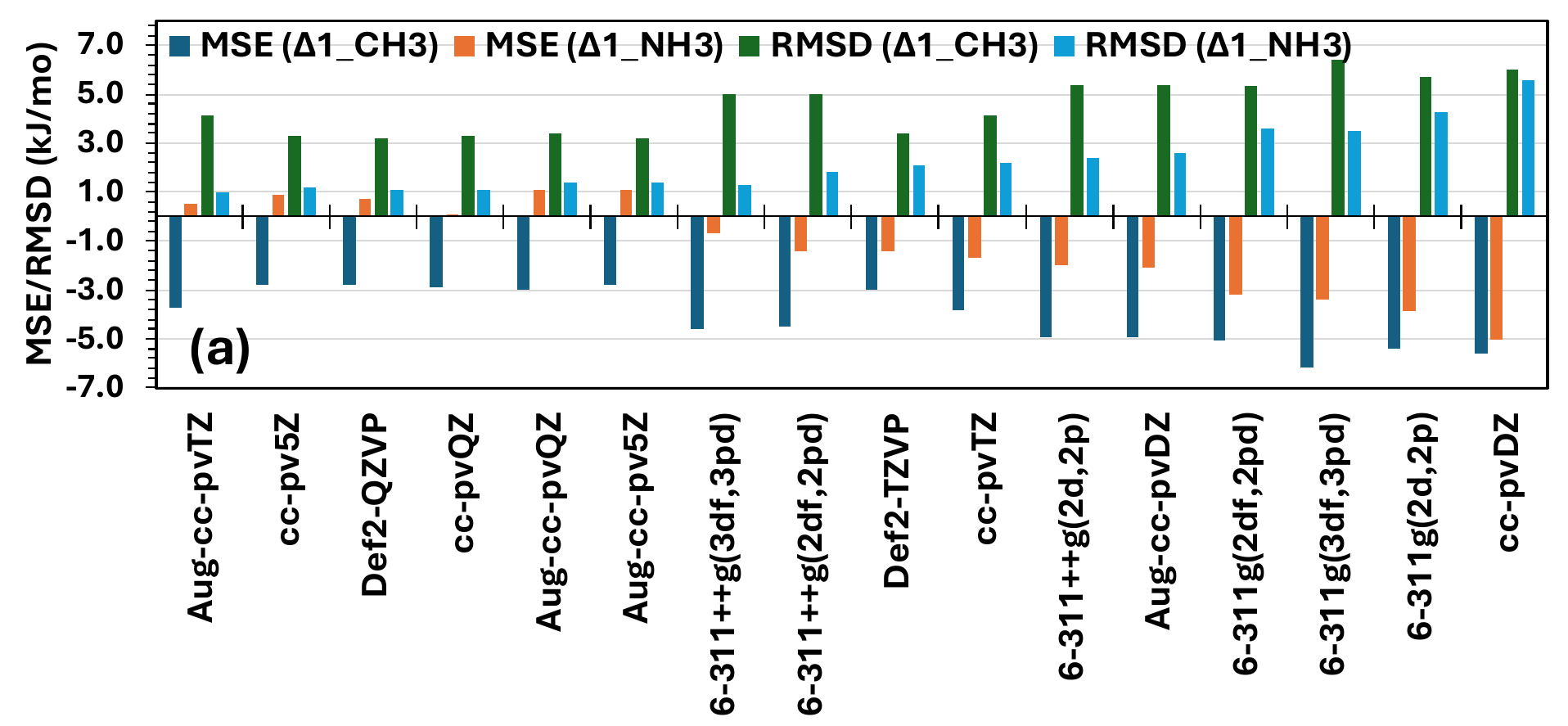}\GPKfig{6.5in}{3.22292in}{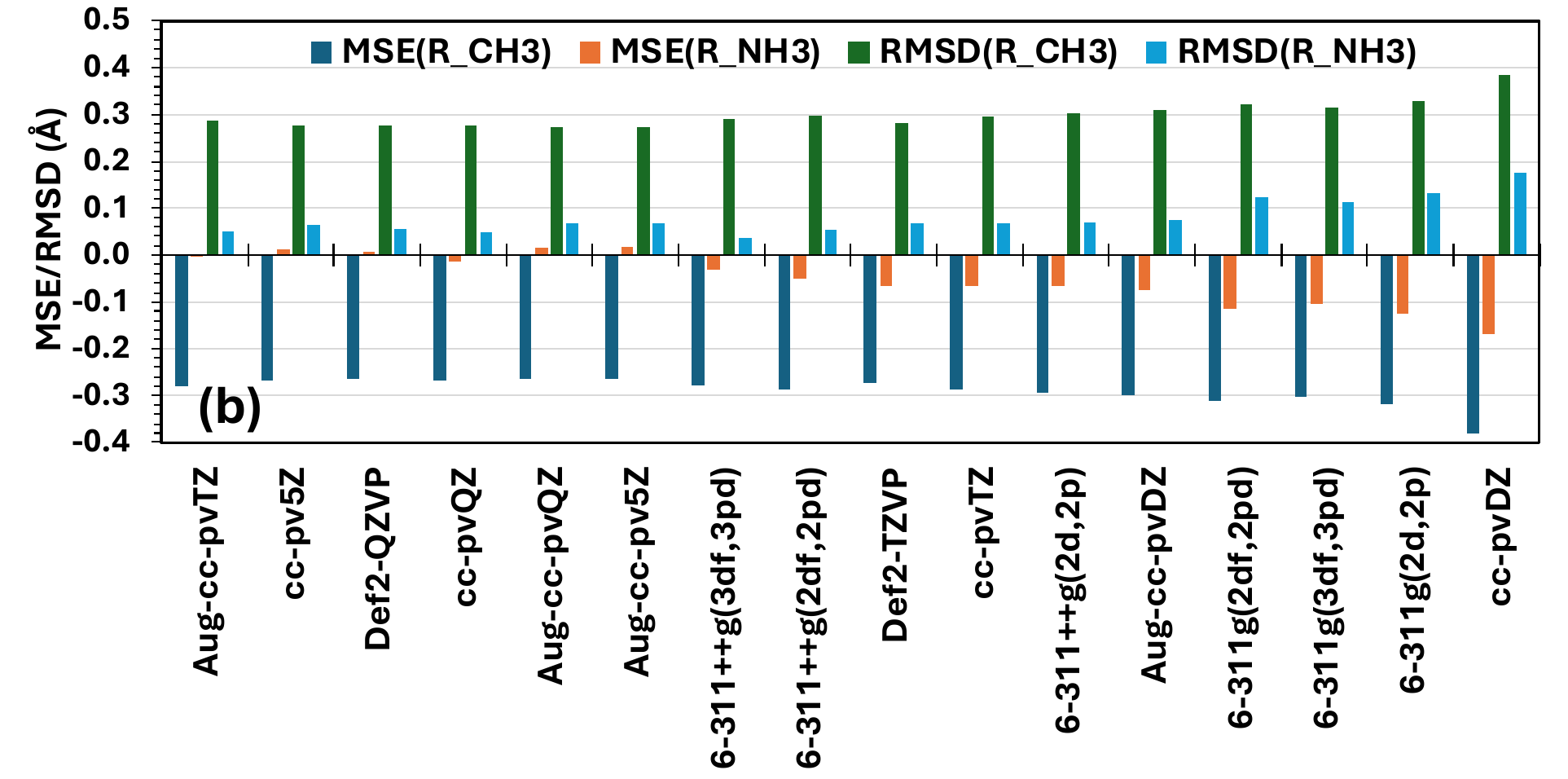}
\par\vspace{-1pt}\caption{Mean signed error (MSE) and root-mean-square deviation (RMSD) for (a) BSSE-uncorrected binding energies (\ensuremath{\Delta}\textsubscript{1}, kJ/mol) and (b) F-bond distances {[}R(F\textperiodcentered{}\textperiodcentered{}\textperiodcentered{}C/N), Å{]} for selected F-bonded complexes DF\textperiodcentered{}\textperiodcentered{}\textperiodcentered{}CH\textsubscript{3}/NH\textsubscript{3} (DF = F\textsubscript{2}, NCF, CNF, NH\textsubscript{3}F\textsuperscript{+}, NH\textsubscript{2}F\textsubscript{2}\textsuperscript{+}), calculated using M06-2X functional with several basis sets. Reference BEs were obtained at DF-RMP2-F12/CBS{[}V(T,Q)Z-F12{]}//DF-RMP2-F12/VDZ-F12, and reference bond lengths at DF-RMP2-F12/VDZ-F12 level. Basis sets are arranged in increasing order of mean absolute deviations (MADs) for BSSE-uncorrected BEs of DF\textperiodcentered{}\textperiodcentered{}\textperiodcentered{}NH\textsubscript{3} complexes. M06-2X appears to perform reliably for CSFB complexes with NH\textsubscript{3}.}\label{fig:mean-signed-error-mse}
\vspace{10pt}
\end{figure}

\begin{table}[!htbp]\centering\setlength{\parskip}{0pt}
\caption{Complete basis set (CBS) limit binding energies (CBS B.E., kJ/\allowbreak{}mol) for F-bonded complexes DF\textperiodcentered{}\textperiodcentered{}\textperiodcentered{}CH\textsubscript{3}/NH\textsubscript{3}, calculated using DF-RMP2-F12[CBS\{V(T,Q)Z-F12\}]//DF-RMP2-F12/\allowbreak{}VDZ-F12 and CCSD(T)/\allowbreak{}CBS//MP2/\allowbreak{}aug-cc-pVDZ (Hobza group method; see main text). Also, BEs from the explicitly correlated Weizmann method W1-F12, using geometries optimized at DF-RMP2-F12/\allowbreak{}VDZ-F12 and M06-2X/\allowbreak{}aug-cc-pVTZ levels are included. BEs without BSSE correction (\ensuremath{\Delta}\textsubscript{1}) and with BSSE correction (\ensuremath{\Delta}\textsubscript{1+BSSE}) are shown for two wave function methods (DF-RMP2-F12 and MP2) and one hybrid meta-GGA (M06-2X). BEs from DF-RMP2-F12[CBS\{V(T,Q)Z-F12\}]//DF-RMP2-F12/\allowbreak{}VDZ-F12 level are used as reference throughout this work.}\label{tab:complete-basis-set-cbs}\vspace{7pt}\nopagebreak
{\fontsize{7.0}{8.0}\fontfamily{ptm}\selectfont\renewcommand{\arraystretch}{1.17}\setlength{\tabcolsep}{1.5pt}
\centerline{\adjustbox{max width=518pt}{\begin{tabular}{|>{\centering\arraybackslash}m{65.1pt}|>{\centering\arraybackslash}m{16.9pt}|>{\centering\arraybackslash}m{16.9pt}|>{\centering\arraybackslash}m{18.5pt}|>{\centering\arraybackslash}m{18.5pt}|>{\centering\arraybackslash}m{16.9pt}|>{\centering\arraybackslash}m{17.0pt}|>{\centering\arraybackslash}m{18.5pt}|>{\centering\arraybackslash}m{16.9pt}|>{\centering\arraybackslash}m{17.1pt}|>{\centering\arraybackslash}m{18.5pt}|>{\centering\arraybackslash}m{65.1pt}|>{\centering\arraybackslash}m{16.9pt}|>{\centering\arraybackslash}m{17.3pt}|>{\centering\arraybackslash}m{18.5pt}|>{\centering\arraybackslash}m{8.0pt}|>{\centering\arraybackslash}m{8.0pt}|>{\centering\arraybackslash}m{17.1pt}|>{\centering\arraybackslash}m{16.9pt}|>{\centering\arraybackslash}m{18.5pt}|>{\centering\arraybackslash}m{16.9pt}|>{\centering\arraybackslash}m{16.9pt}|>{\centering\arraybackslash}m{18.5pt}|}
\specialrule{0.9pt}{0pt}{0pt}
\multicolumn{5}{|>{\centering\arraybackslash}p{147.9pt}|}{\textbf{Opt} \textbf{@DF-RMP2-F12/\allowbreak{}VDZ-F12}} & \multicolumn{3}{>{\centering\arraybackslash}p{58.4pt}|}{\textbf{Opt} \textbf{@MP2/} \textbf{AVDZ}} & \multicolumn{3}{>{\centering\arraybackslash}p{58.5pt}|}{\textbf{Opt} \textbf{@M06-2X/\allowbreak{}AVTZ}} & \multicolumn{6}{>{\centering\arraybackslash}p{148.4pt}|}{\textbf{Opt} \textbf{@DF-RMP2-F12/\allowbreak{}VDZ-F12}} & \multicolumn{3}{>{\centering\arraybackslash}p{58.5pt}|}{\textbf{Opt} \textbf{@MP2/} \textbf{AVDZ}} & \multicolumn{3}{>{\centering\arraybackslash}p{58.3pt}|}{\textbf{Opt} \textbf{@M06-2X/\allowbreak{}AVTZ}} \\* \hline\noalign{\penalty10000}
\multicolumn{1}{|>{\centering\arraybackslash}m{65.1pt}|}{\multirow{2}{65.1pt}{\centering \makebox[\linewidth][c]{\rotatebox{90}{\hspace{2pt}\textbf{ DF\textperiodcentered{}\textperiodcentered{}\textperiodcentered{}CH}\textbf{\textsubscript{3}}\hspace{2pt}}}}} & \multicolumn{2}{>{\centering\arraybackslash}m{36.8pt}|}{\makebox[\linewidth][c]{\rotatebox{90}{\hspace{2pt}\textbf{VDZ-F12}\hspace{2pt}}}} & \multicolumn{2}{>{\centering\arraybackslash}m{40.1pt}|}{\makebox[\linewidth][c]{\rotatebox{90}{\hspace{2pt}\textbf{DF-RMP2-F12/} \textbf{CBS B.E.}\hspace{2pt}}}} & \multicolumn{2}{>{\centering\arraybackslash}m{36.9pt}|}{\makebox[\linewidth][c]{\rotatebox{90}{\hspace{2pt}\textbf{AVDZ}\hspace{2pt}}}} & \makebox[\linewidth][c]{\rotatebox{90}{\hspace{2pt}\textbf{CCSD(T)/\allowbreak{}CBS B.E.}\hspace{2pt}}} & \makebox[\linewidth][c]{\rotatebox{90}{\hspace{2pt}\textbf{\ensuremath{\Delta}}\textbf{\textsubscript{1}}\hspace{2pt}}} & \makebox[\linewidth][c]{\rotatebox{90}{\hspace{2pt}\textbf{\ensuremath{\Delta}}\textbf{\textsubscript{1+BSSE}}\hspace{2pt}}} & \makebox[\linewidth][c]{\rotatebox{90}{\hspace{2pt}\textbf{W1-F12}\hspace{2pt}}} & \multicolumn{1}{>{\centering\arraybackslash}m{65.1pt}|}{\makebox[\linewidth][c]{\rotatebox{90}{\hspace{2pt}\textbf{DF\textperiodcentered{}\textperiodcentered{}\textperiodcentered{}NH}\textbf{\textsubscript{3}}\hspace{2pt}}}} & \multicolumn{2}{>{\centering\arraybackslash}m{37.2pt}|}{\makebox[\linewidth][c]{\rotatebox{90}{\hspace{2pt}\textbf{VDZ-F12}\hspace{2pt}}}} & \multicolumn{2}{>{\centering\arraybackslash}m{29.3pt}|}{\makebox[\linewidth][c]{\rotatebox{90}{\hspace{2pt}\textbf{CBS B.E.}\hspace{2pt}}}} & \multicolumn{3}{>{\centering\arraybackslash}m{47.7pt}|}{\makebox[\linewidth][c]{\rotatebox{90}{\hspace{2pt}\textbf{AVDZ}\hspace{2pt}}}} & \makebox[\linewidth][c]{\rotatebox{90}{\hspace{2pt}\textbf{CCSD(T)/} \textbf{CBS B.E.}\hspace{2pt}}} & \makebox[\linewidth][c]{\rotatebox{90}{\hspace{2pt}\textbf{\ensuremath{\Delta}}\textbf{\textsubscript{1}}\hspace{2pt}}} & \makebox[\linewidth][c]{\rotatebox{90}{\hspace{2pt}\textbf{\ensuremath{\Delta}}\textbf{\textsubscript{1+BSSE}}\hspace{2pt}}} & \makebox[\linewidth][c]{\rotatebox{90}{\hspace{2pt}\textbf{W1-F12}\hspace{2pt}}} \\* \cline{2-23}\noalign{\penalty10000}
\multicolumn{1}{|>{\centering\arraybackslash}m{65.1pt}|}{} & \makebox[\linewidth][c]{\rotatebox{90}{\hspace{2pt}\textbf{\ensuremath{\Delta}}\textbf{\textsubscript{1}}\hspace{2pt}}} & \makebox[\linewidth][c]{\rotatebox{90}{\hspace{2pt}\textbf{\ensuremath{\Delta}}\textbf{\textsubscript{1+BSSE}}\hspace{2pt}}} & \makebox[\linewidth][c]{\rotatebox{90}{\hspace{2pt}\textbf{V(T,Q)Z-F12}\hspace{2pt}}} & \makebox[\linewidth][c]{\rotatebox{90}{\hspace{2pt}\textbf{W1-F12}\hspace{2pt}}} & \makebox[\linewidth][c]{\rotatebox{90}{\hspace{2pt}\textbf{\ensuremath{\Delta}}\textbf{\textsubscript{1}}\hspace{2pt}}} & \makebox[\linewidth][c]{\rotatebox{90}{\hspace{2pt}\textbf{\ensuremath{\Delta}}\textbf{\textsubscript{1+BSSE}}\hspace{2pt}}} & \makebox[\linewidth][c]{\rotatebox{90}{\hspace{2pt}\textbf{AV(T,Q)Z}\hspace{2pt}}} &  &  &  &  & \makebox[\linewidth][c]{\rotatebox{90}{\hspace{2pt}\textbf{\ensuremath{\Delta}}\textbf{\textsubscript{1}}\hspace{2pt}}} & \makebox[\linewidth][c]{\rotatebox{90}{\hspace{2pt}\textbf{\ensuremath{\Delta}}\textbf{\textsubscript{1+BSSE}}\hspace{2pt}}} & \makebox[\linewidth][c]{\rotatebox{90}{\hspace{2pt}\textbf{V(T,Q)Z-F12}\hspace{2pt}}} & \multicolumn{2}{>{\centering\arraybackslash}m{18.6pt}|}{\makebox[\linewidth][c]{\rotatebox{90}{\hspace{2pt}\textbf{W1-F12}\hspace{2pt}}}} & \makebox[\linewidth][c]{\rotatebox{90}{\hspace{2pt}\textbf{\ensuremath{\Delta}}\textbf{\textsubscript{1}}\hspace{2pt}}} & \makebox[\linewidth][c]{\rotatebox{90}{\hspace{2pt}\textbf{\ensuremath{\Delta}}\textbf{\textsubscript{1+BSSE}}\hspace{2pt}}} & \makebox[\linewidth][c]{\rotatebox{90}{\hspace{2pt}\textbf{AV(T,Q)Z}\hspace{2pt}}} &  &  &  \\* \hline\noalign{\penalty10000}
\multicolumn{1}{|>{\centering\arraybackslash}m{65.1pt}|}{\textbf{F}\textbf{\textsubscript{2}}\textbf{\textperiodcentered{}\textperiodcentered{}\textperiodcentered{}CH}\textbf{\textsubscript{3}}} & -3.2 & -3.0 & \textbf{-3.4} & \textbf{-3.1} & -4.6 & -2.4 & \textbf{-3.5} & -7.9 & -7.4 & \textbf{-3.2} & \textbf{ F}\textbf{\textsubscript{2}}\textbf{\textperiodcentered{}\textperiodcentered{}\textperiodcentered{}NH}\textbf{\textsubscript{3}} & -8.0 & -7.6 & \textbf{-7.8} & \multicolumn{2}{>{\centering\arraybackslash}m{18.6pt}|}{\textbf{-7.4}} & -11.9 & -8.0 & \textbf{-7.2} & -6.4 & -6.1 & \textbf{-7.3} \\* \hline\noalign{\penalty10000}
\multicolumn{1}{|>{\centering\arraybackslash}m{65.1pt}|}{\textbf{SF}\textbf{\textsubscript{6}}\textbf{\textperiodcentered{}\textperiodcentered{}\textperiodcentered{}CH}\textbf{\textsubscript{3}}} & -1.6 & -1.5 & \textbf{-1.8} & \textbf{N/\allowbreak{}A} & -4.5 & -1.1 & \textbf{-2.6} & -2.5 & -1.9 & \textbf{N/\allowbreak{}A} & \textbf{SF}\textbf{\textsubscript{6}}\textbf{\textperiodcentered{}\textperiodcentered{}\textperiodcentered{}NH}\textbf{\textsubscript{3}} & -1.7 & -1.4 & \textbf{-1.4} & \multicolumn{2}{>{\centering\arraybackslash}m{18.6pt}|}{\textbf{-1.7}} & -4.0 & -1.3 & \textbf{-1.9} & -1.4 & -1.3 & \textbf{-1.6} \\* \hline\noalign{\penalty10000}
\multicolumn{1}{|>{\centering\arraybackslash}m{65.1pt}|}{\textbf{NCF\textperiodcentered{}\textperiodcentered{}\textperiodcentered{}CH}\textbf{\textsubscript{3}}} & -1.7 & -1.5 & \textbf{-1.7} & \textbf{-1.7} & -2.3 & -1.2 & \textbf{-2.0} & -3.1 & -2.6 & \textbf{-0.8} & \textbf{NCF\textperiodcentered{}\textperiodcentered{}\textperiodcentered{}NH}\textbf{\textsubscript{3}} & -5.5 & -5.0 & \textbf{-5.1} & \multicolumn{2}{>{\centering\arraybackslash}m{18.6pt}|}{\textbf{-5.2}} & -6.0 & -4.4 & \textbf{-5.5} & -4.7 & -4.5 & \textbf{-5.2} \\* \hline\noalign{\penalty10000}
\multicolumn{1}{|>{\centering\arraybackslash}m{65.1pt}|}{\textbf{CNF\textperiodcentered{}\textperiodcentered{}\textperiodcentered{}CH}\textbf{\textsubscript{3}}} & -2.2 & -2.1 & \textbf{-2.4} & \textbf{-2.2} & -3.4 & -1.7 & \textbf{-2.5} & -4.2 & -3.6 & \textbf{-1.7} & \textbf{CNF\textperiodcentered{}\textperiodcentered{}\textperiodcentered{}NH}\textbf{\textsubscript{3}} & -7.2 & -6.7 & \textbf{-6.9} & \multicolumn{2}{>{\centering\arraybackslash}m{18.6pt}|}{\textbf{-6.4}} & -8.4 & -6.1 & \textbf{-6.7} & -5.4 & -5.1 & \textbf{-6.4} \\ \hline
\multicolumn{1}{|>{\centering\arraybackslash}m{65.1pt}|}{\textbf{NF}\textbf{\textsubscript{3}}\textbf{\textperiodcentered{}\textperiodcentered{}\textperiodcentered{}CH}\textbf{\textsubscript{3}}} & -1.6 & -1.4 & \textbf{-1.7} & \textbf{-1.7} & -3.4 & -1.0 & \textbf{-2.1} & -2.7 & -2.3 & \textbf{-0.4} & \textbf{NF}\textbf{\textsubscript{3}}\textbf{\textperiodcentered{}\textperiodcentered{}\textperiodcentered{}NH}\textbf{\textsubscript{3}} & -2.1 & -1.7 & \textbf{-1.9} & \multicolumn{2}{>{\centering\arraybackslash}m{18.6pt}|}{\textbf{-2.0}} & \textbf{b} & \textbf{b} & \textbf{b} & -1.2 & -1.0 & \textbf{-1.6} \\ \hline
\multicolumn{1}{|>{\centering\arraybackslash}m{65.1pt}|}{\textbf{ClF}\textbf{\textsubscript{3}}\textbf{\textperiodcentered{}\textperiodcentered{}\textperiodcentered{}CH}\textbf{\textsubscript{3}}} & -2.3 & -2.1 & \textbf{-2.4} & \textbf{-2.4} & -4.3 & -1.9 & \textbf{-2.9} & -5.2 & -4.7 & \textbf{-1.5} & \textbf{ClF}\textbf{\textsubscript{3}}\textbf{\textperiodcentered{}\textperiodcentered{}\textperiodcentered{}NH}\textbf{\textsubscript{3}} & -4.5 & -4.1 & \textbf{-4.5} & \multicolumn{2}{>{\centering\arraybackslash}m{18.6pt}|}{\textbf{-4.5}} & -7.4 & -4.5 & \textbf{-4.9} & -3.5 & -3.2 & \textbf{-4.5} \\ \hline
\multicolumn{1}{|>{\centering\arraybackslash}m{65.1pt}|}{\textbf{OF}\textbf{\textsubscript{2}}\textbf{\textperiodcentered{}\textperiodcentered{}\textperiodcentered{}CH}\textbf{\textsubscript{3}}} & -1.8 & -1.7 & \textbf{-2.0} & \textbf{-1.9} & -3.8 & -1.2 & \textbf{-2.2} & -4.4 & -4.0 & \textbf{-1.5} & \textbf{OF}\textbf{\textsubscript{2}}\textbf{\textperiodcentered{}\textperiodcentered{}\textperiodcentered{}NH}\textbf{\textsubscript{3}} & -3.5 & -3.1 & \textbf{-3.4} & \multicolumn{2}{>{\centering\arraybackslash}m{18.6pt}|}{\textbf{-3.3}} & -5.6 & -2.9 & \textbf{-3.6} & -2.8 & -2.4 & \textbf{-3.3} \\ \hline
\multicolumn{1}{|>{\centering\arraybackslash}m{65.1pt}|}{\textbf{NF}\textbf{\textsubscript{5}}\textbf{\textperiodcentered{}\textperiodcentered{}\textperiodcentered{}CH}\textbf{\textsubscript{3}}} & -2.9 & -2.7 & \textbf{-3.1} & \textbf{N/\allowbreak{}A} & -4.8 & -2.1 & \textbf{-3.0} & -5.1 & -4.6 & \textbf{N/\allowbreak{}A} & \textbf{NF}\textbf{\textsubscript{5}}\textbf{\textperiodcentered{}\textperiodcentered{}\textperiodcentered{}NH}\textbf{\textsubscript{3}} & -7.6 & -7.1 & \textbf{-7.3} & \multicolumn{2}{>{\centering\arraybackslash}m{18.6pt}|}{\textbf{-7.8}} & -10.0 & -6.7 & \textbf{-8.2} & -6.6 & -6.2 & \textbf{-7.7} \\ \hline
\multicolumn{1}{|>{\centering\arraybackslash}m{65.1pt}|}{\textbf{FCCF\textperiodcentered{}\textperiodcentered{}\textperiodcentered{}CH}\textbf{\textsubscript{3}}} & -1.5 & -1.3 & \textbf{-1.5} & \textbf{-1.5} & -3.6 & -0.9 & \textbf{-2.2} & -2.1 & -1.6 & \textbf{0.2} & \textbf{FCCF\textperiodcentered{}\textperiodcentered{}\textperiodcentered{}NH}\textbf{\textsubscript{3}} & -1.0 & -0.6 & \textbf{-0.6} & \multicolumn{2}{>{\centering\arraybackslash}m{18.6pt}|}{\textbf{-0.8}} & -2.1 & -0.2 & \textbf{-1.2} & -0.6 & -0.5 & \textbf{-0.7} \\ \hline
\multicolumn{1}{|>{\centering\arraybackslash}m{65.1pt}|}{\textbf{CF}\textbf{\textsubscript{3}}\textbf{CCF\textperiodcentered{}\textperiodcentered{}\textperiodcentered{}CH}\textbf{\textsubscript{3}}} & -1.4 & -1.3 & \textbf{-1.5} & \textbf{-1.4} & -2.9 & -0.9 & \textbf{-1.8} & -2.7 & -2.3 & \textbf{N/\allowbreak{}A} & \textbf{CF}\textbf{\textsubscript{3}}\textbf{CCF\textperiodcentered{}\textperiodcentered{}\textperiodcentered{}NH}\textbf{\textsubscript{3}} & -4.1 & -3.6 & \textbf{-3.6} & \multicolumn{2}{>{\centering\arraybackslash}m{18.6pt}|}{\textbf{-3.7}} & -5.5 & -3.2 & \textbf{-3.9} & -3.4 & -3.2 & \textbf{-3.5} \\ \hline
\multicolumn{1}{|>{\centering\arraybackslash}m{65.1pt}|}{\textbf{CF}\textbf{\textsubscript{3}}\textbf{OF\textperiodcentered{}\textperiodcentered{}\textperiodcentered{}CH}\textbf{\textsubscript{3}}} & -2.3 & -2.1 & \textbf{-2.5} & \textbf{N/\allowbreak{}A} & \textbf{b} & \textbf{b} & \textbf{b} & -4.7 & -4.2 & \textbf{-2.2} & \textbf{CF}\textbf{\textsubscript{3}}\textbf{OF\textperiodcentered{}\textperiodcentered{}\textperiodcentered{}NH}\textbf{\textsubscript{3}} & -4.6 & -4.1 & \textbf{-4.3} & \multicolumn{2}{>{\centering\arraybackslash}m{18.6pt}|}{\textbf{-4.1}} & -6.9 & -4.0 & \textbf{-4.2} & -2.3 & -2.1 & \textbf{-3.9} \\ \hline
\multicolumn{1}{|>{\centering\arraybackslash}m{65.1pt}|}{\textbf{F}\textbf{\textsubscript{3}}\textbf{NO\textperiodcentered{}\textperiodcentered{}\textperiodcentered{}CH}\textbf{\textsubscript{3}}} & -2.0 & -1.8 & \textbf{-2.1} & \textbf{-2.1} & -4.1 & -1.3 & \textbf{-2.2} & -3.6 & -3.1 & \textbf{-1.3} & \textbf{F}\textbf{\textsubscript{3}}\textbf{NO\textperiodcentered{}\textperiodcentered{}\textperiodcentered{}NH}\textbf{\textsubscript{3}} & \textbf{b} & \textbf{b} & \textbf{b} & \multicolumn{2}{>{\centering\arraybackslash}m{18.6pt}|}{\textbf{b}} & -5.3 & -2.5 & \textbf{-3.3} & -2.3 & -2.0 & \textbf{-3.0} \\ \hline
\multicolumn{1}{|>{\centering\arraybackslash}m{65.1pt}|}{\textbf{SF}\textbf{\textsubscript{5}}\textbf{\textperiodcentered{}\textperiodcentered{}\textperiodcentered{}CH}\textbf{\textsubscript{3}}} & -1.6 & -1.4 & \textbf{-1.8} & \textbf{N/\allowbreak{}A} & -4.0 & -1.1 & \textbf{-2.5} & -3.0 & -2.6 & \textbf{N/\allowbreak{}A} & \textbf{13) SF}\textbf{\textsubscript{5}}\textbf{\textperiodcentered{}\textperiodcentered{}\textperiodcentered{}NH}\textbf{\textsubscript{3}} & -3.0 & -2.6 & \textbf{-2.7} & \multicolumn{2}{>{\centering\arraybackslash}m{18.6pt}|}{\textbf{N/\allowbreak{}A}} & -5.0 & -2.4 & \textbf{-3.6} & -2.0 & -1.8 & \textbf{N/\allowbreak{}A} \\* \hline\noalign{\penalty10000}
\multicolumn{1}{|>{\centering\arraybackslash}m{65.1pt}|}{\textbf{NH}\textbf{\textsubscript{3}}\textbf{F+\textperiodcentered{}\textperiodcentered{}\textperiodcentered{}CH}\textbf{\textsubscript{3}}} & -9.7 & -9.5 & \textbf{-9.8} & \textbf{-10.2} & -10.7 & -8.9 & \textbf{-10.3} & -15.5 & -15.0 & \textbf{-9.5} & \textbf{NH}\textbf{\textsubscript{3}}\textbf{F+\textperiodcentered{}\textperiodcentered{}\textperiodcentered{}NH}\textbf{\textsubscript{3}} & -34.4 & -33.9 & \textbf{-33.9} & \multicolumn{2}{>{\centering\arraybackslash}m{18.6pt}|}{\textbf{-34.6}} & -36.3 & -33.3 & \textbf{-34.8} & -34.3 & -34.0 & \textbf{-34.5} \\* \hline\noalign{\penalty10000}
\multicolumn{1}{|>{\centering\arraybackslash}m{65.1pt}|}{\textbf{NH}\textbf{\textsubscript{2}}\textbf{F}\textbf{\textsubscript{2}}\textbf{\textsuperscript{+}}\textbf{\textperiodcentered{}\textperiodcentered{}\textperiodcentered{}CH}\textbf{\textsubscript{3}}} & -11.2 & -11.0 & \textbf{-11.4} & \textbf{-11.7} & -12.6 & -10.5 & \textbf{-11.8} & -16.5 & -15.9 & \textbf{-11.0} & \textbf{NH}\textbf{\textsubscript{2}}\textbf{F}\textbf{\textsubscript{2}}\textbf{\textsuperscript{+}}\textbf{\textperiodcentered{}\textperiodcentered{}\textperiodcentered{}NH}\textbf{\textsubscript{3}} & -39.3 & -38.7 & \textbf{-38.8} & \multicolumn{2}{>{\centering\arraybackslash}m{18.6pt}|}{\textbf{-39.3}} & -41.7 & -38.3 & \textbf{-39.5} & -39.0 & -38.5 & \textbf{-39.3} \\* \specialrule{0.9pt}{0pt}{0pt}\noalign{\penalty10000}
\end{tabular}}}}
\par\nopagebreak\vspace{1pt}{\setstretch{1}\setlength{\parskip}{1.5pt}\fontsize{9}{10.5}\selectfont\raggedright\noindent \textsuperscript{b}Convergence error during optimization of respective F-bonded complexes. \textsuperscript{N/\allowbreak{}A}W12-F12 calculations failed to converge.\par}
\end{table}

\begin{table}[!htbp]\centering\setlength{\parskip}{0pt}
{\renewcommand\tablename{TABLE}\caption{BEs calculated at different basis sets VnZ-F12 (n = D, T and Q) by using explicitly correlated method DF-RMP2-F12. CBS limit BEs (CBS B.E. in kJ/\allowbreak{}mol) calculated for F-bonded complexes DF\textperiodcentered{}\textperiodcentered{}\textperiodcentered{}CH\textsubscript{3}/NH\textsubscript{3} using DF-RMP2-F12[CBS\{V(T,Q)Z-F12\}]//DF-RMP2-F12/\allowbreak{}VDZ-F12 level. BEs without BSSE correction (\textbf{\ensuremath{\Delta}}\textbf{\textsubscript{1}}) and those with BSSE-correction (\textbf{\ensuremath{\Delta}}\textbf{\textsubscript{1+BSSE}}) are also included.}\label{tab:bes-calculated-at-different}}\vspace{7pt}\nopagebreak
{\fontsize{8.0}{9.2}\fontfamily{ptm}\selectfont\renewcommand{\arraystretch}{1.17}\setlength{\tabcolsep}{1.5pt}
\centerline{\adjustbox{max width=518pt}{\begin{tabular}{|>{\centering\arraybackslash}m{60.6pt}|>{\centering\arraybackslash}m{17.9pt}|>{\centering\arraybackslash}m{27.8pt}|>{\centering\arraybackslash}m{17.3pt}|>{\centering\arraybackslash}m{28.6pt}|>{\centering\arraybackslash}m{19.7pt}|>{\centering\arraybackslash}m{30.3pt}|>{\centering\arraybackslash}m{41.6pt}|>{\centering\arraybackslash}m{60.6pt}|>{\centering\arraybackslash}m{30.3pt}|>{\centering\arraybackslash}m{30.4pt}|>{\centering\arraybackslash}m{24.1pt}|>{\centering\arraybackslash}m{30.4pt}|>{\centering\arraybackslash}m{24.1pt}|>{\centering\arraybackslash}m{20.8pt}|>{\centering\arraybackslash}m{35.5pt}|}
\specialrule{0.9pt}{0pt}{0pt}
\multicolumn{1}{|>{\centering\arraybackslash}m{60.6pt}|}{} & \multicolumn{7}{>{\centering\arraybackslash}p{201.2pt}|}{\textbf{optimization@DF-RMP2-F12/\allowbreak{}VDZ-F12}} & \multicolumn{8}{>{\centering\arraybackslash}p{277.2pt}|}{\textbf{optimization@DF-RMP2-F12/\allowbreak{}VDZ-F12}} \\* \hline\noalign{\penalty10000}
\multicolumn{1}{|>{\centering\arraybackslash}p{60.6pt}|}{\multirow{2}{60.6pt}{\centering \textbf{DF\textperiodcentered{}\textperiodcentered{}\textperiodcentered{}CH}\textbf{\textsubscript{3}}}} & \multicolumn{2}{>{\centering\arraybackslash}p{48.7pt}|}{\textbf{VDZ-F12}} & \multicolumn{2}{>{\centering\arraybackslash}p{48.9pt}|}{\textbf{VTZ-F12}} & \multicolumn{2}{>{\centering\arraybackslash}p{53.0pt}|}{\textbf{VQZ-F12}} & \textbf{CBS B.E.} & \multicolumn{1}{>{\centering\arraybackslash}p{60.6pt}|}{\multirow{2}{60.6pt}{\centering \textbf{DF\textperiodcentered{}\textperiodcentered{}\textperiodcentered{}NH}\textbf{\textsubscript{3}}}} & \multicolumn{2}{>{\centering\arraybackslash}p{63.7pt}|}{\textbf{VDZ-F12}} & \multicolumn{2}{>{\centering\arraybackslash}p{57.5pt}|}{\textbf{VTZ-F12}} & \multicolumn{2}{>{\centering\arraybackslash}p{47.9pt}|}{\textbf{VQZ-F12}} & \textbf{CBS B.E.} \\* \cline{2-8}\cline{10-16}\noalign{\penalty10000}
\multicolumn{1}{|>{\centering\arraybackslash}m{60.6pt}|}{} & \makebox[\linewidth][c]{\rotatebox{90}{\hspace{2pt}\textbf{\ensuremath{\Delta}}\textbf{\textsubscript{1}}\hspace{2pt}}} & \makebox[\linewidth][c]{\rotatebox{90}{\hspace{2pt}\textbf{\ensuremath{\Delta}}\textbf{\textsubscript{1+BSSE}}\hspace{2pt}}} & \makebox[\linewidth][c]{\rotatebox{90}{\hspace{2pt}\textbf{\ensuremath{\Delta}}\textbf{\textsubscript{1}}\hspace{2pt}}} & \makebox[\linewidth][c]{\rotatebox{90}{\hspace{2pt}\textbf{\ensuremath{\Delta}}\textbf{\textsubscript{1+BSSE}}\hspace{2pt}}} & \makebox[\linewidth][c]{\rotatebox{90}{\hspace{2pt}\textbf{\ensuremath{\Delta}}\textbf{\textsubscript{1}}\hspace{2pt}}} & \makebox[\linewidth][c]{\rotatebox{90}{\hspace{2pt}\textbf{\ensuremath{\Delta}}\textbf{\textsubscript{1+BSSE}}\hspace{2pt}}} & \makebox[\linewidth][c]{\rotatebox{90}{\hspace{2pt}\textbf{V(T,Q)Z-F12}\hspace{2pt}}} &  & \makebox[\linewidth][c]{\rotatebox{90}{\hspace{2pt}\textbf{\ensuremath{\Delta}}\textbf{\textsubscript{1}}\hspace{2pt}}} & \makebox[\linewidth][c]{\rotatebox{90}{\hspace{2pt}\textbf{\ensuremath{\Delta}}\textbf{\textsubscript{1+BSSE}}\hspace{2pt}}} & \makebox[\linewidth][c]{\rotatebox{90}{\hspace{2pt}\textbf{\ensuremath{\Delta}}\textbf{\textsubscript{1}}\hspace{2pt}}} & \makebox[\linewidth][c]{\rotatebox{90}{\hspace{2pt}\textbf{\ensuremath{\Delta}}\textbf{\textsubscript{1+BSSE}}\hspace{2pt}}} & \makebox[\linewidth][c]{\rotatebox{90}{\hspace{2pt}\textbf{\ensuremath{\Delta}}\textbf{\textsubscript{1}}\hspace{2pt}}} & \makebox[\linewidth][c]{\rotatebox{90}{\hspace{2pt}\textbf{\ensuremath{\Delta}}\textbf{\textsubscript{1+BSSE}}\hspace{2pt}}} & \makebox[\linewidth][c]{\rotatebox{90}{\hspace{2pt}\textbf{V(T,Q)Z-F12}\hspace{2pt}}} \\* \hline\noalign{\penalty10000}
\multicolumn{1}{|>{\centering\arraybackslash}m{60.6pt}|}{F\textsubscript{2}\textperiodcentered{}\textperiodcentered{}\textperiodcentered{}CH\textsubscript{3}} & -3.2 & -3.0 & -3.2 & -3.1 & -3.2 & -3.2 & -3.4 & F\textsubscript{2}\textperiodcentered{}\textperiodcentered{}\textperiodcentered{}NH\textsubscript{3} & -8.0 & -7.6 & -7.7 & -7.5 & -7.7 & -7.6 & -7.8 \\* \hline\noalign{\penalty10000}
\multicolumn{1}{|>{\centering\arraybackslash}m{60.6pt}|}{SF\textsubscript{6}\textperiodcentered{}\textperiodcentered{}\textperiodcentered{}CH\textsubscript{3}} & -1.6 & -1.5 & -1.6 & -1.6 & -1.6 & -1.6 & -1.8 & SF\textsubscript{6}\textperiodcentered{}\textperiodcentered{}\textperiodcentered{}NH\textsubscript{3} & -1.7 & -1.4 & -1.5 & -1.3 & -1.4 & -1.4 & -1.4 \\* \hline\noalign{\penalty10000}
\multicolumn{1}{|>{\centering\arraybackslash}m{60.6pt}|}{NCF\textperiodcentered{}\textperiodcentered{}\textperiodcentered{}CH\textsubscript{3}} & -1.7 & -1.5 & -1.6 & -1.6 & -1.7 & -1.6 & -1.7 & NCF\textperiodcentered{}\textperiodcentered{}\textperiodcentered{}NH\textsubscript{3} & -5.5 & -5.0 & -5.2 & -5.0 & -5.1 & -5.1 & -5.1 \\* \hline\noalign{\penalty10000}
\multicolumn{1}{|>{\centering\arraybackslash}m{60.6pt}|}{CNF\textperiodcentered{}\textperiodcentered{}\textperiodcentered{}CH\textsubscript{3}} & -2.2 & -2.1 & -2.2 & -2.1 & -2.2 & -2.2 & -2.4 & CNF\textperiodcentered{}\textperiodcentered{}\textperiodcentered{}NH\textsubscript{3} & -7.2 & -6.7 & -6.9 & -6.7 & -6.8 & -6.8 & -6.9 \\ \hline
\multicolumn{1}{|>{\centering\arraybackslash}m{60.6pt}|}{NF\textsubscript{3}\textperiodcentered{}\textperiodcentered{}\textperiodcentered{}CH\textsubscript{3}} & -1.6 & -1.4 & -1.6 & -1.5 & -1.6 & -1.5 & -1.7 & NF\textsubscript{3}\textperiodcentered{}\textperiodcentered{}\textperiodcentered{}NH\textsubscript{3} & -2.1 & -1.7 & -1.9 & -1.7 & -1.8 & -1.8 & -1.9 \\ \hline
\multicolumn{1}{|>{\centering\arraybackslash}m{60.6pt}|}{ClF\textsubscript{3}\textperiodcentered{}\textperiodcentered{}\textperiodcentered{}CH\textsubscript{3}} & -2.3 & -2.1 & -2.3 & -2.2 & -2.3 & -2.2 & -2.4 & ClF\textsubscript{3}\textperiodcentered{}\textperiodcentered{}\textperiodcentered{}NH\textsubscript{3} & -4.5 & -4.1 & -4.3 & -4.1 & -4.3 & -4.2 & -4.5 \\ \hline
\multicolumn{1}{|>{\centering\arraybackslash}m{60.6pt}|}{OF\textsubscript{2}\textperiodcentered{}\textperiodcentered{}\textperiodcentered{}CH\textsubscript{3}} & -1.8 & -1.7 & -1.8 & -1.7 & -1.8 & -1.8 & -2.0 & OF\textsubscript{2}\textperiodcentered{}\textperiodcentered{}\textperiodcentered{}NH\textsubscript{3} & -3.5 & -3.1 & -3.2 & -3.1 & -3.2 & -3.1 & -3.4 \\ \hline
\multicolumn{1}{|>{\centering\arraybackslash}m{60.6pt}|}{NF\textsubscript{5}\textperiodcentered{}\textperiodcentered{}\textperiodcentered{}CH\textsubscript{3}} & -2.9 & -2.7 & -2.9 & -2.8 & -2.9 & -2.9 & -3.1 & NF\textsubscript{5}\textperiodcentered{}\textperiodcentered{}\textperiodcentered{}NH\textsubscript{3} & -7.6 & -7.1 & -7.3 & -7.1 & -7.3 & -7.2 & -7.3 \\ \hline
\multicolumn{1}{|>{\centering\arraybackslash}m{60.6pt}|}{FCCF\textperiodcentered{}\textperiodcentered{}\textperiodcentered{}CH\textsubscript{3}} & -1.5 & -1.3 & -1.4 & -1.4 & -1.4 & -1.4 & -1.5 & FCCF\textperiodcentered{}\textperiodcentered{}\textperiodcentered{}NH\textsubscript{3} & -1.0 & -0.6 & -0.7 & -0.6 & -0.7 & -0.6 & -0.6 \\ \hline
\multicolumn{1}{|>{\centering\arraybackslash}m{60.6pt}|}{CF\textsubscript{3}CCF\textperiodcentered{}\textperiodcentered{}\textperiodcentered{}CH\textsubscript{3}} & -1.4 & -1.3 & -1.4 & -1.3 & -1.4 & -1.4 & -1.5 & CF\textsubscript{3}CCF\textperiodcentered{}\textperiodcentered{}\textperiodcentered{}NH\textsubscript{3} & -4.1 & -3.6 & -3.7 & -3.6 & -3.7 & -3.6 & -3.6 \\ \hline
\multicolumn{1}{|>{\centering\arraybackslash}m{60.6pt}|}{CF\textsubscript{3}OF\textperiodcentered{}\textperiodcentered{}\textperiodcentered{}CH\textsubscript{3}} & -2.3 & -2.1 & -2.3 & -2.2 & -2.3 & -2.3 & -2.5 & CF\textsubscript{3}OF\textperiodcentered{}\textperiodcentered{}\textperiodcentered{}NH\textsubscript{3} & -4.6 & -4.1 & -4.3 & -4.1 & -4.2 & -4.2 & -4.3 \\ \hline
\multicolumn{1}{|>{\centering\arraybackslash}m{60.6pt}|}{F\textsubscript{3}NO\textperiodcentered{}\textperiodcentered{}\textperiodcentered{}CH\textsubscript{3}} & -2.0 & -1.8 & -2.0 & -1.9 & -2.0 & -1.9 & -2.1 & F\textsubscript{3}NO\textperiodcentered{}\textperiodcentered{}\textperiodcentered{}NH\textsubscript{3} & b & b & b & b & b & b & b \\ \hline
\multicolumn{1}{|>{\centering\arraybackslash}m{60.6pt}|}{SF\textsubscript{5}\textperiodcentered{}\textperiodcentered{}\textperiodcentered{}CH\textsubscript{3}} & -1.6 & -1.4 & -1.6 & -1.5 & -1.6 & -1.5 & -1.8 & SF\textsubscript{5}\textperiodcentered{}\textperiodcentered{}\textperiodcentered{}NH\textsubscript{3} & -3.0 & -2.6 & -2.7 & -2.5 & -2.6 & -2.6 & -2.7 \\* \hline\noalign{\penalty10000}
\multicolumn{1}{|>{\centering\arraybackslash}m{60.6pt}|}{NH\textsubscript{3}F+\textperiodcentered{}\textperiodcentered{}\textperiodcentered{}CH\textsubscript{3}} & -9.7 & -9.5 & -9.7 & -9.6 & -9.7 & -9.7 & -9.8 & NH\textsubscript{3}F+\textperiodcentered{}\textperiodcentered{}\textperiodcentered{}NH\textsubscript{3} & -34.4 & -33.9 & -34.0 & -33.8 & -34.0 & -33.9 & -33.9 \\* \hline\noalign{\penalty10000}
\multicolumn{1}{|>{\centering\arraybackslash}m{60.6pt}|}{NH\textsubscript{2}F\textsubscript{2}\textsuperscript{+}\textperiodcentered{}\textperiodcentered{}\textperiodcentered{}CH\textsubscript{3}} & -11.2 & -11.0 & -11.2 & -11.1 & -11.2 & -11.2 & -11.4 & NH\textsubscript{2}F\textsubscript{2}\textsuperscript{+}\textperiodcentered{}\textperiodcentered{}\textperiodcentered{}NH\textsubscript{3} & -39.3 & -38.7 & -38.9 & -38.7 & -38.8 & -38.7 & -38.8 \\* \specialrule{0.9pt}{0pt}{0pt}\noalign{\penalty10000}
\end{tabular}}}}
\par\nopagebreak\vspace{1pt}{\setstretch{1}\setlength{\parskip}{1.5pt}\fontsize{9}{10.5}\selectfont\raggedright\noindent \textsuperscript{b}Convergence error during optimization of respective F-bonded complexes.\par}
\end{table}

\begin{table}[!htbp]\centering\setlength{\parskip}{0pt}
\caption{Shifts in DF bond distances (\ensuremath{\Delta}R(DF) involved in F-bonded complexes DF\textperiodcentered{}\textperiodcentered{}\textperiodcentered{}CH\textsubscript{3}/NH\textsubscript{3} obtained from DF-RMP2-F12/\allowbreak{}VDZ-F12, MP2/\allowbreak{}aug-cc-pVDZ and M06-2X/\allowbreak{}aug-cc-pVTZ level of theories. Units for all distances are Å.}\label{tab:shifts-in-df-bond}\vspace{7pt}\nopagebreak
{\fontsize{7.0}{8.0}\fontfamily{ptm}\selectfont\renewcommand{\arraystretch}{1.17}\setlength{\tabcolsep}{1.5pt}
\centerline{\adjustbox{max width=518pt}{\begin{tabular}{|>{\centering\arraybackslash}m{61.3pt}|>{\centering\arraybackslash}m{56.8pt}|>{\centering\arraybackslash}m{56.8pt}|>{\centering\arraybackslash}m{43.5pt}|>{\centering\arraybackslash}m{70.2pt}|>{\centering\arraybackslash}m{76.2pt}|>{\centering\arraybackslash}m{69.7pt}|>{\centering\arraybackslash}m{89.5pt}|}
\specialrule{0.9pt}{0pt}{0pt}
\multicolumn{4}{|>{\centering\arraybackslash}m{227.5pt}|}{\textbf{DF\textperiodcentered{}\textperiodcentered{}\textperiodcentered{}CH}\textbf{\textsubscript{3}}} & \multicolumn{4}{>{\centering\arraybackslash}m{314.5pt}|}{\textbf{DF\textperiodcentered{}\textperiodcentered{}\textperiodcentered{}NH}\textbf{\textsubscript{3}}} \\* \hline\noalign{\penalty10000}
\multicolumn{1}{|>{\centering\arraybackslash}m{61.3pt}|}{\multirow{2}{61.3pt}{\centering \textbf{DF\textperiodcentered{}\textperiodcentered{}\textperiodcentered{}CH}\textbf{\textsubscript{3}}}} & \textbf{DF-RMP2-F12/\allowbreak{}VDZ-F12} & \textbf{MP2/\allowbreak{}AVDZ} & \textbf{M06-2X/\allowbreak{}AVTZ} & \multirow{2}{70.2pt}{\centering \textbf{DF\textperiodcentered{}\textperiodcentered{}\textperiodcentered{}NH}\textbf{\textsubscript{3}}} & \textbf{DF-RMP2-F12/\allowbreak{}VDZ-F12} & \textbf{MP2/\allowbreak{}AVDZ} & \textbf{M06-2X/\allowbreak{}AVTZ} \\* \cline{2-4}\cline{6-8}\noalign{\penalty10000}
\multicolumn{1}{|>{\centering\arraybackslash}m{61.3pt}|}{} & \textbf{\ensuremath{\Delta}R} & \textbf{\ensuremath{\Delta}R} & \textbf{\ensuremath{\Delta}R} &  & \textbf{\ensuremath{\Delta}R} & \textbf{\ensuremath{\Delta}R} & \textbf{\ensuremath{\Delta}R} \\* \hline\noalign{\penalty10000}
\multicolumn{1}{|>{\centering\arraybackslash}m{61.3pt}|}{\textbf{1) F}\textbf{\textsubscript{2}}\textbf{\textperiodcentered{}\textperiodcentered{}\textperiodcentered{}CH}\textbf{\textsubscript{3}}} & 0.0066 & 0.0081 & 0.0149 & \textbf{1) F}\textbf{\textsubscript{2}}\textbf{\textperiodcentered{}\textperiodcentered{}\textperiodcentered{}NH}\textbf{\textsubscript{3}} & 0.0141 & 0.0281 & 0.0125 \\* \hline\noalign{\penalty10000}
\multicolumn{1}{|>{\centering\arraybackslash}m{61.3pt}|}{\textbf{2) SF}\textbf{\textsubscript{6}}\textbf{\textperiodcentered{}\textperiodcentered{}\textperiodcentered{}CH}\textbf{\textsubscript{3}}} & 0.0006 & -0.0019 & -0.0006 & \textbf{2) SF}\textbf{\textsubscript{6}}\textbf{\textperiodcentered{}\textperiodcentered{}\textperiodcentered{}NH}\textbf{\textsubscript{3}} & -0.0076 & -0.0082 & -0.0064 \\* \hline\noalign{\penalty10000}
\multicolumn{1}{|>{\centering\arraybackslash}m{61.3pt}|}{\textbf{3) NCF\textperiodcentered{}\textperiodcentered{}\textperiodcentered{}CH}\textbf{\textsubscript{3}}} & -0.0004 & -0.0002 & 0.0004 & \textbf{3) NCF\textperiodcentered{}\textperiodcentered{}\textperiodcentered{}NH}\textbf{\textsubscript{3}} & -0.0025 & -0.0023 & -0.0026 \\* \hline\noalign{\penalty10000}
\multicolumn{1}{|>{\centering\arraybackslash}m{61.3pt}|}{\textbf{4) CNF\textperiodcentered{}\textperiodcentered{}\textperiodcentered{}CH}\textbf{\textsubscript{3}}} & 0.0004 & 0.0010 & 0.0019 & \textbf{4) CNF\textperiodcentered{}\textperiodcentered{}\textperiodcentered{}NH}\textbf{\textsubscript{3}} & -0.0007 & 0.0004 & -0.0001 \\ \hline
\multicolumn{1}{|>{\centering\arraybackslash}m{61.3pt}|}{\textbf{5) NF}\textbf{\textsubscript{3}}\textbf{\textperiodcentered{}\textperiodcentered{}\textperiodcentered{}CH}\textbf{\textsubscript{3}}} & 0.0009 & 0.0011 & -0.0007 & \textbf{5) NF}\textbf{\textsubscript{3}}\textbf{\textperiodcentered{}\textperiodcentered{}\textperiodcentered{}NH}\textbf{\textsubscript{3}} & -0.0074 & -0.0006 & -0.0066 \\ \hline
\multicolumn{1}{|>{\centering\arraybackslash}m{61.3pt}|}{\textbf{6) ClF}\textbf{\textsubscript{3}}\textbf{\textperiodcentered{}\textperiodcentered{}\textperiodcentered{}CH}\textbf{\textsubscript{3}}} & 0.0001 & 0.0024 & 0.0070 & \textbf{6) ClF}\textbf{\textsubscript{3}}\textbf{\textperiodcentered{}\textperiodcentered{}\textperiodcentered{}NH}\textbf{\textsubscript{3}} & -0.0044 & -0.0002 & -0.0023 \\ \hline
\multicolumn{1}{|>{\centering\arraybackslash}m{61.3pt}|}{\textbf{7) OF}\textbf{\textsubscript{2}}\textbf{\textperiodcentered{}\textperiodcentered{}\textperiodcentered{}CH}\textbf{\textsubscript{3}}} & 0.0011 & 0.0015 & 0.0031 & \textbf{7) OF}\textbf{\textsubscript{2}}\textbf{\textperiodcentered{}\textperiodcentered{}\textperiodcentered{}NH}\textbf{\textsubscript{3}} & 0.0003 & 0.0015 & -0.0004 \\ \hline
\multicolumn{1}{|>{\centering\arraybackslash}m{61.3pt}|}{\textbf{8) NF}\textbf{\textsubscript{5}}\textbf{\textperiodcentered{}\textperiodcentered{}\textperiodcentered{}CH}\textbf{\textsubscript{3}}} & 0.0000 & -0.0011 & 0.0030 & \textbf{8) NF}\textbf{\textsubscript{5}}\textbf{\textperiodcentered{}\textperiodcentered{}\textperiodcentered{}NH}\textbf{\textsubscript{3}} & -0.0141 & -0.0156 & -0.0074 \\ \hline
\multicolumn{1}{|>{\centering\arraybackslash}m{61.3pt}|}{\textbf{9) FCCF\textperiodcentered{}\textperiodcentered{}\textperiodcentered{}CH}\textbf{\textsubscript{3}}} & 0.0008 & -0.0001 & 0.0001 & \textbf{9) FCCF\textperiodcentered{}\textperiodcentered{}\textperiodcentered{}NH}\textbf{\textsubscript{3}} & -0.0019 & -0.0023 & -0.0020 \\ \hline
\multicolumn{1}{|>{\centering\arraybackslash}m{61.3pt}|}{\textbf{10) CF}\textbf{\textsubscript{3}}\textbf{CCF\textperiodcentered{}\textperiodcentered{}\textperiodcentered{}CH}\textbf{\textsubscript{3}}} & -0.0005 & -0.0008 & 0.0000 & \textbf{10) CF}\textbf{\textsubscript{3}}\textbf{CCF\textperiodcentered{}\textperiodcentered{}\textperiodcentered{}NH}\textbf{\textsubscript{3}} & -0.0028 & -0.0033 & -0.0027 \\ \hline
\multicolumn{1}{|>{\centering\arraybackslash}m{61.3pt}|}{\textbf{11) CF}\textbf{\textsubscript{3}}\textbf{OF\textperiodcentered{}\textperiodcentered{}\textperiodcentered{}CH}\textbf{\textsubscript{3}}} & 0.0026 & b & 0.0044 & \textbf{11) CF}\textbf{\textsubscript{3}}\textbf{OF\textperiodcentered{}\textperiodcentered{}\textperiodcentered{}NH}\textbf{\textsubscript{3}} & 0.0049 & 0.0075 & 0.0018 \\ \hline
\multicolumn{1}{|>{\centering\arraybackslash}m{61.3pt}|}{\textbf{12) F}\textbf{\textsubscript{3}}\textbf{NO\textperiodcentered{}\textperiodcentered{}\textperiodcentered{}CH}\textbf{\textsubscript{3}}} & 0.0006 & 0.0028 & 0.0002 & \textbf{12) F}\textbf{\textsubscript{3}}\textbf{NO\textperiodcentered{}\textperiodcentered{}\textperiodcentered{}NH}\textbf{\textsubscript{3}} & b & -0.0053 & -0.0125 \\ \hline
\multicolumn{1}{|>{\centering\arraybackslash}m{61.3pt}|}{\textbf{13) SF}\textbf{\textsubscript{5}}\textbf{\textperiodcentered{}\textperiodcentered{}\textperiodcentered{}CH}\textbf{\textsubscript{3}}} & 0.0005 & 0.0002 & -0.0010 & \textbf{13) SF}\textbf{\textsubscript{5}}\textbf{\textperiodcentered{}\textperiodcentered{}\textperiodcentered{}NH}\textbf{\textsubscript{3}} & -0.0075 & -0.0080 & -0.0063 \\* \hline\noalign{\penalty10000}
\multicolumn{1}{|>{\centering\arraybackslash}m{61.3pt}|}{\textbf{14) NH}\textbf{\textsubscript{3}}\textbf{F+\textperiodcentered{}\textperiodcentered{}\textperiodcentered{}CH}\textbf{\textsubscript{3}}} & 0.0046 & 0.0051 & 0.0192 & \textbf{14) NH}\textbf{\textsubscript{3}}\textbf{F+\textperiodcentered{}\textperiodcentered{}\textperiodcentered{}NH}\textbf{\textsubscript{3}} & 0.0082 & 0.0116 & 0.0118 \\* \hline\noalign{\penalty10000}
\multicolumn{1}{|>{\centering\arraybackslash}m{61.3pt}|}{\textbf{15) NH}\textbf{\textsubscript{2}}\textbf{F}\textbf{\textsubscript{2}}\textbf{\textsuperscript{+}}\textbf{\textperiodcentered{}\textperiodcentered{}\textperiodcentered{}CH}\textbf{\textsubscript{3}}} & 0.0031 & 0.0038 & 0.0161 & \textbf{15) NH}\textbf{\textsubscript{2}}\textbf{F}\textbf{\textsubscript{2}}\textbf{\textsuperscript{+}}\textbf{\textperiodcentered{}\textperiodcentered{}\textperiodcentered{}NH}\textbf{\textsubscript{3}} & 0.0039 & 0.0073 & 0.0083 \\* \specialrule{0.9pt}{0pt}{0pt}\noalign{\penalty10000}
\end{tabular}}}}
\par\nopagebreak\vspace{1pt}{\setstretch{1}\setlength{\parskip}{1.5pt}\fontsize{9}{10.5}\selectfont\raggedright\noindent \textsuperscript{b}Convergence error during optimization of respective F-bonded complexes.\par}
\end{table}

\begin{table}[!htbp]\centering\setlength{\parskip}{0pt}
\caption{Deviation (\ensuremath{\delta}R, in Å) and percentage deviation (\%\ensuremath{\delta}R) on the F-bond distances for selected F-bonded complexes DF\textperiodcentered{}\textperiodcentered{}\textperiodcentered{}CH\textsubscript{3}/NH\textsubscript{3} (DF = F\textsubscript{2}, NCF, CNF, NH\textsubscript{3}F\textsuperscript{+}, and NH\textsubscript{2}F\textsubscript{2}\textsuperscript{+}), calculated at several \emph{ab initio} levels of theory relative to the corresponding results obtained at CCSD(T)-F12b/\allowbreak{}VDZ-F12 level.}\label{tab:deviation-r-in-and}\vspace{7pt}\nopagebreak
{\fontsize{8.0}{9.2}\fontfamily{ptm}\selectfont\renewcommand{\arraystretch}{1.12}\setlength{\tabcolsep}{1.5pt}
\centerline{\adjustbox{max width=518pt}{\begin{tabular}{|>{\centering\arraybackslash}m{126.5pt}|>{\centering\arraybackslash}m{21.9pt}|>{\centering\arraybackslash}m{24.9pt}|>{\centering\arraybackslash}m{27.8pt}|>{\centering\arraybackslash}m{28.3pt}|>{\centering\arraybackslash}m{26.0pt}|>{\centering\arraybackslash}m{26.7pt}|>{\centering\arraybackslash}m{31.3pt}|>{\centering\arraybackslash}m{31.7pt}|>{\centering\arraybackslash}m{28.7pt}|>{\centering\arraybackslash}m{64.6pt}|}
\specialrule{0.9pt}{0pt}{0pt}
\multicolumn{11}{|>{\centering\arraybackslash}m{468.4pt}|}{\textbf{DF\textperiodcentered{}\textperiodcentered{}\textperiodcentered{}CH}\textbf{\textsubscript{3}}} \\* \hline\noalign{\penalty10000}
\multicolumn{1}{|>{\centering\arraybackslash}m{126.5pt}|}{\multirow{2}{126.5pt}{\centering \textbf{Methods}}} & \multicolumn{2}{>{\centering\arraybackslash}m{49.8pt}|}{F\textsubscript{2}\textperiodcentered{}\textperiodcentered{}\textperiodcentered{}CH\textsubscript{3}} & \multicolumn{2}{>{\centering\arraybackslash}m{59.0pt}|}{NCF\textperiodcentered{}\textperiodcentered{}\textperiodcentered{}CH\textsubscript{3}} & \multicolumn{2}{>{\centering\arraybackslash}m{55.7pt}|}{CNF\textperiodcentered{}\textperiodcentered{}\textperiodcentered{}CH\textsubscript{3}} & \multicolumn{2}{>{\centering\arraybackslash}m{66.0pt}|}{NH\textsubscript{3}F\textsuperscript{+}\textperiodcentered{}\textperiodcentered{}\textperiodcentered{}CH\textsubscript{3}} & \multicolumn{2}{>{\centering\arraybackslash}m{96.3pt}|}{NH\textsubscript{2}F\textsubscript{2}\textsuperscript{+}\textperiodcentered{}\textperiodcentered{}\textperiodcentered{}CH\textsubscript{3}} \\* \cline{2-11}\noalign{\penalty10000}
\multicolumn{1}{|>{\centering\arraybackslash}m{126.5pt}|}{} & \%\ensuremath{\delta}\textsubscript{R} & \ensuremath{\delta}\textsubscript{R} & \%\ensuremath{\delta}\textsubscript{R} & \ensuremath{\delta}\textsubscript{R} & \%\ensuremath{\delta}\textsubscript{R} & \ensuremath{\delta}\textsubscript{R} & \%\ensuremath{\delta}\textsubscript{R} & \ensuremath{\delta}\textsubscript{R} & \%\ensuremath{\delta}\textsubscript{R} & \ensuremath{\delta}\textsubscript{R} \\* \hline\noalign{\penalty10000}
\multicolumn{1}{|>{\centering\arraybackslash}m{126.5pt}|}{\textbf{ }{\leavevmode\color[HTML]{EE0000}\textbf{1) }}{\leavevmode\color[HTML]{FF0000}\textbf{CCSD(T)-F12b/\allowbreak{}VDZ-F12}}} & 0.0 & 0.0 & 0.0 & 0.0 & 0.0 & 0.0 & 0.0 & 0.0 & 0.0 & 0.0 \\* \hline\noalign{\penalty10000}
\multicolumn{1}{|>{\centering\arraybackslash}m{126.5pt}|}{2) CCSD(T)/\allowbreak{}AVDZ} & -8.2 & -0.234 & -1.7 & -0.058 & -2.9 & -0.095 & -0.5 & -0.015 & -0.6 & -0.017 \\* \hline\noalign{\penalty10000}
\multicolumn{1}{|>{\centering\arraybackslash}m{126.5pt}|}{{\leavevmode\color[HTML]{FF0000}3) CCSD(T)/\allowbreak{}AVTZ}} & \textbf{-0.7} & -0.021 & \textbf{-2.1} & -0.072 & \textbf{-2.6} & -0.083 & \textbf{N/\allowbreak{}A} & N/\allowbreak{}A & \textbf{N/\allowbreak{}A} & N/\allowbreak{}A \\* \hline\noalign{\penalty10000}
\multicolumn{1}{|>{\centering\arraybackslash}m{126.5pt}|}{4) CCSD/\allowbreak{}AVDZ} & 3.1 & 0.088 & 0.4 & 0.013 & -0.6 & -0.019 & 2.5 & 0.072 & 1.9 & 0.055 \\ \hline
\multicolumn{1}{|>{\centering\arraybackslash}m{126.5pt}|}{5) CCSD/\allowbreak{}AVTZ} & \textbf{6.4} & 0.181 & \textbf{0.4} & 0.012 & \textbf{0.1} & 0.004 & \textbf{2.4} & 0.070 & \textbf{2.1} & 0.061 \\ \hline
\multicolumn{1}{|>{\centering\arraybackslash}m{126.5pt}|}{6) DCSD/\allowbreak{}AVDZ} & -3.2 & -0.091 & -0.4 & -0.012 & -1.7 & -0.055 & 1.3 & 0.038 & 1.1 & 0.031 \\ \hline
\multicolumn{1}{|>{\centering\arraybackslash}m{126.5pt}|}{7) DCSD/\allowbreak{}AVTZ} & 1.8 & 0.050 & -0.7 & -0.024 & -1.2 & -0.038 & 2.7 & 0.078 & N/\allowbreak{}A & N/\allowbreak{}A \\ \hline
\multicolumn{1}{|>{\centering\arraybackslash}m{126.5pt}|}{\textbf{8) DF-RMP2-F12/\allowbreak{}VDZ-F12 }} & \textbf{0.7} & \textbf{0.019} & \textbf{0.4} & \textbf{0.013} & \textbf{0.2} & \textbf{0.008} & \textbf{1.7} & \textbf{0.050} & \textbf{1.7} & \textbf{0.048} \\ \hline
\multicolumn{1}{|>{\centering\arraybackslash}m{126.5pt}|}{9) DF-RMP2/\allowbreak{}AVTZ} & -0.3 & -0.009 & -1.1 & -0.037 & -2.1 & -0.067 & 2.5 & 0.073 & 2.0 & 0.059 \\ \hline
\multicolumn{1}{|>{\centering\arraybackslash}m{126.5pt}|}{10) MBPT(2)/\allowbreak{}VTZ} & 4.5 & 0.129 & 1.6 & 0.055 & 1.5 & 0.048 & 2.5 & 0.073 & 2.1 & 0.060 \\ \hline
\multicolumn{1}{|>{\centering\arraybackslash}m{126.5pt}|}{11) SCS-MP2/\allowbreak{}AVTZ} & -0.4 & -0.012 & -1.1 & -0.038 & -2.1 & -0.068 & 2.5 & 0.073 & 2.0 & 0.058 \\ \hline
\multicolumn{1}{|>{\centering\arraybackslash}m{126.5pt}|}{12) MP2/\allowbreak{}AVDZ} & 0.5 & 0.015 & -0.8 & -0.027 & -2.1 & -0.069 & 1.5 & 0.043 & 0.9 & 0.026 \\ \hline
\multicolumn{1}{|>{\centering\arraybackslash}m{126.5pt}|}{13) MP2/\allowbreak{}AVTZ} & 3.8 & 0.107 & -0.9 & -0.032 & -1.6 & -0.053 & 1.5 & 0.044 & 1.1 & 0.033 \\ \hline
\multicolumn{1}{|>{\centering\arraybackslash}m{126.5pt}|}{14) MP2/\allowbreak{}AVQZ} & 3.8 & 0.108 & -0.1 & -0.004 & -0.4 & -0.013 & 2.1 & 0.060 & 1.8 & 0.052 \\ \hline
\multicolumn{1}{|>{\centering\arraybackslash}m{126.5pt}|}{15) MP2/\allowbreak{}AV5Z} & 3.8 & 0.109 & -0.1 & -0.004 & -0.4 & -0.012 & 2.1 & 0.062 & 1.8 & 0.053 \\ \specialrule{0.9pt}{0pt}{0pt}
\end{tabular}}}
\par\nointerlineskip
\centerline{\adjustbox{max width=518pt}{\begin{tabular}{|>{\centering\arraybackslash}m{126.5pt}|>{\centering\arraybackslash}m{21.9pt}|>{\centering\arraybackslash}m{24.9pt}|>{\centering\arraybackslash}m{27.8pt}|>{\centering\arraybackslash}m{28.3pt}|>{\centering\arraybackslash}m{26.0pt}|>{\centering\arraybackslash}m{26.7pt}|>{\centering\arraybackslash}m{31.3pt}|>{\centering\arraybackslash}m{31.7pt}|>{\centering\arraybackslash}m{28.7pt}|>{\centering\arraybackslash}m{64.6pt}|}
\multicolumn{11}{|>{\centering\arraybackslash}m{468.4pt}|}{\textbf{DF\textperiodcentered{}\textperiodcentered{}\textperiodcentered{}NH}\textbf{\textsubscript{3}}} \\* \hline\noalign{\penalty10000}
\multicolumn{1}{|>{\centering\arraybackslash}m{126.5pt}|}{} & \multicolumn{2}{>{\centering\arraybackslash}m{49.8pt}|}{\textbf{F}\textbf{\textsubscript{2}}\textbf{\textperiodcentered{}\textperiodcentered{}\textperiodcentered{}NH}\textbf{\textsubscript{3}}} & \multicolumn{2}{>{\centering\arraybackslash}m{59.0pt}|}{\textbf{NCF\textperiodcentered{}\textperiodcentered{}\textperiodcentered{}NH}\textbf{\textsubscript{3}}} & \multicolumn{2}{>{\centering\arraybackslash}m{55.7pt}|}{\textbf{CNF\textperiodcentered{}\textperiodcentered{}\textperiodcentered{}NH}\textbf{\textsubscript{3}}} & \multicolumn{2}{>{\centering\arraybackslash}m{66.0pt}|}{\textbf{NH}\textbf{\textsubscript{3}}\textbf{F}\textbf{\textsuperscript{+}}\textbf{\textperiodcentered{}\textperiodcentered{}\textperiodcentered{}NH}\textbf{\textsubscript{3}}} & \multicolumn{2}{>{\centering\arraybackslash}m{96.3pt}|}{\textbf{NH}\textbf{\textsubscript{2}}\textbf{F}\textbf{\textsubscript{2}}\textbf{\textsuperscript{+}}\textbf{\textperiodcentered{}\textperiodcentered{}\textperiodcentered{}NH}\textbf{\textsubscript{3}}} \\* \hline\noalign{\penalty10000}
\multicolumn{1}{|>{\centering\arraybackslash}m{126.5pt}|}{} & \%\ensuremath{\delta}\textsubscript{R} & \ensuremath{\delta}\textsubscript{R} & \%\ensuremath{\delta}\textsubscript{R} & \ensuremath{\delta}\textsubscript{R} & \%\ensuremath{\delta}\textsubscript{R} & \ensuremath{\delta}\textsubscript{R} & \%\ensuremath{\delta}\textsubscript{R} & \ensuremath{\delta}\textsubscript{R} & \%\ensuremath{\delta}\textsubscript{R} & \ensuremath{\delta}\textsubscript{R} \\* \hline\noalign{\penalty10000}
\multicolumn{1}{|>{\centering\arraybackslash}m{126.5pt}|}{1) {\leavevmode\color[HTML]{FF0000}\textbf{CCSD(T)-F12b/\allowbreak{}VDZ-F12}}} & 0.0 & 0.0 & 0.0 & 0.0 & 0.0 & 0.0 & 0.0 & 0.0 & 0.0 & 0.0 \\* \hline\noalign{\penalty10000}
\multicolumn{1}{|>{\centering\arraybackslash}m{126.5pt}|}{2) CCSD(T)/\allowbreak{}AVDZ} & -6.6 & -0.174 & -1.0 & -0.030 & -1.6 & -0.047 & -1.9 & -0.050 & -1.8 & -0.047 \\* \hline\noalign{\penalty10000}
\multicolumn{1}{|>{\centering\arraybackslash}m{126.5pt}|}{3) CCSD(T)/\allowbreak{}AVTZ} & \textbf{-0.5} & -0.012 & \textbf{-0.9} & -0.028 & \textbf{-1.0} & -0.029 & \textbf{-0.3} & -0.008 & \textbf{-0.4} & -0.010 \\ \hline
\multicolumn{1}{|>{\centering\arraybackslash}m{126.5pt}|}{4) CCSD/\allowbreak{}AVDZ} & 0.9 & 0.023 & 0.9 & 0.027 & 0.4 & 0.011 & 0.6 & 0.015 & 0.2 & 0.006 \\ \hline
\multicolumn{1}{|>{\centering\arraybackslash}m{126.5pt}|}{5) CCSD/\allowbreak{}AVTZ} & \textbf{5.5} & 0.145 & \textbf{0.5} & 0.017 & \textbf{1.1} & 0.032 & \textbf{1.7} & 0.045 & \textbf{1.5} & 0.039 \\ \hline
\multicolumn{1}{|>{\centering\arraybackslash}m{126.5pt}|}{6) DCSD/\allowbreak{}AVDZ} & -4.0 & -0.105 & 0.4 & 0.012 & -0.6 & -0.017 & -1.1 & -0.028 & -1.1 & -0.029 \\ \hline
\multicolumn{1}{|>{\centering\arraybackslash}m{126.5pt}|}{7) DCSD/\allowbreak{}AVTZ} & 1.7 & 0.045 & 0.3 & 0.009 & 0.0 & 0.000 & 0.4 & 0.011 & -0.2 & -0.005 \\ \hline
\multicolumn{1}{|>{\centering\arraybackslash}m{126.5pt}|}{\textbf{8) DF-RMP2-F12/\allowbreak{}VDZ-F12 }} & \textbf{-0.4} & \textbf{-0.010} & \textbf{0.1} & \textbf{0.003} & \textbf{-0.7} & \textbf{-0.021} & \textbf{0.5} & \textbf{0.014} & \textbf{0.4} & \textbf{0.011} \\ \hline
\multicolumn{1}{|>{\centering\arraybackslash}m{126.5pt}|}{9) DF-RMP2/\allowbreak{}AVTZ} & -1.1 & -0.030 & -0.4 & -0.012 & -0.7 & -0.021 & 0.1 & 0.003 & -0.1 & -0.003 \\ \hline
\multicolumn{1}{|>{\centering\arraybackslash}m{126.5pt}|}{10) MBPT(2)/\allowbreak{}VTZ} & 0.5 & 0.014 & 0.1 & 0.002 & -0.8 & -0.023 & -0.5 & -0.013 & -0.6 & -0.015 \\ \hline
\multicolumn{1}{|>{\centering\arraybackslash}m{126.5pt}|}{11) SCS-MP2/\allowbreak{}AVTZ} & -1.3 & -0.033 & -0.4 & -0.013 & -1.0 & -0.031 & 0.1 & 0.003 & -0.2 & -0.004 \\ \hline
\multicolumn{1}{|>{\centering\arraybackslash}m{126.5pt}|}{12) MP2/\allowbreak{}AVDZ} & -7.0 & -0.184 & -0.5 & -0.016 & -1.9 & -0.057 & -1.4 & -0.037 & -1.3 & -0.035 \\ \hline
\multicolumn{1}{|>{\centering\arraybackslash}m{126.5pt}|}{13) MP2/\allowbreak{}AVTZ} & -1.2 & -0.031 & -0.3 & -0.010 & -0.9 & -0.026 & 0.1 & 0.003 & 0.0 & -0.001 \\* \hline\noalign{\penalty10000}
\multicolumn{1}{|>{\centering\arraybackslash}m{126.5pt}|}{14) MP2/\allowbreak{}AVQZ} & -0.8 & -0.020 & -0.3 & -0.020 & -0.8 & 0.010 & 0.5 & -0.024 & 0.3 & 0.009 \\* \hline\noalign{\penalty10000}
\multicolumn{1}{|>{\centering\arraybackslash}m{126.5pt}|}{15) MP2/\allowbreak{}AV5Z} & -0.7 & -0.019 & -0.3 & -0.008 & -0.8 & -0.023 & 0.5 & 0.013 & 0.4 & 0.011 \\* \specialrule{0.9pt}{0pt}{0pt}\noalign{\penalty10000}
\end{tabular}}}}
\par\nopagebreak\vspace{1pt}{\setstretch{1}\setlength{\parskip}{1.5pt}\fontsize{9}{10.5}\selectfont\raggedright\noindent \textsuperscript{\#}Due to high computational cost, NH\textsubscript{3}F\textsuperscript{+}\textperiodcentered{}\textperiodcentered{}\textperiodcentered{}CH\textsubscript{3} and NH\textsubscript{2}F\textsubscript{2}\textsuperscript{+}\textperiodcentered{}\textperiodcentered{}\textperiodcentered{}CH\textsubscript{3} could not be optimized at CCSD(T)-F12b/\allowbreak{}VTZ-F12; all selected DF\textperiodcentered{}\textperiodcentered{}\textperiodcentered{}CH\textsubscript{3}/NH\textsubscript{3} complexes (DF = F\textsubscript{2}, NCF, CNF, NH\textsubscript{3}F\textsuperscript{+}, and NH\textsubscript{2}F\textsubscript{2}\textsuperscript{+}) were instead optimized at the CCSD(T)-F12b/\allowbreak{}VDZ-F12 level. \textsuperscript{N/\allowbreak{}A}Convergence failure during optimization.\par}
\end{table}

\begin{table}[!htbp]\centering\setlength{\parskip}{0pt}
\caption{Percentage deviation (\%\ensuremath{\delta}\textsubscript{BE}) and deviation (\ensuremath{\delta}\textsubscript{BE}, in kJ/\allowbreak{}mol) on BEs for selected F-bonded complexes DF\textperiodcentered{}\textperiodcentered{}\textperiodcentered{}CH\textsubscript{3}/NH\textsubscript{3} (DF = F\textsubscript{2}, NCF, CNF, NH\textsubscript{3}F\textsuperscript{+}, and NH\textsubscript{2}F\textsubscript{2}\textsuperscript{+}) computed using different \emph{ab initio} methods, relative to reference BEs obtained at CCSD(T)-F12b/\allowbreak{}CBS[V(T,Q)Z-F12]//CCSD(T)-F12b/\allowbreak{}VDZ-F12 (denoted CCSD(T)-F12b/\allowbreak{}CBS in the text).\textsuperscript{\#}}\label{tab:percentage-deviation-be-and}\vspace{7pt}\nopagebreak
{\fontsize{8.0}{9.2}\fontfamily{ptm}\selectfont\renewcommand{\arraystretch}{1.0}\setlength{\tabcolsep}{1.5pt}
\centerline{\adjustbox{max width=518pt}{\begin{tabular}{|>{\centering\arraybackslash}m{126.5pt}|>{\centering\arraybackslash}m{27.4pt}|>{\centering\arraybackslash}m{23.8pt}|>{\centering\arraybackslash}m{27.4pt}|>{\centering\arraybackslash}m{26.8pt}|>{\centering\arraybackslash}m{25.7pt}|>{\centering\arraybackslash}m{25.5pt}|>{\centering\arraybackslash}m{30.7pt}|>{\centering\arraybackslash}m{30.5pt}|>{\centering\arraybackslash}m{28.2pt}|>{\centering\arraybackslash}m{61.6pt}|}
\specialrule{0.9pt}{0pt}{0pt}
\multicolumn{11}{|>{\centering\arraybackslash}m{464.1pt}|}{\textbf{DF\textperiodcentered{}\textperiodcentered{}\textperiodcentered{}CH}\textbf{\textsubscript{3}}} \\* \hline\noalign{\penalty10000}
\multicolumn{1}{|>{\centering\arraybackslash}m{126.5pt}|}{\multirow{2}{126.5pt}{\centering \textbf{Methods}}} & \multicolumn{2}{>{\centering\arraybackslash}m{54.3pt}|}{F\textsubscript{2}\textperiodcentered{}\textperiodcentered{}\textperiodcentered{}CH\textsubscript{3}} & \multicolumn{2}{>{\centering\arraybackslash}m{57.2pt}|}{NCF\textperiodcentered{}\textperiodcentered{}\textperiodcentered{}CH\textsubscript{3}} & \multicolumn{2}{>{\centering\arraybackslash}m{54.2pt}|}{CNF\textperiodcentered{}\textperiodcentered{}\textperiodcentered{}CH\textsubscript{3}} & \multicolumn{2}{>{\centering\arraybackslash}m{64.2pt}|}{NH\textsubscript{3}F\textsuperscript{+}\textperiodcentered{}\textperiodcentered{}\textperiodcentered{}CH\textsubscript{3}} & \multicolumn{2}{>{\centering\arraybackslash}m{92.8pt}|}{NH\textsubscript{2}F\textsubscript{2}\textsuperscript{+}\textperiodcentered{}\textperiodcentered{}\textperiodcentered{}CH\textsubscript{3}} \\* \cline{2-11}\noalign{\penalty10000}
\multicolumn{1}{|>{\centering\arraybackslash}m{126.5pt}|}{} & \%\ensuremath{\delta}\textsubscript{BE} & \ensuremath{\delta}\textsubscript{BE} & \%\ensuremath{\delta}\textsubscript{BE} & \ensuremath{\delta}\textsubscript{BE} & \%\ensuremath{\delta}\textsubscript{BE} & \ensuremath{\delta}\textsubscript{BE} & \%\ensuremath{\delta}\textsubscript{BE} & \ensuremath{\delta}\textsubscript{BE} & \%\ensuremath{\delta}\textsubscript{BE} & \ensuremath{\delta}\textsubscript{BE} \\* \hline\noalign{\penalty10000}
\multicolumn{1}{|>{\centering\arraybackslash}m{126.5pt}|}{1) {\leavevmode\color[HTML]{FF0000}CCSD(T)-F12b/\allowbreak{}VDZ-F12}} & 3.1 & 0.1 & 5.9 & 0.1 & 4.5 & 0.1 & -5.9 & {\leavevmode\color[HTML]{FF0000}-0.6} & -3.4 & -0.4 \\* \hline\noalign{\penalty10000}
\multicolumn{1}{|>{\centering\arraybackslash}m{126.5pt}|}{2) CCSD(T)/\allowbreak{}AVDZ} & -71.9 & {\leavevmode\color[HTML]{FF0000}-2.3} & -47.1 & -0.8 & -59.1 & -1.3 & -11.8 & -1.2 & -13.6 & -1.6 \\* \hline\noalign{\penalty10000}
\multicolumn{1}{|>{\centering\arraybackslash}m{126.5pt}|}{{\leavevmode\color[HTML]{FF0000}3) CCSD(T)/\allowbreak{}AVTZ}} & -21.9 & {\leavevmode\color[HTML]{FF0000}-0.7} & -29.4 & -0.5 & -31.8 & -0.7 & N/\allowbreak{}A & N/\allowbreak{}A & N/\allowbreak{}A & N/\allowbreak{}A \\* \hline\noalign{\penalty10000}
\multicolumn{1}{|>{\centering\arraybackslash}m{126.5pt}|}{4) CCSD/\allowbreak{}AVDZ} & -15.6 & -0.5 & -23.5 & -0.4 & -36.4 & {\leavevmode\color[HTML]{FF0000}-0.8} & 0.0 & 0.0 & -2.5 & -0.3 \\ \hline
\multicolumn{1}{|>{\centering\arraybackslash}m{126.5pt}|}{5) CCSD/\allowbreak{}AVTZ} & 18.8 & {\leavevmode\color[HTML]{FF0000}0.6} & -5.9 & -0.1 & -4.5 & -0.1 & 5.9 & 0.6 & 4.2 & 0.5 \\ \hline
\multicolumn{1}{|>{\centering\arraybackslash}m{126.5pt}|}{6) DCSD/\allowbreak{}AVDZ} & -50.0 & -1.6 & -29.4 & -0.5 & -45.5 & -1.0 & 21.6 & 2.2 & 22.0 & {\leavevmode\color[HTML]{FF0000}2.6} \\ \hline
\multicolumn{1}{|>{\centering\arraybackslash}m{126.5pt}|}{7) DCSD/\allowbreak{}AVTZ} & -6.2 & -0.2 & -11.8 & -0.2 & -18.2 & -0.4 & 15.7 & {\leavevmode\color[HTML]{FF0000}1.6} & N/\allowbreak{}A & N/\allowbreak{}A \\ \hline
\multicolumn{1}{|>{\centering\arraybackslash}m{126.5pt}|}{\textbf{8) DF-RMP2-F12/\allowbreak{}VDZ-F12 }} & 0.0 & 0.0 & 0.0 & 0.0 & 0.0 & 0.0 & 4.9 & 0.5 & 5.1 & {\leavevmode\color[HTML]{FF0000}0.6} \\ \hline
\multicolumn{1}{|>{\centering\arraybackslash}m{126.5pt}|}{\textbf{9) DF-RMP2-F12/\allowbreak{}CBS*}} & -6.2 & -0.2 & 0.0 & 0.0 & -9.1 & -0.2 & 3.9 & {\leavevmode\color[HTML]{FF0000}0.4} & 3.4 & 0.4 \\ \hline
\multicolumn{1}{|>{\centering\arraybackslash}m{126.5pt}|}{\textbf{10) W1-F12}\textbf{\textsuperscript{\$}}} & 3.1 & 0.1 & 0.0 & 0.0 & 0.0 & 0.0 & 0.0 & 0.0 & 0.8 & {\leavevmode\color[HTML]{FF0000}0.1} \\ \hline
\multicolumn{1}{|>{\centering\arraybackslash}m{126.5pt}|}{\textbf{11) CCSD(T)-F12b/\allowbreak{}CBS**}} & 0.0 & 0.0 & 0.0 & 0.0 & 0.0 & 0.0 & 0.0 & 0.0 & 0.8 & {\leavevmode\color[HTML]{FF0000}0.1} \\ \hline
\multicolumn{1}{|>{\centering\arraybackslash}m{126.5pt}|}{12) DF-RMP2/\allowbreak{}AVTZ} & -21.9 & -0.7 & -23.5 & -0.4 & -31.8 & -0.7 & 13.7 & 1.4 & 13.6 & {\leavevmode\color[HTML]{FF0000}1.6} \\ \hline
\multicolumn{1}{|>{\centering\arraybackslash}m{126.5pt}|}{13) MBPT(2)/\allowbreak{}VTZ} & 9.4 & 0.3 & 0.0 & 0.0 & -9.1 & -0.2 & 11.8 & {\leavevmode\color[HTML]{FF0000}1.2} & 10.2 & 1.2 \\ \hline
\multicolumn{1}{|>{\centering\arraybackslash}m{126.5pt}|}{14) SCS-MP2/\allowbreak{}AVTZ} & -21.9 & -0.7 & -23.5 & -0.4 & -31.8 & -0.7 & 13.7 & 1.4 & 13.6 & {\leavevmode\color[HTML]{FF0000}1.6} \\ \hline
\multicolumn{1}{|>{\centering\arraybackslash}m{126.5pt}|}{15) MP2/\allowbreak{}AVDZ} & -43.8 & {\leavevmode\color[HTML]{FF0000}-1.4} & -35.3 & -0.6 & -54.5 & -1.2 & -4.9 & -0.5 & -6.8 & -0.8 \\ \hline
\multicolumn{1}{|>{\centering\arraybackslash}m{126.5pt}|}{16) MP2/\allowbreak{}AVTZ} & -6.2 & -0.2 & -17.6 & -0.3 & -22.7 & {\leavevmode\color[HTML]{FF0000}-0.5} & 0.0 & 0.0 & -0.8 & -0.1 \\ \hline
\multicolumn{1}{|>{\centering\arraybackslash}m{126.5pt}|}{17) MP2/\allowbreak{}AVQZ} & 3.1 & 0.1 & -11.8 & -0.2 & -9.1 & {\leavevmode\color[HTML]{FF0000}-0.2} & 3.9 & 0.4 & 3.4 & 0.4 \\ \hline
\multicolumn{1}{|>{\centering\arraybackslash}m{126.5pt}|}{18) MP2/\allowbreak{}AV5Z} & 6.3 & 0.2 & 0.0 & 0.0 & 0.0 & 0.0 & 7.8 & {\leavevmode\color[HTML]{FF0000}0.8} & 6.8 & 0.8 \\ \specialrule{0.9pt}{0pt}{0pt}
\end{tabular}}}
\par\nointerlineskip
\centerline{\adjustbox{max width=518pt}{\begin{tabular}{|>{\centering\arraybackslash}m{126.5pt}|>{\centering\arraybackslash}m{27.4pt}|>{\centering\arraybackslash}m{23.8pt}|>{\centering\arraybackslash}m{27.4pt}|>{\centering\arraybackslash}m{26.8pt}|>{\centering\arraybackslash}m{25.7pt}|>{\centering\arraybackslash}m{25.5pt}|>{\centering\arraybackslash}m{30.7pt}|>{\centering\arraybackslash}m{30.5pt}|>{\centering\arraybackslash}m{28.2pt}|>{\centering\arraybackslash}m{61.6pt}|}
\multicolumn{11}{|>{\centering\arraybackslash}m{464.1pt}|}{\textbf{DF\textperiodcentered{}\textperiodcentered{}\textperiodcentered{}NH}\textbf{\textsubscript{3}}} \\* \hline\noalign{\penalty10000}
\multicolumn{1}{|>{\centering\arraybackslash}m{126.5pt}|}{} & \multicolumn{2}{>{\centering\arraybackslash}m{54.3pt}|}{\textbf{F}\textbf{\textsubscript{2}}\textbf{\textperiodcentered{}\textperiodcentered{}\textperiodcentered{}NH}\textbf{\textsubscript{3}}} & \multicolumn{2}{>{\centering\arraybackslash}m{57.2pt}|}{\textbf{NCF\textperiodcentered{}\textperiodcentered{}\textperiodcentered{}NH}\textbf{\textsubscript{3}}} & \multicolumn{2}{>{\centering\arraybackslash}m{54.2pt}|}{\textbf{CNF\textperiodcentered{}\textperiodcentered{}\textperiodcentered{}NH}\textbf{\textsubscript{3}}} & \multicolumn{2}{>{\centering\arraybackslash}m{64.2pt}|}{\textbf{NH}\textbf{\textsubscript{3}}\textbf{F}\textbf{\textsuperscript{+}}\textbf{\textperiodcentered{}\textperiodcentered{}\textperiodcentered{}NH}\textbf{\textsubscript{3}}} & \multicolumn{2}{>{\centering\arraybackslash}m{92.8pt}|}{\textbf{NH}\textbf{\textsubscript{2}}\textbf{F}\textbf{\textsubscript{2}}\textbf{\textsuperscript{+}}\textbf{\textperiodcentered{}\textperiodcentered{}\textperiodcentered{}NH}\textbf{\textsubscript{3}}} \\* \hline\noalign{\penalty10000}
\multicolumn{1}{|>{\centering\arraybackslash}m{126.5pt}|}{} & \%\ensuremath{\delta}\textsubscript{BE} & \ensuremath{\delta}\textsubscript{BE} & \%\ensuremath{\delta}\textsubscript{BE} & \ensuremath{\delta}\textsubscript{BE} & \%\ensuremath{\delta}\textsubscript{BE} & \ensuremath{\delta}\textsubscript{BE} & \%\ensuremath{\delta}\textsubscript{BE} & \ensuremath{\delta}\textsubscript{BE} & \%\ensuremath{\delta}\textsubscript{BE} & \ensuremath{\delta}\textsubscript{BE} \\* \hline\noalign{\penalty10000}
\multicolumn{1}{|>{\centering\arraybackslash}m{126.5pt}|}{1) {\leavevmode\color[HTML]{FF0000}CCSD(T)-F12b/\allowbreak{}VDZ-F12}} & -4.0 & -0.3 & -5.8 & -0.3 & -4.7 & -0.3 & -1.4 & {\leavevmode\color[HTML]{FF0000}-0.5} & -1.3 & -0.5 \\* \hline\noalign{\penalty10000}
\multicolumn{1}{|>{\centering\arraybackslash}m{126.5pt}|}{2) CCSD(T)/\allowbreak{}AVDZ} & -57.3 & {\leavevmode\color[HTML]{FF0000}-4.3} & -15.4 & -0.8 & -21.9 & -1.4 & -6.9 & -2.4 & -7.6 & -3.0 \\* \hline\noalign{\penalty10000}
\multicolumn{1}{|>{\centering\arraybackslash}m{126.5pt}|}{{\leavevmode\color[HTML]{FF0000}3) CCSD(T)/\allowbreak{}AVTZ}} & -12.0 & -0.9 & -9.6 & -0.5 & -10.9 & -0.7 & -2.0 & -0.7 & -2.3 & {\leavevmode\color[HTML]{FF0000}-0.9} \\ \hline
\multicolumn{1}{|>{\centering\arraybackslash}m{126.5pt}|}{4) CCSD/\allowbreak{}AVDZ} & -14.7 & {\leavevmode\color[HTML]{FF0000}-1.1} & -3.8 & -0.2 & -10.9 & -0.7 & -1.2 & -0.4 & -2.5 & -1.0 \\ \hline
\multicolumn{1}{|>{\centering\arraybackslash}m{126.5pt}|}{5) CCSD/\allowbreak{}AVTZ} & 17.3 & {\leavevmode\color[HTML]{FF0000}1.3} & 1.9 & 0.1 & 1.6 & 0.1 & 3.5 & 1.2 & 2.5 & 1.0 \\ \hline
\multicolumn{1}{|>{\centering\arraybackslash}m{126.5pt}|}{6) DCSD/\allowbreak{}AVDZ} & -42.7 & {\leavevmode\color[HTML]{FF0000}-3.2} & -5.8 & -0.3 & -15.6 & -1.0 & -4.9 & -1.7 & -5.6 & -2.2 \\ \hline
\multicolumn{1}{|>{\centering\arraybackslash}m{126.5pt}|}{7) DCSD/\allowbreak{}AVTZ} & -1.3 & -0.1 & 0.0 & 0.0 & -4.7 & -0.3 & 0.0 & 0.0 & -1.3 & {\leavevmode\color[HTML]{FF0000}-0.5} \\ \hline
\multicolumn{1}{|>{\centering\arraybackslash}m{126.5pt}|}{\textbf{8) DF-RMP2-F12/\allowbreak{}VDZ-F12 }} & -6.7 & -0.5 & -5.8 & -0.3 & -12.5 & {\leavevmode\color[HTML]{FF0000}-0.8} & 0.6 & 0.2 & 0.0 & 0.0 \\ \hline
\multicolumn{1}{|>{\centering\arraybackslash}m{126.5pt}|}{\textbf{9) DF-RMP2-F12/\allowbreak{}CBS*}} & -4.0 & -0.3 & 1.9 & 0.1 & -7.8 & -0.5 & 2.0 & {\leavevmode\color[HTML]{FF0000}0.7} & 1.3 & 0.5 \\ \hline
\multicolumn{1}{|>{\centering\arraybackslash}m{126.5pt}|}{\textbf{10) W1-F12}\textbf{\textsuperscript{\$}}} & 1.3 & 0.1 & 0.0 & 0.0 & 0.0 & 0.0 & 0.0 & 0.0 & 0.0 & 0.0 \\ \hline
\multicolumn{1}{|>{\centering\arraybackslash}m{126.5pt}|}{\textbf{11) CCSD(T)-F12b/\allowbreak{}CBS**}} & 0.0 & 0.0 & 0.0 & 0.0 & 0.0 & 0.0 & 0.0 & {\leavevmode\color[HTML]{FF0000}0.0} & 0.0 & 0.0 \\ \hline
\multicolumn{1}{|>{\centering\arraybackslash}m{126.5pt}|}{12) DF-RMP2/\allowbreak{}AVTZ} & -13.3 & -1.0 & -7.7 & -0.4 & -17.2 & {\leavevmode\color[HTML]{FF0000}-1.1} & 0.0 & 0.0 & -0.8 & -0.3 \\ \hline
\multicolumn{1}{|>{\centering\arraybackslash}m{126.5pt}|}{13) MBPT(2)/\allowbreak{}VTZ} & -8.0 & -0.6 & -17.3 & -0.9 & -23.4 & {\leavevmode\color[HTML]{FF0000}-1.5} & -2.6 & -0.9 & -3.6 & -1.4 \\ \hline
\multicolumn{1}{|>{\centering\arraybackslash}m{126.5pt}|}{14) SCS-MP2/\allowbreak{}AVTZ} & -13.3 & -1.0 & -7.7 & -0.4 & -17.2 & {\leavevmode\color[HTML]{FF0000}-1.1} & 0.0 & 0.0 & -0.8 & -0.3 \\ \hline
\multicolumn{1}{|>{\centering\arraybackslash}m{126.5pt}|}{15) MP2/\allowbreak{}AVDZ} & -58.7 & {\leavevmode\color[HTML]{FF0000}-4.4} & -15.4 & -0.8 & -31.3 & -2.0 & -4.9 & -1.7 & -6.1 & -2.4 \\ \hline
\multicolumn{1}{|>{\centering\arraybackslash}m{126.5pt}|}{16) MP2/\allowbreak{}AVTZ} & -13.3 & -1.0 & -7.7 & -0.4 & -17.2 & {\leavevmode\color[HTML]{FF0000}-1.1} & 0.0 & 0.0 & -0.8 & -0.3 \\* \hline\noalign{\penalty10000}
\multicolumn{1}{|>{\centering\arraybackslash}m{126.5pt}|}{17) MP2/\allowbreak{}AVQZ} & -9.3 & -0.7 & -3.8 & -0.2 & -12.5 & {\leavevmode\color[HTML]{FF0000}-0.8} & 1.2 & 0.4 & 0.5 & 0.2 \\* \hline\noalign{\penalty10000}
\multicolumn{1}{|>{\centering\arraybackslash}m{126.5pt}|}{18) MP2/\allowbreak{}AV5Z} & -4.0 & -0.3 & 0.0 & 0.0 & -7.8 & -0.5 & 1.7 & {\leavevmode\color[HTML]{FF0000}0.6} & 1.3 & 0.5 \\* \specialrule{0.9pt}{0pt}{0pt}\noalign{\penalty10000}
\end{tabular}}}}
\par\nopagebreak\vspace{1pt}{\setstretch{1}\setlength{\parskip}{1.5pt}\fontsize{9}{10.5}\selectfont\raggedright\noindent \textsuperscript{\#}Due to high computational cost, NH\textsubscript{3}F\textsuperscript{+}\textperiodcentered{}\textperiodcentered{}\textperiodcentered{}CH\textsubscript{3} and NH\textsubscript{2}F\textsubscript{2}\textsuperscript{+}\textperiodcentered{}\textperiodcentered{}\textperiodcentered{}CH\textsubscript{3} could not be optimized at CCSD(T)-F12b/\allowbreak{}VTZ-F12; all selected DF\textperiodcentered{}\textperiodcentered{}\textperiodcentered{}CH\textsubscript{3}/NH\textsubscript{3} complexes (DF = F\textsubscript{2}, NCF, CNF, NH\textsubscript{3}F\textsuperscript{+}, and NH\textsubscript{2}F\textsubscript{2}\textsuperscript{+}) were instead optimized at CCSD(T)-F12b/\allowbreak{}VDZ-F12. \textsuperscript{N/\allowbreak{}A}Convergence failure during optimization. \textsuperscript{\$}Non-relativistic, valence-electron CCSD(T)/\allowbreak{}CBS BEs calculated at CCSD(T)-F12b/\allowbreak{}VDZ-F12 optimized geometries. * CBS limit BEs calculated at DF-RMP2-F12/[CBS\{V(T,Q)Z-F12\}]//DF-RMP2-F12/\allowbreak{}VDZ-F12 level, using Schwenke-type extrapolation. **CBS limit BEs calculated at CCSD(T)-F12b/\allowbreak{}CBS[V(T,Q)Z-F12]//DF-RMP2-F12/\allowbreak{}VDZ-F12 level of theory, using Schwenke-type extrapolation.\par}
\end{table}

\begin{table}[!htbp]\centering\setlength{\parskip}{0pt}
\caption{Percentage deviation (\%\ensuremath{\delta}\textsubscript{BE}) and deviation (\ensuremath{\delta}\textsubscript{BE}, in kJ/\allowbreak{}mol) on BEs for selected F-bonded complexes DF\textperiodcentered{}\textperiodcentered{}\textperiodcentered{}CH\textsubscript{3}/NH\textsubscript{3} (D--F = F\textsubscript{2}, NCF, CNF, NH\textsubscript{3}F\textsuperscript{+}, and NH\textsubscript{2}F\textsubscript{2}\textsuperscript{+}) computed at different ab initio levels, relative to reference BEs obtained at CCSD(T)-F12b/\allowbreak{}VDZ-F12 level.}\label{tab:percentage-deviation-be-and2}\vspace{7pt}\nopagebreak
{\fontsize{8.0}{9.2}\fontfamily{ptm}\selectfont\renewcommand{\arraystretch}{1.12}\setlength{\tabcolsep}{1.5pt}
\centerline{\adjustbox{max width=518pt}{\begin{tabular}{|>{\centering\arraybackslash}m{126.5pt}|>{\centering\arraybackslash}m{28.5pt}|>{\centering\arraybackslash}m{24.8pt}|>{\centering\arraybackslash}m{28.4pt}|>{\centering\arraybackslash}m{28.3pt}|>{\centering\arraybackslash}m{26.6pt}|>{\centering\arraybackslash}m{26.6pt}|>{\centering\arraybackslash}m{32.1pt}|>{\centering\arraybackslash}m{32.0pt}|>{\centering\arraybackslash}m{29.3pt}|>{\centering\arraybackslash}m{66.1pt}|}
\specialrule{0.9pt}{0pt}{0pt}
\multicolumn{11}{|>{\centering\arraybackslash}m{479.1pt}|}{\textbf{DF\textperiodcentered{}\textperiodcentered{}\textperiodcentered{}CH}\textbf{\textsubscript{3}}} \\* \hline\noalign{\penalty10000}
\multicolumn{1}{|>{\centering\arraybackslash}m{126.5pt}|}{\multirow{2}{126.5pt}{\centering \textbf{Methods}}} & \multicolumn{2}{>{\centering\arraybackslash}m{56.2pt}|}{F\textsubscript{2}\textperiodcentered{}\textperiodcentered{}\textperiodcentered{}CH\textsubscript{3}} & \multicolumn{2}{>{\centering\arraybackslash}m{59.8pt}|}{NCF\textperiodcentered{}\textperiodcentered{}\textperiodcentered{}CH\textsubscript{3}} & \multicolumn{2}{>{\centering\arraybackslash}m{56.2pt}|}{CNF\textperiodcentered{}\textperiodcentered{}\textperiodcentered{}CH\textsubscript{3}} & \multicolumn{2}{>{\centering\arraybackslash}m{67.1pt}|}{NH\textsubscript{3}F\textsuperscript{+}\textperiodcentered{}\textperiodcentered{}\textperiodcentered{}CH\textsubscript{3}} & \multicolumn{2}{>{\centering\arraybackslash}m{98.5pt}|}{NH\textsubscript{2}F\textsubscript{2}\textsuperscript{+}\textperiodcentered{}\textperiodcentered{}\textperiodcentered{}CH\textsubscript{3}} \\* \cline{2-11}\noalign{\penalty10000}
\multicolumn{1}{|>{\centering\arraybackslash}m{126.5pt}|}{} & \%\ensuremath{\delta}\textsubscript{BE} & \ensuremath{\delta}\textsubscript{BE} & \%\ensuremath{\delta}\textsubscript{BE} & \ensuremath{\delta}\textsubscript{BE} & \%\ensuremath{\delta}\textsubscript{BE} & \ensuremath{\delta}\textsubscript{BE} & \%\ensuremath{\delta}\textsubscript{BE} & \ensuremath{\delta}\textsubscript{BE} & \%\ensuremath{\delta}\textsubscript{BE} & \ensuremath{\delta}\textsubscript{BE} \\* \hline\noalign{\penalty10000}
\multicolumn{1}{|>{\centering\arraybackslash}m{126.5pt}|}{1) {\leavevmode\color[HTML]{FF0000}CCSD(T)-F12b/\allowbreak{}VDZ-F12}} & 0.0 & 0.0 & 0.0 & 0.0 & 0.0 & 0.0 & 0.0 & 0.0 & 0.0 & 0.0 \\* \hline\noalign{\penalty10000}
\multicolumn{1}{|>{\centering\arraybackslash}m{126.5pt}|}{2) CCSD(T)/\allowbreak{}AVDZ} & -77.4 & -2.4 & -56.3 & -0.9 & -66.7 & -1.4 & -5.6 & -0.6 & -9.8 & -1.2 \\* \hline\noalign{\penalty10000}
\multicolumn{1}{|>{\centering\arraybackslash}m{126.5pt}|}{3) CCSD(T)/\allowbreak{}AVTZ} & -25.8 & -0.8 & -37.5 & -0.6 & -38.1 & -0.8 & N/\allowbreak{}A & N/\allowbreak{}A & N/\allowbreak{}A & N/\allowbreak{}A \\* \hline\noalign{\penalty10000}
\multicolumn{1}{|>{\centering\arraybackslash}m{126.5pt}|}{4) CCSD/\allowbreak{}AVDZ} & -19.4 & -0.6 & -31.3 & -0.5 & -42.9 & -0.9 & 5.6 & 0.6 & 0.8 & 0.1 \\ \hline
\multicolumn{1}{|>{\centering\arraybackslash}m{126.5pt}|}{5) CCSD/\allowbreak{}AVTZ} & 16.1 & 0.5 & -12.5 & -0.2 & -9.5 & -0.2 & 11.1 & 1.2 & 7.4 & 0.9 \\ \hline
\multicolumn{1}{|>{\centering\arraybackslash}m{126.5pt}|}{6) DCSD/\allowbreak{}AVDZ} & -54.8 & -1.7 & -37.5 & -0.6 & -52.4 & -1.1 & 25.9 & 2.8 & 24.6 & 3.0 \\ \hline
\multicolumn{1}{|>{\centering\arraybackslash}m{126.5pt}|}{7) DCSD/\allowbreak{}AVTZ} & -9.7 & -0.3 & -18.8 & -0.3 & -23.8 & -0.5 & 20.4 & 2.2 & N/\allowbreak{}A & N/\allowbreak{}A \\ \hline
\multicolumn{1}{|>{\centering\arraybackslash}m{126.5pt}|}{\textbf{8) DF-RMP2-F12/\allowbreak{}VDZ-F12 }} & \textbf{-3.2} & \textbf{-0.1} & \textbf{-6.2} & \textbf{-0.1} & \textbf{-4.8} & \textbf{-0.1} & \textbf{10.2} & \textbf{1.1} & \textbf{8.2} & \textbf{1.0} \\ \hline
\multicolumn{1}{|>{\centering\arraybackslash}m{126.5pt}|}{9) DF-RMP2/\allowbreak{}AVTZ} & -25.8 & -0.8 & -31.3 & -0.5 & -38.1 & -0.8 & 18.5 & 2.0 & 16.4 & 2.0 \\ \hline
\multicolumn{1}{|>{\centering\arraybackslash}m{126.5pt}|}{10) MBPT(2)/\allowbreak{}VTZ} & 6.5 & 0.2 & -6.2 & -0.1 & -14.3 & -0.3 & 16.7 & 1.8 & 13.1 & 1.6 \\ \hline
\multicolumn{1}{|>{\centering\arraybackslash}m{126.5pt}|}{11) SCS-MP2/\allowbreak{}AVTZ} & -25.8 & -0.8 & -31.3 & -0.5 & -38.1 & -0.8 & 18.5 & 2.0 & 16.4 & 2.0 \\ \hline
\multicolumn{1}{|>{\centering\arraybackslash}m{126.5pt}|}{12) MP2/\allowbreak{}AVDZ} & -48.4 & -1.5 & -43.8 & -0.7 & -61.9 & -1.3 & 0.9 & 0.1 & -3.3 & -0.4 \\ \hline
\multicolumn{1}{|>{\centering\arraybackslash}m{126.5pt}|}{13) MP2/\allowbreak{}AVTZ} & -9.7 & -0.3 & -25.0 & -0.4 & -28.6 & -0.6 & 5.6 & 0.6 & 2.5 & 0.3 \\ \hline
\multicolumn{1}{|>{\centering\arraybackslash}m{126.5pt}|}{14) MP2/\allowbreak{}AVQZ} & 0.0 & 0.0 & -18.8 & -0.3 & -14.3 & -0.3 & 9.3 & 1.0 & 6.6 & 0.8 \\ \hline
\multicolumn{1}{|>{\centering\arraybackslash}m{126.5pt}|}{15) MP2/\allowbreak{}AV5Z} & 3.2 & 0.1 & -6.2 & -0.1 & -4.8 & -0.1 & 13.0 & 1.4 & 9.8 & 1.2 \\ \specialrule{0.9pt}{0pt}{0pt}
\end{tabular}}}
\par\nointerlineskip
\centerline{\adjustbox{max width=518pt}{\begin{tabular}{|>{\centering\arraybackslash}m{126.5pt}|>{\centering\arraybackslash}m{28.5pt}|>{\centering\arraybackslash}m{24.8pt}|>{\centering\arraybackslash}m{28.4pt}|>{\centering\arraybackslash}m{28.3pt}|>{\centering\arraybackslash}m{26.6pt}|>{\centering\arraybackslash}m{26.6pt}|>{\centering\arraybackslash}m{32.1pt}|>{\centering\arraybackslash}m{32.0pt}|>{\centering\arraybackslash}m{29.3pt}|>{\centering\arraybackslash}m{66.1pt}|}
\multicolumn{11}{|>{\centering\arraybackslash}m{479.1pt}|}{\textbf{DF\textperiodcentered{}\textperiodcentered{}\textperiodcentered{}NH}\textbf{\textsubscript{3}}} \\* \hline\noalign{\penalty10000}
\multicolumn{1}{|>{\centering\arraybackslash}m{126.5pt}|}{} & \multicolumn{2}{>{\centering\arraybackslash}m{56.2pt}|}{\textbf{F}\textbf{\textsubscript{2}}\textbf{\textperiodcentered{}\textperiodcentered{}\textperiodcentered{}NH}\textbf{\textsubscript{3}}} & \multicolumn{2}{>{\centering\arraybackslash}m{59.8pt}|}{\textbf{NCF\textperiodcentered{}\textperiodcentered{}\textperiodcentered{}NH}\textbf{\textsubscript{3}}} & \multicolumn{2}{>{\centering\arraybackslash}m{56.2pt}|}{\textbf{CNF\textperiodcentered{}\textperiodcentered{}\textperiodcentered{}NH}\textbf{\textsubscript{3}}} & \multicolumn{2}{>{\centering\arraybackslash}m{67.1pt}|}{\textbf{NH}\textbf{\textsubscript{3}}\textbf{F}\textbf{\textsuperscript{+}}\textbf{\textperiodcentered{}\textperiodcentered{}\textperiodcentered{}NH}\textbf{\textsubscript{3}}} & \multicolumn{2}{>{\centering\arraybackslash}m{98.5pt}|}{\textbf{NH}\textbf{\textsubscript{2}}\textbf{F}\textbf{\textsubscript{2}}\textbf{\textsuperscript{+}}\textbf{\textperiodcentered{}\textperiodcentered{}\textperiodcentered{}NH}\textbf{\textsubscript{3}}} \\* \hline\noalign{\penalty10000}
\multicolumn{1}{|>{\centering\arraybackslash}m{126.5pt}|}{} & \%\ensuremath{\delta}\textsubscript{BE} & \ensuremath{\delta}\textsubscript{BE} & \%\ensuremath{\delta}\textsubscript{BE} & \ensuremath{\delta}\textsubscript{BE} & \%\ensuremath{\delta}\textsubscript{BE} & \ensuremath{\delta}\textsubscript{BE} & \%\ensuremath{\delta}\textsubscript{BE} & \ensuremath{\delta}\textsubscript{BE} & \%\ensuremath{\delta}\textsubscript{BE} & \ensuremath{\delta}\textsubscript{BE} \\* \hline\noalign{\penalty10000}
\multicolumn{1}{|>{\centering\arraybackslash}m{126.5pt}|}{1) {\leavevmode\color[HTML]{FF0000}CCSD(T)-F12b/\allowbreak{}VDZ-F12}} & 0.0 & 0.0 & 0.0 & 0.0 & 0.0 & 0.0 & 0.0 & 0.0 & 0.0 & 0.0 \\* \hline\noalign{\penalty10000}
\multicolumn{1}{|>{\centering\arraybackslash}m{126.5pt}|}{2) CCSD(T)/\allowbreak{}AVDZ} & -51.3 & -4.0 & -9.1 & -0.5 & -16.4 & -1.1 & -5.4 & -1.9 & -6.3 & -2.5 \\* \hline\noalign{\penalty10000}
\multicolumn{1}{|>{\centering\arraybackslash}m{126.5pt}|}{3) CCSD(T)/\allowbreak{}AVTZ} & -7.7 & -0.6 & -3.6 & -0.2 & -6.0 & -0.4 & -0.6 & -0.2 & -1.0 & -0.4 \\ \hline
\multicolumn{1}{|>{\centering\arraybackslash}m{126.5pt}|}{4) CCSD/\allowbreak{}AVDZ} & -10.3 & -0.8 & 1.8 & 0.1 & -6.0 & -0.4 & 0.3 & 0.1 & -1.3 & -0.5 \\ \hline
\multicolumn{1}{|>{\centering\arraybackslash}m{126.5pt}|}{5) CCSD/\allowbreak{}AVTZ} & 20.5 & 1.6 & 7.3 & 0.4 & 6.0 & 0.4 & 4.8 & 1.7 & 3.8 & 1.5 \\ \hline
\multicolumn{1}{|>{\centering\arraybackslash}m{126.5pt}|}{6) DCSD/\allowbreak{}AVDZ} & -37.2 & -2.9 & 0.0 & 0.0 & -10.4 & -0.7 & -3.4 & -1.2 & -4.3 & -1.7 \\ \hline
\multicolumn{1}{|>{\centering\arraybackslash}m{126.5pt}|}{7) DCSD/\allowbreak{}AVTZ} & 2.6 & 0.2 & 5.5 & 0.3 & 0.0 & 0.0 & 1.4 & 0.5 & 0.0 & 0.0 \\ \hline
\multicolumn{1}{|>{\centering\arraybackslash}m{126.5pt}|}{\textbf{8) DF-RMP2-F12/\allowbreak{}VDZ-F12 }} & \textbf{-2.6} & \textbf{-0.2} & \textbf{0.0} & \textbf{0.0} & \textbf{-7.5} & \textbf{-0.5} & \textbf{2.0} & \textbf{0.7} & \textbf{1.3} & \textbf{0.5} \\ \hline
\multicolumn{1}{|>{\centering\arraybackslash}m{126.5pt}|}{9) DF-RMP2/\allowbreak{}AVTZ} & -9.0 & -0.7 & -1.8 & -0.1 & -11.9 & -0.8 & 1.4 & 0.5 & 0.5 & 0.2 \\ \hline
\multicolumn{1}{|>{\centering\arraybackslash}m{126.5pt}|}{10) MBPT(2)/\allowbreak{}VTZ} & -3.8 & -0.3 & -10.9 & -0.6 & -17.9 & -1.2 & -1.1 & -0.4 & -2.3 & -0.9 \\ \hline
\multicolumn{1}{|>{\centering\arraybackslash}m{126.5pt}|}{11) SCS-MP2/\allowbreak{}AVTZ} & -9.0 & -0.7 & -1.8 & -0.1 & -11.9 & -0.8 & 1.4 & 0.5 & 0.5 & 0.2 \\ \hline
\multicolumn{1}{|>{\centering\arraybackslash}m{126.5pt}|}{12) MP2/\allowbreak{}AVDZ} & -52.6 & -4.1 & -9.1 & -0.5 & -25.4 & -1.7 & -3.4 & -1.2 & 4.8 & -1.9 \\ \hline
\multicolumn{1}{|>{\centering\arraybackslash}m{126.5pt}|}{13) MP2/\allowbreak{}AVTZ} & -9.0 & -0.7 & -1.8 & -0.1 & -11.9 & -0.8 & 1.4 & 0.5 & 0.5 & 0.2 \\* \hline\noalign{\penalty10000}
\multicolumn{1}{|>{\centering\arraybackslash}m{126.5pt}|}{14) MP2/\allowbreak{}AVQZ} & -5.1 & -0.4 & 1.8 & 0.1 & -7.5 & -0.5 & 2.6 & 0.9 & 1.8 & 0.7 \\* \hline\noalign{\penalty10000}
\multicolumn{1}{|>{\centering\arraybackslash}m{126.5pt}|}{15) MP2/\allowbreak{}AV5Z} & 0.0 & 0.0 & 5.5 & 0.3 & -3.0 & -0.2 & 3.1 & 1.1 & 2.5 & 1.0 \\* \specialrule{0.9pt}{0pt}{0pt}\noalign{\penalty10000}
\end{tabular}}}}
\par\nopagebreak\vspace{1pt}{\setstretch{1}\setlength{\parskip}{1.5pt}\fontsize{9}{10.5}\selectfont\raggedright\noindent \textsuperscript{\#}Due to high computational cost, NH\textsubscript{3}F\textsuperscript{+}\textperiodcentered{}\textperiodcentered{}\textperiodcentered{}CH\textsubscript{3} and NH\textsubscript{2}F\textsubscript{2}\textsuperscript{+}\textperiodcentered{}\textperiodcentered{}\textperiodcentered{}CH\textsubscript{3} could not be optimized at CCSD(T)-F12b/\allowbreak{}VTZ-F12; all selected DF\textperiodcentered{}\textperiodcentered{}\textperiodcentered{}CH\textsubscript{3}/NH\textsubscript{3} complexes (DF = F\textsubscript{2}, NCF, CNF, NH\textsubscript{3}F\textsuperscript{+}, and NH\textsubscript{2}F\textsubscript{2}\textsuperscript{+}) were instead optimized at CCSD(T)-F12b/\allowbreak{}VDZ-F12. \textsuperscript{N/\allowbreak{}A}Convergence failure during optimization.\par}
\end{table}

\begin{table}[!htbp]\centering\setlength{\parskip}{0pt}
\caption{Deviation (\ensuremath{\delta}\textsubscript{BE}, kJ/mol) and percentage deviation (\%\ensuremath{\delta}\textsubscript{BE}) in BEs for all F-bonded complexes DF\textperiodcentered{}\textperiodcentered{}\textperiodcentered{}CH\textsubscript{3}/NH\textsubscript{3}, calculated at DF-RMP2-F12/VDZ-F12, MP2/AVDZ, and M06-2X, relative to reference values obtained at DF-RMP2-F12/CBS//DF-RMP2-F12/VDZ-F12. Reference BEs from DF-RMP2-F12/CBS//DF-RMP2-F12/VDZ-F12 level are also included.}\label{tab:deviation-be-kj-mol}\vspace{7pt}\nopagebreak
{\fontsize{8.0}{9.2}\fontfamily{ptm}\selectfont\renewcommand{\arraystretch}{1.17}\setlength{\tabcolsep}{1.5pt}
\centerline{\adjustbox{max width=518pt}{\begin{tabular}{|>{\centering\arraybackslash}p{106.7pt}|>{\centering\arraybackslash}p{128.7pt}|>{\centering\arraybackslash}p{27.1pt}|>{\centering\arraybackslash}p{36.0pt}|>{\centering\arraybackslash}p{34.6pt}|>{\centering\arraybackslash}p{39.2pt}|>{\centering\arraybackslash}p{31.2pt}|>{\centering\arraybackslash}p{74.1pt}|}
\specialrule{0.9pt}{0pt}{0pt}
\multicolumn{8}{|>{\centering\arraybackslash}p{498.6pt}|}{\textbf{DF\textperiodcentered{}\textperiodcentered{}\textperiodcentered{}CH}\textbf{\textsubscript{3}}} \\* \hline\noalign{\penalty10000}
\multicolumn{1}{|>{\centering\arraybackslash}m{106.7pt}|}{\multirow{2}{106.7pt}{\centering DF\textperiodcentered{}\textperiodcentered{}\textperiodcentered{}CH\textsubscript{3}}} & BE[DF-RMP2-F12/\allowbreak{}CBS//DF- RMP2-F12/\allowbreak{}VDZ-F12] & \multicolumn{2}{>{\centering\arraybackslash}p{66.0pt}|}{DF-RMP2-F12/\allowbreak{}VDZ-F12} & \multicolumn{2}{>{\centering\arraybackslash}p{76.8pt}|}{MP2/\allowbreak{}AVDZ} & \multicolumn{2}{>{\centering\arraybackslash}p{108.3pt}|}{M06-2X/\allowbreak{}AVTZ (NO BSSE)} \\* \cline{2-8}\noalign{\penalty10000}
\multicolumn{1}{|>{\centering\arraybackslash}p{106.7pt}|}{} & BE & \ensuremath{\delta}\textsubscript{BE} & \% \ensuremath{\delta}\textsubscript{BE} & \ensuremath{\delta}\textsubscript{BE} & \% \ensuremath{\delta}\textsubscript{BE} & \ensuremath{\delta}\textsubscript{BE} & \% \ensuremath{\delta}\textsubscript{BE} \\* \hline\noalign{\penalty10000}
\multicolumn{1}{|>{\centering\arraybackslash}p{106.7pt}|}{1) F\textsubscript{2}\textperiodcentered{}\textperiodcentered{}\textperiodcentered{}CH\textsubscript{3}} & -3.4 & 0.4 & 11.8 & 1.0 & 29.4 & -4.5 & -132.4 \\* \hline\noalign{\penalty10000}
\multicolumn{1}{|>{\centering\arraybackslash}p{106.7pt}|}{2) SF\textsubscript{6}\textperiodcentered{}\textperiodcentered{}\textperiodcentered{}CH\textsubscript{3}} & -1.8 & 0.3 & 16.7 & 0.7 & 38.9 & -0.7 & -38.9 \\* \hline\noalign{\penalty10000}
\multicolumn{1}{|>{\centering\arraybackslash}p{106.7pt}|}{3) NCF\textperiodcentered{}\textperiodcentered{}\textperiodcentered{}CH\textsubscript{3}} & -1.7 & 0.2 & 11.8 & 0.5 & 29.4 & -1.4 & -82.4 \\* \hline\noalign{\penalty10000}
\multicolumn{1}{|>{\centering\arraybackslash}p{106.7pt}|}{4) CNF\textperiodcentered{}\textperiodcentered{}\textperiodcentered{}CH\textsubscript{3}} & -2.4 & 0.3 & 12.5 & 0.7 & 29.2 & -1.8 & -75.0 \\ \hline
\multicolumn{1}{|>{\centering\arraybackslash}p{106.7pt}|}{5) NF\textsubscript{3}\textperiodcentered{}\textperiodcentered{}\textperiodcentered{}CH\textsubscript{3}} & -1.7 & 0.3 & 17.6 & 0.7 & 41.2 & -1.0 & -58.8 \\ \hline
\multicolumn{1}{|>{\centering\arraybackslash}p{106.7pt}|}{6) ClF\textsubscript{3}\textperiodcentered{}\textperiodcentered{}\textperiodcentered{}CH\textsubscript{3}} & -2.4 & 0.3 & 12.5 & 0.5 & 20.8 & -2.8 & -116.7 \\ \hline
\multicolumn{1}{|>{\centering\arraybackslash}p{106.7pt}|}{7) OF\textsubscript{2}\textperiodcentered{}\textperiodcentered{}\textperiodcentered{}CH\textsubscript{3}} & -2.0 & 0.3 & 15.0 & 0.8 & 40.0 & -2.4 & -120.0 \\ \hline
\multicolumn{1}{|>{\centering\arraybackslash}p{106.7pt}|}{8) NF\textsubscript{5}\textperiodcentered{}\textperiodcentered{}\textperiodcentered{}CH\textsubscript{3}} & -3.1 & 0.4 & 12.9 & 1.0 & 32.3 & -2.0 & -64.5 \\ \hline
\multicolumn{1}{|>{\centering\arraybackslash}p{106.7pt}|}{9) FCCF\textperiodcentered{}\textperiodcentered{}\textperiodcentered{}CH\textsubscript{3}} & -1.5 & 0.2 & 13.3 & 0.6 & 40.0 & -0.6 & -40.0 \\ \hline
\multicolumn{1}{|>{\centering\arraybackslash}p{106.7pt}|}{10) CF\textsubscript{3}CCF\textperiodcentered{}\textperiodcentered{}\textperiodcentered{}CH\textsubscript{3}} & -1.5 & 0.2 & 13.3 & 0.6 & 40.0 & -1.2 & -80.0 \\ \hline
\multicolumn{1}{|>{\centering\arraybackslash}p{106.7pt}|}{11) CF\textsubscript{3}OF\textperiodcentered{}\textperiodcentered{}\textperiodcentered{}CH\textsubscript{3}} & -2.5 & 0.4 & 16.0 & N/\allowbreak{}A & N/\allowbreak{}A & -2.2 & -88.0 \\ \hline
\multicolumn{1}{|>{\centering\arraybackslash}p{106.7pt}|}{12) F\textsubscript{3}NO\textperiodcentered{}\textperiodcentered{}\textperiodcentered{}CH\textsubscript{3}} & -2.1 & 0.3 & 14.3 & 0.8 & 38.1 & -1.5 & -71.4 \\ \hline
\multicolumn{1}{|>{\centering\arraybackslash}p{106.7pt}|}{13) SF\textsubscript{5}\textperiodcentered{}\textperiodcentered{}\textperiodcentered{}CH\textsubscript{3}} & -1.8 & 0.4 & 22.2 & 0.7 & 38.9 & -1.2 & -66.7 \\ \hline
\multicolumn{1}{|>{\centering\arraybackslash}p{106.7pt}|}{14) NH\textsubscript{3}F+\textperiodcentered{}\textperiodcentered{}\textperiodcentered{}CH\textsubscript{3}} & -9.8 & 0.3 & 3.1 & 0.9 & 9.2 & -5.7 & -58.2 \\ \hline
\multicolumn{1}{|>{\centering\arraybackslash}p{106.7pt}|}{15) NH\textsubscript{2}F\textsubscript{2}\textsuperscript{+}\textperiodcentered{}\textperiodcentered{}\textperiodcentered{}CH\textsubscript{3}} & -11.4 & 0.4 & 3.5 & 0.9 & 7.9 & -5.1 & -44.7 \\ \hline
\multicolumn{1}{|>{\centering\arraybackslash}p{106.7pt}|}{MSE} & 0.0 & \multicolumn{2}{>{\centering\arraybackslash}p{66.0pt}|}{0.3} & \multicolumn{2}{>{\centering\arraybackslash}p{76.8pt}|}{0.7} & \multicolumn{2}{>{\centering\arraybackslash}p{108.3pt}|}{-2.3} \\ \hline
\multicolumn{1}{|>{\centering\arraybackslash}p{106.7pt}|}{MAD} & 0.0 & \multicolumn{2}{>{\centering\arraybackslash}p{66.0pt}|}{0.3} & \multicolumn{2}{>{\centering\arraybackslash}p{76.8pt}|}{0.7} & \multicolumn{2}{>{\centering\arraybackslash}p{108.3pt}|}{2.3} \\ \hline
\multicolumn{1}{|>{\centering\arraybackslash}p{106.7pt}|}{RMSD} & 0.0 & \multicolumn{2}{>{\centering\arraybackslash}p{66.0pt}|}{0.3} & \multicolumn{2}{>{\centering\arraybackslash}p{76.8pt}|}{0.8} & \multicolumn{2}{>{\centering\arraybackslash}p{108.3pt}|}{2.7} \\* \hline\noalign{\penalty10000}
\multicolumn{1}{|>{\centering\arraybackslash}p{106.7pt}|}{MAX} & 0.0 & \multicolumn{2}{>{\centering\arraybackslash}p{66.0pt}|}{0.4} & \multicolumn{2}{>{\centering\arraybackslash}p{76.8pt}|}{1.0} & \multicolumn{2}{>{\centering\arraybackslash}p{108.3pt}|}{-5.7} \\* \hline\noalign{\penalty10000}
\multicolumn{1}{|>{\centering\arraybackslash}p{106.7pt}|}{MA\%D} & 0.0 & \multicolumn{2}{>{\centering\arraybackslash}p{66.0pt}|}{13.1\%} & \multicolumn{2}{>{\centering\arraybackslash}p{76.8pt}|}{31.1\%} & \multicolumn{2}{>{\centering\arraybackslash}p{108.3pt}|}{75.8\%} \\* \specialrule{0.9pt}{0pt}{0pt}\noalign{\penalty10000}
\end{tabular}}}}
\end{table}

\begin{table}[!htbp]\centering\setlength{\parskip}{0pt}
{\setstretch{1}\fontsize{9.5}{11.5}\selectfont\setlength{\parskip}{0pt}\raggedright\noindent \textbf{Table S7} (Continued)\par}\vspace{-2pt}
{\fontsize{8.0}{9.2}\fontfamily{ptm}\selectfont\renewcommand{\arraystretch}{1.12}\setlength{\tabcolsep}{1.5pt}
\centerline{\adjustbox{max width=518pt}{\begin{tabular}{|>{\centering\arraybackslash}m{112.5pt}|>{\centering\arraybackslash}m{125.8pt}|>{\centering\arraybackslash}m{43.4pt}|>{\centering\arraybackslash}m{55.3pt}|>{\centering\arraybackslash}m{27.1pt}|>{\centering\arraybackslash}m{37.7pt}|>{\centering\arraybackslash}m{30.6pt}|>{\centering\arraybackslash}m{51.5pt}|}
\specialrule{0.9pt}{0pt}{0pt}
\multicolumn{8}{|>{\centering\arraybackslash}m{504.9pt}|}{\textbf{DF\textperiodcentered{}\textperiodcentered{}\textperiodcentered{}NH}\textbf{\textsubscript{3}}} \\* \hline\noalign{\penalty10000}
\multicolumn{1}{|>{\centering\arraybackslash}m{112.5pt}|}{\multirow{3}{112.5pt}{\centering DF\textperiodcentered{}\textperiodcentered{}\textperiodcentered{}NH\textsubscript{3}}} & \multirow{2}{125.8pt}{\centering BE[DF-RMP2-F12/\allowbreak{}CBS//DF-RMP2-F12/\allowbreak{}VDZ-F12]} & \multicolumn{2}{>{\centering\arraybackslash}m{101.7pt}|}{\multirow{2}{101.7pt}{\centering DF-RMP2-F12/\allowbreak{}VDZ-F12}} & \multicolumn{2}{>{\centering\arraybackslash}m{67.8pt}|}{\multirow{2}{67.8pt}{\centering MP2/\allowbreak{}AVDZ}} & \multicolumn{2}{>{\centering\arraybackslash}m{85.1pt}|}{\multirow{2}{85.1pt}{\centering M06-2X/\allowbreak{}AVTZ (NO BSSE)}} \\ 
\multicolumn{1}{|>{\centering\arraybackslash}m{112.5pt}|}{} &  & \multicolumn{2}{>{\centering\arraybackslash}m{101.7pt}|}{} & \multicolumn{2}{>{\centering\arraybackslash}m{67.8pt}|}{} & \multicolumn{2}{>{\centering\arraybackslash}m{85.1pt}|}{} \\* \cline{2-8}\noalign{\penalty10000}
\multicolumn{1}{|>{\centering\arraybackslash}m{112.5pt}|}{} & BE & \ensuremath{\delta}\textsubscript{BE} & \% \ensuremath{\delta}\textsubscript{BE} & \ensuremath{\delta}\textsubscript{BE} & \% \ensuremath{\delta}\textsubscript{BE} & \ensuremath{\delta}\textsubscript{BE} & \% \ensuremath{\delta}\textsubscript{BE} \\* \hline\noalign{\penalty10000}
\multicolumn{1}{|>{\centering\arraybackslash}m{112.5pt}|}{1) F\textsubscript{2}\textperiodcentered{}\textperiodcentered{}\textperiodcentered{}NH\textsubscript{3}} & -7.8 & 0.2 & 2.6 & -0.2 & 2.6 & 1.4 & 17.9 \\* \hline\noalign{\penalty10000}
\multicolumn{1}{|>{\centering\arraybackslash}m{112.5pt}|}{2) SF\textsubscript{6}\textperiodcentered{}\textperiodcentered{}\textperiodcentered{}NH\textsubscript{3}} & -1.4 & 0.0 & 0.0 & 0.1 & -7.1 & 0.0 & 0.0 \\* \hline\noalign{\penalty10000}
\multicolumn{1}{|>{\centering\arraybackslash}m{112.5pt}|}{3) NCF\textperiodcentered{}\textperiodcentered{}\textperiodcentered{}NH\textsubscript{3}} & -5.1 & 0.1 & 2.0 & 0.7 & -13.7 & 0.4 & 7.8 \\ \hline
\multicolumn{1}{|>{\centering\arraybackslash}m{112.5pt}|}{4) CNF\textperiodcentered{}\textperiodcentered{}\textperiodcentered{}NH\textsubscript{3}} & -6.9 & 0.2 & 2.9 & 0.8 & -11.6 & 1.5 & 21.7 \\ \hline
\multicolumn{1}{|>{\centering\arraybackslash}m{112.5pt}|}{5) NF\textsubscript{3}\textperiodcentered{}\textperiodcentered{}\textperiodcentered{}NH\textsubscript{3}} & -1.9 & 0.2 & 10.5 & b & b & 0.7 & 36.8 \\ \hline
\multicolumn{1}{|>{\centering\arraybackslash}m{112.5pt}|}{6) ClF\textsubscript{3}\textperiodcentered{}\textperiodcentered{}\textperiodcentered{}NH\textsubscript{3}} & -4.5 & 0.4 & 8.9 & 0.0 & 0.0 & 1.0 & 22.2 \\ \hline
\multicolumn{1}{|>{\centering\arraybackslash}m{112.5pt}|}{7) OF\textsubscript{2}\textperiodcentered{}\textperiodcentered{}\textperiodcentered{}NH\textsubscript{3}} & -3.4 & 0.3 & 8.8 & 0.5 & -14.7 & 0.6 & 17.6 \\ \hline
\multicolumn{1}{|>{\centering\arraybackslash}m{112.5pt}|}{8) NF\textsubscript{5}\textperiodcentered{}\textperiodcentered{}\textperiodcentered{}NH\textsubscript{3}} & -7.3 & 0.2 & 2.7 & 0.6 & -8.2 & 0.7 & 9.6 \\ \hline
\multicolumn{1}{|>{\centering\arraybackslash}m{112.5pt}|}{9) FCCF\textperiodcentered{}\textperiodcentered{}\textperiodcentered{}NH\textsubscript{3}} & -0.6 & 0.0 & 0.0 & 0.4 & -66.7 & 0.0 & 0.0 \\ \hline
\multicolumn{1}{|>{\centering\arraybackslash}m{112.5pt}|}{10) CF\textsubscript{3}CCF\textperiodcentered{}\textperiodcentered{}\textperiodcentered{}NH\textsubscript{3}} & -3.6 & 0.0 & 0.0 & 0.4 & -11.1 & 0.2 & 5.6 \\ \hline
\multicolumn{1}{|>{\centering\arraybackslash}m{112.5pt}|}{11) CF\textsubscript{3}OF\textperiodcentered{}\textperiodcentered{}\textperiodcentered{}NH\textsubscript{3}} & -4.3 & 0.2 & 4.7 & 0.3 & -7.0 & 2.0 & 46.5 \\ \hline
\multicolumn{1}{|>{\centering\arraybackslash}m{112.5pt}|}{12) F\textsubscript{3}NO\textperiodcentered{}\textperiodcentered{}\textperiodcentered{}NH\textsubscript{3}} & b & b & b & b & b & b & b \\ \hline
\multicolumn{1}{|>{\centering\arraybackslash}m{112.5pt}|}{13) SF\textsubscript{5}\textperiodcentered{}\textperiodcentered{}\textperiodcentered{}NH\textsubscript{3}} & -2.7 & 0.1 & 3.7 & 0.3 & -11.1 & 0.7 & 25.9 \\ \hline
\multicolumn{1}{|>{\centering\arraybackslash}m{112.5pt}|}{14) NH\textsubscript{3}F+\textperiodcentered{}\textperiodcentered{}\textperiodcentered{}NH\textsubscript{3}} & -33.9 & 0.0 & 0.0 & 0.6 & -1.8 & -0.4 & -1.2 \\ \hline
\multicolumn{1}{|>{\centering\arraybackslash}m{112.5pt}|}{15) NH\textsubscript{2}F\textsubscript{2}\textsuperscript{+}\textperiodcentered{}\textperiodcentered{}\textperiodcentered{}NH\textsubscript{3}} & -38.8 & 0.1 & 0.3 & 0.5 & -1.3 & -0.2 & -0.5 \\ \hline
\multicolumn{1}{|>{\centering\arraybackslash}m{112.5pt}|}{MSE} & 0.0 & \multicolumn{2}{>{\centering\arraybackslash}m{101.7pt}|}{0.1} & \multicolumn{2}{>{\centering\arraybackslash}m{67.8pt}|}{0.4} & \multicolumn{2}{>{\centering\arraybackslash}m{85.1pt}|}{0.6} \\ \hline
\multicolumn{1}{|>{\centering\arraybackslash}m{112.5pt}|}{MAD} & 0.0 & \multicolumn{2}{>{\centering\arraybackslash}m{101.7pt}|}{0.1} & \multicolumn{2}{>{\centering\arraybackslash}m{67.8pt}|}{0.4} & \multicolumn{2}{>{\centering\arraybackslash}m{85.1pt}|}{0.7} \\ \hline
\multicolumn{1}{|>{\centering\arraybackslash}m{112.5pt}|}{RMSD} & 0.0 & \multicolumn{2}{>{\centering\arraybackslash}m{101.7pt}|}{0.2} & \multicolumn{2}{>{\centering\arraybackslash}m{67.8pt}|}{0.5} & \multicolumn{2}{>{\centering\arraybackslash}m{85.1pt}|}{0.9} \\* \hline\noalign{\penalty10000}
\multicolumn{1}{|>{\centering\arraybackslash}m{112.5pt}|}{MAX} & 0.0 & \multicolumn{2}{>{\centering\arraybackslash}m{101.7pt}|}{0.4} & \multicolumn{2}{>{\centering\arraybackslash}m{67.8pt}|}{0.8} & \multicolumn{2}{>{\centering\arraybackslash}m{85.1pt}|}{2.0} \\* \hline\noalign{\penalty10000}
\multicolumn{1}{|>{\centering\arraybackslash}m{112.5pt}|}{MA\%D} & 0.0 & \multicolumn{2}{>{\centering\arraybackslash}m{101.7pt}|}{3.4 \%} & \multicolumn{2}{>{\centering\arraybackslash}m{67.8pt}|}{11.7\%} & \multicolumn{2}{>{\centering\arraybackslash}m{85.1pt}|}{16.0\%} \\* \specialrule{0.9pt}{0pt}{0pt}\noalign{\penalty10000}
\end{tabular}}}}
\par\nopagebreak\vspace{1pt}{\setstretch{1}\setlength{\parskip}{1.5pt}\fontsize{9}{10.5}\selectfont\raggedright\noindent \textsuperscript{b}Convergence error during optimization of respective F-bonded complexes.\par}
\end{table}

\begin{table}[!htbp]\centering\setlength{\parskip}{0pt}
\caption{Deviation (\ensuremath{\delta}\textsubscript{R}, Å) and percentage deviation (\%\ensuremath{\delta}\textsubscript{R}) on F-bond distances [R(F\textperiodcentered{}\textperiodcentered{}\textperiodcentered{}C/\allowbreak{}N)] for the F-bonded complexes DF\textperiodcentered{}\textperiodcentered{}\textperiodcentered{}CH\textsubscript{3}/NH\textsubscript{3}, calculated at MP2/\allowbreak{}AVDZ and M06-2X/\allowbreak{}AVTZ, relative to reference values obtained at DF-RMP2-F12/\allowbreak{}VDZ-F12.}\label{tab:deviation-r-and-percentage}\vspace{7pt}\nopagebreak
{\fontsize{8.0}{9.2}\fontfamily{ptm}\selectfont\renewcommand{\arraystretch}{1.17}\setlength{\tabcolsep}{1.5pt}
\centerline{\adjustbox{max width=518pt}{\begin{tabular}{|>{\centering\arraybackslash}p{95.1pt}|>{\centering\arraybackslash}m{151.2pt}|>{\centering\arraybackslash}b{39.5pt}|>{\centering\arraybackslash}b{38.4pt}|>{\centering\arraybackslash}b{39.5pt}|>{\centering\arraybackslash}b{107.3pt}|}
\specialrule{0.9pt}{0pt}{0pt}
\multicolumn{6}{|>{\centering\arraybackslash}p{486.0pt}|}{\textbf{DF\textperiodcentered{}\textperiodcentered{}\textperiodcentered{}CH}\textbf{\textsubscript{3}}} \\* \hline\noalign{\penalty10000}
\multicolumn{1}{|>{\centering\arraybackslash}m{95.1pt}|}{\multirow{2}{95.1pt}{\centering DF\textperiodcentered{}\textperiodcentered{}\textperiodcentered{}CH\textsubscript{3}}} & DF-RMP2-F12/\allowbreak{}VDZ-F12 & \multicolumn{2}{>{\centering\arraybackslash}p{80.9pt}|}{MP2/\allowbreak{}AVDZ} & \multicolumn{2}{>{\centering\arraybackslash}p{149.8pt}|}{M06-2X/\allowbreak{}AVTZ} \\* \cline{2-6}\noalign{\penalty10000}
\multicolumn{1}{|>{\centering\arraybackslash}p{95.1pt}|}{} & R & \ensuremath{\delta}\textsubscript{R} & \% \ensuremath{\delta}\textsubscript{R} & \ensuremath{\delta}\textsubscript{R} & \% \ensuremath{\delta}\textsubscript{R} \\* \hline\noalign{\penalty10000}
\multicolumn{1}{|>{\centering\arraybackslash}p{95.1pt}|}{1) F\textsubscript{2}\textperiodcentered{}\textperiodcentered{}\textperiodcentered{}CH\textsubscript{3}} & 2.869 & -0.004 & -0.1 & -0.221 & -7.7 \\* \hline\noalign{\penalty10000}
\multicolumn{1}{|>{\centering\arraybackslash}p{95.1pt}|}{2) SF\textsubscript{6}\textperiodcentered{}\textperiodcentered{}\textperiodcentered{}CH\textsubscript{3}} & 3.327 & -0.072 & -2.2 & -0.306 & -9.2 \\* \hline\noalign{\penalty10000}
\multicolumn{1}{|>{\centering\arraybackslash}p{95.1pt}|}{3) NCF\textperiodcentered{}\textperiodcentered{}\textperiodcentered{}CH\textsubscript{3}} & 3.428 & -0.040 & -1.2 & -0.410 & -12.0 \\* \hline\noalign{\penalty10000}
\multicolumn{1}{|>{\centering\arraybackslash}p{95.1pt}|}{4) CNF\textperiodcentered{}\textperiodcentered{}\textperiodcentered{}CH\textsubscript{3}} & 3.262 & -0.077 & -2.4 & -0.286 & -8.8 \\ \hline
\multicolumn{1}{|>{\centering\arraybackslash}p{95.1pt}|}{5) NF\textsubscript{3}\textperiodcentered{}\textperiodcentered{}\textperiodcentered{}CH\textsubscript{3}} & 3.295 & -0.048 & -1.5 & -0.293 & -8.9 \\ \hline
\multicolumn{1}{|>{\centering\arraybackslash}p{95.1pt}|}{6) ClF\textsubscript{3}\textperiodcentered{}\textperiodcentered{}\textperiodcentered{}CH\textsubscript{3}} & 3.160 & -0.188 & -5.9 & -0.336 & -10.6 \\ \hline
\multicolumn{1}{|>{\centering\arraybackslash}p{95.1pt}|}{7) OF\textsubscript{2}\textperiodcentered{}\textperiodcentered{}\textperiodcentered{}CH\textsubscript{3}} & 3.212 & -0.040 & -1.2 & -0.284 & -8.8 \\ \hline
\multicolumn{1}{|>{\centering\arraybackslash}p{95.1pt}|}{8) NF\textsubscript{5}\textperiodcentered{}\textperiodcentered{}\textperiodcentered{}CH\textsubscript{3}} & 3.092 & -0.041 & -1.3 & -0.164 & -5.3 \\ \hline
\multicolumn{1}{|>{\centering\arraybackslash}p{95.1pt}|}{9) FCCF\textperiodcentered{}\textperiodcentered{}\textperiodcentered{}CH\textsubscript{3}} & 3.346 & -0.103 & -3.1 & -0.324 & -9.7 \\ \hline
\multicolumn{1}{|>{\centering\arraybackslash}p{95.1pt}|}{10) CF\textsubscript{3}CCF\textperiodcentered{}\textperiodcentered{}\textperiodcentered{}CH\textsubscript{3}} & 3.457 & -0.149 & -4.3 & -0.434 & -12.6 \\ \hline
\multicolumn{1}{|>{\centering\arraybackslash}p{95.1pt}|}{11) CF\textsubscript{3}OF\textperiodcentered{}\textperiodcentered{}\textperiodcentered{}CH\textsubscript{3}} & 3.088 & b & b & -0.197 & -6.4 \\ \hline
\multicolumn{1}{|>{\centering\arraybackslash}p{95.1pt}|}{12) F\textsubscript{3}NO\textperiodcentered{}\textperiodcentered{}\textperiodcentered{}CH\textsubscript{3}} & 3.234 & -0.034 & -1.1 & -0.264 & -8.2 \\ \hline
\multicolumn{1}{|>{\centering\arraybackslash}p{95.1pt}|}{13) SF\textsubscript{5}\textperiodcentered{}\textperiodcentered{}\textperiodcentered{}CH\textsubscript{3}} & 3.329 & -0.087 & -2.6 & -0.308 & -9.3 \\ \hline
\multicolumn{1}{|>{\centering\arraybackslash}p{95.1pt}|}{14) NH\textsubscript{3}F+\textperiodcentered{}\textperiodcentered{}\textperiodcentered{}CH\textsubscript{3}} & 2.947 & -0.007 & -0.2 & -0.254 & -8.6 \\ \hline
\multicolumn{1}{|>{\centering\arraybackslash}p{95.1pt}|}{15) NH\textsubscript{2}F\textsubscript{2}\textsuperscript{+}\textperiodcentered{}\textperiodcentered{}\textperiodcentered{}CH\textsubscript{3}} & 2.941 & -0.022 & -0.7 & -0.228 & -7.8 \\ \hline
\multicolumn{1}{|>{\centering\arraybackslash}p{95.1pt}|}{\textbf{MSE}} & \textbf{0.0} & \multicolumn{2}{>{\centering\arraybackslash}m{80.9pt}|}{\textbf{-0.065}} & \multicolumn{2}{>{\centering\arraybackslash}m{149.8pt}|}{\textbf{-0.287}} \\ \hline
\multicolumn{1}{|>{\centering\arraybackslash}p{95.1pt}|}{\textbf{MAD}} & \textbf{0.0} & \multicolumn{2}{>{\centering\arraybackslash}m{80.9pt}|}{\textbf{0.065}} & \multicolumn{2}{>{\centering\arraybackslash}m{149.8pt}|}{\textbf{0.287}} \\ \hline
\multicolumn{1}{|>{\centering\arraybackslash}p{95.1pt}|}{\textbf{RMSD}} & \textbf{0.0} & \multicolumn{2}{>{\centering\arraybackslash}m{80.9pt}|}{\textbf{0.083}} & \multicolumn{2}{>{\centering\arraybackslash}m{149.8pt}|}{\textbf{0.296}} \\* \hline\noalign{\penalty10000}
\multicolumn{1}{|>{\centering\arraybackslash}p{95.1pt}|}{\textbf{MAX}} & \textbf{0.0} & \multicolumn{2}{>{\centering\arraybackslash}m{80.9pt}|}{\textbf{-0.188}} & \multicolumn{2}{>{\centering\arraybackslash}m{149.8pt}|}{\textbf{-0.434}} \\* \hline\noalign{\penalty10000}
\multicolumn{1}{|>{\centering\arraybackslash}p{95.1pt}|}{\textbf{MA\%D}} & \textbf{0.0} & \multicolumn{2}{>{\centering\arraybackslash}b{80.9pt}|}{\textbf{2.0 \%}} & \multicolumn{2}{>{\centering\arraybackslash}b{149.8pt}|}{\textbf{8.9 \%}} \\* \specialrule{0.9pt}{0pt}{0pt}\noalign{\penalty10000}
\end{tabular}}}}
\end{table}

\begin{table}[!htbp]\centering\setlength{\parskip}{0pt}
{\setstretch{1}\fontsize{9.5}{11.5}\selectfont\setlength{\parskip}{0pt}\raggedright\noindent \textbf{Table S8} (Continued)\par}\vspace{-2pt}
{\fontsize{8.0}{9.2}\fontfamily{ptm}\selectfont\renewcommand{\arraystretch}{1.17}\setlength{\tabcolsep}{1.5pt}
\centerline{\adjustbox{max width=518pt}{\begin{tabular}{|>{\centering\arraybackslash}m{8.0pt}|>{\centering\arraybackslash}m{122.2pt}|>{\centering\arraybackslash}m{114.2pt}|>{\centering\arraybackslash}b{37.8pt}|>{\centering\arraybackslash}b{70.3pt}|>{\centering\arraybackslash}b{43.3pt}|>{\centering\arraybackslash}b{63.3pt}|}
\specialrule{0.9pt}{0pt}{0pt}
\multicolumn{7}{|>{\centering\arraybackslash}m{477.2pt}|}{\textbf{DF\textperiodcentered{}\textperiodcentered{}\textperiodcentered{}NH}\textbf{\textsubscript{3}}} \\* \hline\noalign{\penalty10000}
\multicolumn{2}{|>{\centering\arraybackslash}m{133.2pt}|}{DF\textperiodcentered{}\textperiodcentered{}\textperiodcentered{}NH\textsubscript{3}} & DF-RMP2-F12/\allowbreak{}VDZ-F12 & \multicolumn{2}{>{\centering\arraybackslash}b{111.1pt}|}{MP2/\allowbreak{}AVDZ} & \multicolumn{2}{>{\centering\arraybackslash}b{109.6pt}|}{M06-2X/\allowbreak{}AVTZ} \\* \hline\noalign{\penalty10000}
\multicolumn{2}{|>{\centering\arraybackslash}m{133.2pt}|}{} & R & \ensuremath{\delta}\textsubscript{R} & \%\ensuremath{\delta}\textsubscript{R} & \ensuremath{\delta}\textsubscript{R} & \%\ensuremath{\delta}\textsubscript{R} \\* \hline\noalign{\penalty10000}
\multicolumn{2}{|>{\centering\arraybackslash}m{133.2pt}|}{1) F\textsubscript{2}\textperiodcentered{}\textperiodcentered{}\textperiodcentered{}NH\textsubscript{3}} & 2.616 & -0.174 & -6.7 & -0.020 & -0.8 \\* \hline\noalign{\penalty10000}
\multicolumn{2}{|>{\centering\arraybackslash}m{133.2pt}|}{2) SF\textsubscript{6}\textperiodcentered{}\textperiodcentered{}\textperiodcentered{}NH\textsubscript{3}} & 3.345 & -0.179 & -5.4 & 0.289 & 8.6 \\* \hline\noalign{\penalty10000}
\multicolumn{2}{|>{\centering\arraybackslash}m{133.2pt}|}{3) NCF\textperiodcentered{}\textperiodcentered{}\textperiodcentered{}NH\textsubscript{3}} & 3.160 & -0.019 & -0.6 & 0.045 & 1.4 \\ \hline
\multicolumn{2}{|>{\centering\arraybackslash}m{133.2pt}|}{4) CNF\textperiodcentered{}\textperiodcentered{}\textperiodcentered{}NH\textsubscript{3}} & 2.960 & -0.036 & -1.2 & 0.071 & 2.4 \\ \hline
\multicolumn{2}{|>{\centering\arraybackslash}m{133.2pt}|}{5) NF\textsubscript{3}\textperiodcentered{}\textperiodcentered{}\textperiodcentered{}NH\textsubscript{3}} & 3.209 & b & b & 0.239 & 7.4 \\ \hline
\multicolumn{2}{|>{\centering\arraybackslash}m{133.2pt}|}{6) ClF\textsubscript{3}\textperiodcentered{}\textperiodcentered{}\textperiodcentered{}NH\textsubscript{3}} & 2.965 & -0.202 & -6.8 & -0.041 & -1.4 \\ \hline
\multicolumn{2}{|>{\centering\arraybackslash}m{133.2pt}|}{7) OF\textsubscript{2}\textperiodcentered{}\textperiodcentered{}\textperiodcentered{}NH\textsubscript{3}} & 2.942 & -0.104 & -3.5 & 0.031 & 1.1 \\ \hline
\multicolumn{2}{|>{\centering\arraybackslash}m{133.2pt}|}{8) NF\textsubscript{5}\textperiodcentered{}\textperiodcentered{}\textperiodcentered{}NH\textsubscript{3}} & 2.846 & -0.070 & -2.5 & 0.003 & 0.1 \\ \hline
\multicolumn{2}{|>{\centering\arraybackslash}m{133.2pt}|}{9) FCCF\textperiodcentered{}\textperiodcentered{}\textperiodcentered{}NH\textsubscript{3}} & 3.444 & -0.132 & -3.8 & 0.204 & 5.9 \\ \hline
\multicolumn{2}{|>{\centering\arraybackslash}m{133.2pt}|}{10) CF\textsubscript{3}CCF\textperiodcentered{}\textperiodcentered{}\textperiodcentered{}NH\textsubscript{3}} & 3.227 & -0.074 & -2.3 & 0.182 & 5.6 \\ \hline
\multicolumn{2}{|>{\centering\arraybackslash}m{133.2pt}|}{11) CF\textsubscript{3}OF\textperiodcentered{}\textperiodcentered{}\textperiodcentered{}NH\textsubscript{3}} & 2.869 & -0.106 & -3.7 & 0.063 & 2.2 \\ \hline
\multicolumn{2}{|>{\centering\arraybackslash}m{133.2pt}|}{12) F\textsubscript{3}NO\textperiodcentered{}\textperiodcentered{}\textperiodcentered{}NH\textsubscript{3}} & b & b & b & b & b \\ \hline
\multicolumn{2}{|>{\centering\arraybackslash}m{133.2pt}|}{13) SF\textsubscript{5}\textperiodcentered{}\textperiodcentered{}\textperiodcentered{}NH\textsubscript{3}} & 3.236 & -0.130 & -4.0 & 0.183 & 5.7 \\ \hline
\multicolumn{2}{|>{\centering\arraybackslash}m{133.2pt}|}{14) NH\textsubscript{3}F+\textperiodcentered{}\textperiodcentered{}\textperiodcentered{}NH\textsubscript{3}} & 2.647 & -0.051 & -1.9 & -0.057 & -2.2 \\ \hline
\multicolumn{2}{|>{\centering\arraybackslash}m{133.2pt}|}{15) NH\textsubscript{2}F\textsubscript{2}\textsuperscript{+}\textperiodcentered{}\textperiodcentered{}\textperiodcentered{}NH\textsubscript{3}} & 2.627 & -0.046 & -1.8 & -0.046 & -1.8 \\ \hline
\multicolumn{2}{|>{\centering\arraybackslash}m{133.2pt}|}{\textbf{MSE}} & \textbf{0.0} & \multicolumn{2}{>{\centering\arraybackslash}m{111.1pt}|}{\textbf{-0.102}} & \multicolumn{2}{>{\centering\arraybackslash}m{109.6pt}|}{\textbf{0.082}} \\ \hline
\multicolumn{2}{|>{\centering\arraybackslash}m{133.2pt}|}{\textbf{MAD}} & \textbf{0.0} & \multicolumn{2}{>{\centering\arraybackslash}m{111.1pt}|}{\textbf{0.102}} & \multicolumn{2}{>{\centering\arraybackslash}m{109.6pt}|}{\textbf{0.105}} \\ \hline
\multicolumn{2}{|>{\centering\arraybackslash}m{133.2pt}|}{\textbf{RMSD}} & \textbf{0.0} & \multicolumn{2}{>{\centering\arraybackslash}m{111.1pt}|}{\textbf{0.116}} & \multicolumn{2}{>{\centering\arraybackslash}m{109.6pt}|}{\textbf{0.138}} \\* \hline\noalign{\penalty10000}
\multicolumn{2}{|>{\centering\arraybackslash}m{133.2pt}|}{\textbf{MAX}} & \textbf{0.0} & \multicolumn{2}{>{\centering\arraybackslash}m{111.1pt}|}{\textbf{-0.202}} & \multicolumn{2}{>{\centering\arraybackslash}m{109.6pt}|}{\textbf{0.289}} \\* \hline\noalign{\penalty10000}
\multicolumn{2}{|>{\centering\arraybackslash}m{133.2pt}|}{\textbf{MA\%D}} & \textbf{0.0} & \multicolumn{2}{>{\centering\arraybackslash}b{111.1pt}|}{\textbf{3.4 \%}} & \multicolumn{2}{>{\centering\arraybackslash}b{109.6pt}|}{\textbf{3.3 \% }} \\* \specialrule{0.9pt}{0pt}{0pt}\noalign{\penalty10000}
\end{tabular}}}}
\par\nopagebreak\vspace{1pt}{\setstretch{1}\setlength{\parskip}{1.5pt}\fontsize{9}{10.5}\selectfont\raggedright\noindent \textsuperscript{b}Convergence error during optimization of respective F-bonded complexes.\par}
\end{table}

\begin{table}[!htbp]\centering\setlength{\parskip}{0pt}
\caption{Deviation (\ensuremath{\delta}\textsubscript{\ensuremath{\angle}}, \textdegree{}) and percentage deviation (\%\ensuremath{\delta}\textsubscript{\ensuremath{\angle}}) on F-bond angles [\ensuremath{\angle}DF\textperiodcentered{}\textperiodcentered{}\textperiodcentered{}C/\allowbreak{}N] for the F-bonded complexes DF\textperiodcentered{}\textperiodcentered{}\textperiodcentered{}CH\textsubscript{3}/NH\textsubscript{3}, calculated at MP2/\allowbreak{}AVDZ and M06-2X/\allowbreak{}AVTZ, relative to reference values obtained at DF-RMP2-F12/\allowbreak{}VDZ-F12.}\label{tab:deviation-and-percentage-deviation}\vspace{7pt}\nopagebreak
{\fontsize{8.0}{9.2}\fontfamily{ptm}\selectfont\renewcommand{\arraystretch}{1.12}\setlength{\tabcolsep}{1.5pt}
\centerline{\adjustbox{max width=518pt}{\begin{tabular}{|>{\centering\arraybackslash}m{60.6pt}|>{\centering\arraybackslash}m{60.5pt}|>{\centering\arraybackslash}b{23.6pt}|>{\centering\arraybackslash}b{33.3pt}|>{\centering\arraybackslash}b{21.8pt}|>{\centering\arraybackslash}b{32.1pt}|>{\centering\arraybackslash}m{60.6pt}|>{\centering\arraybackslash}m{58.5pt}|>{\centering\arraybackslash}b{27.8pt}|>{\centering\arraybackslash}b{31.1pt}|>{\centering\arraybackslash}b{18.2pt}|>{\centering\arraybackslash}b{32.0pt}|}
\specialrule{0.9pt}{0pt}{0pt}
\multicolumn{6}{|>{\centering\arraybackslash}m{246.9pt}|}{\textbf{DF\textperiodcentered{}\textperiodcentered{}\textperiodcentered{}CH}\textbf{\textsubscript{3}}} & \multicolumn{6}{>{\centering\arraybackslash}m{243.3pt}|}{\textbf{DF\textperiodcentered{}\textperiodcentered{}\textperiodcentered{}NH}\textbf{\textsubscript{3}}} \\* \hline\noalign{\penalty10000}
\multicolumn{1}{|>{\centering\arraybackslash}m{60.6pt}|}{\textbf{DF\textperiodcentered{}\textperiodcentered{}\textperiodcentered{}CH}\textbf{\textsubscript{3}}} & \textbf{DF-RMP2-F12/\allowbreak{}VDZ-F12} & \multicolumn{2}{>{\centering\arraybackslash}m{59.9pt}|}{\textbf{MP2/\allowbreak{}AVDZ}} & \multicolumn{2}{>{\centering\arraybackslash}m{56.9pt}|}{\textbf{M06-2X/\allowbreak{}AVTZ}} & \textbf{DF\textperiodcentered{}\textperiodcentered{}\textperiodcentered{}NH}\textbf{\textsubscript{3}} & \textbf{DF-RMP2-F12/\allowbreak{}VDZ-F12} & \multicolumn{2}{>{\centering\arraybackslash}m{61.9pt}|}{\textbf{MP2/\allowbreak{}AVDZ}} & \multicolumn{2}{>{\centering\arraybackslash}m{53.2pt}|}{\textbf{M06-2X/\allowbreak{}AVTZ}} \\* \hline\noalign{\penalty10000}
\multicolumn{1}{|>{\centering\arraybackslash}m{60.6pt}|}{} & \ensuremath{\angle} & \ensuremath{\delta}\textsubscript{\ensuremath{\angle}} & \% \ensuremath{\delta}\textsubscript{\ensuremath{\angle}} & \ensuremath{\delta}\textsubscript{\ensuremath{\angle}} & \%\ensuremath{\delta}\textsubscript{\ensuremath{\angle}} &  & \ensuremath{\angle} & \ensuremath{\delta}\textsubscript{\ensuremath{\angle}} & \% \ensuremath{\delta}\textsubscript{\ensuremath{\angle}} & \ensuremath{\delta}\textsubscript{\ensuremath{\angle}} & \% \ensuremath{\delta}\textsubscript{\ensuremath{\angle}} \\ \hline
\multicolumn{1}{|>{\centering\arraybackslash}m{60.6pt}|}{F\textsubscript{2}\textperiodcentered{}\textperiodcentered{}\textperiodcentered{}CH\textsubscript{3}} & 178.8 & 1.2 & 0.7 & 0.8 & 0.4 & F\textsubscript{2}\textperiodcentered{}\textperiodcentered{}\textperiodcentered{}NH\textsubscript{3} & 180.0 & 0.0 & 0.0 & -1.9 & -1.1 \\ \hline
\multicolumn{1}{|>{\centering\arraybackslash}m{60.6pt}|}{SF\textsubscript{6}\textperiodcentered{}\textperiodcentered{}\textperiodcentered{}CH\textsubscript{3}} & 180.0 & -0.2 & -0.1 & -0.9 & -0.5 & SF\textsubscript{6}\textperiodcentered{}\textperiodcentered{}\textperiodcentered{}NH\textsubscript{3} & 180.0 & -2.8 & -1.6 & -3.0 & -1.7 \\ \hline
\multicolumn{1}{|>{\centering\arraybackslash}m{60.6pt}|}{NCF\textperiodcentered{}\textperiodcentered{}\textperiodcentered{}CH\textsubscript{3}} & 180.0 & -0.1 & -0.1 & -0.1 & -0.1 & NCF\textperiodcentered{}\textperiodcentered{}\textperiodcentered{}NH\textsubscript{3} & 180.0 & 0.0 & 0.0 & 0.0 & 0.0 \\ \hline
\multicolumn{1}{|>{\centering\arraybackslash}m{60.6pt}|}{CNF\textperiodcentered{}\textperiodcentered{}\textperiodcentered{}CH\textsubscript{3}} & 179.6 & 0.4 & 0.2 & 0.4 & 0.2 & CNF\textperiodcentered{}\textperiodcentered{}\textperiodcentered{}NH\textsubscript{3} & 178.7 & 1.3 & 0.7 & 1.3 & 0.7 \\ \hline
\multicolumn{1}{|>{\centering\arraybackslash}m{60.6pt}|}{NF\textsubscript{3}\textperiodcentered{}\textperiodcentered{}\textperiodcentered{}CH\textsubscript{3}} & 179.4 & -0.1 & -0.1 & -9.8 & -5.5 & NF\textsubscript{3}\textperiodcentered{}\textperiodcentered{}\textperiodcentered{}NH\textsubscript{3} & 179.1 & b & b & -10.9 & -6.1 \\ \hline
\multicolumn{1}{|>{\centering\arraybackslash}m{60.6pt}|}{ClF\textsubscript{3}\textperiodcentered{}\textperiodcentered{}\textperiodcentered{}CH\textsubscript{3}} & 169.8 & 8.7 & 5.1 & 9.8 & 5.8 & ClF\textsubscript{3}\textperiodcentered{}\textperiodcentered{}\textperiodcentered{}NH\textsubscript{3} & 179.4 & 0.0 & 0.0 & 0.3 & 0.2 \\ \hline
\multicolumn{1}{|>{\centering\arraybackslash}m{60.6pt}|}{OF\textsubscript{2}\textperiodcentered{}\textperiodcentered{}\textperiodcentered{}CH\textsubscript{3}} & 179.7 & -1.0 & -0.6 & -2.3 & -1.3 & OF\textsubscript{2}\textperiodcentered{}\textperiodcentered{}\textperiodcentered{}NH\textsubscript{3} & 176.7 & -0.5 & -0.3 & -19.6 & -11.1 \\ \hline
\multicolumn{1}{|>{\centering\arraybackslash}m{60.6pt}|}{NF\textsubscript{5}\textperiodcentered{}\textperiodcentered{}\textperiodcentered{}CH\textsubscript{3}} & 173.9 & 3.6 & 2.1 & 2.7 & 1.6 & NF\textsubscript{5}\textperiodcentered{}\textperiodcentered{}\textperiodcentered{}NH\textsubscript{3} & 178.8 & 0.1 & 0.1 & -5.5 & -3.1 \\ \hline
\multicolumn{1}{|>{\centering\arraybackslash}m{60.6pt}|}{FCCF\textperiodcentered{}\textperiodcentered{}\textperiodcentered{}CH\textsubscript{3}} & 178.0 & 1.9 & 1.1 & 1.7 & 1.0 & FCCF\textperiodcentered{}\textperiodcentered{}\textperiodcentered{}NH\textsubscript{3} & 179.6 & 0.4 & 0.2 & 0.4 & 0.2 \\ \hline
\multicolumn{1}{|>{\centering\arraybackslash}m{60.6pt}|}{CF\textsubscript{3}CCF\textperiodcentered{}\textperiodcentered{}\textperiodcentered{}CH\textsubscript{3}} & 180.0 & -0.4 & -0.2 & -0.5 & -0.3 & CF\textsubscript{3}CCF\textperiodcentered{}\textperiodcentered{}\textperiodcentered{}NH\textsubscript{3} & 180.0 & 0.0 & 0.0 & 0.0 & 0.0 \\ \hline
\multicolumn{1}{|>{\centering\arraybackslash}m{60.6pt}|}{CF\textsubscript{3}OF\textperiodcentered{}\textperiodcentered{}\textperiodcentered{}CH\textsubscript{3}} & 173.6 & b & b & 3.5 & 2.0 & CF\textsubscript{3}OF\textperiodcentered{}\textperiodcentered{}\textperiodcentered{}NH\textsubscript{3} & 178.8 & -0.9 & -0.5 & -2.8 & -1.6 \\ \hline
\multicolumn{1}{|>{\centering\arraybackslash}m{60.6pt}|}{F\textsubscript{3}NO\textperiodcentered{}\textperiodcentered{}\textperiodcentered{}CH\textsubscript{3}} & 162.7 & 12.4 & 7.6 & 14.4 & 8.9 & F\textsubscript{3}NO\textperiodcentered{}\textperiodcentered{}\textperiodcentered{}NH\textsubscript{3} & b & b & b & b & b \\ \hline
\multicolumn{1}{|>{\centering\arraybackslash}m{60.6pt}|}{SF\textsubscript{5}\textperiodcentered{}\textperiodcentered{}\textperiodcentered{}CH\textsubscript{3}} & 180.0 & -0.8 & -0.4 & -1.5 & -0.8 & SF\textsubscript{5}\textperiodcentered{}\textperiodcentered{}\textperiodcentered{}NH\textsubscript{3} & 179.7 & 0.2 & 0.1 & -7.3 & -4.1 \\ \hline
\multicolumn{1}{|>{\centering\arraybackslash}m{60.6pt}|}{NH\textsubscript{3}F+\textperiodcentered{}\textperiodcentered{}\textperiodcentered{}CH\textsubscript{3}} & 179.4 & 0.6 & 0.3 & -0.7 & -0.4 & NH\textsubscript{3}F+\textperiodcentered{}\textperiodcentered{}\textperiodcentered{}NH\textsubscript{3} & 180.0 & 0.0 & 0.0 & 0.0 & 0.0 \\ \hline
\multicolumn{1}{|>{\centering\arraybackslash}m{60.6pt}|}{NH\textsubscript{2}F\textsubscript{2}\textsuperscript{+}\textperiodcentered{}\textperiodcentered{}\textperiodcentered{}CH\textsubscript{3}} & 176.4 & 0.4 & 0.2 & 2.5 & 1.4 & NH\textsubscript{2}F\textsubscript{2}\textsuperscript{+}\textperiodcentered{}\textperiodcentered{}\textperiodcentered{}NH\textsubscript{3} & 177.3 & 0.9 & 0.5 & 0.8 & 0.5 \\ \hline
\multicolumn{1}{|>{\centering\arraybackslash}m{60.6pt}|}{\textbf{MSE}} & \textbf{0.0} & \multicolumn{2}{>{\centering\arraybackslash}m{59.9pt}|}{\textbf{1.9}} & \multicolumn{2}{>{\centering\arraybackslash}m{56.9pt}|}{\textbf{1.3}} & \textbf{MSE} & \textbf{0.0} & \multicolumn{2}{>{\centering\arraybackslash}m{61.9pt}|}{\textbf{-0.1}} & \multicolumn{2}{>{\centering\arraybackslash}m{53.2pt}|}{\textbf{-3.4}} \\ \hline
\multicolumn{1}{|>{\centering\arraybackslash}m{60.6pt}|}{\textbf{MAD}} & \textbf{0.0} & \multicolumn{2}{>{\centering\arraybackslash}m{59.9pt}|}{\textbf{2.3}} & \multicolumn{2}{>{\centering\arraybackslash}m{56.9pt}|}{\textbf{3.3}} & \textbf{MAD} & \textbf{0.0} & \multicolumn{2}{>{\centering\arraybackslash}m{61.9pt}|}{\textbf{0.5}} & \multicolumn{2}{>{\centering\arraybackslash}m{53.2pt}|}{\textbf{3.8}} \\ \hline
\multicolumn{1}{|>{\centering\arraybackslash}m{60.6pt}|}{\textbf{RMSD}} & \textbf{0.0} & \multicolumn{2}{>{\centering\arraybackslash}m{59.9pt}|}{\textbf{4.2}} & \multicolumn{2}{>{\centering\arraybackslash}m{56.9pt}|}{\textbf{5.4}} & \textbf{RMSD} & \textbf{0.0} & \multicolumn{2}{>{\centering\arraybackslash}m{61.9pt}|}{\textbf{0.9}} & \multicolumn{2}{>{\centering\arraybackslash}m{53.2pt}|}{\textbf{6.6}} \\* \hline\noalign{\penalty10000}
\multicolumn{1}{|>{\centering\arraybackslash}m{60.6pt}|}{\textbf{MAX}} & \textbf{0.0} & \multicolumn{2}{>{\centering\arraybackslash}m{59.9pt}|}{\textbf{12.4}} & \multicolumn{2}{>{\centering\arraybackslash}m{56.9pt}|}{\textbf{14.4}} & \textbf{MAX} & \textbf{0.0} & \multicolumn{2}{>{\centering\arraybackslash}m{61.9pt}|}{\textbf{-2.8}} & \multicolumn{2}{>{\centering\arraybackslash}m{53.2pt}|}{\textbf{-19.6}} \\* \hline\noalign{\penalty10000}
\multicolumn{1}{|>{\centering\arraybackslash}m{60.6pt}|}{\textbf{MA\%D}} & \textbf{0.0} & \multicolumn{2}{>{\centering\arraybackslash}b{59.9pt}|}{\textbf{1.3 \%}} & \multicolumn{2}{>{\centering\arraybackslash}b{56.9pt}|}{\textbf{1.2 \%}} & \textbf{MA\%D} & \textbf{0.0} & \multicolumn{2}{>{\centering\arraybackslash}b{61.9pt}|}{\textbf{0.3 \%}} & \multicolumn{2}{>{\centering\arraybackslash}b{53.2pt}|}{\textbf{2.2 \%}} \\* \specialrule{0.9pt}{0pt}{0pt}\noalign{\penalty10000}
\end{tabular}}}}
\par\nopagebreak\vspace{1pt}{\setstretch{1}\setlength{\parskip}{1.5pt}\fontsize{9}{10.5}\selectfont\raggedright\noindent \textsuperscript{b}Convergence error during optimization of respective F-bonded complexes.\par}
\end{table}

\begin{table}[!htbp]\centering\setlength{\parskip}{0pt}
\caption{BEs (kJ/\allowbreak{}mol) and F-bond distances [R(F\textperiodcentered{}\textperiodcentered{}\textperiodcentered{}C/\allowbreak{}N), Å] for the F-bonded complexes DF\textperiodcentered{}\textperiodcentered{}\textperiodcentered{}CH\textsubscript{3}/NH\textsubscript{3}, calculated at DF-RMP2-F12/\allowbreak{}CBS//DF-RMP2-F12/\allowbreak{}VDZ-F12, DF-RMP2-F12/\allowbreak{}VDZ-F12, MP2/\allowbreak{}AVDZ, and M06-2X/\allowbreak{}AVTZ levels of theory. These data were used to generate the plots in \textbf{F}\textbf{IG.}\textbf{ 3}.}\label{tab:bes-kj-mol-and}\vspace{7pt}\nopagebreak
{\fontsize{7.4}{8.5}\fontfamily{ptm}\selectfont\renewcommand{\arraystretch}{1.17}\setlength{\tabcolsep}{1.5pt}
\centerline{\adjustbox{max width=518pt}{\begin{tabular}{|>{\centering\arraybackslash}m{66.5pt}|>{\centering\arraybackslash}m{17.3pt}|>{\centering\arraybackslash}m{18.6pt}|>{\centering\arraybackslash}m{8.0pt}|>{\centering\arraybackslash}m{9.6pt}|>{\centering\arraybackslash}m{8.0pt}|>{\centering\arraybackslash}m{10.5pt}|>{\centering\arraybackslash}m{8.0pt}|>{\centering\arraybackslash}m{9.3pt}|>{\centering\arraybackslash}m{8.0pt}|>{\centering\arraybackslash}m{10.1pt}|>{\centering\arraybackslash}m{8.0pt}|>{\centering\arraybackslash}m{10.9pt}|>{\centering\arraybackslash}m{19.4pt}|>{\centering\arraybackslash}m{66.5pt}|>{\centering\arraybackslash}m{17.5pt}|>{\centering\arraybackslash}m{19.0pt}|>{\centering\arraybackslash}m{17.3pt}|>{\centering\arraybackslash}m{18.9pt}|>{\centering\arraybackslash}m{17.4pt}|>{\centering\arraybackslash}m{19.4pt}|>{\centering\arraybackslash}m{18.6pt}|>{\centering\arraybackslash}m{19.2pt}|}
\specialrule{0.9pt}{0pt}{0pt}
\multicolumn{14}{|>{\centering\arraybackslash}m{251.1pt}|}{\textbf{DF\textperiodcentered{}\textperiodcentered{}\textperiodcentered{}CH}\textbf{\textsubscript{3}}} & \multicolumn{9}{>{\centering\arraybackslash}m{237.9pt}|}{\textbf{DF\textperiodcentered{}\textperiodcentered{}\textperiodcentered{}NH}\textbf{\textsubscript{3}}} \\* \hline\noalign{\penalty10000}
\multicolumn{1}{|>{\centering\arraybackslash}m{66.5pt}|}{\multirow{2}{66.5pt}{\centering \textbf{DF\textperiodcentered{}\textperiodcentered{}\textperiodcentered{}} \textbf{CH}\textbf{\textsubscript{3}}}} & \multicolumn{2}{>{\centering\arraybackslash}m{38.9pt}|}{BE[DF-RMP2-F12/\allowbreak{}CBS//DF-RMP2-F12/\allowbreak{}VDZ-F12]} & \multicolumn{4}{>{\centering\arraybackslash}m{45.1pt}|}{DF-RMP2-F12/\allowbreak{}VDZ-F12} & \multicolumn{4}{>{\centering\arraybackslash}m{44.4pt}|}{MP2/\allowbreak{}AVDZ} & \multicolumn{3}{>{\centering\arraybackslash}m{44.3pt}|}{M06-2X/\allowbreak{}AVTZ (NO BSSE)} & \textbf{DF\textperiodcentered{}\textperiodcentered{}\textperiodcentered{}NH}\textbf{\textsubscript{3}} & \multicolumn{2}{>{\centering\arraybackslash}m{39.5pt}|}{BE[DF-RMP2-F12/\allowbreak{}CBS//DF-RMP2-F12/\allowbreak{}VDZ-F12]} & \multicolumn{2}{>{\centering\arraybackslash}m{39.2pt}|}{DF-RMP2-F12/\allowbreak{}VDZ-F12} & \multicolumn{2}{>{\centering\arraybackslash}m{39.8pt}|}{MP2/\allowbreak{}AVDZ} & \multicolumn{2}{>{\centering\arraybackslash}m{40.9pt}|}{M06-2X/\allowbreak{}AVTZ (NO BSSE)} \\* \cline{2-23}\noalign{\penalty10000}
\multicolumn{1}{|>{\centering\arraybackslash}m{66.5pt}|}{} & \makebox[\linewidth][c]{\rotatebox{90}{\hspace{2pt}BE\hspace{2pt}}} & \multicolumn{2}{>{\centering\arraybackslash}m{29.6pt}|}{\makebox[\linewidth][c]{\rotatebox{90}{\hspace{2pt}R\hspace{2pt}}}} & \multicolumn{2}{>{\centering\arraybackslash}m{20.6pt}|}{\makebox[\linewidth][c]{\rotatebox{90}{\hspace{2pt}BE\hspace{2pt}}}} & \multicolumn{2}{>{\centering\arraybackslash}m{21.5pt}|}{\makebox[\linewidth][c]{\rotatebox{90}{\hspace{2pt}R\hspace{2pt}}}} & \multicolumn{2}{>{\centering\arraybackslash}m{20.3pt}|}{\makebox[\linewidth][c]{\rotatebox{90}{\hspace{2pt}BE\hspace{2pt}}}} & \multicolumn{2}{>{\centering\arraybackslash}m{21.1pt}|}{\makebox[\linewidth][c]{\rotatebox{90}{\hspace{2pt}R\hspace{2pt}}}} & \makebox[\linewidth][c]{\rotatebox{90}{\hspace{2pt}BE\hspace{2pt}}} & \makebox[\linewidth][c]{\rotatebox{90}{\hspace{2pt}R\hspace{2pt}}} &  & \makebox[\linewidth][c]{\rotatebox{90}{\hspace{2pt}BE\hspace{2pt}}} & \makebox[\linewidth][c]{\rotatebox{90}{\hspace{2pt}R\hspace{2pt}}} & \makebox[\linewidth][c]{\rotatebox{90}{\hspace{2pt}BE\hspace{2pt}}} & \makebox[\linewidth][c]{\rotatebox{90}{\hspace{2pt}R\hspace{2pt}}} & \makebox[\linewidth][c]{\rotatebox{90}{\hspace{2pt}BE\hspace{2pt}}} & \makebox[\linewidth][c]{\rotatebox{90}{\hspace{2pt}R\hspace{2pt}}} & \makebox[\linewidth][c]{\rotatebox{90}{\hspace{2pt}BE\hspace{2pt}}} & \makebox[\linewidth][c]{\rotatebox{90}{\hspace{2pt}R\hspace{2pt}}} \\* \hline\noalign{\penalty10000}
\multicolumn{1}{|>{\centering\arraybackslash}m{66.5pt}|}{\textbf{F}\textbf{\textsubscript{2}}\textbf{\textperiodcentered{}\textperiodcentered{}\textperiodcentered{}CH}\textbf{\textsubscript{3}}} & -3.4 & 2.869 & \multicolumn{2}{>{\centering\arraybackslash}m{20.6pt}|}{-3.0} & \multicolumn{2}{>{\centering\arraybackslash}m{21.5pt}|}{2.869} & \multicolumn{2}{>{\centering\arraybackslash}m{20.3pt}|}{-2.4} & \multicolumn{2}{>{\centering\arraybackslash}m{21.1pt}|}{2.865} & \multicolumn{2}{>{\centering\arraybackslash}m{21.9pt}|}{-7.9} & 2.648 & \textbf{ F}\textbf{\textsubscript{2}}\textbf{\textperiodcentered{}\textperiodcentered{}\textperiodcentered{}NH}\textbf{\textsubscript{3}} & -7.8 & 2.616 & -7.6 & 2.616 & -8.0 & 2.442 & -6.4 & 2.596 \\* \hline\noalign{\penalty10000}
\multicolumn{1}{|>{\centering\arraybackslash}m{66.5pt}|}{\textbf{SF}\textbf{\textsubscript{6}}\textbf{\textperiodcentered{}\textperiodcentered{}\textperiodcentered{}CH}\textbf{\textsubscript{3}}} & -1.8 & 3.327 & \multicolumn{2}{>{\centering\arraybackslash}m{20.6pt}|}{-1.5} & \multicolumn{2}{>{\centering\arraybackslash}m{21.5pt}|}{3.327} & \multicolumn{2}{>{\centering\arraybackslash}m{20.3pt}|}{-1.1} & \multicolumn{2}{>{\centering\arraybackslash}m{21.1pt}|}{3.255} & \multicolumn{2}{>{\centering\arraybackslash}m{21.9pt}|}{-2.5} & 3.021 & \textbf{SF}\textbf{\textsubscript{6}}\textbf{\textperiodcentered{}\textperiodcentered{}\textperiodcentered{}NH}\textbf{\textsubscript{3}} & -1.4 & 3.345 & -1.4 & 3.345 & -1.3 & 3.166 & -1.4 & 3.634 \\* \hline\noalign{\penalty10000}
\multicolumn{1}{|>{\centering\arraybackslash}m{66.5pt}|}{\textbf{NCF\textperiodcentered{}\textperiodcentered{}\textperiodcentered{}CH}\textbf{\textsubscript{3}}} & -1.7 & 3.428 & \multicolumn{2}{>{\centering\arraybackslash}m{20.6pt}|}{-1.5} & \multicolumn{2}{>{\centering\arraybackslash}m{21.5pt}|}{3.428} & \multicolumn{2}{>{\centering\arraybackslash}m{20.3pt}|}{-1.2} & \multicolumn{2}{>{\centering\arraybackslash}m{21.1pt}|}{3.388} & \multicolumn{2}{>{\centering\arraybackslash}m{21.9pt}|}{-3.1} & 3.018 & \textbf{NCF\textperiodcentered{}\textperiodcentered{}\textperiodcentered{}NH}\textbf{\textsubscript{3}} & -5.1 & 3.160 & -5.0 & 3.160 & -4.4 & 3.141 & -4.7 & 3.205 \\* \hline\noalign{\penalty10000}
\multicolumn{1}{|>{\centering\arraybackslash}m{66.5pt}|}{\textbf{CNF\textperiodcentered{}\textperiodcentered{}\textperiodcentered{}CH}\textbf{\textsubscript{3}}} & -2.4 & 3.262 & \multicolumn{2}{>{\centering\arraybackslash}m{20.6pt}|}{-2.1} & \multicolumn{2}{>{\centering\arraybackslash}m{21.5pt}|}{3.262} & \multicolumn{2}{>{\centering\arraybackslash}m{20.3pt}|}{-1.7} & \multicolumn{2}{>{\centering\arraybackslash}m{21.1pt}|}{3.185} & \multicolumn{2}{>{\centering\arraybackslash}m{21.9pt}|}{-4.2} & 2.976 & \textbf{CNF\textperiodcentered{}\textperiodcentered{}\textperiodcentered{}NH}\textbf{\textsubscript{3}} & -6.9 & 2.960 & -6.7 & 2.960 & -6.1 & 2.924 & -5.4 & 3.031 \\ \hline
\multicolumn{1}{|>{\centering\arraybackslash}m{66.5pt}|}{\textbf{NF}\textbf{\textsubscript{3}}\textbf{\textperiodcentered{}\textperiodcentered{}\textperiodcentered{}CH}\textbf{\textsubscript{3}}} & -1.7 & 3.295 & \multicolumn{2}{>{\centering\arraybackslash}m{20.6pt}|}{-1.4} & \multicolumn{2}{>{\centering\arraybackslash}m{21.5pt}|}{3.295} & \multicolumn{2}{>{\centering\arraybackslash}m{20.3pt}|}{-1.0} & \multicolumn{2}{>{\centering\arraybackslash}m{21.1pt}|}{3.247} & \multicolumn{2}{>{\centering\arraybackslash}m{21.9pt}|}{-2.7} & 3.002 & \textbf{NF}\textbf{\textsubscript{3}}\textbf{\textperiodcentered{}\textperiodcentered{}\textperiodcentered{}NH}\textbf{\textsubscript{3}} & -1.9 & 3.209 & -1.7 & 3.209 & b & b & -1.2 & 3.448 \\ \hline
\multicolumn{1}{|>{\centering\arraybackslash}m{66.5pt}|}{\textbf{ClF}\textbf{\textsubscript{3}}\textbf{\textperiodcentered{}\textperiodcentered{}\textperiodcentered{}CH}\textbf{\textsubscript{3}}} & -2.4 & 3.160 & \multicolumn{2}{>{\centering\arraybackslash}m{20.6pt}|}{-2.1} & \multicolumn{2}{>{\centering\arraybackslash}m{21.5pt}|}{3.160} & \multicolumn{2}{>{\centering\arraybackslash}m{20.3pt}|}{-1.9} & \multicolumn{2}{>{\centering\arraybackslash}m{21.1pt}|}{2.972} & \multicolumn{2}{>{\centering\arraybackslash}m{21.9pt}|}{-5.2} & 2.824 & \textbf{ClF}\textbf{\textsubscript{3}}\textbf{\textperiodcentered{}\textperiodcentered{}\textperiodcentered{}NH}\textbf{\textsubscript{3}} & -4.5 & 2.965 & -4.1 & 2.965 & -4.5 & 2.763 & -3.5 & 2.924 \\ \hline
\multicolumn{1}{|>{\centering\arraybackslash}m{66.5pt}|}{\textbf{OF}\textbf{\textsubscript{2}}\textbf{\textperiodcentered{}\textperiodcentered{}\textperiodcentered{}CH}\textbf{\textsubscript{3}}} & -2.0 & 3.212 & \multicolumn{2}{>{\centering\arraybackslash}m{20.6pt}|}{-1.7} & \multicolumn{2}{>{\centering\arraybackslash}m{21.5pt}|}{3.212} & \multicolumn{2}{>{\centering\arraybackslash}m{20.3pt}|}{-1.2} & \multicolumn{2}{>{\centering\arraybackslash}m{21.1pt}|}{3.172} & \multicolumn{2}{>{\centering\arraybackslash}m{21.9pt}|}{-4.4} & 2.928 & \textbf{OF}\textbf{\textsubscript{2}}\textbf{\textperiodcentered{}\textperiodcentered{}\textperiodcentered{}NH}\textbf{\textsubscript{3}} & -3.4 & 2.942 & -3.1 & 2.942 & -2.9 & 2.838 & -2.8 & 2.973 \\ \hline
\multicolumn{1}{|>{\centering\arraybackslash}m{66.5pt}|}{\textbf{NF}\textbf{\textsubscript{5}}\textbf{\textperiodcentered{}\textperiodcentered{}\textperiodcentered{}CH}\textbf{\textsubscript{3}}} & -3.1 & 3.092 & \multicolumn{2}{>{\centering\arraybackslash}m{20.6pt}|}{-2.7} & \multicolumn{2}{>{\centering\arraybackslash}m{21.5pt}|}{3.092} & \multicolumn{2}{>{\centering\arraybackslash}m{20.3pt}|}{-2.1} & \multicolumn{2}{>{\centering\arraybackslash}m{21.1pt}|}{3.051} & \multicolumn{2}{>{\centering\arraybackslash}m{21.9pt}|}{-5.1} & 2.928 & \textbf{NF}\textbf{\textsubscript{5}}\textbf{\textperiodcentered{}\textperiodcentered{}\textperiodcentered{}NH}\textbf{\textsubscript{3}} & -7.3 & 2.846 & -7.1 & 2.846 & -6.7 & 2.776 & -6.6 & 2.849 \\ \hline
\multicolumn{1}{|>{\centering\arraybackslash}m{66.5pt}|}{\textbf{FCCF\textperiodcentered{}\textperiodcentered{}\textperiodcentered{}CH}\textbf{\textsubscript{3}}} & -1.5 & 3.346 & \multicolumn{2}{>{\centering\arraybackslash}m{20.6pt}|}{-1.3} & \multicolumn{2}{>{\centering\arraybackslash}m{21.5pt}|}{3.346} & \multicolumn{2}{>{\centering\arraybackslash}m{20.3pt}|}{-0.9} & \multicolumn{2}{>{\centering\arraybackslash}m{21.1pt}|}{3.243} & \multicolumn{2}{>{\centering\arraybackslash}m{21.9pt}|}{-2.1} & 3.022 & \textbf{FCCF\textperiodcentered{}\textperiodcentered{}\textperiodcentered{}NH}\textbf{\textsubscript{3}} & -0.6 & 3.444 & -0.6 & 3.444 & -0.2 & 3.312 & -0.6 & 3.648 \\ \hline
\multicolumn{1}{|>{\centering\arraybackslash}m{66.5pt}|}{\textbf{CF}\textbf{\textsubscript{3}}\textbf{CCF\textperiodcentered{}\textperiodcentered{}\textperiodcentered{}CH}\textbf{\textsubscript{3}}} & -1.5 & 3.457 & \multicolumn{2}{>{\centering\arraybackslash}m{20.6pt}|}{-1.3} & \multicolumn{2}{>{\centering\arraybackslash}m{21.5pt}|}{3.457} & \multicolumn{2}{>{\centering\arraybackslash}m{20.3pt}|}{-0.9} & \multicolumn{2}{>{\centering\arraybackslash}m{21.1pt}|}{3.308} & \multicolumn{2}{>{\centering\arraybackslash}m{21.9pt}|}{-2.7} & 3.023 & \textbf{CF}\textbf{\textsubscript{3}}\textbf{CCF\textperiodcentered{}\textperiodcentered{}\textperiodcentered{}NH}\textbf{\textsubscript{3}} & -3.6 & 3.227 & -3.6 & 3.227 & -3.2 & 3.153 & -3.4 & 3.409 \\ \hline
\multicolumn{1}{|>{\centering\arraybackslash}m{66.5pt}|}{\textbf{CF}\textbf{\textsubscript{3}}\textbf{OF\textperiodcentered{}\textperiodcentered{}\textperiodcentered{}CH}\textbf{\textsubscript{3}}} & -2.5 & 3.088 & \multicolumn{2}{>{\centering\arraybackslash}m{20.6pt}|}{-2.1} & \multicolumn{2}{>{\centering\arraybackslash}m{21.5pt}|}{3.088} & \multicolumn{2}{>{\centering\arraybackslash}m{20.3pt}|}{b} & \multicolumn{2}{>{\centering\arraybackslash}m{21.1pt}|}{b} & \multicolumn{2}{>{\centering\arraybackslash}m{21.9pt}|}{-4.7} & 2.891 & \textbf{CF}\textbf{\textsubscript{3}}\textbf{OF\textperiodcentered{}\textperiodcentered{}\textperiodcentered{}NH}\textbf{\textsubscript{3}} & -4.3 & 2.869 & -4.1 & 2.869 & -4.0 & 2.763 & -2.3 & 2.932 \\ \hline
\multicolumn{1}{|>{\centering\arraybackslash}m{66.5pt}|}{\textbf{F}\textbf{\textsubscript{3}}\textbf{NO\textperiodcentered{}\textperiodcentered{}\textperiodcentered{}CH}\textbf{\textsubscript{3}}} & -2.1 & 3.234 & \multicolumn{2}{>{\centering\arraybackslash}m{20.6pt}|}{-1.8} & \multicolumn{2}{>{\centering\arraybackslash}m{21.5pt}|}{3.234} & \multicolumn{2}{>{\centering\arraybackslash}m{20.3pt}|}{-1.3} & \multicolumn{2}{>{\centering\arraybackslash}m{21.1pt}|}{3.200} & \multicolumn{2}{>{\centering\arraybackslash}m{21.9pt}|}{-3.6} & 2.970 & \textbf{F}\textbf{\textsubscript{3}}\textbf{NO\textperiodcentered{}\textperiodcentered{}\textperiodcentered{}NH}\textbf{\textsubscript{3}} & b & b & b & b & -2.5 & 2.935 & -2.3 & 3.186 \\ \hline
\multicolumn{1}{|>{\centering\arraybackslash}m{66.5pt}|}{\textbf{ SF}\textbf{\textsubscript{5}}\textbf{\textperiodcentered{}\textperiodcentered{}\textperiodcentered{}CH}\textbf{\textsubscript{3}}} & -1.8 & 3.329 & \multicolumn{2}{>{\centering\arraybackslash}m{20.6pt}|}{-1.4} & \multicolumn{2}{>{\centering\arraybackslash}m{21.5pt}|}{3.329} & \multicolumn{2}{>{\centering\arraybackslash}m{20.3pt}|}{-1.1} & \multicolumn{2}{>{\centering\arraybackslash}m{21.1pt}|}{3.242} & \multicolumn{2}{>{\centering\arraybackslash}m{21.9pt}|}{-3.0} & 3.021 & \textbf{SF}\textbf{\textsubscript{5}}\textbf{\textperiodcentered{}\textperiodcentered{}\textperiodcentered{}NH}\textbf{\textsubscript{3}} & -2.7 & 3.236 & -2.6 & 3.236 & -2.4 & 3.106 & -2.0 & 3.419 \\* \hline\noalign{\penalty10000}
\multicolumn{1}{|>{\centering\arraybackslash}m{66.5pt}|}{\textbf{NH}\textbf{\textsubscript{3}}\textbf{F+\textperiodcentered{}\textperiodcentered{}\textperiodcentered{}CH}\textbf{\textsubscript{3}}} & -9.8 & 2.947 & \multicolumn{2}{>{\centering\arraybackslash}m{20.6pt}|}{-9.5} & \multicolumn{2}{>{\centering\arraybackslash}m{21.5pt}|}{2.947} & \multicolumn{2}{>{\centering\arraybackslash}m{20.3pt}|}{-8.9} & \multicolumn{2}{>{\centering\arraybackslash}m{21.1pt}|}{2.940} & \multicolumn{2}{>{\centering\arraybackslash}m{21.9pt}|}{-15.5} & 2.693 & \textbf{NH}\textbf{\textsubscript{3}}\textbf{F+\textperiodcentered{}\textperiodcentered{}\textperiodcentered{}NH}\textbf{\textsubscript{3}} & -33.9 & 2.647 & -33.9 & 2.647 & -33.3 & 2.596 & -34.3 & 2.590 \\* \hline\noalign{\penalty10000}
\multicolumn{1}{|>{\centering\arraybackslash}m{66.5pt}|}{\textbf{NH}\textbf{\textsubscript{2}}\textbf{F}\textbf{\textsubscript{2}}\textbf{\textsuperscript{+}}\textbf{\textperiodcentered{}\textperiodcentered{}\textperiodcentered{}CH}\textbf{\textsubscript{3}}} & -11.4 & 2.941 & \multicolumn{2}{>{\centering\arraybackslash}m{20.6pt}|}{-11.0} & \multicolumn{2}{>{\centering\arraybackslash}m{21.5pt}|}{2.941} & \multicolumn{2}{>{\centering\arraybackslash}m{20.3pt}|}{-10.5} & \multicolumn{2}{>{\centering\arraybackslash}m{21.1pt}|}{2.919} & \multicolumn{2}{>{\centering\arraybackslash}m{21.9pt}|}{-16.5} & 2.713 & \textbf{NH}\textbf{\textsubscript{2}}\textbf{F}\textbf{\textsubscript{2}}\textbf{\textsuperscript{+}}\textbf{\textperiodcentered{}\textperiodcentered{}\textperiodcentered{}NH}\textbf{\textsubscript{3}} & -38.8 & 2.627 & -38.7 & 2.627 & -38.3 & 2.581 & -39.0 & 2.581 \\* \specialrule{0.9pt}{0pt}{0pt}\noalign{\penalty10000}
\end{tabular}}}}
\par\nopagebreak\vspace{1pt}{\setstretch{1}\setlength{\parskip}{1.5pt}\fontsize{9}{10.5}\selectfont\raggedright\noindent \textsuperscript{b}Convergence error during optimization of respective F-bonded complexes.\par}
\end{table}

\begin{table}[!htbp]\centering\setlength{\parskip}{0pt}
\caption{BSSE-uncorrected BEs (\ensuremath{\Delta}\textsubscript{1}) for selected F-bonded complexes DF\textperiodcentered{}\textperiodcentered{}\textperiodcentered{}CH\textsubscript{3}/NH\textsubscript{3} (DF = F\textsubscript{2}, NCF, CNF, NH\textsubscript{3}F\textsuperscript{+}, and NH\textsubscript{2}F\textsubscript{2}\textsuperscript{+}), calculated using the M06-2X functional with several basis sets, relative to reference BEs obtained at DF-RMP2-F12/\allowbreak{}CBS[V(T,Q)Z-F12]//DF-RMP2-F12/\allowbreak{}VDZ-F12 (denoted DF-RMP2-F12/\allowbreak{}CBS).}\label{tab:bsse-uncorrected-bes-1}\vspace{7pt}\nopagebreak
{\fontsize{7.83}{9.0}\fontfamily{ptm}\selectfont\renewcommand{\arraystretch}{1.17}\setlength{\tabcolsep}{1.5pt}
\centerline{\adjustbox{max width=518pt}{\begin{tabular}{|>{\centering\arraybackslash}m{85.6pt}|>{\centering\arraybackslash}m{21.6pt}|>{\centering\arraybackslash}m{31.3pt}|>{\centering\arraybackslash}m{31.3pt}|>{\centering\arraybackslash}m{38.5pt}|>{\centering\arraybackslash}m{42.1pt}|>{\centering\arraybackslash}m{70.4pt}|>{\centering\arraybackslash}m{21.6pt}|>{\centering\arraybackslash}m{31.3pt}|>{\centering\arraybackslash}m{31.3pt}|>{\centering\arraybackslash}m{38.5pt}|>{\centering\arraybackslash}m{42.1pt}|}
\specialrule{0.9pt}{0pt}{0pt}
\multicolumn{6}{|>{\centering\arraybackslash}m{265.2pt}|}{\textbf{DF\textperiodcentered{}\textperiodcentered{}\textperiodcentered{}CH}\textbf{\textsubscript{3}}} & \multicolumn{6}{>{\centering\arraybackslash}m{250.0pt}|}{\textbf{DF\textperiodcentered{}\textperiodcentered{}\textperiodcentered{}NH}\textbf{\textsubscript{3}}} \\* \hline\noalign{\penalty10000}
\multicolumn{1}{|>{\centering\arraybackslash}m{85.6pt}|}{\multirow{2}{85.6pt}{\centering \textbf{Basis sets}}} & \textbf{F}\textbf{\textsubscript{2}}\textbf{\textperiodcentered{}\textperiodcentered{}\textperiodcentered{}} \textbf{CH}\textbf{\textsubscript{3}} & \textbf{NCF\textperiodcentered{}\textperiodcentered{}\textperiodcentered{}} \textbf{CH}\textbf{\textsubscript{3}} & \textbf{CNF\textperiodcentered{}\textperiodcentered{}\textperiodcentered{}} \textbf{CH}\textbf{\textsubscript{3}} & \textbf{NH}\textbf{\textsubscript{3}}\textbf{F}\textbf{\textsuperscript{+}}\textbf{\textperiodcentered{}\textperiodcentered{}\textperiodcentered{}} \textbf{CH}\textbf{\textsubscript{3}} & \textbf{NH}\textbf{\textsubscript{2}}\textbf{F}\textbf{\textsubscript{2}}\textbf{\textsuperscript{+}}\textbf{\textperiodcentered{}\textperiodcentered{}\textperiodcentered{}} \textbf{CH}\textbf{\textsubscript{3}} & \multirow{2}{70.4pt}{\centering \textbf{Basis sets}} & \textbf{F}\textbf{\textsubscript{2}}\textbf{\textperiodcentered{}\textperiodcentered{}\textperiodcentered{}} \textbf{NH}\textbf{\textsubscript{3}} & \textbf{NCF\textperiodcentered{}\textperiodcentered{}\textperiodcentered{}} \textbf{NH}\textbf{\textsubscript{3}} & \textbf{CNF\textperiodcentered{}\textperiodcentered{}\textperiodcentered{}} \textbf{NH}\textbf{\textsubscript{3}} & \textbf{NH}\textbf{\textsubscript{3}}\textbf{F}\textbf{\textsuperscript{+}}\textbf{\textperiodcentered{}\textperiodcentered{}\textperiodcentered{}} \textbf{NH}\textbf{\textsubscript{3}} & \textbf{NH}\textbf{\textsubscript{2}}\textbf{F}\textbf{\textsubscript{2}}\textbf{\textsuperscript{+}}\textbf{\textperiodcentered{}\textperiodcentered{}\textperiodcentered{}} \textbf{NH}\textbf{\textsubscript{3}} \\* \cline{2-6}\cline{8-12}\noalign{\penalty10000}
\multicolumn{1}{|>{\centering\arraybackslash}m{85.6pt}|}{} & \ensuremath{\Delta}\textsubscript{1} & \ensuremath{\Delta}\textsubscript{1} & \ensuremath{\Delta}\textsubscript{1} & \ensuremath{\Delta}\textsubscript{1} & \ensuremath{\Delta}\textsubscript{1} &  & \ensuremath{\Delta}\textsubscript{1} & \ensuremath{\Delta}\textsubscript{1} & \ensuremath{\Delta}\textsubscript{1} & \ensuremath{\Delta}\textsubscript{1} & \ensuremath{\Delta}\textsubscript{1} \\* \hline\noalign{\penalty10000}
\multicolumn{1}{|>{\centering\arraybackslash}m{85.6pt}|}{VDZ} & -11.5 & -4.5 & -5.5 & -17.2 & -18.1 & VDZ & -12.7 & -7.8 & -8.7 & -41.6 & -46.7 \\* \hline\noalign{\penalty10000}
\multicolumn{1}{|>{\centering\arraybackslash}m{85.6pt}|}{AVDZ} & -9.3 & -3.5 & -5.1 & -17.2 & -18.2 & AVDZ & -9.6 & -5.6 & -7.3 & -37.8 & -42.6 \\* \hline\noalign{\penalty10000}
\multicolumn{1}{|>{\centering\arraybackslash}m{85.6pt}|}{VTZ} & -8.3 & -3.5 & -4.4 & -15.3 & -16.3 & VTZ & -8.6 & -5.9 & -6.8 & -37.3 & -42.3 \\* \hline\noalign{\penalty10000}
\multicolumn{1}{|>{\centering\arraybackslash}m{85.6pt}|}{AVTZ} & -7.9 & -3.1 & -4.2 & -15.5 & -16.5 & AVTZ & -6.4 & -4.7 & -5.4 & -34.3 & -39.0 \\ \hline
\multicolumn{1}{|>{\centering\arraybackslash}m{85.6pt}|}{VQZ} & -7.1 & -2.6 & -3.6 & -14.5 & -15.4 & VQZ & -6.5 & -5.0 & -5.7 & -35.2 & -39.7 \\ \hline
\multicolumn{1}{|>{\centering\arraybackslash}m{85.6pt}|}{AVQZ} & -7.0 & -2.6 & -3.6 & -14.7 & -15.6 & AVQZ & -5.6 & -4.4 & -5.0 & -33.7 & -38.2 \\ \hline
\multicolumn{1}{|>{\centering\arraybackslash}m{85.6pt}|}{V5Z} & -7.1 & -2.4 & -3.3 & -14.5 & -15.5 & V5Z & -5.8 & -4.5 & -5.1 & -34.1 & -38.7 \\ \hline
\multicolumn{1}{|>{\centering\arraybackslash}m{85.6pt}|}{AV5Z} & -6.8 & -2.4 & -3.3 & -14.6 & -15.5 & AV5Z & -5.5 & -4.4 & -5.0 & -33.9 & -38.4 \\ \hline
\multicolumn{1}{|>{\centering\arraybackslash}m{85.6pt}|}{Def2-TZVP} & -7.2 & -2.7 & -3.7 & -14.6 & -15.4 & Def2-TZVP & -8.0 & -5.7 & -6.5 & -37.2 & -42.0 \\ \hline
\multicolumn{1}{|>{\centering\arraybackslash}m{85.6pt}|}{Def2-QZVP} & -7.1 & -2.4 & -3.3 & -14.4 & -15.3 & Def2-QZVP & -6.1 & -4.6 & -5.3 & -34.3 & -38.9 \\ \hline
\multicolumn{1}{|>{\centering\arraybackslash}m{85.6pt}|}{6-311g(2d,2p)} & -10.1 & -4.8 & -5.8 & -17.0 & -18.1 & 6-311g(2d,2p) & -11.0 & -7.5 & -8.8 & -39.8 & -44.8 \\ \hline
\multicolumn{1}{|>{\centering\arraybackslash}m{85.6pt}|}{6-311++g(2d,2p)} & -9.4 & -3.6 & -4.8 & -17.1 & -18.2 & 6-311++g(2d,2p) & -9.3 & -5.9 & -7.2 & -37.5 & -42.5 \\ \hline
\multicolumn{1}{|>{\centering\arraybackslash}m{85.6pt}|}{6-311g(2df,2pd)} & -9.7 & -4.7 & -5.7 & -16.5 & -17.6 & 6-311g(2df,2pd) & -10.3 & -7.2 & -8.4 & -39.0 & -43.8 \\ \hline
\multicolumn{1}{|>{\centering\arraybackslash}m{85.6pt}|}{6-311++g(2df,2pd)} & -8.9 & -3.5 & -4.7 & -16.6 & -17.7 & 6-311++g(2df,2pd) & -8.6 & -5.6 & -6.9 & -36.7 & -41.6 \\ \hline
\multicolumn{1}{|>{\centering\arraybackslash}m{85.6pt}|}{6-311g(3df,3pd)} & -11.4 & -5.5 & -6.7 & -17.5 & -18.5 & 6-311g(3df,3pd) & -11.0 & -7.8 & -8.8 & -38.6 & -43.1 \\* \hline\noalign{\penalty10000}
\multicolumn{1}{|>{\centering\arraybackslash}m{85.6pt}|}{6-311++g(3df,3pd)} & -9.2 & -3.5 & -4.7 & -16.5 & -17.6 & 6-311++g(3df,3pd) & -8.2 & -4.9 & -6.2 & -36.0 & -40.7 \\* \hline\noalign{\penalty10000}
\multicolumn{1}{|>{\centering\arraybackslash}m{85.6pt}|}{\textbf{DF-RMP2-F12/\allowbreak{}CBS}} & \textbf{-3.4} & \textbf{-1.7} & \textbf{-2.4} & \textbf{-9.8} & \textbf{-11.4} & \textbf{DF-RMP2-F12/\allowbreak{}CBS} & \textbf{-7.8} & \textbf{-5.1} & \textbf{-6.9} & \textbf{-33.9} & \textbf{-38.8} \\* \specialrule{0.9pt}{0pt}{0pt}\noalign{\penalty10000}
\end{tabular}}}}
\end{table}

\begin{table}[!htbp]\centering\setlength{\parskip}{0pt}
\caption{BSSE-corrected binding energies (\ensuremath{\Delta}\textsubscript{1+BSSE}, kJ/\allowbreak{}mol) for selected F-bonded complexes DF\textperiodcentered{}\textperiodcentered{}\textperiodcentered{}CH\textsubscript{3}/NH\textsubscript{3} (DF = F\textsubscript{2}, NCF, CNF, NH\textsubscript{3}F\textsuperscript{+}, and NH\textsubscript{2}F\textsubscript{2}\textsuperscript{+}), calculated using the M06-2X functional with several basis sets, relative to reference binding energies obtained at DF-RMP2-F12/\allowbreak{}CBS[V(T,Q)Z-F12]//DF-RMP2-F12/\allowbreak{}VDZ-F12 (denoted DF-RMP2-F12/\allowbreak{}CBS).}\label{tab:bsse-corrected-binding-energies}\vspace{7pt}\nopagebreak
{\fontsize{8.0}{9.2}\fontfamily{ptm}\selectfont\renewcommand{\arraystretch}{1.17}\setlength{\tabcolsep}{1.5pt}
\centerline{\adjustbox{max width=518pt}{\begin{tabular}{|>{\centering\arraybackslash}m{87.6pt}|>{\centering\arraybackslash}p{25.1pt}|>{\centering\arraybackslash}p{32.0pt}|>{\centering\arraybackslash}p{32.0pt}|>{\centering\arraybackslash}p{41.0pt}|>{\centering\arraybackslash}p{43.0pt}|>{\centering\arraybackslash}m{74.0pt}|>{\centering\arraybackslash}p{28.0pt}|>{\centering\arraybackslash}p{32.3pt}|>{\centering\arraybackslash}p{32.3pt}|>{\centering\arraybackslash}p{39.8pt}|>{\centering\arraybackslash}p{44.9pt}|}
\specialrule{0.9pt}{0pt}{0pt}
\multicolumn{6}{|>{\centering\arraybackslash}m{275.7pt}|}{\textbf{DF\textperiodcentered{}\textperiodcentered{}\textperiodcentered{}CH}\textbf{\textsubscript{3}}} & \multicolumn{6}{>{\centering\arraybackslash}m{266.3pt}|}{\textbf{DF\textperiodcentered{}\textperiodcentered{}\textperiodcentered{}NH}\textbf{\textsubscript{3}}} \\* \hline\noalign{\penalty10000}
\multicolumn{1}{|>{\centering\arraybackslash}m{87.6pt}|}{\multirow{2}{87.6pt}{\centering \textbf{Basis sets}}} & \textbf{F}\textbf{\textsubscript{2}}\textbf{\textperiodcentered{}\textperiodcentered{}\textperiodcentered{}} \textbf{CH}\textbf{\textsubscript{3}} & \textbf{NCF\textperiodcentered{}\textperiodcentered{}\textperiodcentered{}} \textbf{CH}\textbf{\textsubscript{3}} & \textbf{CNF\textperiodcentered{}\textperiodcentered{}\textperiodcentered{}} \textbf{CH}\textbf{\textsubscript{3}} & \textbf{NH}\textbf{\textsubscript{3}}\textbf{F}\textbf{\textsuperscript{+}}\textbf{\textperiodcentered{}\textperiodcentered{}\textperiodcentered{}} \textbf{CH}\textbf{\textsubscript{3}} & \textbf{NH}\textbf{\textsubscript{2}}\textbf{F}\textbf{\textsubscript{2}}\textbf{\textsuperscript{+}}\textbf{\textperiodcentered{}\textperiodcentered{}\textperiodcentered{}} \textbf{CH}\textbf{\textsubscript{3}} & \multirow{2}{74.0pt}{\centering \textbf{Basis sets}} & \textbf{F}\textbf{\textsubscript{2}}\textbf{\textperiodcentered{}\textperiodcentered{}\textperiodcentered{}} \textbf{NH}\textbf{\textsubscript{3}} & \textbf{NCF\textperiodcentered{}\textperiodcentered{}\textperiodcentered{}} \textbf{NH}\textbf{\textsubscript{3}} & \textbf{CNF\textperiodcentered{}\textperiodcentered{}\textperiodcentered{}} \textbf{NH}\textbf{\textsubscript{3}} & \textbf{NH}\textbf{\textsubscript{3}}\textbf{F}\textbf{\textsuperscript{+}}\textbf{\textperiodcentered{}\textperiodcentered{}\textperiodcentered{}} \textbf{NH}\textbf{\textsubscript{3}} & \textbf{NH}\textbf{\textsubscript{2}}\textbf{F}\textbf{\textsubscript{2}}\textbf{\textsuperscript{+}}\textbf{\textperiodcentered{}\textperiodcentered{}\textperiodcentered{}} \textbf{NH}\textbf{\textsubscript{3}} \\* \cline{2-6}\cline{8-12}\noalign{\penalty10000}
\multicolumn{1}{|>{\centering\arraybackslash}m{87.6pt}|}{} & {\fontsize{6.6}{7.5}\selectfont \ensuremath{\Delta}\textsubscript{1+BSSE}} & {\fontsize{6.6}{7.5}\selectfont \ensuremath{\Delta}\textsubscript{1+BSSE}} & {\fontsize{6.6}{7.5}\selectfont \ensuremath{\Delta}\textsubscript{1+BSSE}} & {\fontsize{6.6}{7.5}\selectfont \ensuremath{\Delta}\textsubscript{1+BSSE}} & {\fontsize{6.6}{7.5}\selectfont \ensuremath{\Delta}\textsubscript{1+BSSE}} &  & {\fontsize{6.6}{7.5}\selectfont \ensuremath{\Delta}\textsubscript{1+BSSE}} & {\fontsize{6.6}{7.5}\selectfont \ensuremath{\Delta}\textsubscript{1+BSSE}} & {\fontsize{6.6}{7.5}\selectfont \ensuremath{\Delta}\textsubscript{1+BSSE}} & {\fontsize{6.6}{7.5}\selectfont \ensuremath{\Delta}\textsubscript{1+BSSE}} & {\fontsize{6.6}{7.5}\selectfont \ensuremath{\Delta}\textsubscript{1+BSSE}} \\* \hline\noalign{\penalty10000}
\multicolumn{1}{|>{\centering\arraybackslash}m{87.6pt}|}{VDZ} & -7.7 & -2.0 & -3.3 & -14.9 & -16.0 & VDZ & -8.7 & -4.2 & -5.5 & -38.3 & -43.5 \\* \hline\noalign{\penalty10000}
\multicolumn{1}{|>{\centering\arraybackslash}m{87.6pt}|}{AVDZ} & -8.4 & -2.9 & -4.0 & -16.2 & -17.1 & AVDZ & -8.2 & -4.8 & -5.9 & -36.3 & -40.9 \\* \hline\noalign{\penalty10000}
\multicolumn{1}{|>{\centering\arraybackslash}m{87.6pt}|}{VTZ} & -7.2 & -2.7 & -3.7 & -14.5 & -15.5 & VTZ & -7.1 & -4.7 & -5.7 & -35.9 & -40.9 \\* \hline\noalign{\penalty10000}
\multicolumn{1}{|>{\centering\arraybackslash}m{87.6pt}|}{AVTZ} & -7.4 & -2.6 & -3.6 & -15.0 & -15.9 & AVTZ & -6.1 & -4.5 & -5.1 & -34.0 & -38.5 \\ \hline
\multicolumn{1}{|>{\centering\arraybackslash}m{87.6pt}|}{VQZ} & -6.4 & -2.2 & -3.2 & -14.1 & -14.9 & VQZ & -5.7 & -4.4 & -5.1 & -34.4 & -39.0 \\ \hline
\multicolumn{1}{|>{\centering\arraybackslash}m{87.6pt}|}{AVQZ} & -6.6 & -2.3 & -3.2 & -14.4 & -15.2 & AVQZ & -5.4 & -4.3 & -4.9 & -33.6 & -38.1 \\ \hline
\multicolumn{1}{|>{\centering\arraybackslash}m{87.6pt}|}{V5Z} & -6.6 & -2.2 & -3.2 & -14.4 & -15.4 & V5Z & -5.4 & -4.3 & -4.9 & -33.9 & -38.5 \\ \hline
\multicolumn{1}{|>{\centering\arraybackslash}m{87.6pt}|}{AV5Z} & -6.7 & -2.3 & -3.3 & -14.5 & -15.4 & AV5Z & -5.5 & -4.4 & -4.9 & -33.8 & -38.4 \\ \hline
\multicolumn{1}{|>{\centering\arraybackslash}m{87.6pt}|}{Def2-TZVP} & -6.6 & -2.2 & -3.2 & -14.2 & -15.0 & Def2-TZVP & -6.3 & -4.7 & -5.5 & -35.5 & -40.3 \\ \hline
\multicolumn{1}{|>{\centering\arraybackslash}m{87.6pt}|}{Def2-QZVP} & -6.5 & -2.3 & -3.1 & -14.3 & -15.2 & Def2-QZVP & -5.4 & -4.4 & -5.0 & -33.9 & -38.5 \\ \hline
\multicolumn{1}{|>{\centering\arraybackslash}m{87.6pt}|}{6-311g(2d,2p)} & -7.5 & -3.0 & -4.1 & -15.4 & -16.6 & 6-311g(2d,2p) & -7.9 & -4.9 & -6.5 & -37.4 & -42.4 \\ \hline
\multicolumn{1}{|>{\centering\arraybackslash}m{87.6pt}|}{6-311++g(2d,2p)} & -8.1 & -3.0 & -4.1 & -15.9 & -16.8 & 6-311++g(2d,2p) & -7.6 & -5.0 & -5.9 & -35.6 & -40.4 \\ \hline
\multicolumn{1}{|>{\centering\arraybackslash}m{87.6pt}|}{6-311g(2df,2pd)} & -7.1 & -2.9 & -4.0 & -14.8 & -16.0 & 6-311g(2df,2pd) & -7.2 & -4.6 & -6.1 & -36.5 & -41.3 \\ \hline
\multicolumn{1}{|>{\centering\arraybackslash}m{87.6pt}|}{6-311++g(2df,2pd)} & -7.6 & -2.9 & -3.9 & -15.4 & -16.3 & 6-311++g(2df,2pd) & -6.9 & -4.7 & -5.6 & -34.8 & -39.5 \\ \hline
\multicolumn{1}{|>{\centering\arraybackslash}m{87.6pt}|}{6-311g(3df,3pd)} & -6.9 & -2.9 & -3.9 & -15.1 & -16.3 & 6-311g(3df,3pd) & -6.0 & -4.3 & -5.4 & -35.0 & -39.5 \\* \hline\noalign{\penalty10000}
\multicolumn{1}{|>{\centering\arraybackslash}m{87.6pt}|}{6-311++g(3df,3pd)} & -7.4 & -2.9 & -3.9 & -15.3 & -16.2 & 6-311++g(3df,3pd) & -6.6 & -4.6 & -5.5 & -34.8 & -39.4 \\* \hline\noalign{\penalty10000}
\multicolumn{1}{|>{\centering\arraybackslash}m{87.6pt}|}{\textbf{DF-RMP2-F12/\allowbreak{}CBS}} & \textbf{-3.4} & \textbf{-1.7} & \textbf{-2.4} & \textbf{-9.8} & \textbf{-11.4} & \textbf{DF-RMP2F12/\allowbreak{}CBS} & \textbf{-7.8} & \textbf{-5.1} & \textbf{-6.9} & \textbf{-33.9} & \textbf{-38.8} \\* \specialrule{0.9pt}{0pt}{0pt}\noalign{\penalty10000}
\end{tabular}}}}
\end{table}

\begin{table}[!htbp]\centering\setlength{\parskip}{0pt}
\caption{Deviations on BSSE-uncorrected BEs (\ensuremath{\delta}\ensuremath{\Delta}\textsubscript{1}, kJ/\allowbreak{}mol) for selected F-bonded complexes DF\textperiodcentered{}\textperiodcentered{}\textperiodcentered{}CH\textsubscript{3}/NH\textsubscript{3} (DF = F\textsubscript{2}, NCF, CNF, NH\textsubscript{3}F\textsuperscript{+}, and NH\textsubscript{2}F\textsubscript{2}\textsuperscript{+}), calculated using the M06-2X functional with several basis sets, relative to reference BEs obtained at DF-RMP2-F12/\allowbreak{}CBS[V(T,Q)Z-F12]//DF-RMP2-F12/\allowbreak{}VDZ-F12 (denoted DF-RMP2-F12/\allowbreak{}CBS). The corresponding BEs are provided in \mbox{\textbf{Table \ref{tab:bsse-uncorrected-bes-1}}}.}\label{tab:deviations-on-bsse-uncorrected}\vspace{7pt}\nopagebreak
{\fontsize{8.0}{9.2}\fontfamily{ptm}\selectfont\renewcommand{\arraystretch}{1.17}\setlength{\tabcolsep}{1.5pt}
\centerline{\adjustbox{max width=518pt}{\begin{tabular}{|>{\centering\arraybackslash}m{87.6pt}|>{\centering\arraybackslash}p{22.1pt}|>{\centering\arraybackslash}p{32.0pt}|>{\centering\arraybackslash}p{32.0pt}|>{\centering\arraybackslash}p{40.6pt}|>{\centering\arraybackslash}p{43.0pt}|>{\centering\arraybackslash}m{76.4pt}|>{\centering\arraybackslash}p{23.7pt}|>{\centering\arraybackslash}p{32.3pt}|>{\centering\arraybackslash}p{32.3pt}|>{\centering\arraybackslash}p{39.6pt}|>{\centering\arraybackslash}p{50.5pt}|}
\specialrule{0.9pt}{0pt}{0pt}
\multicolumn{6}{|>{\centering\arraybackslash}m{272.3pt}|}{\textbf{DF\textperiodcentered{}\textperiodcentered{}\textperiodcentered{}CH}\textbf{\textsubscript{3}}} & \multicolumn{6}{>{\centering\arraybackslash}m{269.7pt}|}{\textbf{DF\textperiodcentered{}\textperiodcentered{}\textperiodcentered{}NH}\textbf{\textsubscript{3}}} \\* \hline\noalign{\penalty10000}
\multicolumn{1}{|>{\centering\arraybackslash}m{87.6pt}|}{\multirow{2}{87.6pt}{\centering \textbf{Basis sets}}} & \textbf{F}\textbf{\textsubscript{2}}\textbf{\textperiodcentered{}\textperiodcentered{}\textperiodcentered{}} \textbf{CH}\textbf{\textsubscript{3}} & \textbf{NCF\textperiodcentered{}\textperiodcentered{}\textperiodcentered{}} \textbf{CH}\textbf{\textsubscript{3}} & \textbf{CNF\textperiodcentered{}\textperiodcentered{}\textperiodcentered{}} \textbf{CH}\textbf{\textsubscript{3}} & \textbf{NH}\textbf{\textsubscript{3}}\textbf{F}\textbf{\textsuperscript{+}}\textbf{\textperiodcentered{}\textperiodcentered{}\textperiodcentered{}} \textbf{CH}\textbf{\textsubscript{3}} & \textbf{NH}\textbf{\textsubscript{2}}\textbf{F}\textbf{\textsubscript{2}}\textbf{\textsuperscript{+}}\textbf{\textperiodcentered{}\textperiodcentered{}\textperiodcentered{}} \textbf{CH}\textbf{\textsubscript{3}} & \multirow{2}{76.4pt}{\centering \textbf{Basis sets}} & \textbf{F}\textbf{\textsubscript{2}}\textbf{\textperiodcentered{}\textperiodcentered{}\textperiodcentered{}} \textbf{NH}\textbf{\textsubscript{3}} & \textbf{NCF\textperiodcentered{}\textperiodcentered{}\textperiodcentered{}} \textbf{NH}\textbf{\textsubscript{3}} & \textbf{CNF\textperiodcentered{}\textperiodcentered{}\textperiodcentered{}} \textbf{NH}\textbf{\textsubscript{3}} & \textbf{NH}\textbf{\textsubscript{3}}\textbf{F}\textbf{\textsuperscript{+}}\textbf{\textperiodcentered{}\textperiodcentered{}\textperiodcentered{}} \textbf{NH}\textbf{\textsubscript{3}} & \textbf{NH}\textbf{\textsubscript{2}}\textbf{F}\textbf{\textsubscript{2}}\textbf{\textsuperscript{+}}\textbf{\textperiodcentered{}\textperiodcentered{}\textperiodcentered{}} \textbf{NH}\textbf{\textsubscript{3}} \\* \cline{2-6}\cline{8-12}\noalign{\penalty10000}
\multicolumn{1}{|>{\centering\arraybackslash}m{87.6pt}|}{} & \ensuremath{\delta}\ensuremath{\Delta}\textsubscript{1} & \ensuremath{\delta}\ensuremath{\Delta}\textsubscript{1} & \ensuremath{\delta}\ensuremath{\Delta}\textsubscript{1} & \ensuremath{\delta}\ensuremath{\Delta}\textsubscript{1} & \ensuremath{\delta}\ensuremath{\Delta}\textsubscript{1} &  & \ensuremath{\delta}\ensuremath{\Delta}\textsubscript{1} & \ensuremath{\delta}\ensuremath{\Delta}\textsubscript{1} & \ensuremath{\delta}\ensuremath{\Delta}\textsubscript{1} & \ensuremath{\delta}\ensuremath{\Delta}\textsubscript{1} & \ensuremath{\delta}\ensuremath{\Delta}\textsubscript{1} \\* \hline\noalign{\penalty10000}
\multicolumn{1}{|>{\centering\arraybackslash}m{87.6pt}|}{VDZ} & -8.1 & -2.8 & -3.1 & -7.4 & -6.7 & VDZ & -4.9 & -2.7 & -1.8 & -7.7 & -7.9 \\* \hline\noalign{\penalty10000}
\multicolumn{1}{|>{\centering\arraybackslash}m{87.6pt}|}{AVDZ} & -5.9 & -1.8 & -2.7 & -7.4 & -6.8 & AVDZ & -1.8 & -0.5 & -0.4 & -3.9 & -3.8 \\* \hline\noalign{\penalty10000}
\multicolumn{1}{|>{\centering\arraybackslash}m{87.6pt}|}{VTZ} & -4.9 & -1.8 & -2.0 & -5.5 & -4.9 & VTZ & -0.8 & -0.8 & 0.1 & -3.4 & -3.5 \\* \hline\noalign{\penalty10000}
\multicolumn{1}{|>{\centering\arraybackslash}m{87.6pt}|}{AVTZ} & -4.5 & -1.4 & -1.8 & -5.7 & -5.1 & AVTZ & 1.4 & 0.4 & 1.5 & -0.4 & -0.2 \\ \hline
\multicolumn{1}{|>{\centering\arraybackslash}m{87.6pt}|}{VQZ} & -3.7 & -0.9 & -1.2 & -4.7 & -4.0 & VQZ & 1.3 & 0.1 & 1.2 & -1.3 & -0.9 \\ \hline
\multicolumn{1}{|>{\centering\arraybackslash}m{87.6pt}|}{AVQZ} & -3.6 & -0.9 & -1.2 & -4.9 & -4.2 & AVQZ & 2.2 & 0.7 & 1.9 & 0.2 & 0.6 \\ \hline
\multicolumn{1}{|>{\centering\arraybackslash}m{87.6pt}|}{V5Z} & -3.7 & -0.7 & -0.9 & -4.7 & -4.1 & V5Z & 2.0 & 0.6 & 1.8 & -0.2 & 0.1 \\ \hline
\multicolumn{1}{|>{\centering\arraybackslash}m{87.6pt}|}{AV5Z} & -3.4 & -0.7 & -0.9 & -4.8 & -4.1 & AV5Z & 2.3 & 0.7 & 1.9 & 0.0 & 0.4 \\ \hline
\multicolumn{1}{|>{\centering\arraybackslash}m{87.6pt}|}{Def2-TZVP} & -3.8 & -1.0 & -1.3 & -4.8 & -4.0 & Def2-TZVP & -0.2 & -0.6 & 0.4 & -3.3 & -3.2 \\ \hline
\multicolumn{1}{|>{\centering\arraybackslash}m{87.6pt}|}{Def2-QZVP} & -3.7 & -0.7 & -0.9 & -4.6 & -3.9 & Def2-QZVP & 1.7 & 0.5 & 1.6 & -0.4 & -0.1 \\ \hline
\multicolumn{1}{|>{\centering\arraybackslash}m{87.6pt}|}{6-311g(2d,2p)} & -6.7 & -3.1 & -3.4 & -7.2 & -6.7 & 6-311g(2d,2p) & -3.2 & -2.4 & -1.9 & -5.9 & -6.0 \\ \hline
\multicolumn{1}{|>{\centering\arraybackslash}m{87.6pt}|}{6-311++g(2d,2p)} & -6.0 & -1.9 & -2.4 & -7.3 & -6.8 & 6-311++g(2d,2p) & -1.5 & -0.8 & -0.3 & -3.6 & -3.7 \\ \hline
\multicolumn{1}{|>{\centering\arraybackslash}m{87.6pt}|}{6-311g(2df,2pd)} & -6.3 & -3.0 & -3.3 & -6.7 & -6.2 & 6-311g(2df,2pd) & -2.5 & -2.1 & -1.5 & -5.1 & -5.0 \\ \hline
\multicolumn{1}{|>{\centering\arraybackslash}m{87.6pt}|}{6-311++g(2df,2pd)} & -5.5 & -1.8 & -2.3 & -6.8 & -6.3 & 6-311++g(2df,2pd) & -0.8 & -0.5 & 0.0 & -2.8 & -2.8 \\ \hline
\multicolumn{1}{|>{\centering\arraybackslash}m{87.6pt}|}{6-311g(3df,3pd)} & -8.0 & -3.8 & -4.3 & -7.7 & -7.1 & 6-311g(3df,3pd) & -3.2 & -2.7 & -1.9 & -4.7 & -4.3 \\* \hline\noalign{\penalty10000}
\multicolumn{1}{|>{\centering\arraybackslash}m{87.6pt}|}{6-311++g(3df,3pd)} & -5.8 & -1.8 & -2.3 & -6.7 & -6.2 & 6-311++g(3df,3pd) & -0.4 & 0.2 & 0.7 & -2.1 & -1.9 \\* \hline\noalign{\penalty10000}
\multicolumn{1}{|>{\centering\arraybackslash}m{87.6pt}|}{\textbf{DF-RMP2-F12/\allowbreak{}CBS}} & 0.0 & 0.0 & 0.0 & 0.0 & 0.0 & \textbf{DF-RMP2-F12/\allowbreak{}CBS} & 0.0 & 0.0 & 0.0 & 0.0 & 0.0 \\* \specialrule{0.9pt}{0pt}{0pt}\noalign{\penalty10000}
\end{tabular}}}}
\end{table}

\begin{table}[!htbp]\centering\setlength{\parskip}{0pt}
\caption{Deviations on BSSE-corrected binding energies (\ensuremath{\delta}\ensuremath{\Delta}\textsubscript{1+BSSE}, kJ/\allowbreak{}mol) for selected F-bonded complexes DF\textperiodcentered{}\textperiodcentered{}\textperiodcentered{}CH\textsubscript{3}/NH\textsubscript{3} (DF = F\textsubscript{2}, NCF, CNF, NH\textsubscript{3}F\textsuperscript{+}, and NH\textsubscript{2}F\textsubscript{2}\textsuperscript{+}), calculated using the M06-2X functional with several basis sets, relative to reference BEs obtained at DF-RMP2-F12/\allowbreak{}CBS[V(T,Q)Z-F12]//DF-RMP2-F12/\allowbreak{}VDZ-F12 (denoted DF-RMP2-F12/\allowbreak{}CBS). BEs at the corresponding basis sets are given in \mbox{\textbf{Table \ref{tab:bsse-corrected-binding-energies}}}.}\label{tab:deviations-on-bsse-corrected}\vspace{7pt}\nopagebreak
{\fontsize{8.0}{9.2}\fontfamily{ptm}\selectfont\renewcommand{\arraystretch}{1.17}\setlength{\tabcolsep}{1.5pt}
\centerline{\adjustbox{max width=518pt}{\begin{tabular}{|>{\centering\arraybackslash}m{87.6pt}|>{\centering\arraybackslash}p{26.9pt}|>{\centering\arraybackslash}p{32.0pt}|>{\centering\arraybackslash}p{32.0pt}|>{\centering\arraybackslash}p{39.7pt}|>{\centering\arraybackslash}p{43.0pt}|>{\centering\arraybackslash}m{73.6pt}|>{\centering\arraybackslash}p{27.1pt}|>{\centering\arraybackslash}p{32.0pt}|>{\centering\arraybackslash}p{32.0pt}|>{\centering\arraybackslash}p{39.4pt}|>{\centering\arraybackslash}p{46.7pt}|}
\specialrule{0.9pt}{0pt}{0pt}
\multicolumn{6}{|>{\centering\arraybackslash}m{276.2pt}|}{\textbf{DF\textperiodcentered{}\textperiodcentered{}\textperiodcentered{}CH}\textbf{\textsubscript{3}}} & \multicolumn{6}{>{\centering\arraybackslash}m{265.8pt}|}{\textbf{DF\textperiodcentered{}\textperiodcentered{}\textperiodcentered{}NH}\textbf{\textsubscript{3}}} \\* \hline\noalign{\penalty10000}
\multicolumn{1}{|>{\centering\arraybackslash}m{87.6pt}|}{\multirow{2}{87.6pt}{\centering \textbf{Basis sets}}} & \textbf{F}\textbf{\textsubscript{2}}\textbf{\textperiodcentered{}\textperiodcentered{}\textperiodcentered{}} \textbf{CH}\textbf{\textsubscript{3}} & \textbf{NCF\textperiodcentered{}\textperiodcentered{}\textperiodcentered{}} \textbf{CH}\textbf{\textsubscript{3}} & \textbf{CNF\textperiodcentered{}\textperiodcentered{}\textperiodcentered{}} \textbf{CH}\textbf{\textsubscript{3}} & \textbf{NH}\textbf{\textsubscript{3}}\textbf{F}\textbf{\textsuperscript{+}}\textbf{\textperiodcentered{}\textperiodcentered{}\textperiodcentered{}} \textbf{CH}\textbf{\textsubscript{3}} & \textbf{NH}\textbf{\textsubscript{2}}\textbf{F}\textbf{\textsubscript{2}}\textbf{\textsuperscript{+}}\textbf{\textperiodcentered{}\textperiodcentered{}\textperiodcentered{}} \textbf{CH}\textbf{\textsubscript{3}} & \multirow{2}{73.6pt}{\centering \textbf{Basis sets}} & \textbf{F}\textbf{\textsubscript{2}}\textbf{\textperiodcentered{}\textperiodcentered{}\textperiodcentered{}} \textbf{NH}\textbf{\textsubscript{3}} & \textbf{NCF\textperiodcentered{}\textperiodcentered{}\textperiodcentered{}} \textbf{NH}\textbf{\textsubscript{3}} & \textbf{CNF\textperiodcentered{}\textperiodcentered{}\textperiodcentered{}} \textbf{NH}\textbf{\textsubscript{3}} & \textbf{NH}\textbf{\textsubscript{3}}\textbf{F}\textbf{\textsuperscript{+}}\textbf{\textperiodcentered{}\textperiodcentered{}\textperiodcentered{}} \textbf{NH}\textbf{\textsubscript{3}} & \textbf{NH}\textbf{\textsubscript{2}}\textbf{F}\textbf{\textsubscript{2}}\textbf{\textsuperscript{+}}\textbf{\textperiodcentered{}\textperiodcentered{}\textperiodcentered{}} \textbf{NH}\textbf{\textsubscript{3}} \\* \cline{2-6}\cline{8-12}\noalign{\penalty10000}
\multicolumn{1}{|>{\centering\arraybackslash}m{87.6pt}|}{} & {\fontsize{6.6}{7.5}\selectfont \ensuremath{\delta}\ensuremath{\Delta}\textsubscript{1+BSSE}} & {\fontsize{6.6}{7.5}\selectfont \ensuremath{\delta}\ensuremath{\Delta}\textsubscript{1+BSSE}} & {\fontsize{6.6}{7.5}\selectfont \ensuremath{\delta}\ensuremath{\Delta}\textsubscript{1+BSSE}} & {\fontsize{6.6}{7.5}\selectfont \ensuremath{\delta}\ensuremath{\Delta}\textsubscript{1+BSSE}} & {\fontsize{6.6}{7.5}\selectfont \ensuremath{\delta}\ensuremath{\Delta}\textsubscript{1+BSSE}} &  & {\fontsize{6.6}{7.5}\selectfont \ensuremath{\delta}\ensuremath{\Delta}\textsubscript{1+BSSE}} & {\fontsize{6.6}{7.5}\selectfont \ensuremath{\delta}\ensuremath{\Delta}\textsubscript{1+BSSE}} & {\fontsize{6.6}{7.5}\selectfont \ensuremath{\delta}\ensuremath{\Delta}\textsubscript{1+BSSE}} & {\fontsize{6.6}{7.5}\selectfont \ensuremath{\delta}\ensuremath{\Delta}\textsubscript{1+BSSE}} & {\fontsize{6.6}{7.5}\selectfont \ensuremath{\delta}\ensuremath{\Delta}\textsubscript{1+BSSE}} \\* \hline\noalign{\penalty10000}
\multicolumn{1}{|>{\centering\arraybackslash}m{87.6pt}|}{VDZ} & -4.3 & -0.3 & -0.9 & -5.1 & -4.6 & VDZ & -0.9 & 0.9 & 1.4 & -4.4 & -4.7 \\* \hline\noalign{\penalty10000}
\multicolumn{1}{|>{\centering\arraybackslash}m{87.6pt}|}{AVDZ} & -5.0 & -1.2 & -1.6 & -6.4 & -5.7 & AVDZ & -0.4 & 0.3 & 1.0 & -2.4 & -2.1 \\* \hline\noalign{\penalty10000}
\multicolumn{1}{|>{\centering\arraybackslash}m{87.6pt}|}{VTZ} & -3.8 & -1.0 & -1.3 & -4.7 & -4.1 & VTZ & 0.7 & 0.4 & 1.2 & -2.0 & -2.1 \\* \hline\noalign{\penalty10000}
\multicolumn{1}{|>{\centering\arraybackslash}m{87.6pt}|}{AVTZ} & -4.0 & -0.9 & -1.2 & -5.2 & -4.5 & AVTZ & 1.7 & 0.6 & 1.8 & -0.1 & 0.3 \\ \hline
\multicolumn{1}{|>{\centering\arraybackslash}m{87.6pt}|}{VQZ} & -3.0 & -0.5 & -0.8 & -4.3 & -3.5 & VQZ & 2.1 & 0.7 & 1.8 & -0.5 & -0.2 \\ \hline
\multicolumn{1}{|>{\centering\arraybackslash}m{87.6pt}|}{AVQZ} & -3.2 & -0.6 & -0.8 & -4.6 & -3.8 & AVQZ & 2.4 & 0.8 & 2.0 & 0.3 & 0.7 \\ \hline
\multicolumn{1}{|>{\centering\arraybackslash}m{87.6pt}|}{V5Z} & -3.2 & -0.5 & -0.8 & -4.6 & -4.0 & V5Z & 2.4 & 0.8 & 2.0 & 0.0 & 0.3 \\ \hline
\multicolumn{1}{|>{\centering\arraybackslash}m{87.6pt}|}{AV5Z} & -3.3 & -0.6 & -0.9 & -4.7 & -4.0 & AV5Z & 2.3 & 0.7 & 2.0 & 0.1 & 0.4 \\ \hline
\multicolumn{1}{|>{\centering\arraybackslash}m{87.6pt}|}{Def2-TZVP} & -3.2 & -0.5 & -0.8 & -4.4 & -3.6 & Def2-TZVP & 1.5 & 0.4 & 1.4 & -1.6 & -1.5 \\ \hline
\multicolumn{1}{|>{\centering\arraybackslash}m{87.6pt}|}{Def2-QZVP} & -3.1 & -0.6 & -0.7 & -4.5 & -3.8 & Def2-QZVP & 2.4 & 0.7 & 1.9 & 0.0 & 0.3 \\ \hline
\multicolumn{1}{|>{\centering\arraybackslash}m{87.6pt}|}{6-311g(2d,2p)} & -4.1 & -1.3 & -1.7 & -5.6 & -5.2 & 6-311g(2d,2p) & -0.1 & 0.2 & 0.4 & -3.5 & -3.6 \\ \hline
\multicolumn{1}{|>{\centering\arraybackslash}m{87.6pt}|}{6-311++g(2d,2p)} & -4.7 & -1.3 & -1.7 & -6.1 & -5.4 & 6-311++g(2d,2p) & 0.2 & 0.1 & 1.0 & -1.7 & -1.6 \\ \hline
\multicolumn{1}{|>{\centering\arraybackslash}m{87.6pt}|}{6-311g(2df,2pd)} & -3.7 & -1.2 & -1.6 & -5.0 & -4.6 & 6-311g(2df,2pd) & 0.6 & 0.5 & 0.8 & -2.6 & -2.5 \\ \hline
\multicolumn{1}{|>{\centering\arraybackslash}m{87.6pt}|}{6-311++g(2df,2pd)} & -4.2 & -1.2 & -1.5 & -5.6 & -4.9 & 6-311++g(2df,2pd) & 0.9 & 0.4 & 1.3 & -0.9 & -0.7 \\ \hline
\multicolumn{1}{|>{\centering\arraybackslash}m{87.6pt}|}{6-311g(3df,3pd)} & -3.5 & -1.2 & -1.5 & -5.3 & -4.9 & 6-311g(3df,3pd) & 1.8 & 0.8 & 1.5 & -1.1 & -0.7 \\* \hline\noalign{\penalty10000}
\multicolumn{1}{|>{\centering\arraybackslash}m{87.6pt}|}{6-311++g(3df,3pd)} & -4.0 & -1.2 & -1.5 & -5.5 & -4.8 & 6-311++g(3df,3pd) & 1.2 & 0.5 & 1.4 & -0.9 & -0.6 \\* \hline\noalign{\penalty10000}
\multicolumn{1}{|>{\centering\arraybackslash}m{87.6pt}|}{\textbf{DF-RMP2-F12/\allowbreak{}CBS}} & 0.0 & 0.0 & 0.0 & 0.0 & 0.0 & \textbf{DF-RMP2F12/\allowbreak{}CBS} & 0.0 & 0.0 & 0.0 & 0.0 & 0.0 \\* \specialrule{0.9pt}{0pt}{0pt}\noalign{\penalty10000}
\end{tabular}}}}
\end{table}

\begin{table}[!htbp]\centering\setlength{\parskip}{0pt}
\caption{F-bond angles [\ensuremath{\angle}DF\textperiodcentered{}\textperiodcentered{}\textperiodcentered{}A, \textdegree{}] for selected F-bonded complexes DF\textperiodcentered{}\textperiodcentered{}\textperiodcentered{}AY (DF = F\textsubscript{2}, NCF, CNF, NH\textsubscript{3}F\textsuperscript{+}, NH\textsubscript{2}F\textsubscript{2}\textsuperscript{+}; AY = CH\textsubscript{3}, NH\textsubscript{3}; A = C, N), calculated at the M06-2X level using different basis sets, including VDZ = cc-pVDZ, AVDZ = Aug-cc-pVDZ, VTZ = cc-pVTZ, AVTZ = Aug-cc-pVTZ, VQZ = cc-pVQZ, AVQZ = Aug-cc-pVQZ, V5Z = cc-pV5Z, AV5Z = Aug-cc-pV5Z.}\label{tab:f-bond-angles-df}\vspace{7pt}\nopagebreak
{\fontsize{8.0}{9.2}\fontfamily{ptm}\selectfont\renewcommand{\arraystretch}{1.17}\setlength{\tabcolsep}{1.5pt}
\centerline{\adjustbox{max width=518pt}{\begin{tabular}{|>{\centering\arraybackslash}m{10.1pt}|>{\centering\arraybackslash}m{60.3pt}|>{\centering\arraybackslash}p{8.6pt}|>{\centering\arraybackslash}m{21.5pt}|>{\centering\arraybackslash}m{23.8pt}|>{\centering\arraybackslash}m{23.8pt}|>{\centering\arraybackslash}m{23.8pt}|>{\centering\arraybackslash}m{23.8pt}|>{\centering\arraybackslash}m{23.8pt}|>{\centering\arraybackslash}m{23.8pt}|>{\centering\arraybackslash}m{18.6pt}|>{\centering\arraybackslash}m{22.4pt}|>{\centering\arraybackslash}m{22.4pt}|>{\centering\arraybackslash}m{26.1pt}|>{\centering\arraybackslash}m{23.8pt}|>{\centering\arraybackslash}m{23.8pt}|>{\centering\arraybackslash}m{23.8pt}|>{\centering\arraybackslash}m{23.8pt}|>{\centering\arraybackslash}m{29.9pt}|>{\centering\arraybackslash}m{29.8pt}|}
\specialrule{0.9pt}{0pt}{0pt}
\multicolumn{3}{|>{\centering\arraybackslash}m{84.9pt}|}{\textbf{DF\textperiodcentered{}\textperiodcentered{}\textperiodcentered{}CH}\textbf{\textsubscript{3}}\textbf{/NH}\textbf{\textsubscript{3}}} & \multicolumn{1}{>{\centering\arraybackslash}m{21.5pt}|}{\makebox[\linewidth][c]{\rotatebox{90}{\hspace{2pt}\textbf{vDZ}\hspace{2pt}}}} & \multicolumn{1}{>{\centering\arraybackslash}m{23.8pt}|}{\makebox[\linewidth][c]{\rotatebox{90}{\hspace{2pt}\textbf{AvDZ}\hspace{2pt}}}} & \multicolumn{1}{>{\centering\arraybackslash}m{23.8pt}|}{\makebox[\linewidth][c]{\rotatebox{90}{\hspace{2pt}\textbf{vTZ}\hspace{2pt}}}} & \multicolumn{1}{>{\centering\arraybackslash}m{23.8pt}|}{\makebox[\linewidth][c]{\rotatebox{90}{\hspace{2pt}\textbf{AvTZ}\hspace{2pt}}}} & \multicolumn{1}{>{\centering\arraybackslash}m{23.8pt}|}{\makebox[\linewidth][c]{\rotatebox{90}{\hspace{2pt}\textbf{vQZ}\hspace{2pt}}}} & \multicolumn{1}{>{\centering\arraybackslash}m{23.8pt}|}{\makebox[\linewidth][c]{\rotatebox{90}{\hspace{2pt}\textbf{AvQZ}\hspace{2pt}}}} & \multicolumn{1}{>{\centering\arraybackslash}m{23.8pt}|}{\makebox[\linewidth][c]{\rotatebox{90}{\hspace{2pt}\textbf{V5Z}\hspace{2pt}}}} & \multicolumn{1}{>{\centering\arraybackslash}m{18.6pt}|}{\makebox[\linewidth][c]{\rotatebox{90}{\hspace{2pt}\textbf{AV5Z}\hspace{2pt}}}} & \multicolumn{1}{>{\centering\arraybackslash}m{22.4pt}|}{\makebox[\linewidth][c]{\rotatebox{90}{\hspace{2pt}\textbf{Def2-TZVP}\hspace{2pt}}}} & \multicolumn{1}{>{\centering\arraybackslash}m{22.4pt}|}{\makebox[\linewidth][c]{\rotatebox{90}{\hspace{2pt}\textbf{Def2-QZVP}\hspace{2pt}}}} & \multicolumn{1}{>{\centering\arraybackslash}m{26.1pt}|}{\makebox[\linewidth][c]{\rotatebox{90}{\hspace{2pt}\textbf{6-311g(2d,2p)}\hspace{2pt}}}} & \multicolumn{1}{>{\centering\arraybackslash}m{23.8pt}|}{\makebox[\linewidth][c]{\rotatebox{90}{\hspace{2pt}\textbf{6-311++g(2d,2p)}\hspace{2pt}}}} & \multicolumn{1}{>{\centering\arraybackslash}m{23.8pt}|}{\makebox[\linewidth][c]{\rotatebox{90}{\hspace{2pt}\textbf{6-311g(2df,2pd)}\hspace{2pt}}}} & \multicolumn{1}{>{\centering\arraybackslash}m{23.8pt}|}{\makebox[\linewidth][c]{\rotatebox{90}{\hspace{2pt}\textbf{6-311++g(2df,2pd)}\hspace{2pt}}}} & \multicolumn{1}{>{\centering\arraybackslash}m{23.8pt}|}{\makebox[\linewidth][c]{\rotatebox{90}{\hspace{2pt}\textbf{6-311g(3df,3pd)}\hspace{2pt}}}} & \multicolumn{1}{>{\centering\arraybackslash}m{29.9pt}|}{\makebox[\linewidth][c]{\rotatebox{90}{\hspace{2pt}\textbf{6-311++g(3df,3pd)}\hspace{2pt}}}} & \multicolumn{1}{>{\centering\arraybackslash}m{29.8pt}|}{\makebox[\linewidth][c]{\rotatebox{90}{\hspace{2pt}\textbf{DF-RMP2-F12/\allowbreak{}VDZ-F12}\hspace{2pt}}}} \\* \hline\noalign{\penalty10000}
\multicolumn{1}{|>{\centering\arraybackslash}m{10.1pt}|}{\multirow{5}{10.1pt}{\centering \makebox[\linewidth][c]{\rotatebox{90}{\hspace{2pt}\textbf{DF\textperiodcentered{}\textperiodcentered{}\textperiodcentered{}CH}\textbf{\textsubscript{3}}\hspace{2pt}}}}} & \textbf{F}\textbf{\textsubscript{2}}\textbf{\textperiodcentered{}\textperiodcentered{}\textperiodcentered{}CH}\textbf{\textsubscript{3}} & \ensuremath{\angle} & 180.0 & 180.0 & 180.0 & 179.6 & 180.0 & 180.0 & 180.0 & 180.0 & 180.0 & 180.0 & 180.0 & 180.0 & 180. & 180. & 180.0 & 180.0 & 178.8 \\* \cline{2-20}\noalign{\penalty10000}
\multicolumn{1}{|>{\centering\arraybackslash}m{10.1pt}|}{} & \textbf{NCF\textperiodcentered{}\textperiodcentered{}\textperiodcentered{}CH}\textbf{\textsubscript{3}} & \ensuremath{\angle} & 180.0 & 180.0 & 180.0 & 179.9 & 180.0 & 180.0 & 180.0 & 180.0 & 180.0 & 180.0 & 180.0 & 180.0 & 180.0 & 180.0 & 180.0 & 179.9 & 180.0 \\* \cline{2-20}\noalign{\penalty10000}
\multicolumn{1}{|>{\centering\arraybackslash}m{10.1pt}|}{} & \textbf{CNF\textperiodcentered{}\textperiodcentered{}\textperiodcentered{}CH}\textbf{\textsubscript{3}} & \ensuremath{\angle} & 180.0 & 180.0 & 180.0 & 180.0 & 180.0 & 180.0 & 180.0 & 180.0 & 180.0 & 180.0 & 180.0 & 180.0 & 180.0 & 180.0 & 180.0 & 179.9 & 179.6 \\* \cline{2-20}\noalign{\penalty10000}
\multicolumn{1}{|>{\centering\arraybackslash}m{10.1pt}|}{} & \textbf{NH}\textbf{\textsubscript{3}}\textbf{F}\textbf{\textsuperscript{+}}\textbf{\textperiodcentered{}\textperiodcentered{}\textperiodcentered{}CH}\textbf{\textsubscript{3}} & \ensuremath{\angle} & 179.8 & 179.2 & 179.6 & 178.7 & 179.8 & 180.0 & 180.0 & 180.0 & 180.0 & 180.0 & 179.9 & 179.6 & 179.3 & 180.0 & 180.0 & 179.7 & 179.4 \\* \cline{2-20}\noalign{\penalty10000}
\multicolumn{1}{|>{\centering\arraybackslash}m{10.1pt}|}{} & \textbf{NH}\textbf{\textsubscript{2}}\textbf{F}\textbf{\textsubscript{2}}\textbf{\textsuperscript{+}}\textbf{\textperiodcentered{}\textperiodcentered{}\textperiodcentered{}CH}\textbf{\textsubscript{3}} & \ensuremath{\angle} & 179.4 & 179.3 & 179.5 & 178.9 & 175.9 & 175.7 & 179.0 & 177.8 & 176.2 & 176.2 & 178.8 & 179.0 & 178.8 & 179.3 & 178.6 & 179.1 & 176.4 \\* \specialrule{0.9pt}{0pt}{0pt}\noalign{\penalty10000}
\end{tabular}}}
\par\nointerlineskip
\centerline{\adjustbox{max width=518pt}{\begin{tabular}{|>{\centering\arraybackslash}m{10.1pt}|>{\centering\arraybackslash}m{60.3pt}|>{\centering\arraybackslash}p{8.6pt}|>{\centering\arraybackslash}m{21.5pt}|>{\centering\arraybackslash}m{23.8pt}|>{\centering\arraybackslash}m{23.8pt}|>{\centering\arraybackslash}m{23.8pt}|>{\centering\arraybackslash}m{23.8pt}|>{\centering\arraybackslash}m{23.8pt}|>{\centering\arraybackslash}m{23.8pt}|>{\centering\arraybackslash}m{18.6pt}|>{\centering\arraybackslash}m{22.4pt}|>{\centering\arraybackslash}m{22.4pt}|>{\centering\arraybackslash}m{26.1pt}|>{\centering\arraybackslash}m{23.8pt}|>{\centering\arraybackslash}m{23.8pt}|>{\centering\arraybackslash}m{23.8pt}|>{\centering\arraybackslash}m{23.8pt}|>{\centering\arraybackslash}m{29.9pt}|>{\centering\arraybackslash}m{29.8pt}|}
\multicolumn{1}{|>{\centering\arraybackslash}m{10.1pt}|}{\multirow{5}{10.1pt}{\centering \makebox[\linewidth][c]{\rotatebox{90}{\hspace{2pt}\textbf{DF\textperiodcentered{}\textperiodcentered{}\textperiodcentered{}NH}\textbf{\textsubscript{3}}\hspace{2pt}}}}} & \textbf{F}\textbf{\textsubscript{2}}\textbf{\textperiodcentered{}\textperiodcentered{}\textperiodcentered{}NH}\textbf{\textsubscript{3}} & \ensuremath{\angle} & 180.0 & 180.0 & 180.0 & 180.0 & 180.0 & 180.0 & 180.0 & 180.0 & 180.0 & 180.0 & 180.0 & 180.0 & 180.0 & 180.0 & 180.0 & 180.0 & 180.0 \\* \cline{2-20}\noalign{\penalty10000}
\multicolumn{1}{|>{\centering\arraybackslash}m{10.1pt}|}{} & \textbf{NCF\textperiodcentered{}\textperiodcentered{}\textperiodcentered{}NH}\textbf{\textsubscript{3}} & \ensuremath{\angle} & 180.0 & 180.0 & 180.0 & 180.0 & 180.0 & 180.0 & 180.0 & 180.0 & 180.0 & 180.0 & 180.0 & 180.0 & 180.0 & 180.0 & 180.0 & 180.0 & 180.0 \\* \cline{2-20}\noalign{\penalty10000}
\multicolumn{1}{|>{\centering\arraybackslash}m{10.1pt}|}{} & \textbf{CNF\textperiodcentered{}\textperiodcentered{}\textperiodcentered{}NH}\textbf{\textsubscript{3}} & \ensuremath{\angle} & 180.0 & 180.0 & 180.0 & 180.0 & 180.0 & 180.0 & 180.0 & 180.0 & 180.0 & 180.0 & 180.0 & 180.0 & 180.0 & 180.0 & 180.0 & 180.0 & 178.7 \\* \cline{2-20}\noalign{\penalty10000}
\multicolumn{1}{|>{\centering\arraybackslash}m{10.1pt}|}{} & \textbf{NH}\textbf{\textsubscript{3}}\textbf{F}\textbf{\textsuperscript{+}}\textbf{\textperiodcentered{}\textperiodcentered{}\textperiodcentered{}NH}\textbf{\textsubscript{3}} & \ensuremath{\angle} & 179.9 & 179.8 & 180.0 & 180.0 & 180.0 & 180.0 & 180.0 & 180.0 & 180.0 & 180.0 & 179.8 & 180.0 & 179.7 & 179.8 & 180.0 & 180.0 & 180.0 \\* \cline{2-20}\noalign{\penalty10000}
\multicolumn{1}{|>{\centering\arraybackslash}m{10.1pt}|}{} & \textbf{NH}\textbf{\textsubscript{2}}\textbf{F}\textbf{\textsubscript{2}}\textbf{\textsuperscript{+}}\textbf{\textperiodcentered{}\textperiodcentered{}\textperiodcentered{}NH}\textbf{\textsubscript{3}} & \ensuremath{\angle} & 179.3 & 178.0 & 179.4 & 178.1 & 176.8 & 178.2 & 178.2 & 178.3 & 178.0 & 178.1 & 179.8 & 178.7 & 179.6 & 178.3 & 179.3 & 178.2 & 177.3 \\* \specialrule{0.9pt}{0pt}{0pt}\noalign{\penalty10000}
\end{tabular}}}}
\end{table}

\end{document}